\documentclass{article}
\usepackage{ifthen}
\usepackage[dvips]{graphicx}

\def\be{\begin{eqnarray}}
\def\ee{\end{eqnarray}}
\def\nn{\nonumber}
\def\bb{\bf}

\def\circn{\circ n}

\def\notsubset{\not\subset}
\def\p{\ }

\newcommand\FIGEPS[4]{
\ifthenelse{\equal{#3}{}}
{\begin{figure}\begin{center}\includegraphics[width=200pt,height=200pt,draft]
{#1.eps}\caption{\footnotesize#2}\label{#1}\end{center}\end{figure}}
{\begin{figure}\begin{center}\includegraphics[scale=0.5]{#1.eps}
\caption{\footnotesize#4}\label{#1}\end{center}\end{figure}}
}

\hoffset= - 1.0in

\begin{document}

\hfill /hepth-0501235

\hfill{ITEP-TH-02/05}\\

\centerline{\Large{Algebraic Geometry of Discrete Dynamics
}}
\centerline{\Large{The case of one variable}}

\bigskip

\centerline{\it V.Dolotin and A.Morozov}

\centerline {ITEP, Moscow and JINR, Dubna; Russia}

\bigskip

\bigskip

\centerline{ABSTRACT}

\bigskip

We argue that discrete dynamics
has natural links to the theory of analytic functions.
Most important, bifurcations and chaotic dynamical properties
are related to intersections of algebraic varieties.
This paves the way to identification of
boundaries of Mandelbrot sets with discriminant
varieties in moduli spaces, which are the central objects
in the worlds of chaos and order (integrability) respectively.
To understand and exploit this relation one needs first to develop
the theory of discrete dynamics as a solid branch of algebraic
geometry, which so far did not pay enough attention to iterated maps.
The basic object to study in this context is Julia sheaf over
the universal Mandelbrot set. The base has a charateristic
combinatorial structure, which can be revealed by resultant analysis
and represented by a basic graph. Sections
(Julia sets) are contractions of a unit disc, related to the action
of Abelian $\bb{Z}$ group on the unit circle. Their singularities
(bifurcations) are located at the points of the universal
discriminant variety.

\bigskip

\bigskip

\newpage

\tableofcontents

\newpage

\section{Introduction}

This paper is devoted to the structure of Mandelbrot set -- a
remarkable and important personage of the modern theoretical
physics, related to chaos and fractals and simultaneously to
analytical functions, Riemann surfaces and phase transitions. This
makes Mandelbrot set  one of the bridges connecting the worlds
of chaos and order (integrability). At the same time the Mandelbrot
set is a very simple object, allowing detailed {\it experimental}
investigation with the help of the easily available computer
programs and profound {\it theoretical} analysis with the help of
undergraduate-level mathematics. This makes Mandelbrot set a
wonderful subject of physical investigation: one needs to look for
theoretical explanations of experimentally observable properties of
the object and theoretical predictions can be immediately verified
by the easily affordable new experiments. In this paper we just lift
the cover of Pandora box: if interested, the reader can easily
continue along any of the research lines, which are mentioned below,
or, digging a little deeper into the experimental data, find and
resolve a lot of new puzzles, -- all this without any special
mathematical background. Remarkably, when thinking about this simple
subject one appears in the close vicinity of modern problems of the
"high theory": the subject is not only simple, it is also deep.

Mandelbrot set shows up in the study of trajectories (world lines)
behavior change under a change of the laws of motion. Generally
speaking, trajectory depends on two types of data: on the motion
law (e.g. on the choice of the Hamiltonian) and on the initial
(boundary) conditions. The pattern of motion (the phase portrait of
the system) can be rather sophisticated: trajectories can tend to
fixed points, to limit cycles, to strange attractors and go away
from the similar types of structures, some trajectories can go to
infinity; trajectories can wrap around each other, knot and unknot
etc. In this paper the information about the phase portrait will
be collected in {\bf Julia sets}, the structure of these sets
encodes information about the world lines dependence on the initial
conditions. If the law of motion is changed, the phase portrait =
Julia set is also deformed. These changes can be pure quantitative
(due to the variations of the shape of trajectories, of foci and
limit cycles positions) and -- sometime -- qualitative (looking as
reshuffling of the phase portrait, creation or elimination of the
fixed points, cycles, attractors etc). The {\bf Mandelbrot set} or
its {\bf boundary}, to be precise, is exactly the set of all the
points in the laws-of-motion space (say in the space of
Hamiltonians) where the qualitative reshufflings (phase transitions)
take place. Therefore the problems about the Mandelbrot set (or,
better, about the set of the phase portraits, associated with
various laws of motion, i.e. about the {\bf "sheaf of Julia sets
over the Mandelbrot set"}) are the typical problems about the
structure of the {\bf space of theories} (laws of motion), about
what changes and what remains intact in transition from one
theory to another. In other words, these are typical problems from
the {\it string theory}\footnote{
As often happens, the name is pure historical and refers to a concrete
set of models, which were first successfully approached from such
direction. Better name could be "the theory of theories"
(sometime "abbreviated" to pretentious "theory of everything") or,
most adequate, "generic quantum field theory". Unfortunately,
these names are already associated with narrower fields of research,
where accents are put on somewhat different issues.
} -- the science which studies the various
sets of similar physical models and studies the change of the
experimentally measurable characteristics with the change of the
model. Moreover, in the case of the Mandelbrot set one can
immediately address the problems from a difficult branch of
string theory, which nowadays attracts a good deal of attention, --
from {\it landscape theory}, which wonders {\it how often} is a
given property encountered in the given set of models (i.e.
associates with the various physical quantities, say, scattering
amplitudes or mass spectra, the {\it measures} on the space of
theories, which specify how frequent are the amplitudes of the given
type or masses of the given values in the models of the given type).
One of the lessons, taught by the study of Mandelbrot sets, is that
even if the {\bf phase transitions} are relatively rare in the space
of theories, the phase transition points are not distributed in
a uniform way, they tend to condense and form clusters of higher
dimensions, up to codimension one, which actually separate --
the naively un-separated -- domains in the space of theories and
make this space disconnected. This lesson is important for the theory
of effective actions, for problems of their analytical continuations,
for construction of special functions ($\tau$-functions) which can
describe these effective actions, and for many other -- conceptual
and practical -- applications of string theory. Dynamics, essential
for this type of applications of Mandelbrot-set theory, is not
obligatory the ordinary dynamics in physical space-time, often the
relevant one is dynamics in the space of coupling constants, i.e. the
(generalized) renormalization group flow.

In this paper we restrict consideration to {\it discrete}
dynamics {\it of a single variable}.
This restriction preserves most essential
properties of the subject, but drastically simplifies computer
simulations and mathematical formalism (for example, substitutes
the study of arbitrary analytic functions by that of polynomials or
the study of arbitrary spectral surfaces of arbitrary dimension by
that of the ordinary Riemann surfaces).
However, throughout the text we use every chance
to show directions of possible generalizations. Getting rid of the
infinite-dimensional algebras (loop algebras), which would be
unavoidable in considerations of continuous dynamics, we concentrate
instead on {\it another} infinity: unification of various one-dimensional
Mandelbrot sets, shown in numerous pictures below and associated
with particular 1-dimensional families of evolution operators,
into a total (infinite-dimensional) {\bf universal Mandelbrot set},
shown symbolically in Fig.\ref{sausages} at the very end of the paper.
Understanding the structure of this set (which has a pure
cathegorial 
nature of the {\bf universal discriminant variety}) can be one of the
intermediate targets of the theory. The notions of {\it discriminant}
and {\it resultant} are assumed familiar from the courses
on general algebra and are actively used already in the Introduction.
Still, if necessary, all the necessary definitions and properties
can be found in s.4.12 of the present paper (the section itself is
devoted to generalizations from polynomials to arbitrary analytic
functions).

The theory of deterministic dynamical systems \cite{dysy},
\be
\dot x^i = \beta^i(x),
\label{dys}
\ee
is one of the main chapters of theoretical and mathematical physics.
One of the goals of the theory is to classify the types of motion --
from completely integrable to fully ergodic, --
and understand their dependence
on the choice of the vector field $\beta(x)$ and initial
conditions. Discrete dynamics, in the avatar of Poincare map,
\be
x^i \longrightarrow f^i(x), \nn \\
\beta^i(f(x)) = \frac{\partial f^i}{\partial x^j}\beta^j(x),
\ee
naturally arises in attempts to develop such a theory and
plays the principal role in computer experiments (which -- in the
absence of adequate theory -- remain
the main source of information about dynamical systems).

In quantum field/string theory (QFT) the most important (but by no
means unique) appearance of the problem (\ref{dys}) is in the role
of renormalization group (RG) flow. As usual, QFT relates (\ref{dys}) to
a problem, formulated in terms of effective actions, this time to
a Callan-Symanzik equation for a function ${\cal F}(x,\varphi)$
of the {\it coupling constants} $x^i$ and additional variable $\varphi$
(having the meaning of logarithm of background {\it field})
\be
\beta^i(x)\frac{\partial{\cal F}}{\partial x^i} +
\frac{\partial{\cal F}}{\partial \varphi} = 0.
\label{CS}
\ee
In other contexts solutions to this equation are known as
{\it characteristics} \cite{char} of the problem (\ref{dys}):
in the case of one variable the solution is arbitrary
function of $\varphi - s(x)$ or of $x(s=\varphi)$, its $\varphi$ dependence
is induced by the action of operator $\exp \left(-\varphi
\sum_i\beta^i\partial_i\right)$.
Until recently only simple types of motions (attraction and repulsion points)
were considered in the context of RG theory, but today this
restriction is no longer fashionable (see, for example, \cite{cycl}
for the first examples of RG with periodic behaviour and
\cite{tH}-\cite{charg} for further generalizations).
In this context the discrete dynamics is associated with the
original Kadanoff-style formulation of renormalization group,
and Callan-Symanzik equation (\ref{CS}) becomes a finite-difference
equation.\footnote{
This discrete equation looks like
$$
{\cal F}_{\pm}(f^{\circ n}(x),\varphi \pm n) = {\cal F}_\pm(x,\varphi)
$$
Its generic solution is arbitrary function of any of its
particular solutions.
Formally, such particular solutions are given by the
differences $\varphi \mp \hat n(x)$,
where $\hat n(x)$ is the "time", needed to reach the point
$x$, i.e. a solution of the equation $f^{\circ n}(x_0) = x$
for given $f$ and $x_0$.
In other words, particular solutions are formally provided by
the $\varphi$-th image and $\varphi$-th preimage
of $f(x)$,
$$
{\cal F}_\pm(x,\varphi) = f^{\circ (\mp\varphi)}(x)
$$
For example, if $f(x) = x^c$ and $f^\circ (x) = x^{c^n}$, then the
generic solution to is an
arbitrary function of the variable $c^{\mp\varphi}\log x$.

Consideration of the present paper provides  peculiar
$\varphi$-independent solutions (zero-modes), associated with
periodic $f$-orbits. For example, the "orbit $\delta$-function"
$$
{\cal F}_\pm(x,\varphi) = F_n'(x)\delta(F_n(x))
$$
is a solution of such type
(see eq.(\ref{derivFn}) below) --
a generalization of continuous-case zero-mode
$$
{\cal F}(x,\varphi) = \det\left(\frac{\partial\beta}{\partial x}\right)
\delta(\beta(x)).
$$
At least in this sense  periodic
$f$-orbits are  discrete analogues of non-trivial
zeroes of $\beta$-function, in continuum limit they can turn into
limit cycles and strange attractors.
}
Bifurcations of dynamical behaviour of trajectories
with the change of $\beta(x)$ in (\ref{dys}) are associated with
transitions between different branches of effective actions
\cite{amm1}.
One of the most intriguing points here is that the theory of
effective actions is actually the one of {\it integrable systems}
-- effective actions usually are $\tau$-functions,
RG flows are multi-directional (since the boundary of integration
domain in functional integral can be varied in many different ways)
and related to Whitham dynamics \cite{Whith}.
Obvious existence of non-trivial RG flows (describing, for example,
how a fractal variety changes with the change of scale)
-- along with many other pieces of evidence -- implies
existence of a deep relation between the worlds of order and chaos
(represented by $\tau$-functions and dynamical systems respectively
and related by the -- yet underdeveloped -- theory of effective actions).
Description of effective actions (prepotentials)
${\cal F}(\varphi,x)$ in situations when (\ref{dys}) exhibits
chaotic motion remains an important open problem.
Of course, one does not expect to express such ${\cal F}$ through
ordinary elementary or special functions, but we are going to
claim in this paper that the adequate language can still be
searched for in the framework of algebraic geometry, in which
the theory of dynamical systems can be naturally embedded, at least
in the discrete case.

\bigskip

The goal of this paper is to overview the connection between
the {\it discrete} dynamics of one {\it complex} variable \cite{Cody},
which studies the properties of the iterated map
\be
x \rightarrow f(x) \rightarrow  f^{\circ 2}(x) :\ = f(f(x))
\rightarrow \ldots \rightarrow  f^{\circ n}(x)
\rightarrow \ldots,
\ee
and the algebraic properties of the (locally) analytic function
$f(x) = \sum_k a_kx^k$, considered as an element of the
(infinite-dimensional) space ${\cal M}$ of all analytic functions
$\bb{C} \rightarrow \bb{C}$ on the complex plane $\bb{C}$.
We attribute the apparent complexity of the Julia and Mandelbrot
sets, associated with the pattern of orbit bifurcations,
to the properties of well defined (but not yet well-studied)
algebraic subspaces in moduli space:
discriminant variety ${\cal D}$ and spaces of shifted $n$-iterated maps
${\cal M}_{\circn}$
\be
F_n(x;f) = f^{\circ n}(x) - x.
\label{Fndef}
\ee
In other words, we claim that Julia and Mandelbrot
sets can be defined as objects in algebraic geometry and studied by
theoretical methods, not only by computer experiments.

We remind, that the boundary of
Julia set $\partial J(f)$ for a given $f$ is the union of all
{\it unstable} periodic orbits of $f$ in $C$, i.e. a subset in the union
${\cal O}(f)$ off all periodic orbits, perhaps, {\it complex},
which, in turn, is nothing but
the set of all roots of the iterated maps $F_n(x;f)$ for all $n$.
The boundary of {\it algebraic} Mandelbrot set
$\partial M(\mu) \subset \mu \subset{\cal M}$ consists of all functions
$f$ from a
given family $\mu \subset {\cal M}$, where the stability of orbits changes.
However, there is no one-to-one correspondence between Mandelbrot sets and
bifurcations of Julia sets, since the latter
can also occur when something happens to {\it unstable} orbits and
some of such events do not necessarily involve {\it stable} orbits,
controlled by the Mandelbrot set.
In order to get rid of the reference to somewhat subtle notion of stability
(which also involves {\it real} instead of {\it complex-analytic}
constraints), we suggest to reveal the algebraic structure of Julia and
Mandelbrot sets and then use it is a definition of their algebraic
counterparts.
It is obvious that boundaries
of Mandelbrot sets are related to discriminants,
$\partial M = {\cal D} \subset {\cal M}$ and can actually be defined
without reference to stability.
The same is true also for Julia sets, which can be defined
in terms of {\it grand orbits} of $f$,
which characterize not only the images of points after the action of $f$,
but also their pre-images.
The bifurcations of so defined algebraic Julia sets occur
at the boundary of what
we call the {\it grand Mandelbrot set}, a straightforward extension of
Mandelbrot set, also representable as discriminant variety.
Its boundary consists of points in the space of maps ({\it moduli space}),
where the structure of periodic orbits (not obligatory {\it stable}) and
their pre-orbits changes.
Hypothetically algebraic Julia sets coincide with the ordinary ones (at
least for some appropriate choice of stability criterium).

\bigskip

To investigate  the algebraic structure of Julia and
Mandelbrot sets we suggest to reformulate the
problem in terms of representation theory.

An analytic function as a map $f:\bb{C}\to\bb{C}$ defines an action of
$\bb{Z}$ on $\bb{C}$: $n\mapsto \underbrace{f\circ\ldots\circ f}_n$.
For $f$ with real coefficients this action has two invariant subsets
$\bb{R}$ and $\bb{C}-\bb{R}$.
Often in discrete dynamics one considers only the action on $\bb{R}$.
However, some peculiar properties of this action (like period doubling,
stability etc.) become transparent only if we consider it as a special case of
the action of $f$ with complex coefficients and no distinguished subsets
(like $\bb{R}$) in $\bb{C}$.

The set of periodic orbits of this action can be considered as a
representation of Abelian group $\bb{Z}$, which depends on the shape of $f$.
One can study these representations, starting
from generic representation, which corresponds to generic position of
$f$ in the space ${\cal M}$.
Other representations result from merging of
the orbits of generic representation,
what happens when $f$ belongs to certain discriminant subsets
${\cal D}_{n}^* = ({\cal D} \cap {\cal M}_{n})^* \subset {\cal M}$.
These discriminant sets are algebraic and have grading,
related to the order of the corresponding merging orbits.
The boundary of Mandelbrot set $\partial M(\mu)$
for a one-parametric family of maps $\mu \subset {\cal M}$
is related to
the union of the countable family of algebraic sets of increasing degree,
$\cup_{n=0}^\infty \left({\cal D}\cap \mu_{\circn}\right)$,
and this results in the "fractal" structure of Mandelbrot set.
This approach allows us to introduce the notion of
discriminant in the (infinite-dimensional) space ${\cal M}$
of analytic functions as the
inverse limit of a sequence of finite-dimensional expressions corresponding
to discriminants in the spaces of polynomials with increasing degree.

On the preimages
of periodic orbits (on bounded grand orbits) the above
$\bb{Z}$-action is not invertible. The structure of the set of preimages
of a given order $k$ has special bifurcations which are not expressible in
terms of bifurcations of periodic orbits. This implies the
existence of a hierarchy of discriminant sets in the space of coefficients
(secondary Mandelbrot sets).
The union of secondary Mandelbrot sets for all $k$ presumably defines all
the bifurcations of Julia sets. We call this union the Grand Mandelbrot
set.

\bigskip

Mentioned above is just one of many possible "definitions" of Julia
and Mandelbrot sets \cite{defs}.
Their interesting (fractal) properties look universal and not
sensitive to particular way they are introduced, what clearly
implies that some canonical universal structure stands behind.
The claim of the present paper is that this canonical structure is
algebraic and is nothing but the intersection of iterated maps with the
universal discriminant and resultant varieties
${\cal D} \subset {\cal M}$ and ${\cal R} \subset {\cal M}\times{\cal M}$,
i.e. the appropriately defined pull-back ${\cal R}^* \subset {\cal M}$.
Different definitions of Mandelbrot
sets are just different vies/projections/sections of ${\cal R}^*$.
The study of Mandelbrot sets is actually the study of ${\cal R}^*$,
which is a complicated but well-defined problem of algebraic
geometry. Following this line of thinking we suggest to substitute
intuitive notions of Julia and Mandelbrot sets by their much better
defined {\it algebraic} counterparts.
The details of these
definitions can require modifications when one starts proving, say,
existence- or equivalence theorems, but the advantage of such
approach is the very possibility to formulate and prove theorems.

\bigskip

Our main conclusions about the structure of Mandelbrot and Julia sets
are collected in section \ref{summ}.
There are no properties of these fractal sets which could not be
discovered and
explained by pure algebraic methods. However, the adequate part of algebraic
theory is not enough developed and requires more attention and work. In the
absence of developed theory -- and even developed language -- a lot of
facts can be better understood through examples than through general
theorems. We present both parts of the story -- theory and examples --
in parallel.\footnote{
To investigate examples, we used powerful discriminant and resultant
facilities of MAPLE and a wonderful {\it Fractal Explorer} (FE) program
\cite{FE}. In particular, the pictures of Mandelbrot and Julia sets
for $1$-parameter families of maps in this paper are generated with the
help of this program.
}
For the sake of convenience we begin in the next section from
listing the main relevant notions and their inter-relations.

\newpage

\section{Notions and notation}

\subsection{Objects, associated with the space $\bb{X}$}

\noindent

$\bullet$ {\bf Phase space $\bb{X}$.}
In principle, in order to define the Mandelbrot space only
$\bb{X}$ only needs to be a topological space. However,
to make powerful algebraic machinery applicable,
$\bb{X}$ should be an algebraically
closed field: the complex plane $\bb{C}$ suits all needs,
while the real line $\bb{R}$ does not.
We usually assume that $\bb{X} = \bb{C}$, but algebraic
construction is easily extendable to $p$-adic fields and
to Galois field. In fact, further generalizations are straightforward:
what is really important is the
ring structure (to define maps from $\bb{X}$ into itself
as polynomials and series)
and a kind of Bezout theorem (allowing to parameterize the
maps by collections of their roots).\footnote{
In order to preserve these properties,
multidimensional discrete dynamics can be introduced
on phase space $\bb{X}^m$ in the following way. The $x$-variables
are substituted by $m$-component vectors $x_1,\ldots,x_m$ (or even
$m+1$-component if the space is $\bb{CP}^m$ and
affine parameterization is used). The relevant maps
$\ f:\ \bb{X}^m \rightarrow \bb{X}^m\ $ are defined by
tensor coefficients $a_{i;\vec k}$,
$$
f_i(x) = \sum_{k_1,\ldots,k_m = 0}^\infty
a_{i;k_1,\ldots,k_m}x_1^{k_1}\ldots x_m^{k_m},
\ \ \ i=1,\ldots,m
$$
instead of (\ref{mapf}).
Then one can introduce and study discriminants, resolvents,
Mandelbrot and Julia sets just in the same way as we are going
to do in one-dimensional situation ($m=1$).
For relevant generalization of discriminant see \cite{Doldis}.
Concrete
multidimensional examples should be examined by this technique, but
this is beyond the scope of the present paper.
}

\bigskip

$\bullet$ {\bf A map $f:\ \bb{X} \rightarrow \bb{X}$} can be represented
as finite (polynomial) or infinite series
\be
f(x) = \sum_{k=0}^\infty a_k x^k.
\label{mapf}
\ee
The series can be assumed convergent at least in some domain of $\bb{X}$
(if the notion of convergency is defined on $\bb{X}$).
Some statements in this paper are extendable even to formal series.

\bigskip

$\bullet$ {\bf $f$-orbit} ${O}(x;f) \subset \bb{X}$ of a point
$x \in \bb{X}$ is a set of all images,
$$\{x,f(x),f^{\circ 2}(x), \ldots, f^{\circ n}(x),\ldots\}.$$
It is convenient to agree that for $n=0$ $\ f^{\circ 0}(x) = x$.
The orbit is {\bf periodic} of order $n$ if $n$ is the smallest positive
integer for which $f^{\circ n}(x) = x$.

\bigskip

$\bullet$
{\bf Grand $f$-orbit} ${GO}(x;f) \subset \bb{X}$ of $x$ includes also all
the pre-images: all points $x'$, such that $f^{\circ k}(x') = x$
for some $k$. There can be many different $x'$ for given $x$ and $k$.
The {\it orbit} is finite, if for some $k\geq 0$ and $n\geq 1$
$f^{\circ(n+k)}(x) = f^{\circ k}(x)$. If $k$ and $n$ are the smallest
for which this property holds, we say that $x$ belongs to the $k$-th
preimage of a periodic orbit of order $n$ (which consists of the
$n$ points $\{f^{\circ k}(x), \ldots, f^{\circ (n+k-1)}(x)\}$). The
corresponding {\it grand orbit} (i.e. the one which ends up in a finite
orbit) is called {\bf bounded grand orbit} (BGO), it looks like a graph,
with infinite trees attached to a loop of length $n$.
Grand orbit defines the ramification structure of functional-inverse
of the map $f$ and of associated Riemann surface.
Different branches\footnote{
For a rooted tree we call {\bf branch} any path, connecting the
root and some highest-level vertex. For infinite trees, like most
grand orbits, the branches have infinite length.
}
of the grand-orbit tree define different branches of the prepotential
(the discrete analogue of ${\cal F}$ in eq.(\ref{CS})).

\FIGEPS{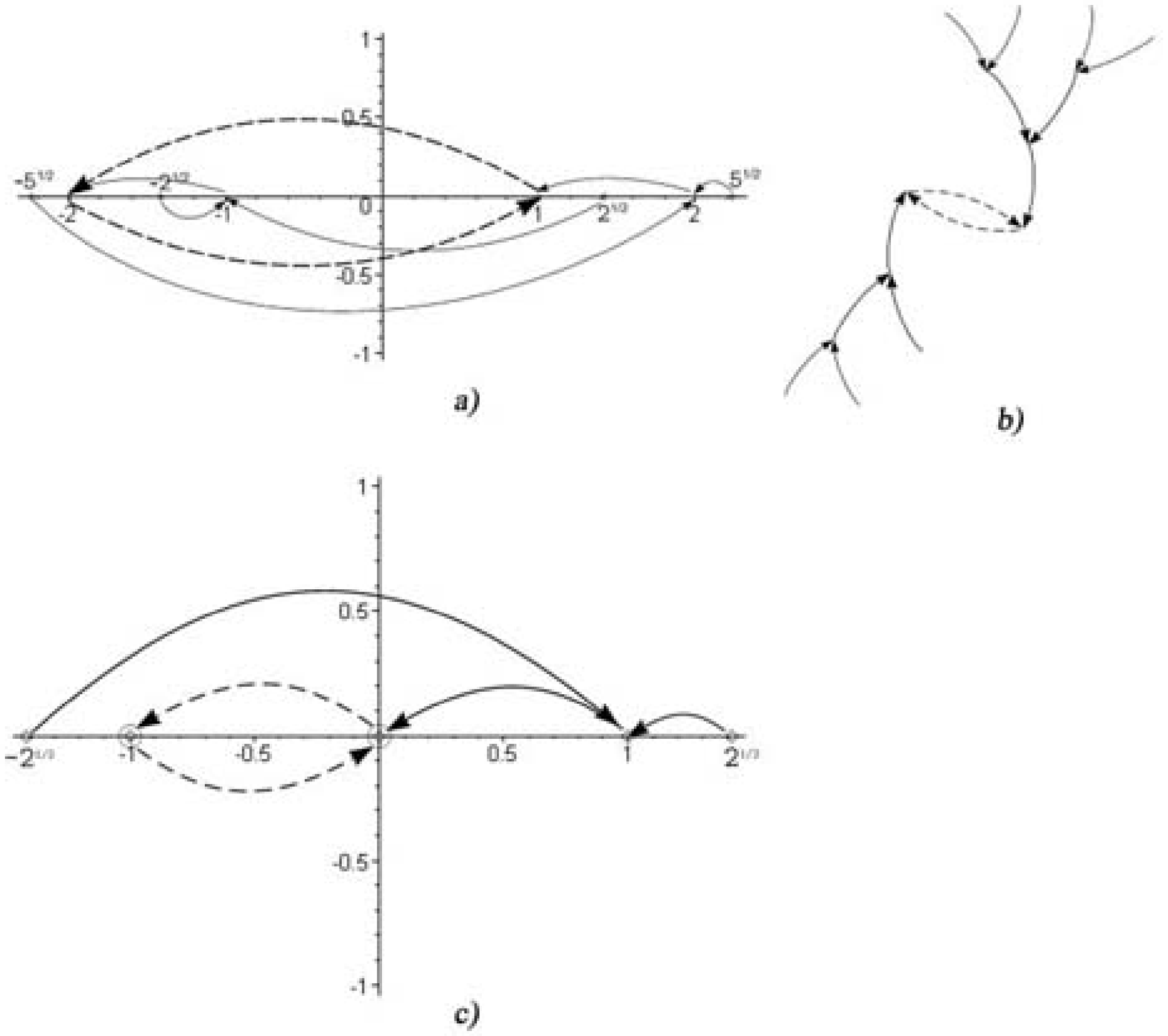}{}
{391,336}
{Typical view of a bounded grand orbit of order $n=2$
(thus the loop is of length $2$) for $f(x)$ which is a quadratic
polynomial (thus two arrows enter every vertex).
(a) Example of generic case: $f(x) = x^2 - 3$,
shown together with its embedding into the complex-$x$ plane.
(b) The same example, $f(x) = x^2 - 3$, only {\it internal}
tree structure is shown.
(c) Example of degenerate case: $f(x) = x^2-1$ (only one arrow
enters $-1$, its second preimage is infinitesimally close to $0$).}

\FIGEPS{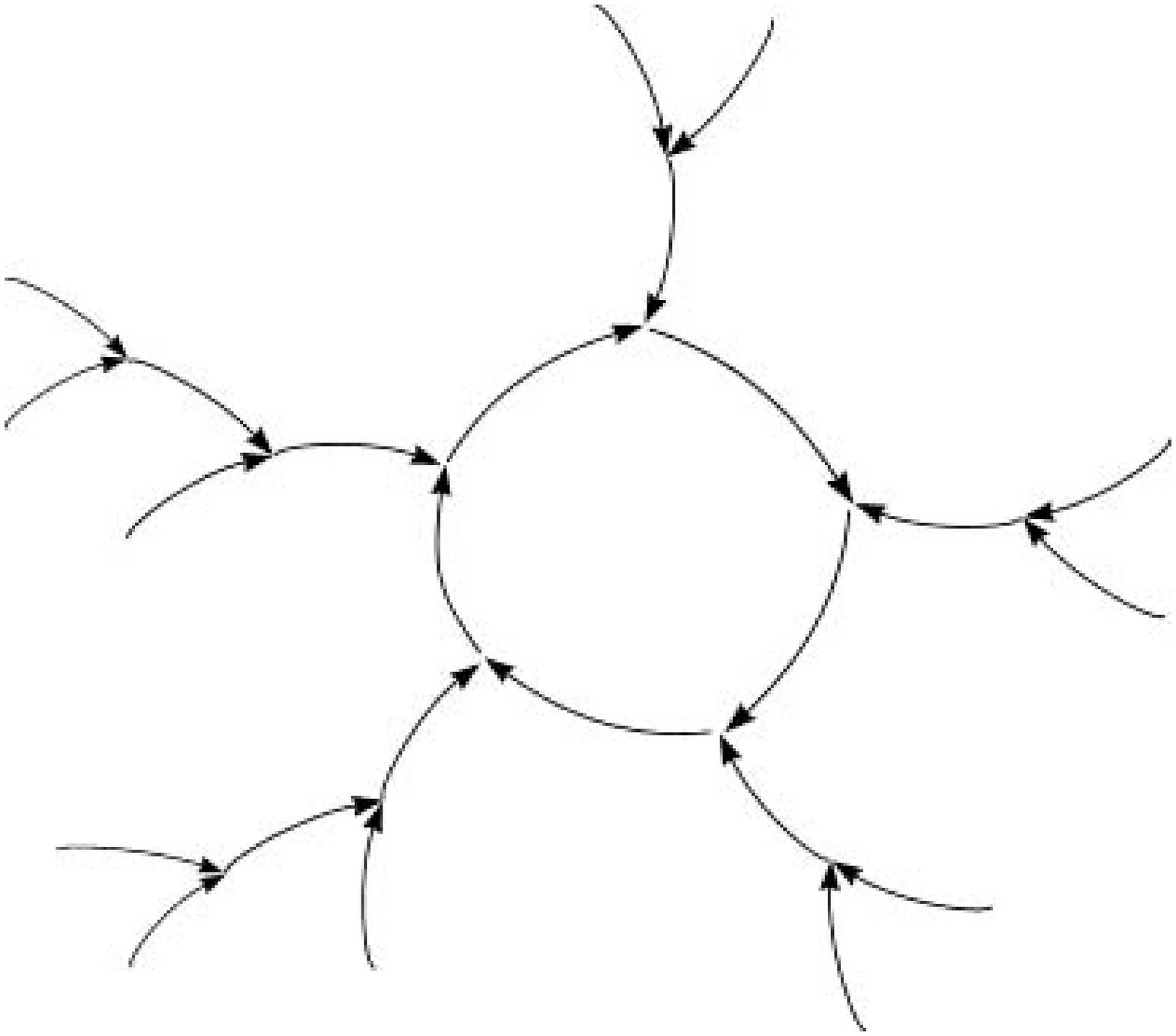}{}
{220,194}
{A non-degenerate bounded grand orbit of order $n=5$ for cubic
$f(x)$. Only internal structure is shown, without
embedding into $\bb{X}$.}

\bigskip

$\bullet$ The union ${\cal O}_{n}(f) \subset \bb{X}$
of all periodic $f$-orbits of order $n$,
the union ${\cal O}_{n,k}(f) \subset \bb{X}$ of all their $k$-preimages
and the unions
${\cal O}(f) =
\cup_{n=1}^\infty {\cal O}_{n}(f) \subset \bb{X}$
of all periodic orbits and
${\cal GO}(f) =
\cup_{k=0,n=1}^\infty {\cal O}_{n,k}(f) \subset \bb{X}$
of all {\it bounded} grand orbits.
The sets ${\cal O}(f) \subset {\cal GO}(f)$ can be smaller than $\bb{X}$
(since there are also non-periodic orbits and thus unbounded grand orbits).

\bigskip

$\bullet$ The sets ${\cal S}_{n}(f) \subset \bb{X}$ and
${\cal S}_{n,k}(f) \subset \bb{X}$ of all roots of the
functions
\be
F_{n}(x;f) = f^{\circ n}(x) - x
\label{Fndef2}
\ee
and
\be
F_{n,k}(x;f) = f^{\circ (n+k)}(x) - f^{\circ k}(x) =
F_{n+k}(x;f) - F_k(x;f) = F_n\left(x;f^{\circ k}\right).
\label{Fnkdef}
\ee
All periodic orbits
${\cal O}_m(f) \subset {\cal S}_n(f)$
whenever $m$ is a divisor of $n$ (in particular, when $m=n$).
Similarly, ${\cal O}_{m,k}(f) \subset {\cal S}_{n,k}(f)$.
This property allows to substitute the study of (bounded) periodic
(grand) orbits by the study of roots of maps $F_{n,k}(x;f)$.
Reshufflings (mergings and decompositions) of (grand) orbits occur
when the roots coincide, i.e. when the map becomes degenerate.
This property allows to substitute the study of the phase
space $\bb{X}$ (where the orbits and roots are living) by the study
of the space ${\cal M}$ of functions, where the varieties of
degenerate functions are known as discriminants ${\cal D}$.

\bigskip

$\bullet$
${\cal O}_n(f) \subset \bb{X}$ is a set of zeroes of function $G_n(x;f)$,
which is an irreducible-\-over-\-any-\-field divisor
of reducible function $F_n(x;f)$,
\be
F_n(x;f) = \prod_{m|n} G_m(x;f),
\ee
where the product is over all divisors $m$ of $n$, including $m=1$ and
$m=n$.
The  functions $G_n$ are more adequate for our purposes than $F_n$, but
the $F_n$ are much easier to define and deal with in sample calculations.

\bigskip

$\bullet$  {\bf Stable periodic orbit} of order $n$ is defined by
conditions
\be
\left\{
\begin{array}{l}
|(f^{\circ n})'(x)| < 1 \ \  {\rm i.e.}\ \ |F_n'(x) + 1| < 1 \\
G_n(x) = 0
\label{stabcon}
\end{array}\right.
\ee
for all its points. This definition is self-consistent because
\be
(f^{\circ n})'(x) = \prod_{k=0}^{n-1} f'(f^{\circ k}(x)) =
\prod_{z\in {\rm orbit}} f'(z),
\ee
i.e. the l.h.s. is actually independent of the choice of $x\in O_n(f)$.
Other periodic orbits are called {\bf unstable}.
Bounded grand orbits are called stable and unstable when periodic
orbits at their ends are stable and unstable respectively.

\bigskip

$\bullet$ The set of stable periodic orbits ${\cal O}_+(f)$
and its complement: the state of unstable periodic orbits ${\cal O}_-(f)$.
By definition
${\cal O}_+(f)\cup {\cal O}_-(f) = {\cal O}(f)$.

\bigskip

$\bullet$ {\bf Julia set} $J(f) \in \bb{X}$ is attraction domain of
stable periodic orbits of $f$, i.e. the set of points
$x \in \bb{X}$ with orbits $O(x;f)$, approaching the set of stable orbits:
$\forall \varepsilon >0$ $\exists k$ and $\exists z_k\in {\cal O}_+(f)$:
$|f^{\circ k}(x) - z_k| < \varepsilon$.
This definition refers to non-algebraic structures like continuity.
However, the boundary $\partial J(f)$ is an almost algebraic object,
since it actually consists of all the unstable periodic orbits and
their grand orbits, which are dense in $\partial J(f)$:
\be
\partial J(f) = \overline{{\cal O}_-(f)} = \overline{{\cal GO}_-(f)}
\ee
Moreover, {\it each} grand orbit "originates in the vicinity"
of ${\cal O}_-(f)$: generically, every branch of the grand-orbit tree
is associated with -- "originates at" -- a particular orbit from
${\cal O}_-(f)$, and when ${\cal O}_+(f) = \emptyset$ this is
a one-to-one correspondence.
If ${\cal O}_+(f) \neq \emptyset$,
the "future" of the grand orbit can be
three-fold: it can terminate in a periodic orbit, stable or unstable,
approach (tend to) some stable orbit or go to infinity (which can
actually be considered as a "reservoir" of additional stable orbits).
If ${\cal O}_+(f) = \emptyset$, the situation is different: strange
attractors can also occur.
Description of $\partial J(f)$ in terms of orbits works even when
${\cal O}_+(f) = \emptyset$, but then it is not a boundary of anything,
the Julia set $J(f)$ itself is not defined.

For a better description of Julia sets see summary in s.\ref{summ}
and s.\ref{Juexa} below.

\subsection{Objects, associated with the space ${\cal M}$}

\noindent

$\bullet$ The {\bf moduli space ${\cal M}$} of all maps.
In principle it can be as big as the set of formal series (\ref{mapf}),
i.e. can be considered as the space of the coefficients $\{a_k\}$.
Some pieces of theory, however, require additional structures, accordingly
one can restrict ${\cal M}$ to sets of
continuous, smooth, locally analytic or any
other convenient sets of maps. We assume that the maps are single-valued,
and their functional
inverses can be described in terms of trees, perhaps, of
infinite valence.
The safest (but clearly over-restricted)
choice is a subspace ${\cal P} \subset {\cal M}$ of all polynomials
with complex coefficients,
which, in turn, can be decomposed into spaces of polynomials of particular
degrees $d$: ${\cal P} = \cup_d^\infty {\cal P}_d$.
One can also consider much smaller subspaces: families $\mu(\vec c) \subset
{\cal M}$, with $a_k(\vec c)$ in (\ref{mapf})
parameterized by several complex
parameters $\vec c = (c_1,c_2,\ldots)$.
Most examples in the literature deal with one-parametric families.

\bigskip

$\bullet$
The moduli subspace ${\cal M}_{\circn} \subset {\cal M}$ of {\it shifted}
$n$-iterated maps, consisting of all maps $F: \ X \longrightarrow \ X$
which have the form (\ref{Fndef2})
for some $f(x)$. The shift by $x$ is important.
Eq.(\ref{Fndef2}) defines canonical mappings
$$
\hat I_n:\ {\cal M} \longrightarrow {\cal M}_{\circn},
$$
which can be also written in terms of the coefficients: for $f(x)$
parameterized as in eq.(\ref{mapf}), and
$$
F_n(x;f) = f^{\circ n}(x) - x = \sum_{k=0}^\infty a_k^{(n)} x^k
$$
we can define $\hat I_n$ as an algebraic map
\be
\hat I_n:\ \{a_k\} \ \longrightarrow\ \{a_k^{(n)}\}
\label{Imap}
\ee
where all $a_k^{(n)}$ are polynomials of $\{a_l\}$.\footnote{
In order to avoid confusion we state
explicitly that $a_k^{(0)} = 0$ and $a_k^{(1)} = a_k - \delta_{k,1}$.
Also ${\cal M}_{\circ 1} = {\cal M}$, but in general analogous relation
is not true for subsets $\mu \subset {\cal M}$: $\mu_{\circ 1}$ does
not obligatory coincide with $\mu$, it differs from it by a shift by $x$.
}
Inverse map $\hat I_n^{-1}$ is defined on ${\cal M}_{\circn}$ only, not on
entire ${\cal M}$, but is not single-valued. In what follows we denote
the pull-back of $\hat I_n$ by {\it double} star,
thus $(\hat I_n f)^{**} = f$ (while $\hat I_n^{-1}
(\hat I_n f)$ can contain a set of other functions in addition to $f$,
all mapped into one and the same point $\hat I_n f$ of ${\cal M}_{\circn}$).

The map $\hat I_n$ can be restricted to polynomials of  given degree, then
\be
\hat I_n:\ {\cal P}_d \longrightarrow {\cal P}_{d^n}
\label{IPd}
\ee
embeds the $(d+1)_{\bb{C}}$-dimensional space into the
$(d^n+1)_{\bb{C}}$ one as an algebraic variety.\footnote{
For example, $({\cal P}_2)_{\circ 2}$ is embedded into
${\cal P}_4$ as algebraic variety:
$$
8A^2(D+1) - 4ABC + B^3 = 0,
$$
$$
64A^2 (4AB^2E + B^3 - 4A^2(D+1)^2)^3 = B^3(16A^2(D+1)-B^3)^3
$$
This follows from
$$
Ax^4 + Bx^3 + Cx^2 + Dx + E =
(ax^2+bx + c)^{\circ 2} -x =
$$
$$
= a^3x^4 + 2a^2b x^3 + (ab^2 + ab + 2a^2c)x^2 +
(2abc + b^2-1) x + (ac^2 + bc + c).
$$
Note, that because of the shift by $x$ one should be careful with
projectivization of this variety.
}
Like in (\ref{IPd}),
often $\mu_{\circn} = \hat I_n(\mu) \notsubset \mu$.
One can, however, use a pull-back to cure this situation:
$\left[{\cal B}\cap\mu_{\circn}\right]^{**} \subset \mu$ for any
${\cal B} \subset {\cal M}$.

In the studies of {\it grand orbits} more general moduli subspaces
are involved: ${\cal M}_{\circ (n,k)} \subset {\cal M}$, consisting of
all maps of the form (\ref{Fnkdef}) for some $f$.

\bigskip

$\bullet$ Similarly, the functions $G_n(x;f)$ define another
canonical set of mappings,
$$
\hat J_n:\ {\cal M} \longrightarrow {\cal M}_{n}.
$$
into subspaces ${\cal M}_{n}\subset {\cal M}$, consisting of all
functions which have the form $G_n(x;f)$ for some $f$.
These are, however, somewhat less explicit varieties than
${\cal M}_{\circn}$, because such are the functions $G_n(x;f)$ --
defined as irreducible constituents of $F_n(x;f)$. Still, they
are quite explicit, say, when $f$ are polynomials of definite
degree, and they are more adequate to describe the structures,
relevant for discrete dynamics.
The $J$-pullback will be denoted by a single star.

\bigskip

$\bullet$ {\bf Discriminant variety} ${\cal D} \subset {\cal M}$ consists of
all {\it degenerate} maps $f$, i.e. such that $f(x)$ and its derivative
$f'(x)$ have a common zero. It is defined by the equation
$$
D(f) = 0
$$
in ${\cal M}$, where $D(f)$ is the square of the Van-der-Monde product
of roots differences:
$$
{\rm for}\ \ f(x) \sim \prod_k (x-r_k)\nn \\
D(f) \sim \prod_{k \neq l} (r_k - r_l).
$$
Restrictions of discriminant onto the spaces
of polynomials of given degree,
${\cal D}({\cal P}_d) = {\cal D}\cap {\cal P}_d$,
are algebraic varieties in ${\cal P}_d$, because $D(f)$ is a polynomial
(of degree $2d-1$)
of the coefficients $\{a_k\}$ in representation (\ref{mapf}) of $f$
(while roots themselves are not polynomial in $\{a_k\}$ the square of
Van-der-Monde product is, basically, by Vieta's theorem).
Discriminant variety ${\cal D}$
has singularities of various codimensions $k$ in ${\cal D}$,
associated with mergings of $k+2$ roots of $f$.

\bigskip

$\bullet$ {\bf Resultant} variety ${\cal R} \subset {\cal M}\times
{\cal M}$ consists of pairs of maps $f(x)$, $g(x)$ which have
common zero. It is defined by the equation
$$
R(f,g) = 0,
$$
and
\be
{\rm for}\ \ f(x) \sim \prod_k (x-r_k) \ \
{\rm and} \ \ \  g(x) \sim \prod_l (x-s_l) \nn \\
R(f,g) \sim \prod_{k,l} (r_k - s_l).
\nn
\ee
Again, if $f$ and $g$ are polynomials, resultant is a polynomial
in coefficients of $f$ and $g$.
Resultant variety has singularities when more that one pair of roots
coincide. Discriminant can be considered as appropriate
resultants' derivative at diagonal $f=g$.
Higher resultants varieties in ${\cal M}\times\ldots\times{\cal M}$
are also important for our purposes, but will not be considered in
the present paper.

\FIGEPS{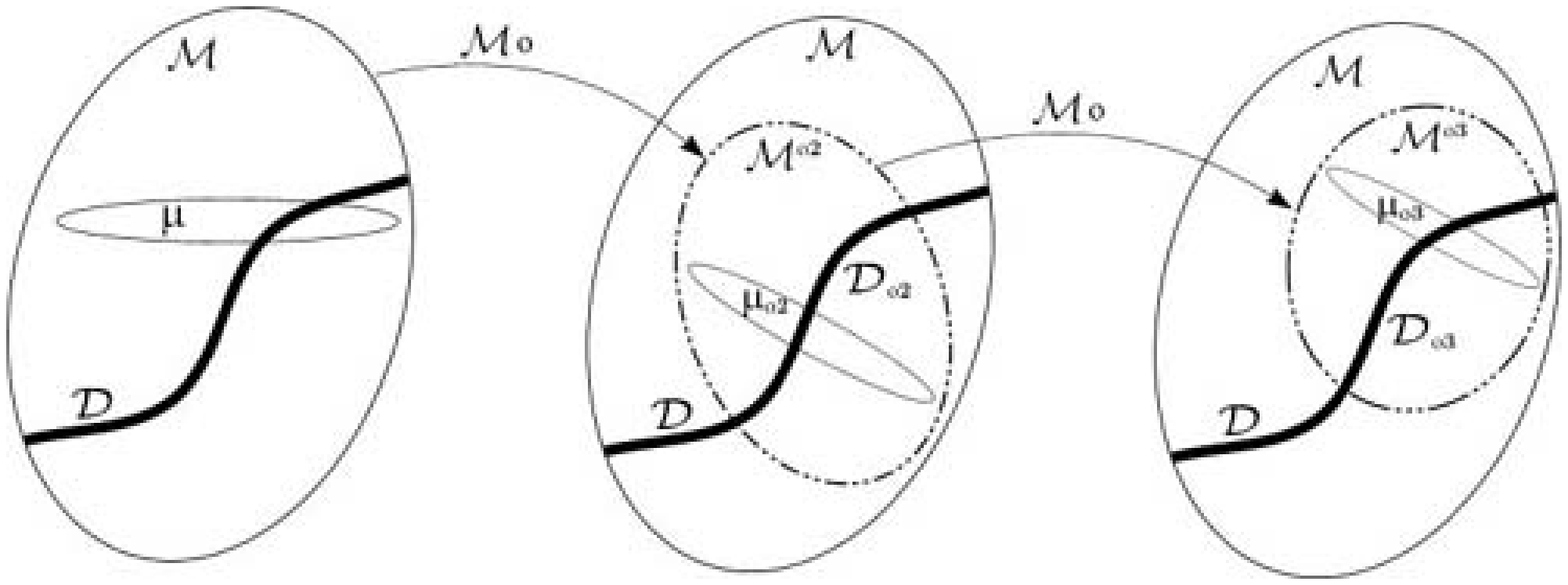}{}
{380,140}
{Schematic view of inter-relations between
the subsets ${\cal M}_{\circn}$, $\mu_{\circn}$ and
${\cal D}_{\circn}$.
}

\bigskip

$\bullet$
Intersections of ${\cal D}$ and ${\cal R}$ with subspaces
like ${\cal M}_{\circn}$ and ${\cal M}_n$, i.e.
${\cal D}_{\circn} = {\cal D}\cap {\cal M}_{\circn}$,
${\cal R}_{\circ (m,n)} = {\cal R} \cap ({\cal M}_{\circ m} \times
{\cal M}_{\circ n})$;
${\cal D}_n = {\cal D}\cap {\cal M}_n$,
${\cal R}_{m,n} = {\cal R} \cap ({\cal M}_m \times
{\cal M}_{m})$
consist of disconnected components (see Fig.\ref{McircninM}).
$\hat I$-pullbacks ${\cal D}_{\circn}^{**},\ {\cal R}_{\circ (m,n)}^{**}
\subset {\cal M}$ and $\hat J$-pullbacks
\be
{\cal D}_n^* = \left\{ f:\ D(G_n(f))=0 \right\} \subset {\cal M}, \nn \\
{\cal R}_{m,n}^* =
\left\{f:\ R(G_m(f),G_n(f))=0 \right\} \subset {\cal M}
\nn
\ee
also consist of numerous components, disjoint and touching.
Particular families $\mu \subset {\cal M}$ intersect these singular
varieties and provide particular sections $\partial M(\mu)$
of this generic structure: the {\bf universal discriminant}
(or resultant) {\bf variety}
$$
{\cal D}^* = {\cal R}^* = \bigcup_{n=1}^\infty {\cal R}_n =
\overline{\bigcup_{n,m=1}^\infty {\cal R}^*_{m,n}
\bigcup_{n=1}^\infty {\cal D}^*_n}
$$
which is the boundary of the
{\bf Universal Mandelbrot set} $\partial M({\cal M})$.

\bigskip

$\bullet$ For a given family of maps $\mu \subset {\cal M}$
the {\bf boundary of {\it algebraic}  Mandelbrot set}
$\partial M(\mu) \subset \mu$ is defined as a $\hat J$-pullback
\be
\partial M(\mu) = \cup_{n=1}^\infty
({\cal D}_n^*\cap \mu) =
\left( {\cal D} \cap (\cup_{n=1}^\infty \mu_n)\right)^* = \nn \\ =
\mu\cup \left(\cup_{n=1}^\infty
\left\{ f:\ D(G_n(f))=0\right\}\right)
\label{partialMandel}
\ee

\bigskip

$\bullet$
Varieties ${\cal D}^{**}$, ${\cal R}^{**}$, ${\cal D}^*$,
${\cal R}^*$ -- and thus the boundary of
algebraic Mandelbrot space $\partial M$ --
can be instead considered as pure topological objects in ${\cal M}$
independent of any additional algebraic structures,
needed to define $F_n$, $G_n$, discriminant and resultant
varieties ${\cal D}$ and ${\cal R}$. For example,
${\cal D}^{**}_1 = {\cal D}^*_1$ consists of all maps
$f \in {\cal M}$ with two coincident fixed points.
Higher components ${\cal D}^*_n \subset {\cal D}^{**}_n$
consist of maps $f$, with two coincident fixed points of their
$n$-th iteration (for ${\cal D}^{**}_n$) and such that any
lower iteration does not have coincident points
(for ${\cal D}^{*}_n$). Resultants consist of maps $f$
with coinciding fixed points of their different iterations.
Fixed points and iterated maps are
pure categorial notions, while to define coincident points
one can make use of topological structure:
two different fixed points merge under continuous deformation
of $f$ in ${\cal M}$.

\bigskip

$\bullet$
{\bf Stability domain} of the periodic order-n orbits $S_n \subset {\cal M}$
is defined by the system (\ref{stabcon}): if a root $x$ of $G_n(f)$
is substituted in to inequality, it becomes a restriction on the
shape of $f$, i.e. defines a domain in ${\cal M}$. This domain is
highly singular and disconnected. For a family $\mu \subset {\cal M}$
we get a section $S_n(\mu) = S_n\cap \mu$, also singular and
disconnected.

$\bullet$
{\bf Mandelbrot set} $M({\cal M}) \subset {\cal M}$
is a union of all stability domains with different $n$:
\be
M = \cup_{n=1}^\infty S_n, \ \ \ M(\mu) = \cup_{n=1}^\infty
S_n(\mu) = M\cap \mu.
\label{Mandel}
\ee
Definitions (\ref{Mandel}) and (\ref{partialMandel})
leave obscure most of the structure of Mandelbrot set and its
boundary. Moreover, even consistency of these two definitions
is not obvious.\footnote{
It deserves emphasizing that from algebraic perspective there
is an  essential difference between the Mandelbrot
and Julia sets themselves and their boundaries.
Boundaries are pure algebraic (or, alternatively, pure
topological) objects, while entire spaces depend on stability
criteria, which, for example, break complex analyticity and
other nice properties present in the description of the boundaries.
}
For better description of Mandelbrot set see s.\ref{summ} below.

\bigskip

$\bullet$ Above definitions,
at least the algebraic part of the story,
can be straightforwardly extended from orbits to grand orbits
and from maps $F_n(f)$ and $G_n(f)$ to $F_{n,k}(f)$ and $G_{n,k}(f)$.
The maps $G_{n,k}(x;f)$, which define $k$-th pre-orbits of periodic
order-n orbits are not obligatory {\it irreducible} constituents of
$F_{n,k}(x;f)$ (what is the case for $k=0$). They are instead related
to peculiar map
$$
z:\ {\cal M} \rightarrow \bb{X}
$$
associating with every $f \in {\cal M}$ the points $x \in \bb{X}$
with degenerate pre-image:
$$
z_f = \{z:\ D(f(x)-z) = 0\}.
$$
Generalizations of Mandelbrot sets to $k\geq 1$ are called
{\bf secondary Mandelbrot sets}. The {\it Grand Mandelbrot
set} is the union of secondary sets with all $k \geq 0$.
Bifurcations of {\it Julia sets} with the variation of $f$ inside
$\mu \subset {\cal M}$ are captured by the structure of
Grand Mandelbrot set.

\bigskip

We see that, the future theory of dynamical systems should include the
study of two purely algebro-geometric objects:

(i) the universal discriminant and resultant varieties
${\cal D} \subset {\cal M}$
and ${\cal R} \subset {\cal M}\times {\cal M}$,

(ii) the map
$\mu \longrightarrow \check\mu = \cup_{n=1}^\infty \mu_{n}$ and the
pull-backs ${\cal D}^* \subset {\cal M}$ and ${\cal R}^* \subset {\cal M}$.

It should investigate intersections of ${\cal D}$ and $\check\mu$ and
${\cal R}$ with $\check\mu\times\check\mu$
and consider further generalizations: to multi-dimensional phase spaces
$\bb{X}$ and to continuous iteration numbers $n$.
It is also interesting to understand how far one can move
with description of ${\cal R}^*$ and universal Mandelbrot space by pure
topological methods, without the ring and other auxiliary
algebraic structures.

\subsection{Combinatorial objects
\label{coo}}

$\bullet$ {\bf Divisors tree} $t[n]$ is a finite rooted tree.
Number $n$ stands at the root, and it is connected to all
$\tau(n)-1$ of its divisors $k|n$ (including $k=1$, but
excluding $k=n$). Each vertex $k$ is further connected to all
$\tau(k)-1$ of {\it its} divisors and so on.
The links are labeled by ratios $m = n/k\geq 2$ and
$m_i = k_i/k_{i+1}\geq 2$. Number $n$ is equal to product of
$m$'s along every branch. See examples in Fig.\ref{divitree}.
The number $\tau(n)$ of divisors of $n = p_1^{a_1}\ldots p_k^{a_k}$
is equal to $\tau(n) = (a_1+1)\ldots (a_k+1)$.
A generating function (Dirchlet function) has asymptotic
\be
D(x) = \sum_{n\leq x} \tau(n) \sim x\log x + (2C-1) + O(\sqrt{x}),
\label{Dirfo}
\ee
$C$ -- Euler constant.

\FIGEPS{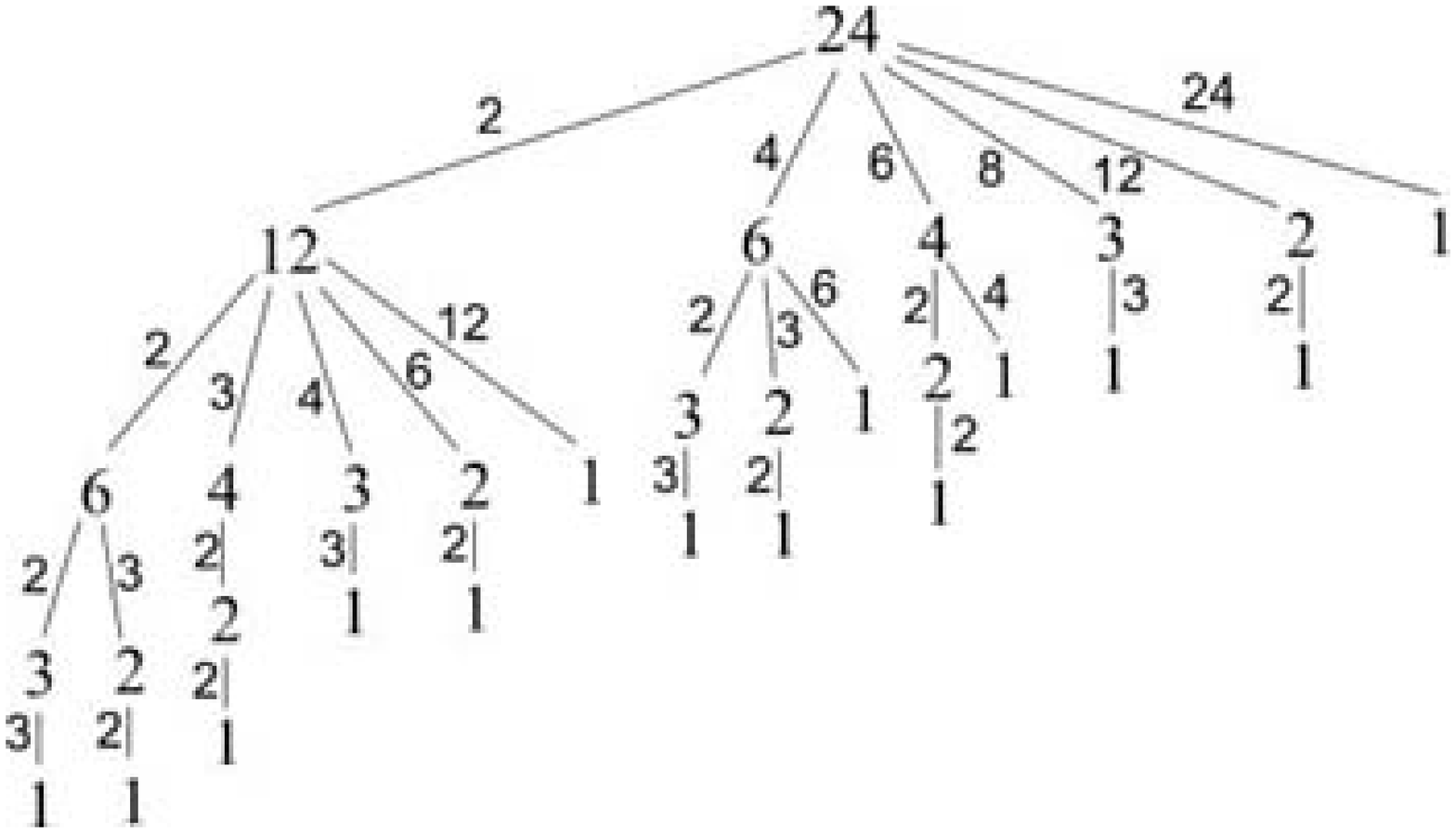}{}
{300,170}
{Divisors tree $t[24]$. It contains divisor trees for all its divisors:
$t[12]$, $t[6]$, $t[4]$, $t[3]$, $t[2]$, $t[1]$, some in several
copies.}

\FIGEPS{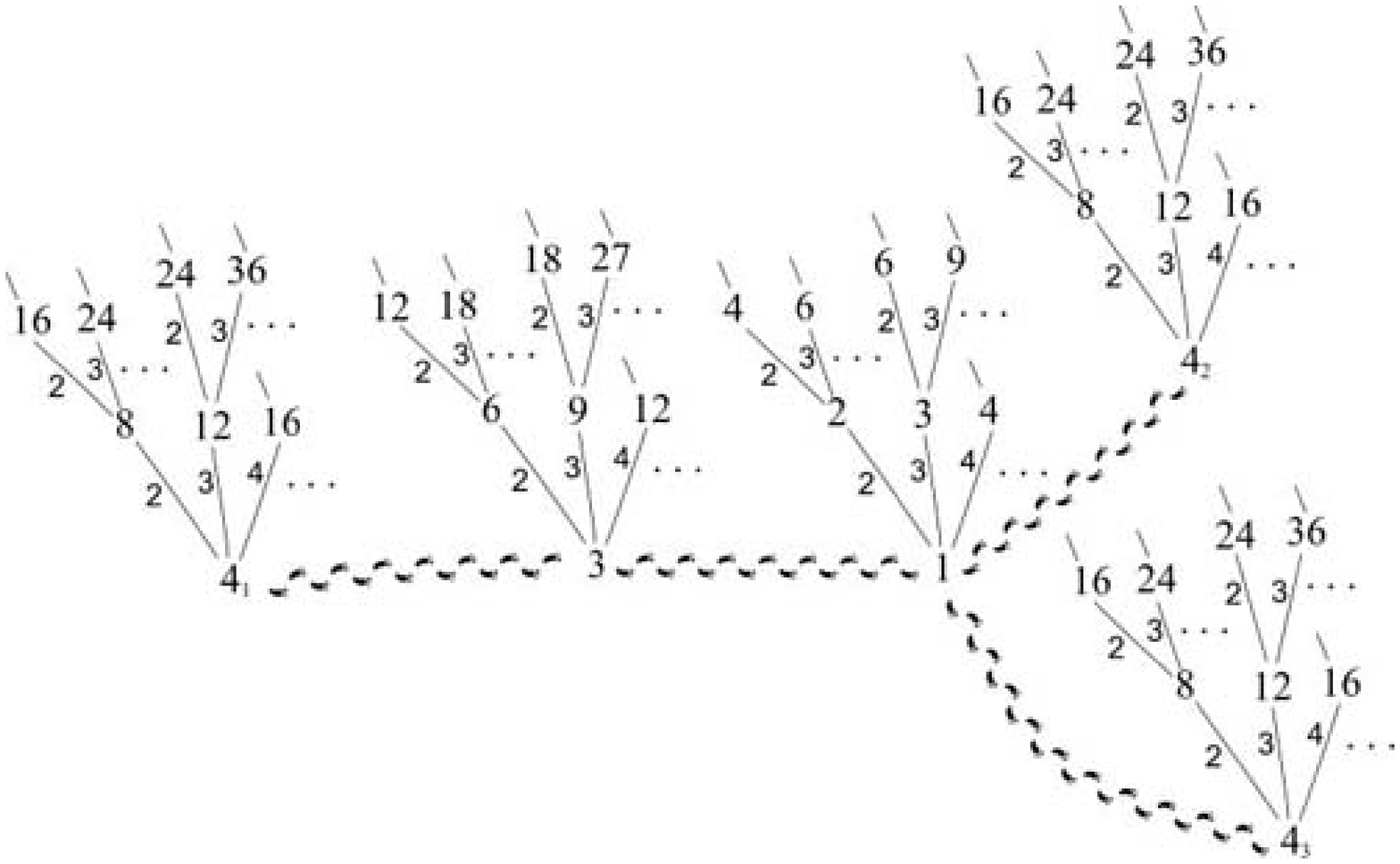}{}
{400,246}
{A piece of the forest of multipliers trees. Trails are shown by "steps".}

\bigskip

$\bullet$ {\bf Multipliers tree} $T_n$
is a rooted tree of infinite valence.
At the root stands the number $n$. The branches at every level $p$ are
labeled by positive integers $m_p = 1,2,3,4,\ldots$.
Every vertex at $p$-th level is connected to the root by a path
$m_1,m_2,\ldots,m_p$ and the product $nm_1m_2\ldots m_p$ stands in it.
The {\bf basic forest} $T = \cup_{n=1}T_n$.
A number of times ${B}(n)$ the number $n$ occurs in the forest $T$ is
equal to the number of branches in its {\it divisors tree}.

The numbers $\tau_p(n)$
of ways to represent $n$ as a product of $p$ integers,
i.e. the numbers of times $n$ appears at the $p-th$ level of
{\it multipliers tree}, are described by the generating function
$$
D_p(x) = \sum_{n\leq x} \tau_p(n) =
\frac{1}{2\pi i}\int_{c-i\infty}^{c + i\infty}
\frac{ds}{s} x^s\zeta^p(s),
$$
$\zeta(s) = \sum_{m=1}^\infty m^{-s}$ is Riemann's $zeta$-function.
Note that (\ref{Dirfo}) describes $\tau(n) = \tau_2(n)$.

\bigskip

$\bullet$ {\bf Basic graph} ${\cal B}$ is obtained from the
multipliers tree $T_1$ by identification of all vertices with the
same numbers, so that the vertices of ${\cal B}$ are in
one-to-one correspondence with all natural numbers. Since
$T_1$ was a rooted tree, the graph ${\cal B}$ is directed:
all links are arrows. There are $\tau(n)-1$ arrows entering
the vertex  $n$ and infinitely many arrows which exit it and
lead to points $mn$ with all natural $m$.

\FIGEPS{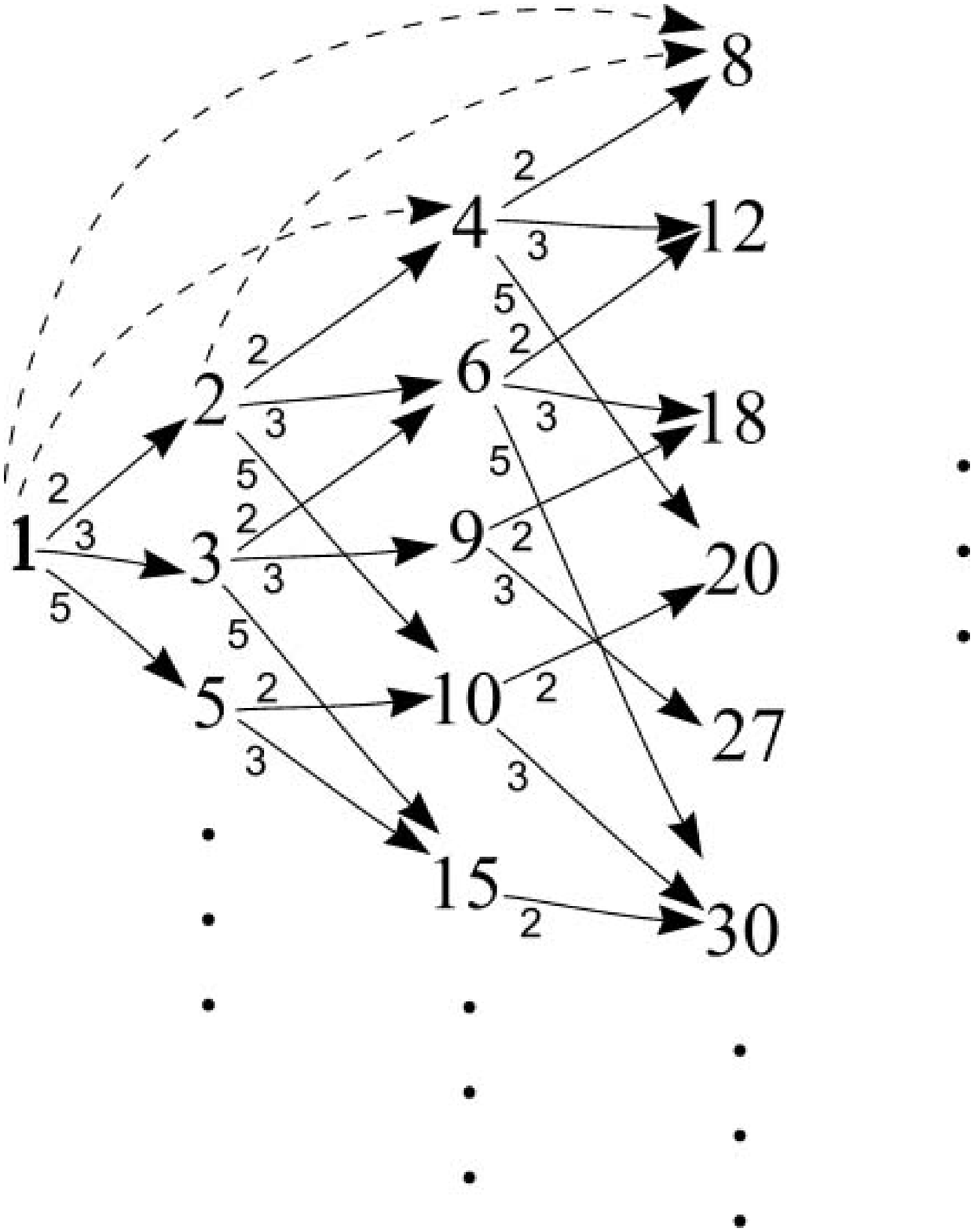}{}
{220,279}
{A fragment of the basic graph.
Only the generating arrows, associated with primes, are shown
(to avoid overloading the picture). All compositions of arrows
should be added as separate arrows (for example, arrows connect
$1$ to $4$, $6$, $8$, $9$ and all other integers).}

\subsection{Relations between the notions}

Relations are summarized in the following table.

{\footnotesize
\be
\hspace{-2cm}
\begin{array}{cccc}
& \underline{{\rm objects\ in}\ \bb{X}}  &  &
\underline{{\rm objects\ in}\ {\cal M}} \\
& & & \\
& & & \\
& {\rm map}\ f & & {\rm moduli\ space}\ {\cal M} \\
& \downarrow &  & \cup \\
& {\rm iterated\ maps}\  F_n(f)& &
     {\rm spaces}\ {\cal M}_{\circn} \\
&  \downarrow & & \\
& F_n(f) = \prod_{k|n} G_k(f) & & {\rm spaces}\ {\cal M}_{k} \\
&  \downarrow & &  \\
& n-{\rm periodic}\ f-{\rm orbits}\ = & & \\
& =\ {\rm roots\ of}\ G_n(f)  & &  \\
&  \downarrow & & \\
& {\rm coincident\ roots\ of}\ G_n(f) & = & \{D\{G_n(f)\} = 0\} \\
& =\ {\rm reshuffling\ of\ orbits} & = &
      {\rm discriminant\ varieties}\ =\\
& & & =\ \partial({\rm Mandelbrot\ set}\ M)  \\
& & & \\
\ \ \ \ \ \ \ \swarrow &&\searrow \ \ \ \ \ \ \ &\\
& & & \\
& &{\rm maps}\  F_{n,k}(f) & {\rm spaces}\ {\cal M}_{\circ(n,k)} \\
& & & \\
{\rm unstable\ periodic\ orbits}\ =& &
        {\rm bounded\ grand}\ f-{\rm orbits} & \\
=\ \partial({\rm Julia\ set}\ J(f)) \ \stackrel{?}{=}        & &  =
      \ {\rm roots\ of}\ F_{n,k}(f) & \\
\partial({\rm algebraic\ Julia\ set}\ J_A(f)) & \stackrel{?}{=} &
      {\rm infinite\ preimage\ of\ BGO}  & \\
  \downarrow & &  \downarrow & \\
& & {\rm coincident\ roots\ of}\ F_{n,k}(f) &=\ \ \{D\{F_{n,k}(f)\} = 0\} \\
{\rm reshuffling\ of}\ J_A(f) & \stackrel{?}{=} &
      {\rm resuffling\ of\ BGO}\ = &
      \partial({\rm grand\ Mandelbrot\ set})  \\
& & &  \\
\end{array}
\nn
\ee
}

\newpage

\section{Summary
\label{summ}}

In this section we briefly summarize our main claims, concerning the
structure of Mandelbrot and Julia sets. They are naturally splited
in three topics.

\subsection{Orbits and grand orbits}

This part of the story includes:

-- The theory of orbit and pre-orbit functions
$G_n(x;f)$ and $G_{n,k}(x;f)$.

-- Classification of orbits, pre-orbits and grand orbits.

-- Intersections (bifurcations) of orbits and
degenerations of grand orbits.

-- Discriminant and resultant analysis, reduced discriminants $d_n$ and
resultants $r_{m,n}$ intersections
of resultant and discriminant varieties.

All these subjects -- to different level of depth --
are considered in s.\ref{pot} below and illustrated
by examples in ss.\ref{Exa} and \ref{Exa2}.
Systematic theory is still lacking.

\subsection{Mandelbrot sets}

Figs.\ref{0Man4},a-b
show three hierarchical levels of Mandelbrot set: the structure of
individual component ($M_1$ in Fig.\ref{0Man4},a); existence of
infinitely many other components (like $M_{3,\alpha}$ in Fig.\ref{0Man4},b),
which look {\it practically} the same after appropriate rescaling;
trails connecting different components $M_{k\alpha}$ with $M_1$,
which are densely populated by other $M_{m\beta}$ (well seen in the
same Fig.\ref{0Man4},b).
These structures (except for the self-similarity property)
are universal: they reflect the structure of the universal Mandelbrot
set, of which the boundary is the universal resultant variety ${\cal R}^*$.
This means that they do not depend on particular choice of the family $\mu$
(above figures are drawn for $f_c = x^4 + c$, which is in no way
distinguished, as is obvious from similar pictures
for some other families, presented in ss.\ref{Exa} and \ref{Exa2}).
The only characteristics which depend on the family
are the sets of indices $\alpha$, labeling degeneracies:
different domains with the same place in algebraic structures.
Algebraic structure reflects intersection properties of different
subvarieties ${\cal R}_{m,n}^*$, while $\alpha$'s parameterize the
intersection subvarieties. To classify the $\alpha$-parameters one should
introduce and study higher resultants and multiparametric families
$\mu \subset {\cal M}$, what is straightforward,
but (except for introductory examples)
is left beyond the scope of the present paper.

The main property of the resultant varieties ${\cal R}^*_{m,n}$,
responsible for the structure of Mandelbrot set, is that
they are non-trivial in real codimension two
only if $n$ is divisible by $m$ or vice versa.
All the rest follows from above-described relation between
the resultant varieties and the boundary of Mandelbrot set.
Stability domains $S_n$ and Mandelbrot sets $M$ are made out of the same
disk-like\footnote{
In the case of multi-parametric families $\mu \subset {\cal M}$,
when dimension of components of Mandelbrot set $M(\mu)$ is equal to
$\ {\rm dim}\ \mu\ $, "disc" may be not a very adequate name. In fact,
these components look more like interiors of cylinders and tori rather
than balls.
}
building blocks, but connect these
blocks in two different ways.
Mandelbrot set and its "forest with trails" structure reflect
the universal ($\mu$-independent) labeling (ordering)
of these {\bf elementary domains} (denoted by $\sigma$
below), while their sizes, locations are self-similarity properties
(which depend on the choice of $\mu$) are
dictated by stability equations.

\FIGEPS{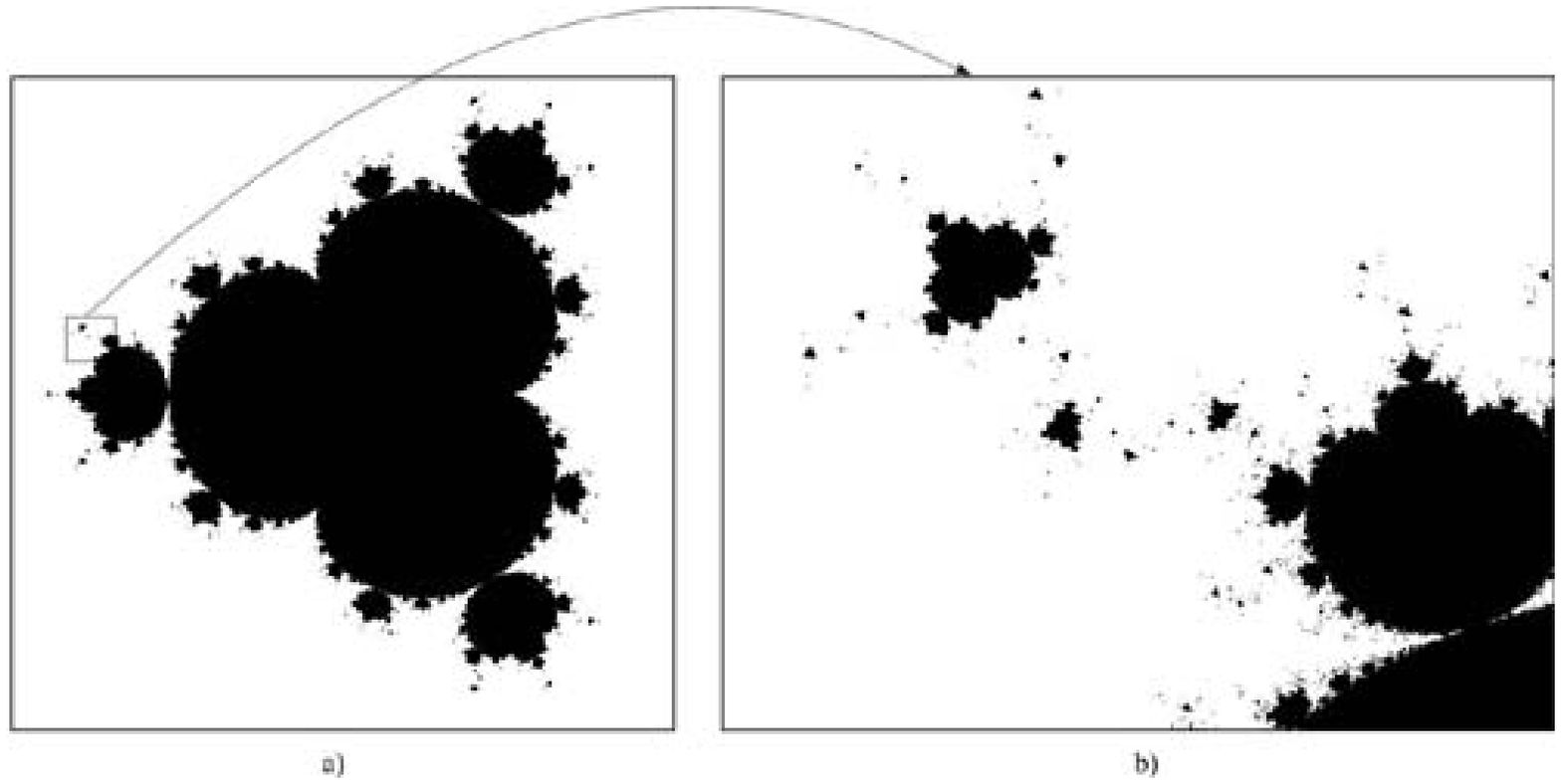}{}
{450,224}
{a) General view
of the Mandelbrot set for the family $f_c(x) = x^4 + c$.
Actually, only the $M_1$ component is well seen. Arrow points
at another component, $M_{3,\alpha}$, shown unlarged in b). The structure of the trail between
$M_1$ and $M_{3,\alpha}$ (almost unobservable in this picture)
is also seen in b). $Z_3$-symmetry of the picture
is due to invariance of equations like $x^4 + c = x$ under the
transformation $x \rightarrow e^{2\pi i/3}x$,
$c \rightarrow e^{2\pi i/3}c$. Associated Julia set has
$Z_4$ symmetry, see Fig.\ref{JuliaMix4}. b) Enlarged component $M_{3,\alpha}$, included in the box in a). It {\it looks} like exact copy of $M_1$ in a). In fact similarity is approximate, but the deviations
are damped by the ratio of sizes, $r_3/r_1 \ll 1$.
In this unlarged picture also the trail is seen between $M_1$ and
$M_{3,\alpha}$ and other $M_{k\alpha}$ component in it.}

\subsubsection{Forest structure}

For any family of maps $\mu$ Mandelbrot set $M(\mu)$ has the
following hierarchical structure, see Fig.\ref{0Man4},a
(for particular $\mu$ some
components -- or, better to say, their intersections with $\mu$
-- can be empty):

$M(\mu)$ consists of infinitely many disconnected components
(of which only one is clearly seen in Fig.\ref{0Man4},a and some
others are revealed by zooming in Fig.\ref{0Man4},b),
$$
M(\mu) = \bigcup_{k,\alpha} M_{k\alpha}(\mu),\nn \\
M_{k\alpha} \cap M_{l\beta} = \emptyset,\ \ {\rm if}\
k\neq l\ {\rm or}\ \alpha\neq \beta,
$$
labeled by natural number $k$ and an extra index
$\alpha$, belonging to the $\mu$-dependent set
$\nu_k(\mu)$. For polynomial $f$ the set is finite, its size
will be denoted by $|\nu_k(\mu)|$.

Every component $M_{k\alpha}(\mu)$ is a union of elementary domains,
each with topology of a disc (interior of a $(2D-1)_{\bb{R}}$-dimensional
sphere if $\mu$ is a $D_{\bb{C}}$-parametric family),
which form a tree-like structure.
Vertices of the tree $T_{k\alpha}$ -- the {\bf skeleton of $M_{k\alpha}$} --
are associated with "centers"
of elementary domains, a link connects two vertices whenever the two
corresponding domains have a common point -- and there can be at most
one such point, where the two domains actually {\it touch} each other.
The skeleton is a rooted tree, moreover, it is actually a {\it covering}
of the multipliers tree $T_k$, introduced in s.\ref{coo}.
The difference is that
every branch at the level $p$ carrying the label $m_p$ has multiplicity
$|\nu^{(p)}_{m_p}(\mu)|$ and thus an extra label
$\alpha_p \in \nu^{(p)}_{m_p}(\mu)$.
The union of skeleton trees is the skeleton of the Mandelbrot set --
the {\bf Mandelbrot forest}
$$
T(M) = \bigcup_{k,\alpha} T_{k\alpha}.
$$

Thus
$$
M_{k\alpha} = \bigcup_{p=0}^\infty M^{(p)}_{k\alpha}
$$
$$
M^{(p)}_{k\alpha} = \bigcup_{
\stackrel{m_1,\ldots,m_p = 1}
{\alpha_1,\ldots,\alpha_p}}^\infty
\sigma^{(p)}\left[\left.
\begin{array}{ccc}
m_1 & \ldots & m_p\\ \alpha_1 & \ldots & \alpha_p
\end{array}\right| k\alpha\right]
$$
and, finally, $M(\mu) = M \cap \mu$,
$$
M = \bigcup_{k,\alpha} \bigcup_{p=0}^\infty
\left(\bigcup_{
\stackrel{m_1,\ldots,m_p = 1}
{\alpha_1,\ldots,\alpha_p}}^\infty
\sigma^{(p)}\left[\left.
\begin{array}{ccc}
m_1 & \ldots & m_p\\ \alpha_1 & \ldots & \alpha_p
\end{array}\right| k\alpha\right]\right).
$$
The elementary domains are
$\ \sigma^{(p)}\left[\left.
\begin{array}{ccc}
m_1 & \ldots & m_p\\ \alpha_1 & \ldots & \alpha_p
\end{array}\right| k\alpha\right]\ $
and two of them touch at exactly one point,
whenever they are connected by a link in the powerful tree
$T_{k\alpha}$, i.e.
when $p'=p+1$, and $m'_i = m_i,$ $\alpha'_i = \alpha_i$
for all $i = 1,\ldots,p$.

\subsubsection{Relation to resultants and discriminants}

The touching point (map $f \in \mu$) belongs to the resultant variety
${\cal R}^*_{n,n'}$, i.e. is a root of $R(G_n(f),G_{n'}(f)) = 0$,
with $n = km_1\ldots m_p$ and $n' = km_1\ldots m_pm'_{p+1} =
nm'_{p+1}$. The boundaries of all elementary domains $\sigma^{(p)}$
with $p>0$ are smooth
(unless $\mu$ crosses a singularity of ${\cal R}$),
while $\partial\sigma^{(0)}[k\alpha]$ has a cusp,
located at discriminant variety ${\cal D}^*_k$, i.e. at $f \in \mu$,
satisfying $D(G_k(f))=0$ ($\sigma^{(0)}[k\alpha]$ itself has a
peculiar cardioid form). It is clear that the bigger the family $\mu$,
the more intersections it has with the resultant and discriminant
varieties, thus the bigger are the sizes of the sets $\nu^{(p)}(\mu)$.
For entire ${\cal M}$ the indices $\alpha_p$ get continuous and parameterize
the entire pullbacks of the resultant and discriminant varieties.

Every elementary domain $\sigma^{(p)}$ touches a single domain of the
lower level $p-1$ and infinitely many domains of the next level $p+1$
(which are labeled by all integer $m = m_{p+1}\geq 2$ and
$\alpha = \alpha_{p+1} \in
\nu^{(p+1)}_{m}(\mu)$). The touching points -- belonging to
zeroes of $R(G_n,G_{nm})$ with all possible $m = m_{p+1}$ are actually
dense in the boundary $\partial\sigma^{(p)}$, i.e. the boundary can
be considered as a closure of the sets of zeroes.

\subsubsection{Relation to stability domains}

As already mentioned, every $\ \sigma^{(p)}\left[\left.
\begin{array}{ccc}
m_1 & \ldots & m_p\\ \alpha_1 & \ldots & \alpha_p
\end{array}\right| k\alpha\right]\ $
is characterized by an integer $n = km_1\ldots m_p$.
Sometime, when other parameters are inessential, we even
denote it by $\sigma_n^{(p)}$.
Stability domain $S_n(\mu)$, defined by the conditions (\ref{stabcon})
is a union of all $\sigma(\mu)$ with the same $n$.
This implies, that $S_n$ is actually a sum over all
branches of {\it divisor tree} $t[n]$, introduced in s.\ref{coo}.
Then $p$ is the length of the branch at $k$ is its end-link.
Other links carry numbers $m_1,\ldots,m_p$.
The only new
ingredient is addition of extra parameters $\alpha$ at every step.
This means that we actually need $\alpha$-decorated divisor trees
(for every family $\mu$), which we denote $\tilde t[n](\mu)$
and imply that links at $p$-th
level carry pairs $m_p,\alpha_p$ with $\alpha_p \in \nu^{(p)}_{m_p}(\mu)$,
and sum over decorated trees imply summation over $\alpha$'s.
In other words, the elementary domains $\sigma$ are actually
labeled by branches of the decorated trees, $B\tilde t$, and
\be
S_n = \bigcup_{B\tilde t[n]}\sigma_n[B\tilde t] = \nn \\ =
\bigcup_{p=0}^\infty\ \bigcup_{\stackrel {k|n}
{\alpha\in \nu_k}}
\left(
\bigcup_{\stackrel{m_1,\ldots,m_p:\
n = km_1\ldots m_p}{\alpha_1\in\nu^{(1)}_{m_1},\ldots,
\alpha_p\in\nu^{(p)}_{m_p}}}
\sigma^{(p)}\left[\left.
\begin{array}{ccc}
m_1 & \ldots & m_p\\ \alpha_1 & \ldots & \alpha_p
\end{array}\right| k\alpha\right]\right)
\label{Sndec}
\ee
It remains an interesting problem to prove {\it directly} that
eqs.(\ref{stabcon}) imply decomposition (\ref{Sndec}).

\subsubsection{Critical points and locations of elementary
domains}

Let $w_f$ denote critical points of the map $f(x)$:
$$f'(w_f) = 0.$$
Then $(f^{\circ n})'(w_f) = 0$, i.e. the critical points always
belong to some {\it stable} orbit, see (\ref{stabcon}). This orbit
is periodic of order $n$ provided
\be
G_n(w_f;f) = 0.
\label{criporbit}
\ee
This is an equation on the map $f$ and its
solutions  define points $f$ in the family $\mu \subset
{\cal M}$, which lie inside stability domain $S_n$, and, actually,
inside certain elementary domains
$\ \sigma^{(p)}\left[\left.
\begin{array}{ccc}
m_1 & \ldots & m_p\\ \alpha_1 & \ldots & \alpha_p
\end{array}\right| k\alpha\right]\ $.
In fact this is a one-to-one correspondence:
solutions to eq.(\ref{criporbit}) enumerate all the
elementary domains.

This implies the following pure algebraic description of
the powerful forest of $M(\mu)$ (formed by the
coverings $T_{k\alpha}$ of the multipliers trees $T_k$).
All vertices of the graph are labeled by solutions of
(\ref{criporbit}) and accordingly carry indices $n$.
Links are labeled by roots of the resultants $R(G_m(f),G_n(f))$,
with appropriate $m$ and $n$ (standing at the ends of the link).
$\alpha$-parameters serve to enumerate different solutions of
(\ref{criporbit}) and different roots of the same resultants.

Critical points $w_f$ also play a special role in the study of
bifurcations of grand orbits and Julia sets, see s.\ref{graob}.

\subsubsection{Perturbation theory and
approximate self-similarity of Mandelbrot set \label{summsim}}

For concrete families $\mu \subset {\cal M}$ a kind of approximated
perturbation theory can be developed in the vicinity of solutions
$f = f_0$ to (\ref{criporbit}).
Namely, one can expand equations (\ref{stabcon}), which define the
shape of stability domain (and thus of the elementary domains
$\sigma[k\alpha]$),
in the small vicinity of the point
$(x,f) = (w_{f_0},f_0)$ and, assuming that the deviation is small,
substitute original exact equations by their approximation for
small deviations. Approximate equations have
a {\it universal} form, depending only on the symmetry of the problem.
This explains why in many examples (where this method is accurate
enough) all the components $M_{k\alpha}$ of Mandelbrot set {\it look}
approximately the same, i.e. why the Mandelbrot set is approximately
self-similar (see Fig.\ref{0Man4},b) and help to classify the
components of stability domain. The same method can be used to investigate
the {\it shape} (not just structure) of Julia sets.
See in sec.\ref{xdExa} below
an example of application of
this procedure to the map families $f_c = x^d + c$.

\subsubsection{Trails in the forest}

The last level of hierarchy in the structure of Mandelbrot set is
represented by {\bf trails}, see Fig.\ref{0Man4},b.
Despite the components $M_{k\alpha}$ do not intersect (do not have
common points), they are linked by a tree-like system of trails:
each $M_{k\alpha}$ is connected to $M_1$ by a single trail
$\tau_{k\alpha}$, which is densely populated by some other components
$M_{l\beta}$. The trail structure is exhaustively described by triangle
{\bf embedding matrix},
$\tau = \{\tau[k\alpha,l\beta]\}$ with unit entry
when one trail is inside another, $\tau_{l\beta} \subset \tau_{k\alpha}$,
and zero entry otherwise.
If locations of all $M_{k\alpha}$ in parameter-$\vec c$ space
are known (for example, evaluated with the help of the perturbation
theory from s.\ref{summsim}), then embedding matrix fully describes
the trails. It is unclear whether any equations in parameter space
$\mu$ can be written, which define the shape of trails.

Embedding matrix seems to be universal, i.e. does not depend on the choice
of the family $\mu$. If true, this means that the trail structure is indeed
a pertinent characteristic of Mandelbrot/resultant variety, not
of its section by $\mu$. However, as usual, the universal structure is
partly hidden because of the presence of non-universal degeneracies,
labeled by $\alpha$-parameters
(at our level of consideration, ignoring higher resultants): concrete
trail systems $\tau(\mu)$ are {\it coverings} of presumably-\-universal
system.

Trail structure requires further investigation and does not get much
attention in the present paper.

\subsection{Sheaf of Julia sets over moduli space}

The structure of Julia set $J(f)$ is also hierarchical and it depends on
the position of the map $f$ in Mandelbrot set, especially on its
location in  Mandelbrot forest, i.e.
on the set of parameters $\{k\alpha,p, m_i\alpha_i\}$, labeling the
elementary domain $\ \sigma^{(p)}\left[\left.
\begin{array}{ccc}
m_1 & \ldots & m_p\\ \alpha_1 & \ldots & \alpha_p
\end{array}\right| k\alpha\right]\ $ which contains $f$.
Since in this paper we do not
classify $\alpha$-dependencies, only the numbers
$k$ (labeling connected component),  $p$ (labeling the level)
and $m_1,\ldots,m_p$ will be interpreted in terms of $J(f)$ structure.

Julia set is a continuous deformation of a set of discs (balls),
which -- the set -- depends on $k$; increase of $p$ by every unit
causes gluing of infinitely many points at the boundary,
at every point exactly $m_p$ points are glued together.
Every elementary domain is associated with a set of stable orbits
(often there is just one), contained
{\it inside} the Julia set together with their grand orbits.
All other (i.e. unstable) periodic orbits and their grand orbits
belong to the boundary of Julia set and almost each particular grand
orbit fills this boundary densely.
At the touching point between two adjacent
elementary domains the stable orbit approaches the boundary $\partial J$
from inside $J$ (perhaps, it is better to say that the boundary
deforms and some of its points -- by groups -- approach the orbit
which lies  {\it inside} $J$),
intersect some unstable orbit and "exchange stability" with it.
Their grand orbits also intersect.
The ratio of orders of these two orbits is $m_p$ and every $m_p$
points of unstable {\it grand} orbit merge with every point of the
stable one, thus strapping the disc at infinitely many
places (at all points of the merging grand orbits) and pushing its
sectors into bubble-like shoots, see Fig.\ref{JuliaMix4}.
As $f$ transfers to the elementary domain at level $p+1$, the
new stable orbit (the one of bigger order)
quits the boundary and immerses into Julia set,
while the orbit of smaller order, which is now unstable, remains
at its boundary: the singularities, created when the grand orbits crossed,
can not disappear.

One can say that every link of the forest describes a phase transition
of Julia set between neighboring elementary domains. An order parameter
of this transition is the distance between approaching stable and unstable
orbits on one side, looking as the contact length between emerging components
of Julia set, and the angle between these components
on the other side, when they are already separated except for
a single common point. Both quantities
vanish at transition point (i.e. when
$f$ is at the touching point between two domains in the Mandelbrot
set). If we move around the transition point in ${\cal M}$, there
is a non-trivial monodromy (the Julia set gets twisted), but on the
way one should obligatory pass through the complement of Mandelbrot
set, where the discs -- which form Julia set -- disappear, only the
boundary remains, Julia set has no "body", only boundary, and exact
definition of monodromy is hard to give.

Julia sets can change structure not only when stable orbit intersects with
unstable one -- what happens when the corresponding
elementary domains $\sigma$ of Mandelbrot
set touch each other, -- but also when unstable grand orbits at the boundary
$\partial J(f)$
cross or degenerate. Such events are not reflected in
the Mandelbrot set itself: a bigger {\bf Grand Mandelbrot set}, which
contains all zeroes of $D(F_{n,k})$ and $R(F_{n,k},F_{m,l})$ on its
boundary, should be introduced to capture all the bifurcations of Julia
set.

See s.\ref{Juexa} for some more details about Julia sets
and their bifurcations.

\FIGEPS{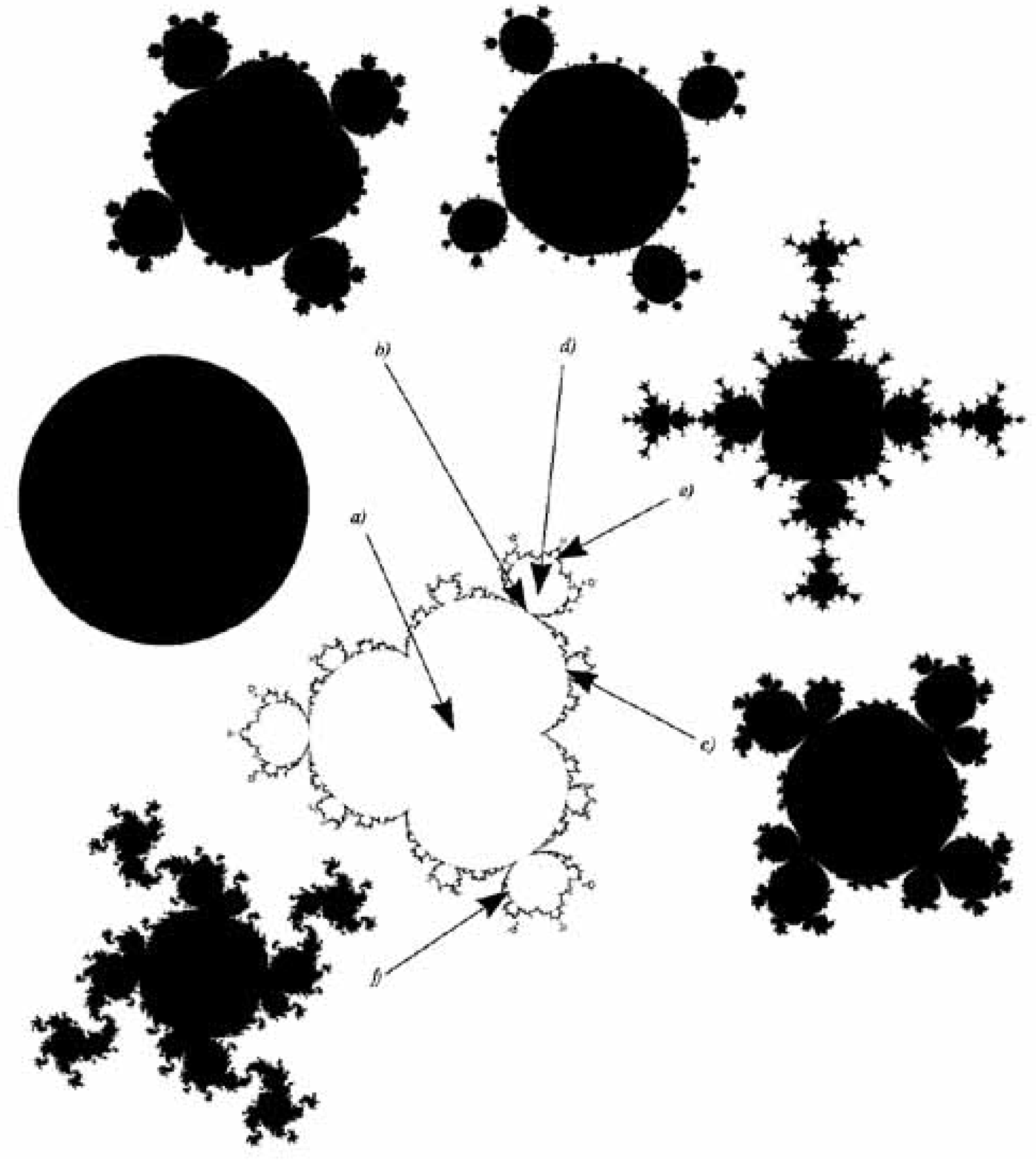}{}{450,497}
{The typical forms of the Julia set
for the map $f_c = x^4 + c$ and $c \in M_1$.
a) $c$ in the center of $\sigma^{(0)}_1$ ($c = 0$). $Z_4$
symmetry of the picture is due to the symmetry of the map
$f_c$. Associated Mandelbrot set has symmetry $Z_3$, see
Fig.\ref{0Man4},a.
b) $c$ at the touching point $\sigma^{(0)}_1\cup \sigma^{(1)}_2$
($c =\frac{5}{16}2^{1/3}(1+i\sqrt{3})$).
Infinitely many {\it pairs} of points on the unit
circle (a) are identified to provide this pattern.
c) $c$ at another touching point $\sigma^{(0)}_1\cup \sigma^{(1)}_3$
($c =\frac{1}{8}(9+i\,111\sqrt{3})^{1/3}$).
Infinitely many {\it triples} of points on the unit
circle (a) are identified to provide this pattern.
d) $c$ inside $\sigma^{(1)}_2$ ($c =0.45+i\,0.8$). This is a further
deformation of (a) in the direction of (b). Once appeared in (b)
the merging points at the boundary and straps of the disc caused
by this merging of boundary points do not disappear, but angles
between the external bubbles increase from the zero value which
they have in (b). They will turn into $2\pi/2$ and bridges
become needles when $c$ reaches the boundary of $\sigma^{(1)}_2$,
say an intersection point $\sigma^{(1)}_2\cup \sigma^{(2)}_4$ (e)
or $\sigma^{(1)}_2\cup \sigma^{(2)}_6$ (f).
e) $c$ at the touching point $\sigma^{(1)}_2\cup \sigma^{(2)}_4$
($c\approx 0.56+0.97i$).
f) $c$ at the touching point $\sigma^{(1)}_2\cup \sigma^{(2)}_6$
($c\approx 0.309-0.862i$). A mixture of merging {\it pairs} and
{\it triples} is clearly seen. The bridges for pairs became needles,
while triplets are touching. Angles between triplets will increase
as $c$ goes inside $\sigma^{(2)}_6$ and reach $2\pi/3$ at its
boundary, and so on.
Arrows point the positions of Julia sets shown in
Fig.\ref{JuliaMix4} in the Mandelbrot set. The "Julia sheaf"
is obtained by "hanging" the corresponding Julia set over each
point of the Mandelbrot set. At the boundary of Mandelbrot set
the Julia sets are reshuffled and the task of the theory is to
describe this entire variety (the sheaf) and all its properties,
both for the universal Mandelbrot set and multi-dimensional
Julia sets associated with multicomponent maps, and for the
particular sections, like the one-parametric family of
single-component quartic maps, $f(x;c) = x^4 + c$ shown in this
particular picture.}

\newpage

\section{Fragments of theory
\label{pot}}

\subsection{Orbits and reduction theory of iterated maps
\label{ortim}
}

In discrete dynamics trajectories are substituted by {\it orbits}
and closed trajectories -- by periodic orbits $O_n$
of the {\it finite} order $n$.
Different orbits have no common points,
and $f$ acts on each $O_n$ by cyclic permutation of points.
All points of every $O_n$ belong to the set
${\cal S}_n(f)$ of roots of the function
$$
F_n(x;f) = f^{\circ n}(x) - x.
$$
The map $f$ generates the action of cyclic group $\hat f(n)$
on ${\cal S}_n(f)$, and this set is decomposed into orbits
of $\hat f(n)$ of the orders $k$, which are divisors of $n$
(all $O_n$ with $k=n$ are among them).
If $n$ is divisible by $k$, then $F_n(x)$ is divisible by $F_k(x)$,
and in general $F_n$ decomposes into irreducible constituents:
\be
F_n(x;f) = \prod_{k|n}^{\tau(n)} G_k(x;f)
\label{rnk}
\ee
with a product over all possible $\tau(n)$
divisors $k$ of $n$ (including $k=1$ and $k=n$).
The number of periodic orbits of order $n$ is equal to
$N_n(f)/n$, where $N_n(f)$ is the number of roots of $G_n(x;f)$.
Irreducibility means that $G_k$ are not decomposable into
simpler constituents in an $\bb{X}$- and $f$-independent way
(of course, when $G_k(x)$ are polynomials they can be decomposed
into products of monomials over $\bb{C}$, but this decomposition
will not take place over $\bb{R}$).

Among the tasks of the theory is the study of
reducibility of the sets ${\cal S}_n(f)$, the ways
they decompose into orbits and the ways this decomposition
changes under the deformations of $f$ within $\mu \subset {\cal M}$.
Since the story is essentially about the roots of
functions, it gets much more transparent in the {\it complex}
setting than in the {\it real} one. Also, the entire theory is
naturally a generalization from the case of the polynomial
functions $f \in {\cal P} \subset {\cal M}$.

Obviously, for $f \in {\cal P}_d \subset {\cal P}$,
$F_n \in {\cal P}_{d^n}$, and $G_k(x)$ in (\ref{rnk})
is a polynomial of certain degree $N_k(d)$.
If $d = p$ is prime, then the group
$\hat f(n) = \bb{Z}_n$ and is isomorphic to the Galois group
over $F_p$ of the cyclotomic polynomial $x^{p^n}-x$
(while the Galois group of $f$ itself over $\bb{C}$ is trivial).
If instead $n=p$ is prime, then $d$ out of the $d^p$ roots of $F_p(x)$
are also the roots of $F_1(x)$, i.e. are invariant points (orbits of
order one) of $f$, while the remaining $d^p-d$ roots decompose into
$\frac{d^p-d}{p}$ orbits of order $n=p$.
According to (\ref{rnk}) the numbers $N_n(d)$ for
$f \in {\cal P}_d$ can be defined recursively: from
$$
\sum_{k|n}^{\tau(n)} N_k(d) = {\rm deg}\ F_n = d^n
$$
it follows, that
$$
N_n(d) = d^n - \sum_{\stackrel{k|n}{k<n}} N_k(d).
$$
The lowest numbers $N_n(d)$ are:
\be
N_1(d) = d, \nn \\
N_2(d) = d^2-d = d(d-1), \nn \\
N_3(d) = d^3-d = (d-1)d(d+1), \nn \\
N_4(d) = d^4 - d^2 = d^2(d^2-1) = (d-1)d^2(d+1), \nn \\
N_5(d) = d^5 - d, \nn \\
N_6(d) = (d-1)d(d+1)(d^3+d-1), \nn \\
N_7(d) = d^7-d, \nn \\
N_8(d) = d^8-d^4 = d^4(d^4-1), \nn \\
N_9(d) = d^9 - d^3 = d^3(d^6-1), \nn \\
N_{10}(d) = d(d^4-1)(d^5+d-1), \nn \\
\ldots \nn
\ee
For $n$ prime, $n=p$, $N_p(d) = d^p-d = d(d^{p-1}-1)$,
and small Fermat theorem guarantees that $N_p(d)$ is
divisible by $p$. Further,
\be
N_{p^k}(d) = d^{p^k} -
d^{p^{k-1}} = d^{p^{k-1}}\left(d^{(p-1)p^{k-1}} -1\right),
\nn \\
N_{p_1p_2}(d) = d^{p_1p_2} - d^{p_1} - d^{p_2} + d, \nn
\ee
{\rm in\ particular},
\be
N_{2p}(d) = (d^{p} - d)(d^p+d-1), \nn \\
N_{3p}(d) = (d^{p} - d)(d^{2p}+d^{p+1} + d^2 -1). \nn
\ee

\subsection{Bifurcations and discriminants:
from real to complex}

Usually considerable part of discrete dynamics concentrates
on the {\it real} part of above story, i.e. deals with
the functions $f$, which map the real line
$\bb{R} \subset \bb{C}$ into itself. Such functions form a
subspace ${\cal M}^{(\bb{R})}$ in ${\cal M}$.
The subset ${\cal P}^{(\bb{R})} \subset {\cal M}^{(\bb{R})}$ consists
of all polynomials with real coefficients.
In variance with the complex situation, different
polynomials $F_n$ in the same ${\cal P}_{n^d}^{(\bb{R})}$ can have
different numbers of {\it real} roots, thus even the
size of the set ${\cal S}_n^{(R)}(f)$ can change when $f$ is
varying inside ${\cal M}^{(R)}$. In fact, the orbits of
$f \in {\cal M}^{(R)}$ are either entirely real,
$O \subset \bb{R}$ or entirely  complex,
$O \subset \bb{C}-\bb{R}$. In the latter case the orbit is
either self-conjugate or there is a complex conjugate orbit
$\bar O \neq O$. The number of real roots of $F_n(x)$,
i.e. the size of the set ${\cal S}_n^{(\bb{R})}(f)$, can change
whenever a self-conjugate orbit or a pair of conjugate orbits,
that was complex, becomes real or vice versa, a real
orbit turns into a complex self-adjoint or a pair of conjugate
orbits. This is generalization of the well known
phenomenon, when a pair of complex-conjugate {\it roots} of
a polynomial from ${\cal P}^{(\bb{R})}$ becomes real, or
vice versa, when a pair of real roots collide and go away
into the complex domain -- just in the case of peculiar
polynomials $F_n(x)$ from ${\cal M}_{\circn}$
this happens at once with entire orbits,
not just with {\it pairs} of roots.
In the case of polynomial's roots, the points in
${\cal P}^{(\bb{R})}$, where some roots can migrate between the
real line and complex domain, belong to {\it discriminant}
varieties ${\cal D}_d \subset {\cal P}_d$, consisting
of polynomials with at least a pair of coincident roots
(i.e. such, that $P(x)$ and $P'(x)$ have at least one common
root).\footnote{The simplest example of non-trivial
discriminant variety is ${\cal D}_2$ -- the quadric
$b^2-4ac=0$ in the space ${\cal P}_2$ of quadratic polynomials
$ax^2+bx+c$. In accordance with (\ref{discf}) below
$$
b^2-4ac = -\frac{1}{a}
\det
\left(
\begin{array}{ccc}
a&b&c\\ 2a&b&0 \\ 0&2a&b
\end{array}
\right)
$$
If quadratic polynomial is considered as a quadric in $\bb{CP}^1$,
$Q(x,y) = ax^2 + bxy + cy^2$, discriminant is a determinant of
$2\times 2$ matrix with two lines formed by the coefficients of
$\frac{\partial Q}{\partial x}$ and $\frac{\partial Q}{\partial y}$:
$$
b^2-4ac = -\det
\left(
\begin{array}{cc}
2a&b \\ b&2c
\end{array}
\right)
$$
}
The same is true for the orbits of $f$:
all kinds of reshuffling  of orbits
take place when $F_n \in {\cal D} \subset {\cal M}$.
Note, that in the last statement neither the discriminant varieties
nor the maps (and their orbits)  are restricted to
{\it polynomials} with {\it real} coefficients, everything is
equally well defined at least for entire
space of {\it complex} polynomials ${\cal P}$. Moreover, a
generalization should exist to entire space ${\cal M}$ of analytic
(not obligatory polynomial) functions, see s.\ref{analyt} below.

Coefficients $a_k^{(n)}$ of
\be
F_n(x) = f^{\circ n}(x) - x = \sum_{k=0}^\infty a^{(n)}_k x^k \nn
\ee
are polynomials of the coefficients $a_k$ of $f(x) =
\sum_{k=0}^\infty a_kx^k$,
thus we have an algebraic map (\ref{Imap})
between the spaces of coefficients
$\hat I_n:\ \{a\} \rightarrow \left\{a^{(n)}(a)\right\}$.
Then the space ${\cal M}$ of coefficients $\{a\}$ of $f$
is divided into connected
components by the $\hat I_n$-pullback of the discriminant variety
${\cal D}_{\circn}$.
Given a path $a(t)$ in the space ${\cal M}$, its image
$\hat I_n(a(t))$ can cross
${\cal D}_{\circn} = {\cal D}\cap {\cal M}_{\circn}$
at a generic (non-singular) point $\hat I_n(a(t_0))$.
$F_n(x;t)$ will have different number of {\it real} roots
before and after the crossing (i.e. for
$t'<t_0<t''$ in a neighborhood of $t_0$).
This change in the number of real roots is called {\bf bifurcation}
in the theory
of dynamical systems. Let the number of real roots $m'$ for
$t'<t_0$ be less then the number $m''$ for $t''>t_0$. If a new root
$x_i(t'')$ belongs to an $f$-orbit $O$, then
{\it all} the elements of $O$ belong to the $f$-invariant domain
$\bb{C}-\bb{R}$ for $t'<t_0$ and "simultaneously come" to the
$f$-invariant domain $\bb{R}$ at $t=t_0$.
Since elements of $O$ and its conjugate $\bar O$ come to the real
axis $\bb{R}$ simultaneously, one can say, that real roots "are born
in pairs" at $t=t_0$.

\subsection{Discriminants and resultants for iterated maps
\label{redi}}

General comments on the definitions of discriminants and resultants
(including the non-polynomials case) are collected in s.\ref{disc}
below. However, for iterated maps these quantities are highly reducible.

Discriminant variety is defined by the equation
$$
D(f) = 0
$$
where, for polynomial $f$,
$D(f)$ is the polynomial of the coefficients $\{a\}$,
see eq.(\ref{discres}) below. Similarly, $D(F_n)$ is a polynomial
of the coefficients $\left\{a^{(n)}(a)\right\}$ and
the resultant $R(F_n,F_m)$ is a polynomial of coefficients
$\left\{a^{(n)}(a)\right\}$ and $\left\{a^{(m)}(a)\right\}$.

Since function $F_n(x)$ is reducible over any field (e.g. over $\bb{R}$
as well as over $\bb{C}$), see (\ref{rnk}), the resultant factorization
rule (\ref{resfact}) implies:
\be
D(F_n) = \prod_{k|n} D(G_k) \prod_{\stackrel{k,l|n}{k>l}}
R^2(G_k,G_l).
\label{DDDR2}
\ee
However, this is only the beginning of the story.
Despite $G_k(x)$ are irreducible constituents
of $F_n(x)$, the resultants $R(G_k,G_{k/m})$ and
discriminants $D(G_k)$ are still reducible:
\be
R(G_k,G_{l}) = r^l(G_k,G_{l}),\ \ l<k, \nn \\
D(G_k) = d^k(G_k) \prod_{m>1}^n R^{m-1}(G_k,G_{k/m})
 = d^k(G_k) \prod_{l|k} r^{k-l}(G_k,G_l), \nn \\
{\rm thus}\ \ \ D(F_n) = \prod_{k|n} \left(d^k(G_k)
\prod_{l|k} r^{k+l}(G_k,G_l)\right).
\label{Dd}
\ee
Indeed, whenever a resultant vanishes,
a root of some $G_{k/m}$ coincides with that of $G_k$ and then
-- since roots of $G_k$ form orbits of order $k$ --
the $k$ roots should
merge by groups of $m$ into $l = k/m$ roots of $G_{k/m}$,
and there are
exactly $l$ such groups. In other words,
whenever a group of $m$ roots of $G_k$
merges with a root of $G_{k/m}$, so do $l = k/m$ other groups,
and $R(G_k,G_{k/m})$ is an $l$-th power of an irreducible quantity,
named $r(G_k,G_{k/m})$ is (\ref{Dd}).

In a little more detail, if
$\alpha_1,\ldots,\alpha_k$ with $k=lm$
are roots of $G_k$ and $\beta_1,\ldots, \beta_l$
are those of $G_{k/m}$, then $\alpha_1 = \beta_1$ implies, say, that
$$
\alpha_1=\ldots=\alpha_m = \beta_1,$$ $$
\alpha_{m+1}=\ldots=\alpha_{2m} = \beta_2,$$ $$ \ldots $$ $$
\alpha_{m(l-1)+1} = \ldots \alpha_{lm} = \beta_l.
$$
Then in the vicinity of this point
$$
R(G_k,G_l) \sim \prod_{i,s}(\alpha_i - \beta_s) \sim
\prod_{s=1}^l \left(\prod_{i = (s-1)m+1}^{sm} (\alpha_i - \beta_s)\right)
$$
(factors, which do not vanish, are omitted).
Internal products are polynomials of the coefficients of $f$ with
{\it first}-order zeroes (the roots themselves are not polynomial in these
coefficients!), and only external product enters our calculus
and provides power $l$ in the first line of eq.(\ref{Dd}).

Similarly,
$$
D(G_k) \sim \prod_{i < j} (\alpha_i-\alpha_j)^2 \sim
$$ $$ \sim
\prod_{r,s=1}^l \left(\prod_{\stackrel{i,j=(s-1)m+1}{i<j}}^{sm}
(\alpha_i-\alpha_j)^2\right)
= \prod_{s=1}^l \left(\prod_{i = (s-1)m+1}^{sm}
(\alpha_i - \beta_s)\right)^{m-1}
$$
what implies that zero of $D(G_k)$ is of order $l(m-1) = k-m$.

The remaining constituent $d(G_k)$ of discriminant $D(G_k)$ describes
intersections among order-$k$ orbits. In this case, whenever two points
of two orbits coincide, so do -- pairwise -- the $k-1$ other points.
Thus the corresponding zero is of order $k$, and irreducible quantity
$d(G_k)$ enters $D(G_k)$ in $k$-th power.

For examination of examples in ss.\ref{Exa} and \ref{Exa2} it is
convenient to have explicit versions of (\ref{Dd}) for a few lowest $n$
and $k$. It is also convenient to use condensed notation:
$$
d_k = d(G_k), \ \ \ \ r_{kl} = r(G_k,G_l),
$$
$r_{kl} = 1$ unless $l$ is divisor of $k$, $l|k$ (or vice versa,
$k$ is divisor of $l$). Then
\be
\begin{array}{ccccc}
k & &  D(F_k)   &&  D(G_k)  \\
&\ \ &&\ & \\
1 && d_1 && d_1 \\
2 && d_1d_2^2r_{21}^3 && d_2^2r_{21}   \\
3 && d_1d_3^3 r_{31}^4 && d_3^3r_{31}^2 \\
4 && d_1d_2^2d_4^4 r_{41}^5 r_{42}^6 r_{21}^3 &&
            d_4^4 r_{41}^3 r_{42}^2 \\
5 && d_1d_5^5 r_{51}^6 && d_5^5 r_{51}^4 \\
6 && d_1d_2^2d_3^3d_6^6 r_{61}^7r_{62}^8r_{63}^9 r_{21}^3r_{31}^4 &&
            d_6^6 r_{61}^5r_{62}^4r_{63}^3 \\
&& \ldots &&
\end{array} \nn
\ee

\bigskip

Decomposition of discriminant imply that, as $f$ is varied,
new roots can emerge in different ways, when different components
of ${\cal D}^*$ are crossed. If the new roots are born when the
component of $R(G_n,G_{k})$ is crossed, where the
orbits of orders $n$ and $k$ intersect, then they appear at
positions of the previously existing roots of $G_k$.
If this happens at real line, than the phenomenon is
known as {\bf period doubling} (it is {\it doubling},
since when more than two new
roots occur at the place of one, they are necessarily complex).

\subsection{Period-doubling and beyond}

The simplest example of orbit reshuffling with the change of
the map $f$ is the period-doubling bifurcation
\cite{dpb}, which can be described as follows.
Let $x_0$ be an invariant point of $f$, i.e. $F_1(x_0)=f(x_0)-x_0 = 0$.
Let us see what happens to an infinitesimally close point
$x_0 + \epsilon$.
\be
f(x_0 + \epsilon) = x_0 + f'(x_0)\epsilon + \ldots, \nn \\
f^{\circ 2}(x_0 + \epsilon) = x_0 + \left[f'(x_0)\right]^2\epsilon +
\ldots, \nn \\
\ldots \nn
\ee
We see that the necessary condition for $x_0 + \epsilon$
with infinitesimally small but non-vanishing $\epsilon$
to be a root of $F_1$ or $F_2$
is $f'(x_0) = 1$ or $f'(x_0) = \pm 1$ respectively (i.e. $F_1$ or
$F_2$ should be degenerate at the point $x_0$).
The period-doubling bifurcation
corresponds to the case $f'(x_0) = -1$, i.e. the map $f$ is such that
$F_2(x)$, but not $F_1(x)$, becomes degenerate, then in the vicinity
of the corresponding stable point (the common zero $x_0$ of $F_2$
and $F_2'$) a new orbit of order $2$ can emerge.
If, more generally, $x_0$ is a root of some other $F_n$,
$F_n(x_0) = 0$, i.e.
describes some (perhaps, reducible, if $n$ is not a simple number)
$f$-orbit of order (period) n, then the same
reasoning can be repeated for $f^{\circ n}$ instead of $f$:
$$
F_n(x_0 + \epsilon) = F'_n(x_0)\epsilon + \ldots
$$
and
$$
F_{2n}(x_0 + \epsilon) = F'_{2n}(x_0)\epsilon + \ldots
$$
Derivative $F_{2n}'(x_0) = F_n'(x_0)( F_n'(x_0) + 2)$.
Period-doubling bifurcation occurs for $f$ with the property that
$(f^{\circ n})'(x_0) = -1$ or $F_n'(x_0) = -2$.

\bigskip

The period-doubling bifurcation, though very important, is not
the only one possible: new orbits can emerge in other ways as well.

First of all, {\it doubling} is relevant only in the case of maps
with real coefficients. In fully complex situation one can encounter
the situations when a higher power $f'(x_0)^k = 1$, $k>2$ (and all
lower powers of $f'(x_0) \neq 1$): then we have the bifurcation when
the period increases by a factor of $k$ and the new orbit
emerges in the vicinity of original one (which loses stability, but
survives).

Second, the new orbits can emerge "sporadically" at "empty places",
with no relation to the previously existing orbits and no obvious
criterium to warn about their appearance. The only reason for them
to occur is the crossing between discriminant variety ${\cal D}$
and moduli spaces ${\cal M}_{\circn}$, ${\cal M}_n$ of iterated maps
or their irreducible constituents.

\subsection{Stability and Mandelbrot set}

Above consideration of period-doubling bifurcation
implies introduction of the notion of "stable orbits"
in the following way:
the point $x$ of the orbit of order $n$, i.e. satisfying
$G_n(x;f)=0$, is called "stable" if
$$
|(f^{\circ n})'(x)|\le 1
$$
and "unstable" otherwise.
Since
\be
(f^{\circ n})'(x)=\prod_{k=1}^n f'\left(f^{\circ k} (x)\right) =
\prod_{{\rm all}\ z\ \in\ {\rm orbit}} f'(z)
\label{derivFn}
\ee
all points of the orbit are simultaneously either stable or unstable.

The {\bf Julia set} $J(f) \subset \bb{X}$ is attraction domain of
stable periodic orbits
in $\bb{X}$. Its boundary $\partial J(f)$ consists of all unstable
periodic orbits
and their grand orbits.

Excluding $x$ from the pair of stability conditions
\be
\left\{\begin{array}{cc}
G_n(x;f) = 0,  \\
|F'_n(x;f) + 1| < 1,
\end{array}\right. \nn
\ee
we obtain a stability domain $S_n$ of the order-$n$ orbits
in moduli space ${\cal M}$.
All zeroes of reduced discriminants $d(G_n)$ and
resultants $r(G_n,G_{mn})$
with arbitrary $m\geq 1$ lie at the boundary of $S_n$.
This property is used in our description of
{\bf Mandelbrot set} in s.\ref{summ}.

Mandelbrot set describes exchanges
of stability between pairs of orbits:
at the boundary of Mandelbrot set stable
orbits intersect with unstable ones,
stable become unstable, while unstable
become stable. Since stable orbits lie
{\it inside} Julia set, and unstable
ones -- on its boundary, this causes
reshuffling between interior and the
boundary of $J(f)$. Since all this actually
happens with entire {\it grand}
orbits, reshuffling involves infinitely many
points and looks like a
fractalization of the boundary.
In fact, this does not exhaust all possible
bifurcations of Julia set
$J(f)$: they can be also caused by crossings and degenerations of
unstable orbits (with no reference to the stable ones).
In order to study
bifurcations of Julia sets one should include into consideration
the {\it pre-images} $O_{n,s}$ of periodic orbits
of $f$, associated with the roots of the functions
\be
F_{n,s}(x) :\ = f^{\circ (n+s)}(x) - f^{\circ s}(x) =
F_n\left(f^{\circ s}(x)\right)
\ee
and study their reducibility properties.
Examples in secs.\ref{Exa} and \ref{Exa2}
below demonstrate that consideration of orbit's pre-images,
discriminants $D(F_{n,s})$ and resultants
$R(F_{n,s},F_{k,r})$ can indeed capture the bifurcations of $J(f)$,
which are overlooked by consideration of orbits alone.
Presumably, it provides {\it complete} theory of Julia sets in terms
Grand Mandelbrot set, which characterizes reshufflings of all orbits,
stable and unstable, and is fully algebraic, does not refer
to additional stability structure.

\subsection{Towards the theory of Julia sets
\label{graob}}

\subsubsection{Grand orbits and algebraic Julia sets}

For a given (algebraically closed) field $\bb{X}$ each series
$f=\sum_{n=0}^d a_nx^n,\ a_n\in\bb{X}$,
defines a map $\bb{X}\stackrel{f}{\to}\bb{X}$ and
thus a (pre-)order $f(x)\succ x$ on $\bb{X}$. Then the set of points of
$\bb{X}$ splits into connected components with respect to $\succ$ ("grand
orbits"):
we say that $x_1,x_2$ belong to the same {\bf grand orbit}, iff $x_1\succ
x,x_2\succ x$ for some $x\in\bb{X}$, i.e. $f^{\circ n}(x_1)=f^{\circ
m}(x_2)$ for some $n$ and $m$. In particular, for $n=0, m=1$ the set of
points $x_2$ satisfying $x_1=f(x_2)$ is the $f$-preimage of a given $x_1$.
So a grand orbit for generic point $x\in\bb{X}$
can be represented by an oriented tree with the
valency of each given vertex $x$ equal to $d+1$, where $d$ is the number of
roots of the equation $f(z)=x$. For a given $x$ the set $x^+$ of points
$x'\succ x$ will be called the {\bf orbit} of $x$.

However, if $x \in \bb{X}$ is a root of some $F_n(x;f)$, then the
{\it orbit} becomes a closed loop of finite length (order) $n$, and the
grand orbit $GO$ is a $d+1$-valent graph (see Fig.), obtained by gluing
vertices at a distance $n$ on a certain totally ordered chain of the general
position tree. For a pair of elements $x_\alpha,x_\beta$ on a periodic
orbit of order $n$ we have
$O_n =x^+_\alpha=x^+_\beta$, moreover both $x_\alpha\succ x_\beta$ and
$x_\beta\succ x_\alpha$ is true, so for the points of $O_n$
the relation $\succ$ is just a pre-order. But for each $x\in O_n$
there is a subset $x^-\subset GO$,
called the {\bf pre-orbit} of $x$, being a well-ordered tree
rooted at $x$. The
set of points $z\in x^-$ for which $f^{\circ s}(z)=x$ will be denoted
$x^{-s}$ (not to be confused with a negative power!),
so that $x^-=\cup_s x^{-s}$. Each set $x^{-s}$ belongs to the roots of
$$
F_{n,s}(x)= F_{n+s}(x)-F_s(x)=f^{\circ (n+s)}(x)-f^{\circ s}(x).
$$

Then for each $n$ we have a finite set of grand orbits with $n$-periodic
"bases" $O_n$. Then we may characterize the initial map $f$ by the structure
of this discrete set of data.

In particular, taking $z\in x^{-}$ we can regard $z^{-}\subset x^{-}$ as a
sequence of maps we can take its inverse limit, i.e. the set of sequences
$\{z_i\}$ where $z_0=z, z_i=f(z_{i+1})$ (or $z_i\in x^{-i}$). For
different $z',z''$ (which may in general belong to different grand orbits as
well) their preorbits $(z')^{-},(z'')^-$ are not only isomorphic as ordered
sets, but have isomorphic inverse limits. In particular, the set of limit
points for $\lim\limits_\leftarrow(z')^-$ and
$\lim\limits_\leftarrow(z'')^-$ in $\bb{F}$ coincide.
We call them {\it the algebraic Julia set} of $f$:
$$
J_A(f) = \lim\limits_\leftarrow (x)^-\ \ \ \forall x\in \bb{X}.
$$

\bigskip

{\bf Hypothesis:} $J_A(f)$ coincides with $J(f)$.

\subsubsection{From algebraic to ordinary Julia set}

In order to explain our expectation that algebraic Julia sets are
related to conventional ones, we formulate two more hypotheses.

A map $f$ defines in $\bb{X}$ two important subsets:
the unions ${\cal O}_+(f)$ and ${\cal O}_-(f)$ of all stable
and unstable periodic orbits of all orders -- both are countable
sets of points in $\bb{X}$.

\bigskip

{\bf Hypothesis:} For almost all $f \in {\cal M}$
the $f$-orbit of almost any point $x \in \bb{X}$
approaches ${\cal O}_+(f)$. Moreover, it approaches exactly one concrete
stable orbit of particular order $n(x)$, which is a characteristic
of the point $x$.

\bigskip

The pre-orbit tree of almost any point $x \in \bb{X}$ has infinitely many
branches. Going backwards along particular branch we approach its
inverse limit, or {\it origin}.

\bigskip

{\bf Hypothesis:} For almost all  $f \in {\cal M}$ and $x \in \bb{X}$
the origins of almost all {\it periodic} branches of grand $f$-orbit of $x$
belong to  ${\cal O}_-(f)$. Moreover, different periodic branches usually
originate at different unstable orbits from ${\cal O}_-(f)$
and almost every orbit in ${\cal O}_-(f)$ is an origin of
some branches of $f$-pre-orbit of $x$.

\bigskip

In other words, almost each grand orbit originates
at the closure of {\it entire}
${\cal O}_-(f)$ and terminates at (tends to) a particular orbit
in ${\cal O}_+(f)$. If these hypotheses are true, one and the same set
${\cal O}_-(f)$ is an origin of almost all grand orbits (not obligatory
bounded). It is this set (or its closure, to be precise) that we
call the boundary of the algebraic Julia set,
$\partial J_A(f) = \overline{{\cal O}_-(f)}$.
Periodic branches and unstable periodic orbits play the same role as
periodic sequences (rational numbers) in the space of all sequences (real
numbers), see s.\ref{Juexa} for some more details.

{\it Bounded} grand orbits reach (not just tend to) a periodic
orbit, however, in this case it can belong to ${\cal O}_-(f)$, not
obligatory to ${\cal O}_+(f)$: bounded grand orbits are not in generic
position in what concerns the "future".
Still, they are not distinguished from the point of view of the "past"
and can be used to study $J_A(f)$.
Bounded grand orbits are convenient to deal with, because they
consist of roots of $F_{n,s}$ (with all $s$) and
can be studied in pure algebraic terms.

\subsubsection{Bifurcations of Julia set}

If we deform $f$ (change the coefficients
of the corresponding series) then
the set of grand orbits move in $\bb{F}$
and may undergo the following two
structural changes:

(i) merging of two distinct periodic orbits $O',O''$,
which happens as soon as
any pair of elements from these orbits merge,

(ii) merging of elements of the same pre-orbit $x^-$.

This merging of components of grand orbits
results into splitting of $J_A(f)$
into disjoint components.

Case (i) corresponds to merging of the roots of $F_n$ and
ii) corresponds to
merging of the roots of $F_{n,s}$. This means that $f$ hits the
discriminants ${\cal D}_{\circn} = {\cal D}(F_n)$ and ${\cal D}(F_{n,s})$
with $s\geq 1$ respectively, i.e. that some functions $F_n(x;f)$ or
$F_{n,s}(x;f)$ become degenerate.
If $x$ is a multiple root of $F_n$ then $F_n(x)=0$ and $F'_n(x)=(f^{\circ
n}(x)-x)'=0$ so $(f^{\circ n}(x))'=1$. Since $n$-periodic $x$ is a
stationary point for the map $f^{\circ n}$, then the last equation means
that the points of the merging orbits $O'_n$ and $O_n''$
change their stability type.
Thus in the case when we can speak about convergence in $\bb{X}$,
the language of "stability"\ is related (equivalent?)
to that of "discriminants", but we will stick to
the latter since it allows to handle also the case ii) and the case
of the arbitrary field $\bb{X}$ as well (which does not need to be
full and have continuous and differentiable functions defined).

In the case (ii) some point of the grand orbit has degenerate preimage
(two or more preimages coincide). A point $z_f\in \bb{X}$ has degenerate
$f$-preimage when discriminant
\be
D(f(x)-z_f) = 0.
\label{zeq}
\ee
This equation defines an important (multi-valued) map
$$
z:\ {\cal M} \longrightarrow \bb{X},
$$
associating a set of points $\{z_f\}$ -- solutions to (\ref{zeq}) --
with every map $f \in {\cal M}$. Bifurcations of the type (ii) occur
whenever some bounded grand orbit crosses this set, i.e. when
$$
F_{n,s}(z_f;f) = 0
$$
for some $n$ and $k$.

The fact, that all bifurcations of bounded grand orbits are either of the
type (i) or of the type (ii), implies that discriminants $D(F_{n,s})$ are
products of discriminants $D(F_n)$ and the new canonical functions
$F_{n,r}(z)$ on ${\cal M}$, with $r\leq s$.
Similarly, the resultants $R(F_{n,s},F_{k,r})$ are made from $R(F_n,F_k)$
and various $F_{m,t}(z)$ with $m|n,k$ and $t\leq s,r$.
Actual expressions are somewhat more involved, because the functions
$F_{n,s}(z)$ are highly reducible (see s.\ref{discago} below)
and their different components
enter differently into formulas for particular resultants.

For description of these reductions it is important that
the map $z_f$ is intimately related to the critical points
$w_f$ of $f$. Indeed, eq.(\ref{zeq}) implies that
$f(x) - z_f$ and $f'(x)$ have a common zero.
The zeroes of derivative $f'(x)$ are critical points $w_f$:
$$
f'(w_f) = 0,
$$
and therefore
$$
z_f = f(w_f),
$$
i.e. there is a one-to-one correspondence between the points $z_f$ and
$w_f$. Moreover, from the definition of $F_{n,s}$ it follows that
\be
F_{n,s}(z_f) = F_{n,s+1}(w_f).
\label{kz1f}
\ee

\subsection{On discriminant analysis for grand orbits
\label{discago}}

\subsubsection{Decomposition formula for $F_{n,s}(x;f)$}

Each zero of $F_{k,r}(x)$ satisfies
$f^{\circ(k+r)}(x) = f^{\circ r}(x)$. Applying
$f^{\circ k}$ to both sides of this equation, we obtain
$f^{\circ(2k+r)}(x) = f^{\circ(k+r)}(x) = f^{\circ r}(x)$.
Repeating this procedure several times we get
$f^{\circ(mk+r)}(x) = f^{\circ r}(x)$ and, finally, applying
$f^{\circ(s-r)}$ we  obtain
$f^{\circ(mk+s)}(x) = f^{\circ s}(x)$ for any $s\geq r$
and any $m$. This means that $F_{k,r}(x)$ is a divisor
of any $F_{n,s}$, provided $k|n$ and $s\geq r$.
Consequently, similarly to (\ref{rnk}) we have
the following decomposition of $F_{n,s}(x)$ into
irreducible (for generic field $\bb{X}$ and map $f$)
components:
\be
F_{n,s}(x;f) =
\prod_{k|n}^{\tau(n)} \prod_{r=0}^s G_{k,r}(x;f)
\label{rnkrat}
\ee
The irreducible function $G_{k,0}(x;f) = G_k(x;f)$ appeared already in
(\ref{rnk}).
In particular, (\ref{rnkrat}) states that
\be
F_{1,1} = G_1G_{1,1}, \nn\\
F_{1,2} = G_1G_{1,1}G_{1,2}, \nn \\
\ldots \nn \\
F_{1,s} = G_1G_{1,1}G_{1,2}\ldots G_{1,s},\nn\\
\ldots \nn
\ee

\be
F_{2,s} = (G_1 G_{1,1} G_{1,2}\ldots G_{1,s})
(G_2 G_{2,1} G_{2,2}\ldots G_{2,s}), \nn \\
F_{3,s} = (G_1 G_{1,1} G_{1,2}\ldots G_{1,s})
(G_3 G_{3,1} G_{3,2}\ldots G_{3,s}), \nn \\
F_{4,s} = (G_1 G_{1,1} G_{1,2}\ldots G_{1,s})
(G_2 G_{2,1} G_{2,2}\ldots G_{2,s})
(G_4 G_{4,1} G_{4,2}\ldots G_{4,s}), \nn \\
\ldots \nn
\ee

\subsubsection{Irreducible constituents of
discriminants and resultants}

A direct analogue of (\ref{DDDR2}) expresses
discriminants $D(F_{n,s})$ through $D(G_{k,r})$
and the resultants $R(G_{k,r},G_{k',r'})$.
A less trivial thing is expression of these
quantities through elementary constituents, which
are, as already predicted, the familiar from
eq.(\ref{Dd})
irreducible discriminants and resultants
$d_k = d(G_k)$ and $r_{nk} = r(G_n,G_k)$,
as well as the new quantities, which are
irreducible components $w_{k,r}(f)$ of
\be
W_{n,s}(f) = \prod_{w_f} G_{n,s}(w_f;f)
\label{Wdef}
\ee
-- the products of all values of $G_{r,s}(x;f)$
at all critical points $w_f$ of $f$,
$f'(w_f) = 0$.\footnote{
We hope that the use of the same letter $w$ for critical
points $w_f \in \bb{X}$ and irreducible components
$w_{k,r}(f) \in {\cal M}$ will not cause too much
confusion.
}
Note, that $G$, not $F$, enters the definition of
$W$ in (\ref{Wdef}), still after substitution of
peculiar values of $x = w_f$ this quantity {\it often}
gets further reducible, the first few reductions are:

\be
W_{n}(f) = w_n(f)\ \ ({\rm irreducible}),\nn \\ \nn \\
W_{1,s}(f) = w_1(f)w_{1,s}(f), \nn \\ \nn \\
\left\{ \begin{array}{l}
W_{2,2r}(f) = w_{2,2r}(f)\ \ ({\rm irreducible}),\\
W_{2,2r+1}(f) = w_2(f)w_{2,2r+1}(f),\\
\end{array}\right.\nn \\ \nn \\
\left\{ \begin{array}{l}
W_{3,3r}(f) = w_{3,3r}(f)\ \ ({\rm irreducible}),\\
W_{3,3r+1}(f) = w_3(f)w_{3,3r+1}(f), \\
W_{3,3r+2}(f) = w_{3,3r+2}(f)\ \ ({\rm irreducible}),\\
\end{array}\right.\nn \\ \nn \\
\left\{ \begin{array}{l}
W_{4,4r}(f) = w_{4,4r}(f)\ \ ({\rm irreducible}),\\
W_{4,4r+1}(f) = w_4(f)w_{4,4r+1}(f),  \\
W_{4,4r+2}(f) = w_{4,4r+2}(f)\ \ ({\rm irreducible}),\\
W_{4,4r+3}(f) = w_{4,4r+3}(f)\ \ ({\rm irreducible}),\\
\end{array}\right.\nn \\
\ldots \nn
\ee
(we actually checked most of these statements only for $n+s\leq 6$).
Presumably, in general
\be
\left\{ \begin{array}{l}
W_{n,s} = w_{n,s}\ \ {\rm for}\ s\neq 1\ {\rm mod}\ n
\ \ ({\rm irreducible}),\\
W_{n,nr+1} = w_nw_{n,nr+1}.
\end{array}\right.
\label{wsmalldef}
\ee

\subsubsection{Discriminant analysis at the level $(n,s) = (1,1)$:
basic example}

Let us begin the proof of (\ref{wsmalldef})
from the simplest case of $W_{1,1}$.
The key point is that
\be
G_{1,1}(x) = f'(x)\ {\rm mod}\ G_1(x).
\label{G11f}
\ee
Then, according to (\ref{resresu}), the residue
$R(G_{1,1},G_1)$ can be expressed as a product over
all critical points $w_f$, i.e. the roots of $f'(x)$:
\be
r_{1|1,1} = R(G_1,G_{1,1}) \sim R(G_1, f') \sim
\prod_w G_1(w) = W_1(f) = w_1(f).
\label{r111}
\ee
Eq.(\ref{r111}) is the first result of grand orbit
discriminant calculus. It relies upon the relation
(\ref{G11f}), actually, on the fact that $f'(x)$
is residual of $G_{1,1}(x)$ division by $G_1(x)$:
then, though $G_{1,1}(x)$ is not divisible by $G_1(x)$
as a function of $x$, $ W_1(f) = \prod_w G_1(w)$ divides
$W_{1,1}(f) = \prod_w G_{1,1}(w)$ as a functional of $f$.
In order to see that
\be
\frac{G_{1,1}(x) - f'(x)}{G_1(x)} =
\frac{\frac{f(f(x)) - f(x)}{f(x) - x} - f'(x)}{f(x) - x}
\label{G11G1}
\ee
is non-singular at all zeroes of
$G_1(x) = F_1(x) = f(x) - x$
it is enough to consider an infinitesimal variation
of such root, $x = x_1 + \chi$, $f(x_1) = x_1$,
and expand all functions at $x_1$ up to the second order
in $\chi$:
$$
G_1(x) = F_1(x) = f(x) - x = (f'(x_1) - 1)\chi + \frac{1}{2}
f''(x_1)\chi^2 + \ldots =
$$ $$
= (f'(x_1) - 1)\chi\left(
1 + \chi \frac{f''(x_1)}{2(f'(x_1) - 1)}
+ O(\chi^2)\right),
$$\\ $$
F_{1,1}(x) = f'(x_1)(f'(x_1)-1)\chi +
\frac{1}{2}f''(x_1)(f'(x_1)^2 + f'(x_1) -1)\chi^2
+ O(\chi^3) =
$$ $$
\ \ \ = (f'(x_1) - 1)\chi\left(
f'(x_1) +  \frac{f''(x_1)\chi}{2(f'(x_1) - 1)}
(f'(x_1)^2 + f'(x_1) -1)
+ O(\chi^2)\right),
$$\\ $$
G_{1,1}(x) = \frac{F_{1,1}(x)}{G_1(x)} =
$$ $$
\ \ = \left(
f'(x_1) +
\frac{f''(x_1)(f'(x_1)^2 + f'(x_1) -1)\chi}{2(f'(x_1) - 1)}
+ O(\chi^2)\right)
\left(1 -  \frac{f''(x_1)\chi}{2(f'(x_1) - 1)}
+ O(\chi^2)\right) =
$$ $$
\ \ = f'(x_1) + \frac{1}{2}\chi f''(x_1)(f'(x_1) + 1)
+ O(\chi^2),
$$
so that (\ref{G11G1}) becomes
$$
\frac{f'(x_1) + \frac{1}{2}\chi f''(x_1)(f'(x_1) + 1)
- f'(x_1) - \chi f''(x_1) + O(\chi^2)}
{(f'(x_1) - 1)\chi + O(\chi^2)}
= f''(x_1) + O(\chi)
$$
and is finite at $\chi = 0$. This proves
divisibility of $W_{1,1}(f)$ and allows to introduce its
irreducible constituent $w_{1,1}(f)$:
\be
W_1(f) = \prod_{w_f} G_1(w_f) = w_1(f), \nn \\
W_{1,1}(f) = \prod_{w_f} G_{1,1}(w_f) =
w_1(f) w_{1,1}(f). \nn
\ee
It enters expression for discriminant $D(G_{1,1})$:
\be
D(G_{1,1}) \sim d_1^{d-1} w_{1,1}
\label{DG11}
\ee
i.e. is actually the irreducible part of this discriminant:
\be
d_{1,1} \sim w_{1,1}.
\label{d11}
\ee
Power $d-1$ in which $d_1 = d(G_1)$ enters (\ref{DG11})
depends on the degree $d$ of the map $f(x)$ (i.e. $f(x)$ is
assumed to be a polynomial of degree $d$).

Eqs.(\ref{d11}) and (\ref{r111}),
\be
r_{1|1,1} \sim w_1, \nn \\
d_{1,1} \sim w_{1,1} \nn
\ee
together with
\be
R(G_{1,1},G_n) = 1,\ \ {\rm for}\ n>1 \nn
\ee
are the outcome of discriminant/resultant analysis
in the sector of $G_1$ and $G_{1,1}$, responsible for
fixed points and their first pre-images.

\subsubsection{Sector $(n,s) = (1,s)$}

For a fixed point $x_1 = f(x_1)$
we denote $f' = f'(x_1)$, $f'' = f''(x_1)$. Then
for $x = x_1 + \chi$ we have:
$$
F_n(x) = \chi({f'}^n - 1)\left(1 + \frac{1}{2}\chi f''
\frac{{f'}^{(n-1)}}{f'-1} + O(\chi^2)\right)
$$
and
$$
G_k(x) = g_k(f')\left(1 + \frac{1}{2}\chi f''h_k(f')
+ O(\chi^2)\right),
$$
where $g_k(\beta)$ are irreducible circular polynomials
(which will appear again in s.\ref{Exa2}) and $h_k(\beta)$
are more sophisticated (with the single exception of
$h_1$ they are also polynomials).
The first several polynomials are:
$$
\begin{array}{ccc}
g_1(\beta) = \beta - 1 &\ & h_1g_1(\beta) = 1 \\
g_2(\beta) = \beta + 1 &\ & h_2(\beta) = 1 \\
g_3(\beta) = \beta^2+\beta + 1 &\ & h_3(\beta) = \beta + 1 \\
g_4(\beta) = \beta^2 + 1 &\ & h_4(\beta) = \beta(\beta + 1) \\
g_5(\beta) = \beta^4 +\beta^3+\beta^2+\beta+1 &\ &
       h_5(\beta) = \beta^3+\beta^2+\beta+1 \\
g_6(\beta) = \beta^2-\beta + 1 &\ &
       h_6(\beta) = \beta^4 + \beta^3 + \beta^2 - 1 \\
&\ldots &
\end{array}
$$
Making use of $F_{n,s} = F_{n+s} - F_s$
and of decomposition formula (\ref{rnkrat}),
it is straightforward to deduce:
$$
G_{1,1}(x) = \frac{F_2-F_1}{G_1} =
 f' + \frac{1}{2}\chi f''(f'+1) + O(\chi^2),
$$ $$
\frac{G_{1,1}(x)-f'(x)}{G_1(x)} = \frac{1}{2}
\frac{f'+1-2}{f'-1}f'' + O(\chi) = \frac{1}{2}f'' + O(\chi),
$$
as we already know.
Here and below $f'(x) = f' + \chi f'' + O(\chi^2)$.
Further, recursively,
$$
G_{1,s}(x) = \frac{F_{s+1}-F_S}{G_1G_{1,1}\ldots G_{1,s-1}} =
 f' + \frac{1}{2}\chi f''(f'+1){f'}^{s-1} + O(\chi^2),
$$ $$
\frac{G_{1,s}(x)-f'(x)}{G_1(x)} = \frac{1}{2}
\frac{{f'}^s+{f'}^{s-1}-2}{f'-1}f'' + O(\chi) =
\frac{1}{2}f''({f'}^{s-1} + 2{f'}^{s-2}+\ldots +2) + O(\chi),
$$
i.e. $G_{1,s}(x) = f'(x)\ {\rm mod}\ G_1(x)$ and, as generalization
of (\ref{G11f}) and (\ref{r111}), we obtain for all $s$
$$
r_{1|1,s} =
R(G_1,G_{1,s}) \sim R(G_1,f') \sim \prod_{w_f}G_1(w_f;f) = w_1.
$$

\subsubsection{Sector $(n,s) = (2,s)$}

For consideration of the sector $(n,s) = (2,s)$ we need to
consider the vicinity of another stable point $x_2$, which
belongs to the orbit of order two, i.e. satisfies
$f(f(x_2))=x_2$, but $\tilde x_2 = f(x_2)\neq x_2$. Denote
$f' = f'(x_2)$, $\tilde{f}' = f'(\tilde x_2) = f'(f(x_2))$,
$f'' = f''(x_2)$, $\tilde{f}'' = f''(\tilde x_2) =
f''(f(x_2))$.
For $x = x_2 + \chi$ we have:
$$
F_1(x) = G_1(x)
= (\tilde x_2-x_2) + \chi(f'-1) + \frac{1}{2}\chi^2 f'' +
O(\chi^3),
$$ $$
F_2(x) = G_1(x)G_2(x)
= \chi\left(f'\tilde{f}'-1\right) + \frac{1}{2}\chi^2
\left(f''\tilde{f}' + \tilde{f}''{f'}^2\right) + O(\chi^3),
$$ $$
F_3(x) = G_1(x)G_3(x)
= (\tilde x_2-x_2) + \chi\left(f'^2\tilde f'-1\right) +
\frac{1}{2}\chi^2\left(f''f'\tilde{f}' + \tilde{f}'' {f'}^3 +
f''(f'\tilde f')^2\right) + O(\chi^3),
$$ $$\ldots $$
Then
$$
G_{2,1}(x) = \frac{F_{2,1}}{G_1G_2G_{1,1}} =
\frac{F_1(F_3-F_1)}{(F_2-F_1)F_2} =
$$ $$
= -\frac{(\tilde x_2-x_2) + \chi(f'-1) + \frac{1}{2}\chi^2 f'' +
O(\chi^3)}
{(\tilde x_2-x_2) - \chi(\tilde f'-1)f'
- \frac{1}{2}\chi^2
\left(f''(\tilde{f}'-1) + \tilde{f}''{f'}^2\right)  +
O(\chi^3)}\cdot
$$ $$
\cdot \frac{\chi\left(f'\tilde f'-1\right)f' +
\frac{1}{2}\chi^2\left(\tilde{f}'' {f'}^3 +
f''\left((f'\tilde f')^2+ f'\tilde{f}' -1\right)\right) +
O(\chi^3)}
{\chi\left(f'\tilde{f}'-1\right) + \frac{1}{2}\chi^2
\left(f''\tilde{f}' + \tilde{f}''{f'}^2\right) + O(\chi^3)
} =
$$ $$
= -\left(1 + \frac{\chi(f'\tilde f'-1)}{\tilde x_2-x_2}
+ O(\chi^2)\right)
\left(f' + \frac{1}{2}\chi f''(f'\tilde f'+1) +
O(\chi^2)\right),
$$
so that
$$
G_{21}(x) + f'(x) = -\chi(f'\tilde f' -1)\left(\frac{1}{2}f'' +
\frac{f'}{\tilde x_2-x_2}\right) + O(\chi^2)
$$
and
$$
\frac{G_{21}(x) + f'(x)}{G_2(x)} =
-\left(\frac{1}{2}f'' +
\frac{f'}{\tilde x_2-x_2}\right) + O(\chi)
$$
is finite at $\chi = 0$. Therefore
$$
r_{2|2,1} = R(G_2,G_{2,1}) \sim -R(G_2,f') \sim
\prod_{w_f} G_2(w_f) = W_2(f) = w_2(f).
$$

\subsubsection{Summary}

In the same way one can consider other quantities,
complete the derivation of (\ref{wsmalldef}) and
deduce the formulas for resultants and discriminants.
Like in (\ref{DG11}), these decomposition
formulas depend explicitly on degree $d$ of the map $f(x)$.
In what follows $$\# s = (d-1)d^{s-1}$$ denotes the number
of $s$-level pre-images of a point on a periodic orbit,
which do not belong to the orbit. Also, $w_n(f)$ enter
all formulas multiplied by peculiar $d$-dependent factors:
$$
\begin{array}{cccc}
\tilde w_n(f) && d=2 & d=3 \\
&&&\\
\tilde w_1 = d^d w_1, & & 2^2=4, & 3^3 = 27, \\
\tilde w_2 = d^{d(d-1)} w_2, && 2^2 = 4, & 3^6 = 729, \\
\tilde w_3 = d^{d(d^2-1)} w_3, && 2^6 = 64, & 3^{24}, \\
\tilde w_4 = d^{d^2(d^2-1)} w_4, && 2^{12} = 4096, & 3^{72} \\
\tilde w_5 = d^{d(d^4-1)} w_5, && 2^{30}, &  3^{240}\ ? \\
\tilde w_6 = d^{d(d^2-1)(d^3+d-1)} w_6, && 2^{54}, &3^{7176}\ ? \\
\ldots &&&
\end{array}
$$
(the last two columns in this table contain the values of
numerical factors for $d=2$ and $d=3$, question marks label
the cases which were {\it not} verified by explicit MAPLE simulations).
Obviously, these factors are made from degree
$N_d(n)$ of the map $G_n(x)$, which was considered in
s.\ref{ortim}:
\be
\tilde w_n(f) = d^{N_d(n)}w_n(f).
\label{wtildew}
\ee

Non-trivial (i.e. not identically unit) resultants are:
$$
R(G_m, G_{n,s}) = \left\{ \begin{array}{ccc}
           \tilde w_n && {\rm if} \ m=n, \\
            1 && {\rm if}\ m\neq n
                 \end{array}\right.
$$
and
$$
R(G_{k,s}, G_{n,s'}) = \left\{\begin{array}{ccc}
             R(G_k,G_n) = r_{k,n}^{k\# s} &&
                   {\rm if}\ s'=s,\  k<n\ ({\rm actually},\ k|n) \\
             \tilde w_n^{\# s} && {\rm if}\ s< s'
                  \end{array} \right.
$$
or, in the form of a table:

{\footnotesize
$$
\begin{tabular}{|c|cccc|cccc|cccc|}
\hline
&&&& &&&& &&&&\\
ns&11&12&13&14 &21&22&23&24 &31&32&33&34\\
&&&& &&&& &&&&\\
\hline
&&&& &&&& &&&&\\
1& $\tilde w_1$& $\tilde w_1$& $\tilde w_1$& $\tilde w_1$&
    1 & 1 & 1 & 1 &   1 & 1 & 1& 1 \\
2 &  1 & 1 & 1 & 1 &
      $\tilde w_2$& $\tilde w_2$&$\tilde w_2$&
      $\tilde w_2$&  1 & 1 & 1 & 1  \\
3 &  1 & 1 & 1 & 1 &  1 & 1 & 1 & 1 &
     $\tilde w_3$& $\tilde w_3$&
     $\tilde w_3$& $\tilde w_3$ \\
4 &  1 & 1 & 1 & 1 &  1 & 1 & 1 & 1 &  1 & 1 & 1 & 1 \\
5 &  1 & 1 & 1 & 1 &  1 & 1 & 1 & 1 &  1 & 1 & 1 & 1 \\
\ldots & & & & & & & & & & & & \\
\hline \hline
&&&& &&&& &&&&\\
11 & -&$\tilde w_1^{\# 1}$& $\tilde w_1^{\# 1}$& $\tilde w_1^{\# 1}$&
       $r_{12}^{\# 1}$ & 1&1&1& $ r_{13}^{\# 1}$&1&1&1 \\
12 & $\tilde w_1^{\# 1}$&-&$\tilde w_1^{\# 2}$&$\tilde w_1^{\# 2}$&
  1&$r_{12}^{\# 2}$ &1&1& 1&$r_{13}^{\# 2}$&1&1 \\
13 & $\tilde w_1^{\# 1}$&$\tilde w_1^{\# 2}$&-&$\tilde w_1^{\# 3}$&
  1&1& $r_{12}^{\# 3}$&1& 1&1&$r_{13}^{\# 3}$&1 \\
14 &$\tilde w_1^{\# 1}$&$\tilde w_1^{\# 2}$&$\tilde w_1^{\# 3}$&-&
  1&1&1&$r_{12}^{\# 4}$& 1&1&1&$r_{13}^{\# 4}$ \\
15 &$\tilde w_1^{\# 1}$&$\tilde w_1^{\# 2}$&
    $\tilde w_1^{\# 3}$&$\tilde w_1^{\# 4}$&
  1&1&1&1& 1&1&1&1 \\
16 &$\tilde w_1^{\# 1}$&$\tilde w_1^{\# 2}$&
    $\tilde w_1^{\# 3}$&$\tilde w_1^{\# 4}$&
  1&1&1&1& 1&1&1&1 \\
\ldots & & & & & & & & & & & & \\
\hline
&&&& &&&& &&&&\\
21 & $r_{12}^{\# 1}$&1&1&1&
   -&$\tilde w_2^{\# 1}$&$\tilde w_2^{\# 1}$&
   $\tilde w_2^{\# 1}$&  1&1&1&1 \\
22 &1&$r_{12}^{\# 2}$&1&1&
    $\tilde w_2^{\# 1}$&-&$\tilde w_2^{\# 2}$&
    $\tilde w_2^{\# 2}$& 1&1&1&1 \\
23 &1&1&$r_{12}^{\# 3}$&1& $\tilde w_2^{\# 1}$&
         $\tilde w_2^{\# 2}$&-&$\tilde w_2^{\# 3}$& 1&1&1&1\\
24 &1&1&1&$r_{12}^{\# 4}$& $\tilde w_2^{\# 1}$&
         $\tilde w_2^{\# 2}$&$\tilde w_2^{\# 3}$&-& 1&1&1&1\\
25 &1&1&1&1& $\tilde w_2^{\# 1}$&$\tilde w_2^{\# 2}$&
         $\tilde w_2^{\# 3}$&$\tilde w_2^{\# 4}$& 1&1&1&1\\
26 &1&1&1&1& $\tilde w_2^{\# 1}$&$\tilde w_2^{\# 2}$&
         $\tilde w_2^{\# 3}$&$\tilde w_2^{\# 4}$& 1&1&1&1\\
\ldots & & & & & & & & & & & & \\
\hline
&&&& &&&& &&&&\\
31 &$r_{13}^{\# 1}$&1&1&1& 1&1&1&1&
      -&$\tilde w_3^{\# 1}$&$\tilde w_3^{\# 1}$&
         $\tilde w_3^{\# 1}$ \\
32 &1&$r_{13}^{\# 2}$&1&1& 1&1&1&1& $\tilde w_3^{\# 1}$&-&
   $\tilde w_3^{\# 2}$&$\tilde w_3^{\# 2}$\\
33 &1&1&$r_{13}^{\# 3}$&1& 1&1&1&1&
      $\tilde w_3^{\# 1}$&$\tilde w_3^{\# 2}$&-&
                  $\tilde w_3^{\# 3}$\\
34 &1&1&1&$r_{13}^{\# 4}$& 1&1&1&1& $\tilde w_3^{\# 1}$&
      $\tilde w_3^{\# 2}$&$\tilde w_3^{\# 3}$&-\\
35 &1&1&1&1& 1&1&1&1& $\tilde w_3^{\# 1}$&
      $\tilde w_3^{\# 2}$&$\tilde w_3^{\# 3}$&
          $\tilde w_3^{\# 4}$\\
\ldots & & & & & & & & & & & & \\
\hline
&&&& &&&& &&&&\\
41 &$r_{14}^{\# 1}$&1&1&1& $(r_{24}^2)^{\# 1}$&1&1&1& 1&1&1&1\\
42 &1&$r_{14}^{\# 2}$&1&1& 1&$(r_{24}^2)^{\# 2}$&1&1& 1&1&1&1\\
43 &1&1&$r_{14}^{\# 3}$&1& 1&1&$(r_{24}^2)^{\# 3}$&1& 1&1&1&1\\
44 &1&1&1&$r_{14}^{\# 3}$& 1&1&1&$(r_{24}^2)^{\# 4}$& 1&1&1&1\\
45 &1&1&1&1& 1&1&1&1& 1&1&1&1\\
\ldots & & & & & & & & & & & & \\
\hline
&&&& &&&& &&&&\\
51 &$r_{15}^{\# 1}$&1&1&1& 1&1&1&1& 1&1&1&1\\
52 &1&$r_{15}^{\# 2}$&1&1& 1&1&1&1& 1&1&1&1\\
53 &1&1&$r_{15}^{\# 3}$&1& 1&1&1&1& 1&1&1&1\\
54 &1&1&1&$r_{15}^{\# 4}$& 1&1&1&1& 1&1&1&1\\
55 &1&1&1&1& 1&1&1&1& 1&1&1&1\\
\ldots & & & & & & & & & & & & \\
\hline
&&&& &&&& &&&&\\
61 &$r_{16}^{\# 1}$&1&1&1& $(r_{26}^2)^{\# 1}$&1&1&1&
                        $(r_{36}^3)^{\# 1}$&1&1&1\\
62 &1&$r_{16}^{\# 2}$&1&1& 1&$(r_{26}^2)^{\# 2}$&1&1&
                             1&$(r_{36}^3)^{\# 2}$&1&1\\
63 &1&1&$r_{16}^{\# 3}$&1& 1&1&$(r_{26}^2)^{\# 3}$&1&
                          1&1&$(r_{36}^3)^{\# 3}$&1\\
64 &1&1&1&$r_{16}^{\# 4}$& 1&1&1&$(r_{26}^2)^{\# 4}$&
                             1&1&1&$(r_{36}^3)^{\# 4}$\\
65 &1&1&1&1& 1&1&1&1& 1&1&1&1\\
\ldots & & & & & & & & & & & & \\
\hline
\end{tabular}
$$
}

Similarly, for discriminants:
\be
D(G_{n,s}) \sim
D(G_n)^{\# s} w_{11}^{\# s/\# 1}
\prod_{r=2}^s W_{n,r}^{\# s/\# r} = \nn \\ =
\left(d_n^n\prod_{\stackrel{k|n}{k<n}}r_{k,n}^{n-k}\right)^{\# s}
w_n^{\delta(n,s)}\prod_{r=1}^s w_{n,r}^{\# s/\# r}
\label{DGns}
\ee
Exponents in this expression are $\# s = (d-1)d^{s-1}$,
$\# s/\# r = d^{s-r}$, and
$\delta(n,s) = d^{r-1}\frac{d^{nr'}-1}{d^n-1} =
\frac{d^{s-1}-d^{r-1}}{d^n-1}$
for $s=nr'+r$, $r',r>0$ (if such $r'$ and $r$
do not exist, i.e. $s\leq n$, $\delta(n,s) = 0$).
The first few values of $\delta(n,s)$ are listed in the table
(stars substitute too lengthy expressions):

{\footnotesize
$$
\left(\begin{tabular}{c|ccccccccccccc}
\hline
&&&&&&&&&&\\
$n\ \backslash\ s\ $ &1&2&3&4&5&6&7&8&9&10&& \ldots \\
&&&&&&&&&&&&\\
\hline
&&&&&&&&&&&&\\
1 &0 &1 &$d+1$& $d^2+d+1$& $d^3+d^2+d+1$
 &*&*&*&*&*&&\\
2 &0&0&1&$d$&$d^2+1$&$d^3+d$&$d^4+d^2+1$&$d^5+d^3+d$&*&*&&\\
3 &0&0&0&1&$d$&$d^2$&$d^3+1$&$d^4+d$&$d^5+d^2$&$d^6+d^3+1$&&\\
4 &0&0&0&0&1&$d$&$d^2$&$d^3$&$d^4+1$&$d^5+d$&&\\
5 &0&0&0&0&0&1&$d$&$d^2$&$d^3$&$d^4$&&\\
6 &0&0&0&0&0&0&1&$d$&$d^2$&$d^3$&&\\
7 &0&0&0&0&0&0&0&1&$d$&$d^2$&&\\
8 &0&0&0&0&0&0&0&0&1&$d$&&\\
9 &0&0&0&0&0&0&0&0&0&1&&\\
10 &0&0&0&0&0&0&0&0&0&0&&\\
\ldots &&&&&&&&&&\\
&&&&&&&&&&\\
\hline
\end{tabular} \right)
$$
}

The first few discriminants are listed in the table
($D_n = D(G_n)$, not to be confused with irreducible $d_n$;
the last two columns contain numerical factors
for $d=2$ and $d=3$\footnote{
Numerical factors are equal to
$d^{d^s(sd-s-1)\#_n(d)}$, where the sequences
$\#_n(d)$ are  $1,1,3,6,15,\ldots$ for $d=2$ and $1,2,\ldots$ for $d=3$.
One can observe that $\#_n(d) = {\rm deg}_c G_n(x;x^d+c)$, but the reason
for this, as well as the very origin of numerical factors,
here and in eq.(\ref{wtildew}), remain obscure.
}):

{\footnotesize
$$
\begin{array}{ccccccc}
ns &&D(G_{ns})&&&d=2&d=3\\
&&&&&&\\
11 &D_1^{d-1}w_{11}&=& d_1^{d-1}w_{11} && 1& -3 \\
12 &D_1^{d(d-1)}w_{11}^dW_{12}&=&
         d_1^{d(d-1)}w_1w_{11}^dw_{12} && 2^4& 3^{27} \\
13 &D_1^{d^2(d-1)}w_{11}^{d^2}W_{12}^dW_{13}&=&
         d_1^{d^2(d-1)}w_1^{d+1}w_{11}^{d^2}w_{12}^dw_{13}
         && 2^{16}& 3^{135} \\
14 &D_1^{d^3(d-1)}w_{11}^{d^3}W_{12}^{d^2}W_{13}^{d}W_{14} &=&
         d_1^{d^3(d-1)}w_1^{d^2+d+1}w_{11}^{d^3}w_{12}^{d^2}w_{13}^d
         w_{14}  && 2^{48} & 3^{567} \\
15 &D_1^{d^4(d-1)}w_{11}^{d^4}W_{12}^{d^3}W_{13}^{d^2}W_{14}^dW_{15}
         &=&
         d_1^{d^4(d-1)}w_1^{d^3+d^2+d+1}w_{11}^{d^4}w_{12}^{d^3}
         w_{13}^{d^2}w_{14}^dw_{15}  && 2^{128} & ? \\
&&&&&&\\
21 &D_2^{d-1}W_{21} &=& d_2^{2(d-1)}r_{12}^{d-1}w_{21} && 1 & 3^6\\
22 &D_2^{d(d-1)}W_{21}^dW_{22}&=& d_2^{2d(d-1)}r_{12}^{d(d-1)}
         w_{21}^dw_{22} && 2^4 & 3^{54} \\
23 &D_2^{d^2(d-1)}W_{21}^{d^2}W_{22}^{d}W_{23}&=&
           d_2^{2d^2(d-1)}r_{12}^{d^2(d-1)}
           w_{2}w_{21}^{d^2}w_{22}^dw_{23} && 2^{16} & ? \\
24 &D_2^{d^3(d-1)}W_{21}^{d^3}W_{22}^{d^2}W_{23}^{d}W_{24}&=&
           d_2^{2d^3(d-1)}r_{12}^{d^3(d-1)}
      w_{2}^dw_{21}^{d^3}w_{22}^{d^2}w_{23}^dw_{24} && 2^{48} & ? \\
25 &D_2^{d^4(d-1)}W_{21}^{d^4}W_{22}^{d^3}W_{23}^{d^2}W_{24}^dW_{25}
        &=&     d_2^{2d^4(d-1)}r_{12}^{d^4(d-1)}
      w_{2}^{d^2+1}w_{21}^{d^4}w_{22}^{d^3}w_{23}^{d^2}w_{24}^dw_{25}
      && 2^{128} & ? \\
&&&&&&\\
31 &D_3^{d-1}W_{31} &=& d_3^{3(d-1)}r_{13}^{2(d-1)}w_{31} && 1 & 3^{24}\\
32 &D_3^{d(d-1)}W_{31}^dW_{32}&=& d_3^{3d(d-1)}r_{13}^{2d(d-1)}
         w_{31}^dw_{32} && 2^{12} & ? \\
33 &D_3^{d^2(d-1)}W_{31}^{d^2}W_{32}^{d}W_{33}&=&
           d_3^{3d^2(d-1)}r_{13}^{2d^2(d-1)}
           w_{31}^{d^2}w_{32}^dw_{33} && 2^{48} & ? \\
34 &D_3^{d^3(d-1)}W_{31}^{d^3}W_{32}^{d^2}W_{33}^{d}W_{34}&=&
           d_3^{3d^3(d-1)}r_{13}^{2d^3(d-1)}
      w_{3}w_{31}^{d^3}w_{32}^{d^2}w_{33}^dw_{34} && 2^{144} & ? \\
35 &D_3^{d^4(d-1)}W_{31}^{d^4}W_{32}^{d^3}W_{33}^{d^2}W_{34}^dW_{35}
        &=&     d_3^{3d^4(d-1)}r_{13}^{2d^4(d-1)}
      w_{3}^dw_{31}^{d^4}w_{32}^{d^3}w_{33}^{d^2}w_{34}^dw_{35}
      && 2^{384} & ? \\
&&&&&&\\
41 &D_4^{d-1}W_{41} &=& d_4^{4(d-1)}r_{14}^{3(d-1)}r_{24}^{2(d-1)}
            w_{41} && 1 & ?\\
42 &D_4^{d(d-1)}W_{41}^dW_{42}&=& d_4^{4d(d-1)}r_{14}^{3d(d-1)}
         r_{24}^{2d(d-1)}
         w_{41}^dw_{42} && 2^{24} & ? \\
43 &D_4^{d^2(d-1)}W_{41}^{d^2}W_{42}^{d}W_{43}&=&
           d_4^{4d^2(d-1)}r_{14}^{3d^2(d-1)}r_{24}^{2d^2(d-1)}
           w_{41}^{d^2}w_{42}^dw_{43} && 2^{96} & ? \\
44 &D_4^{d^3(d-1)}W_{41}^{d^3}W_{42}^{d^2}W_{43}^{d}W_{44}&=&
           d_4^{4d^3(d-1)}r_{14}^{3d^3(d-1)}r_{24}^{2d^3(d-1)}
      w_{41}^{d^3}w_{42}^{d^2}w_{43}^dw_{44} && ? & ? \\
45 &D_4^{d^4(d-1)}W_{41}^{d^4}W_{42}^{d^3}W_{43}^{d^2}W_{44}^dW_{45}
        &=&     d_4^{4d^4(d-1)}r_{14}^{3d^4(d-1)}r_{24}^{2d^4(d-1)}
      w_{4}w_{41}^{d^4}w_{42}^{d^3}w_{43}^{d^2}w_{44}^dw_{45}
      && ? & ? \\
&&&&&&\\
51 &D_5^{d-1}W_{51} &=& d_5^{5(d-1)}r_{15}^{4(d-1)}w_{51} && 1 & ? \\
52 &D_5^{d(d-1)}W_{51}^dW_{52}&=& d_5^{5d(d-1)}r_{15}^{4d(d-1)}
         w_{51}^dw_{52} && 2^{60} & ? \\
&&&&&&\\
61 &D_6^{d-1}W_{61} &=& d_6^{6(d-1)}r_{16}^{5(d-1)}
         r_{26}^{4(d-1)}r_{36}^{3(d-1)}       w_{61} && 1 & ? \\
62 &D_6^{d(d-1)}W_{61}^dW_{62}&=& d_6^{6d(d-1)}r_{16}^{5d(d-1)}
          r_{26}^{4d(d-1)}r_{36}^{3d(d-1)}
         w_{21}^dw_{22} && ? & ? \\
\ldots &&&&&&\\
\end{array}
$$
}

\subsubsection{On interpretation of $w_{n,k}$}

Responsible for degenerations of pre-orbits are
intersections with the set $\{z_f\} = f(\{w_f\})$,
which is $f$-image of the set of critical points of $f$.
Let us describe what happens when $\{z_f\}$ is crossed
by pre-orbits of different levels.

\bigskip

$\bullet$ $\{G_n(f)=0\} \bigcap \{z_f\} \neq \emptyset$.

Let $A$ be the intersection point, denote its pre-images
along the periodic orbit through $A^{-1}, A^{-2}, \ldots,
A^{-n} = A$, and the pre-images of level $s$ on the
pre-image tree through $B^{-s}_{i_1\ldots i_{\# s}}(A)$,
where indices $i$ can be further ordered according to the
tree structure. Points on the level $s$ of the tree,
rooted at pre-image $A^{-l}$ will be denoted through
$B^{-s}_{i_1\ldots i_{\# s}}(A^{-l})$.
If $A\in \{z_f\}$, then two of its pre-images coincide.
There are two possibilities: either these coinciding
preimages belong to pre-orbit tree, say $B^{-1}_1(A) = B^{-1}_2(A)$,
or one of them lie on the orbit, say $B^{-1}_1(A) = A^{-1}$.
In the first case $w_{n,1}(f) = 0$, in the second case
$w_n(f) = 0$, and the fact that only two such possibilities
exist is reflected in decomposition formula
$$\prod_{z_f}G_n(z_f) = \prod_{w_f} G_{n,1}(w_f) =
W_{n,1}(f) = w_n(f)w_{n,1}(f)$$
(products at the l.h.s. are needed to build up quantities,
which depend on the coefficients of $f(x)$, not on their
irrational combinations, entering expressions for individual
points $z_f$ and $w_f$ -- the latter are roots of $f'(x)$).
Let us analyze these two cases in more detail.

\FIGEPS{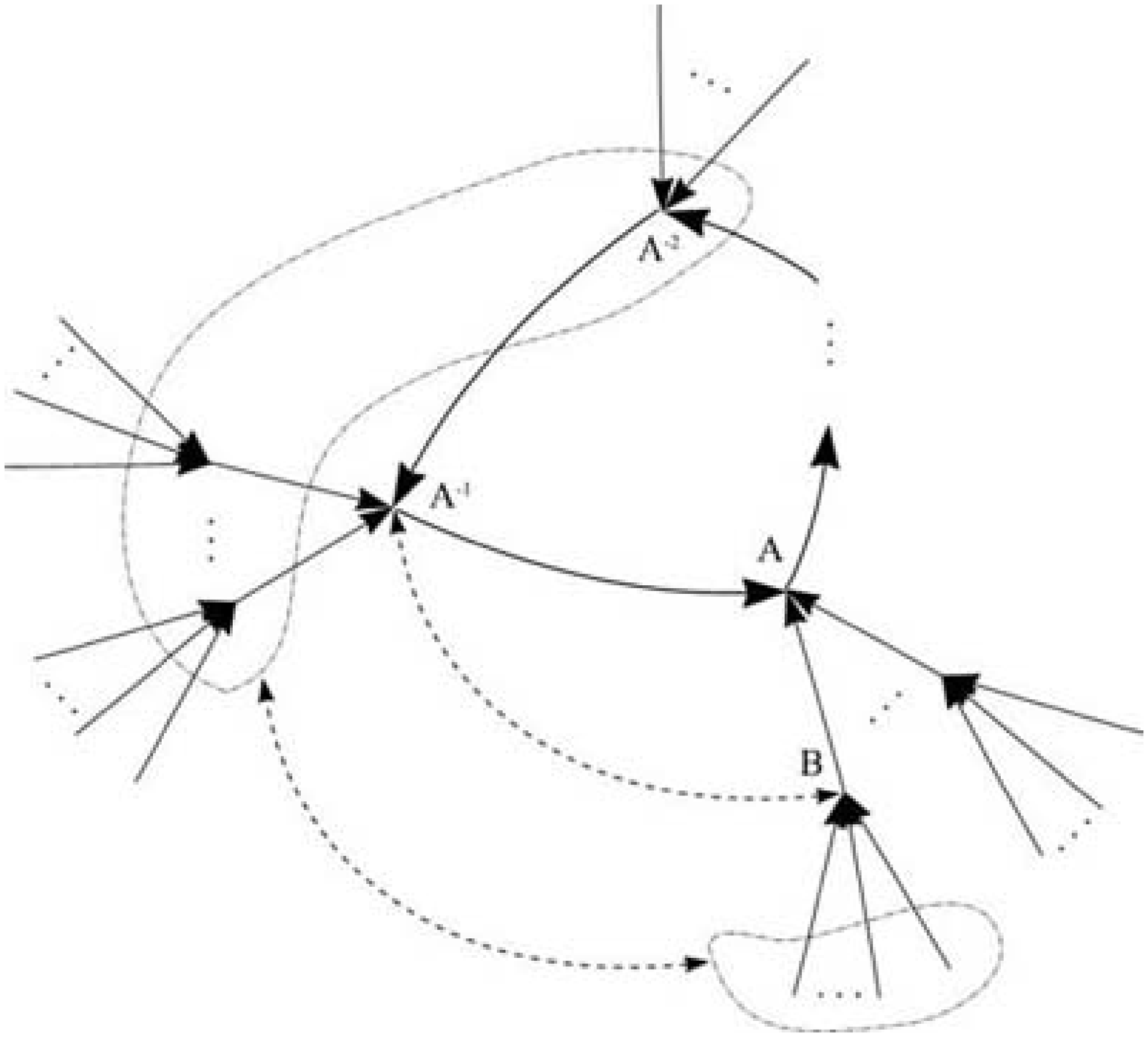}{}
{325,293}
{A periodic orbit of
order $n$ and grand-orbit trees, rooted at the points of the orbit.
$w_n(f)$ vanishes when (i) a point $A$ on the orbit coincides with an
"irreducible point" from the set $\{z_f\}$, which is $f$-image of
the set of critical points $\{w_f\}$, and (ii)
when one of the coinciding first pre-images of $A$ is also a point on
the orbit. Arrows show the pre-image
points of different levels,
which coincide when $A \in \{z_f\}$ and $w_n(f)=0$, i.e. when $A^{-1}=B$.}
\bigskip

$\circ$ $w_n(f)=0$. See Fig.\ref{wnf0}.

Since $B^{-1}_1(A) = A^{-1}$ exactly one point from $\{G_{n,1}=0\}$
coincides with exactly one point on $\{G_n=0\}$, thus the resultant
has simple zero, $$R(G_{n},G_{n,1}) = r_{n|n,1} = \tilde w_n \sim w_n.$$
Numeric $d$-dependent factor, distinguishing between $\tilde w_n$
and $w_n$ requires separate explanation.

Pre-images of $B^{-1}_1(A)$ -- the set of $d$ (different!) points
$B^{-2}_1(A),\ldots,B^{-2}_d(A)$ from $\{G_{n,2}=0\}$
should coincide pairwise with pre-images
of $A^{-1}$, which are $d-1$ points $B^{-1}_1(A^{-1}), \ldots,
B^{-1}_{d-1}(A^{-1})$ from $\{G_{n,1}=0\}$
and the point $A^{-2}$ on the orbit, i.e. from $\{G_{n}=0\}$.
Thus the corresponding resultants will have zeroes of orders
$\# 1 = d-1$ and one respectively:
$$R(G_{n,1},G_{n,2}) = r_{n,1|n,2}^{\# 1} \sim w_n^{\# 1}$$ and
$$R(G_{n},G_{n,2}) = r_{n|n,2} \sim w_n$$

Next pre-images of $A^{-1}$ consist of $A^{-3} \in \{G_n = 0\}$,
$\# 1 = d-1$ points $B^{-1}(A^{-2}) \in \{G_{n,1}=0\}$ and
$\# 2 = d(d-1)$ points $B^{-1}(A^{-2}) \in \{G_{n,2}=0\}$, while
those of $B^{-1}_1$ are $\# 3/\# 1 = d^2$ points from $\{B^{-3}(A)\}$.
Each of $d^2$ points from the $A^{-1}$ second pre-image should
coincide with one of the $d^2$ points from that of $B^{-1}_1$.
This provides relations
$$R(G_{n,2},G_{n,3}) = r_{n,2|n,3}^{\# 2} \sim w_n^{\# 2},$$
$$R(G_{n,1},G_{n,3}) = r_{n,1|n,3}^{\# 1} \sim w_n^{\# 1}$$ and
$$R(G_{n},G_{n,3}) = r_{n|n,3} \sim w_n$$.

Continuing along the same line, we deduce:
$$R(G_{n,r},G_{n,s}) = r_{n,r|n,s}^{\# r} \sim w_n^{\# r},\ \
{\rm for}\ r<s$$ and
$$R(G_{n},G_{n,s}) = r_{n|n,s} \sim w_n$$.

Actually, this exhausts the set of non-trivial (non-unit)
resultants for orbits of coincident periods $n$, but does not
exhaust the possible appearances of $w_n$: it will show up again
in discriminants, see below.

\bigskip

$\circ$ $w_{n,1}(f)=0$. See Fig.\ref{wn1f0}.

In this case we deal with a single tree, rooted at
$A \in G_n \cap \{z_f\}$ and thus we can safely omit reference
to $A$ in $B(A)$. Thus, the starting point is
$B^{-1}_1 = B^{-1}_2$. (In particular, there is nothing to
discuss in the case of $d=2$, when $\# 1 = 1$, there are no
{\it a priori} different points at the level $s=1$, which could
occasionally coincide, and all $w_{n,1} = 1$.)
We can conclude, that discriminant $D(G_{n,1}) \sim w_{n,1}$.

This is not the full description of discriminant, because
it also contains contributions coming from intersections of different
periodic orbits, which cause the corresponding points on pre-orbit
trees to coincide as well. These factors were already evaluated
in s.\ref{redi} above, in application to discriminant
of $G_{n,s}$ they should just be raised to the power $\# s$,
counting the number of pairs of points which are forced to coincide
at level $s$ when their roots on the orbits merge.
Thus finally
$$D(G_{n,1}) \sim D(G_n)^{\# 1}w_{n,1} = w_{n,1}
\left(d_n^n \prod_{\stackrel{k|n}{k<n}}r_{k,n}^{n-k}\right)^{\# 1},$$
and the remaining undetermined factor is just a constant on ${\cal M}$,
which depends on $d$, but not on $f$.
The same argument explains the expressions for non-trivial resultants
$$R(G_{n,s},G_{k,s}) = R(G_n,G_k)^{\# s} = r_{n,k}^{k\# s},\ \
k|n,\ k<n.$$

Considering next pre-images of $B^{-1}_1$ and $B^{-1}_2$, we obtain
at level $s$ the two coincident sets of $\# s/\# 1 = d^{s-1}$ points
each and thus $D(G_{n,s}) \sim w_{n,1}^{d^{s-1}}D(G_n)^{\# s}$.
However, many
more factors enter in the expression for higher discriminants:
all $w_{n,r}$ with $r<s$ contribute, and, remarkably,
$w_n$ also contributes when $s>n$ -- in accordance with
decomposition formula (\ref{wsmalldef}). To see all this we should return
to the very beginning and consider the intersections of {\it pre-orbits}
with the set $\{z_f\}$.

\FIGEPS{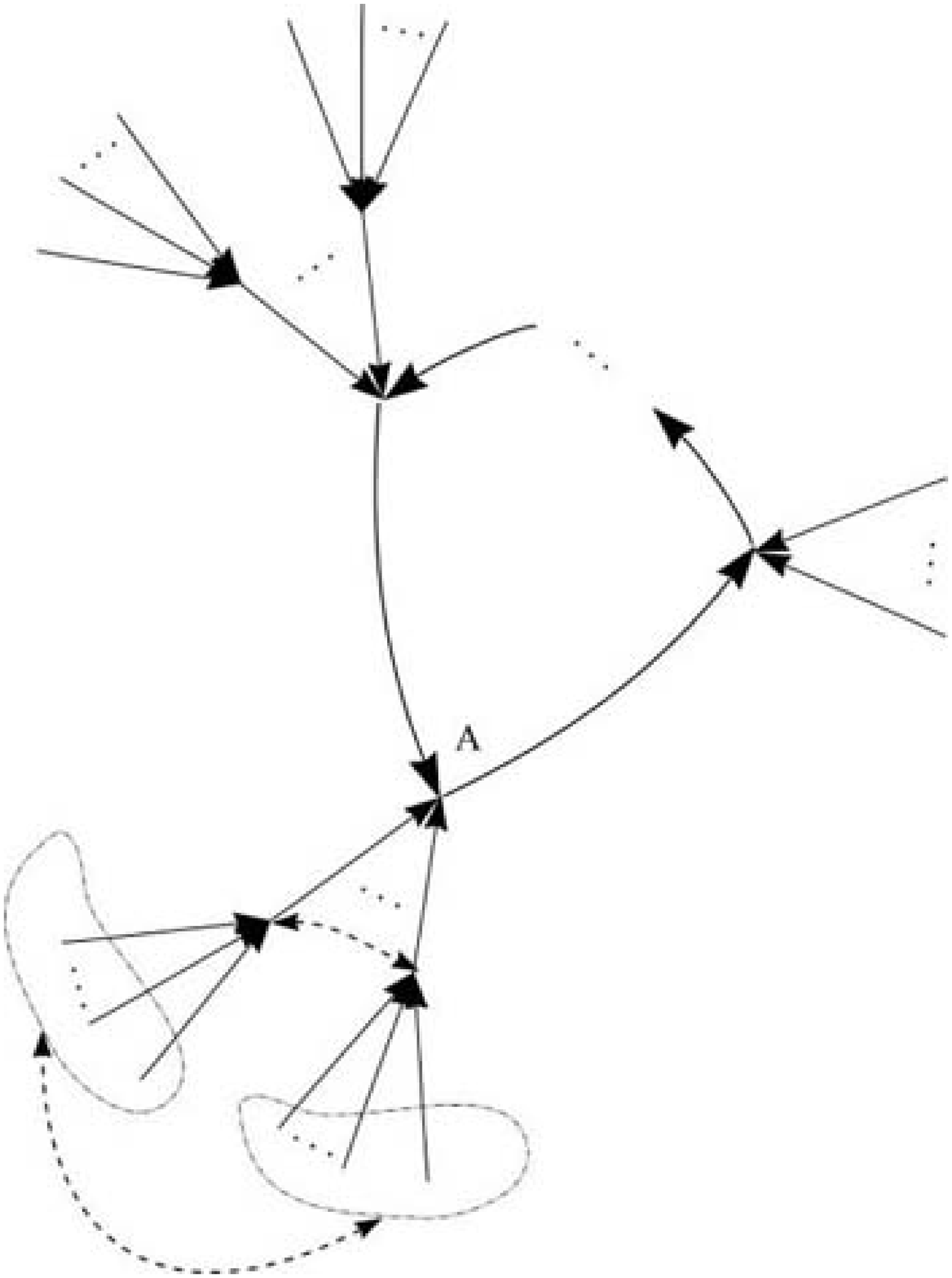}{}
{270,363}
{The same periodic orbit of
order $n$ and the same point $A$, intersecting $\{z_f\}$ as in the
previous Fig.\ref{wnf0}. $w_{n,1}(f)$ vanishes when
(i) a point $A$ coincides with a point from $\{z_f\}$,  and (ii)
when both coinciding first pre-images of $A$ belong to pre-image tree,
rooted at $A$. Arrows show the pre-image
points of different levels,
which coincide when $A \in \{z_f\}$ and $w_{n,1}(f)=0$.}

\FIGEPS{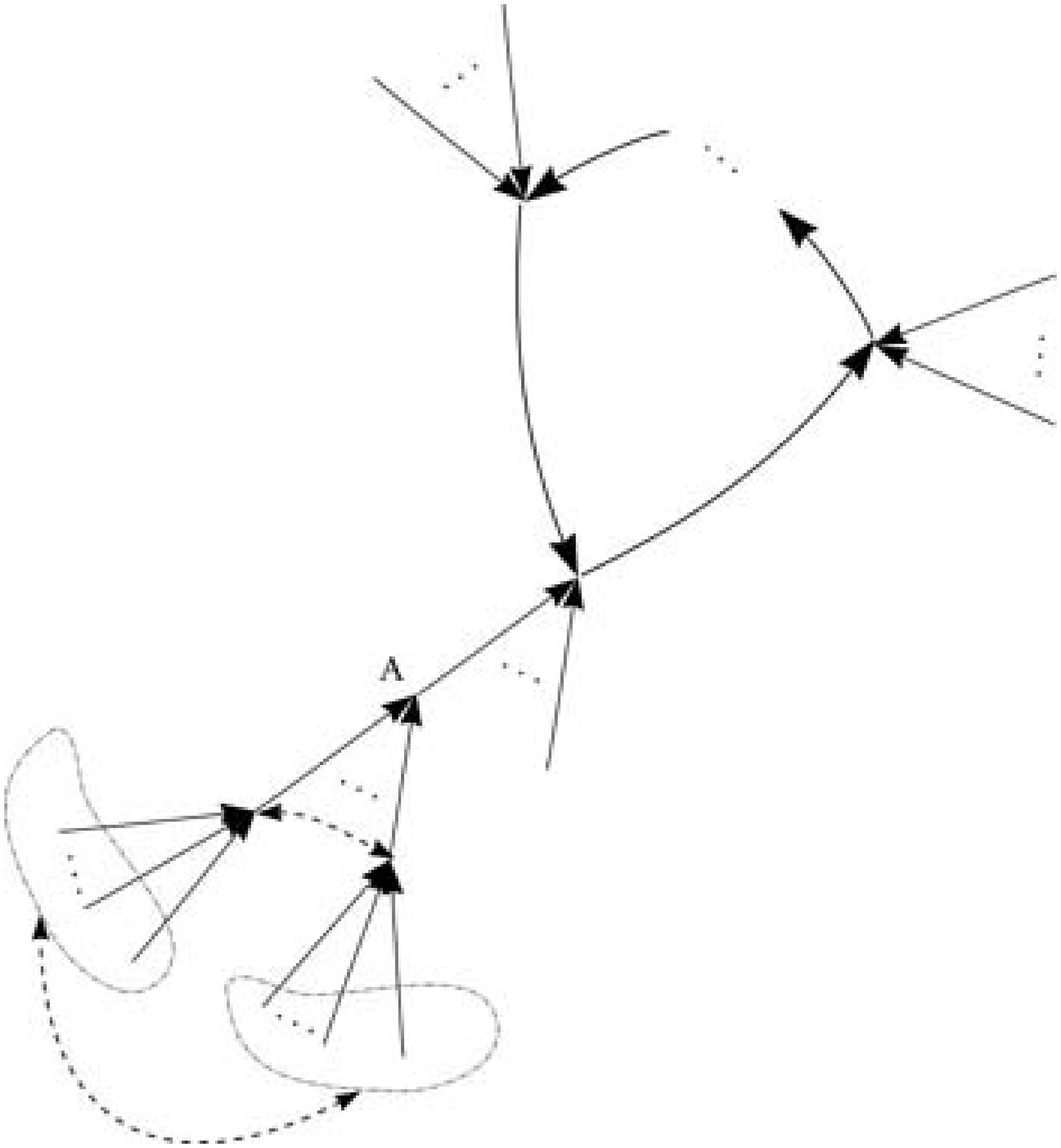}{}
{319,345}
{A point which intersects $\{z_f\} = f(\{w_f\}$ belongs
to the $r-1$-th level of the grand orbit tree, $r>1$.
Arrows show the pre-image points of different levels,
which coincide in this situation.}

\bigskip

$\bullet$ $\{G_{n,r-1}(f)=0\} \bigcap \{z_f\} \neq \emptyset$,
$r>1$.
See Fig.\ref{wn2f0}.

Let $B^{-(r-1)}_1 \in \{z_f\}$. This means that some two preimages
of this point coincide, say $B^{-r}_1 = B^{-r}_2$,
and at level $s \geq r$
there will be two coinciding sets, of $\# s/\# r = d^{s-r}$ points each.
Since, $G_{n,r-1}(z_f;f) = G_{n,r-1}(f(w_f);f) \sim G_{n,r}(w_f;f)$,
all this happens when $W_{n,r}(f) = 0$, and we conclude that
$D(G_{n,s}) \sim W_{n,r}^{d^{s-r}}$ for any $r\leq s$.
We can further use (\ref{wsmalldef}) to substitute $W_{n,r}(f)$
by $w_{n,r}(f)$ for $r\neq 1 \ {\rm mod}\ n$ and by
$w_{n,r}(f)w_n(f)$ for $r=1 \ {\rm mod}\ n$.
Collecting everything together we finally reproduce (\ref{DGns}):
$$
D(G_{n,s}) \sim D(G_n)^{\# s}
w_{11}^{\# s/\# 1} \prod_{r=2}^s W_{n,r}^{\# s/\# r} =
$$ $$ =
\left(d_n^n \prod_{\stackrel{k|n}{k<n}}
r_{k,n}^{n-k}\right)^{(d-1)d^{s-1}}
w_n^{\delta(n,s)} \prod_{r=1}^s w_{n,r}^{d^{s-r}}
$$
Index $\delta(n,s)$ is obtained by summation over all
$1<r\leq s$, such that $r = 1 \ {\rm mod}\ n$, of the weights
$\# s/\# r = d^{s-r}$. Each term in this sum corresponds to
the tree, warped $\frac{n}{r-1}$ times
around the periodic orbit, see Fig.\ref{warpedtree}.
For $s=nr'+r$ and $r',r>0$ we have
$$\delta(n,s) = d^{r-1}\frac{d^{nr'}-1}{d^n-1} =
\frac{d^{s-1}-d^{r-1}}{d^n-1}$$
If such $r'$ and $r$
do not exist, i.e. if $s\leq n$, $\delta(n,s) = 0$.

\FIGEPS{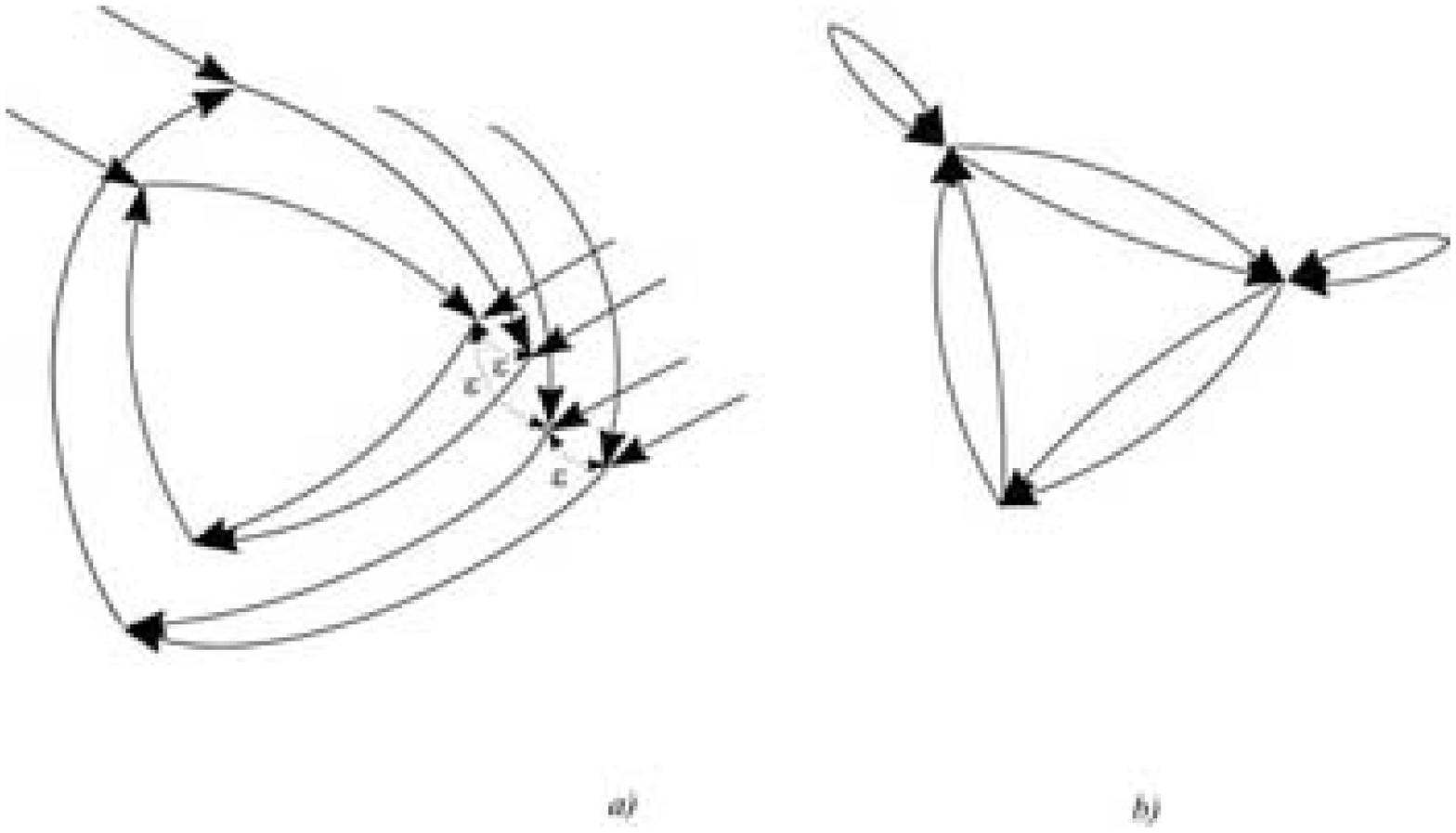}{}
{454,258}
{a) Pre-orbit tree, warped around the periodic
orbit of order $n=3$. b) The limit degenerate orbit with multiplicities.}

\subsection{Combinatorics of discriminants and resultants}

Systematic analysis of orbits of the map $f: \ \bb{X}
\rightarrow \bb{X}$ includes determination of
the following characteristics, used in our presentation
in s.\ref{summ}:

\bigskip

$\bullet$ Number $N_n(f)/n$ of periodic orbits of order $n$,
which is $n$ times smaller than the number $N_n(f)$ of roots
of $G_n(x;f)$, $$N_n(f) = {\rm deg}_x[G_n].$$

$\bullet$ Number $S(n;f)$ of different combinations of
periodic orbits of order $n$ which {\it can} be
stable for some $f$. (It deserves emphasizing, that some orbits are never
stable: for $f=x^2+c$ only one of the two fixed points -- orbits of order
one -- has non-vanishing stability domain in the plane of complex $c$.)

$\bullet$ Number $\tilde {\cal N}_n(f)$ of elementary components $\sigma$
in stability domain $S_n$. It is also useful to count separately
the numbers ${\cal N}_n^{(p)}(f)$ of elementary components $\sigma$
at level $p$,
$$\tilde {\cal N}_n(f) = \sum_{p=0}^\infty {\cal N}_n^{(p)}(f).$$

$\bullet$ Number ${\cal N}_n(f)$ of solutions to the equation
$G_n(x=0,f) = 0$, considered as an equation for $f$.
Solutions of this equation for particular
families of maps $\mu \subset {\cal M}$
are also important, because they
define some points inside the $\sigma$-domains
(and thus -- approximately --
the location of domains). If the family $\mu$ is one-parametric and
parameter is $c$, then
$${\cal N}_n(f) = {\rm deg}_c[G_n].$$
Presumably, $\tilde S(n;f) = \tilde{\cal N}_n(f) = {\cal N}_n(f)$.

$\bullet$ Number of zeroes of reduced
discriminant $d_n = d(G_n)$, introduced
in s.\ref{redi}. For one-parametric families of maps $f$ the number
of zeroes is equal to $\ {\rm deg}_c[d_n]$.
Zeroes of $d_n$ are associated with self-intersections of
periodic orbits and define the cusps on various components.
Usually the number $q$ of cusps depends only on the level of $p$
of the component $\sigma_n^{(p)}$ (and not on $n$ and other
parameters like $m_i$ and $\alpha_i$).

$\bullet$ Number of zeroes of reduced resultants
$r_{n,k} = r(G_n,G_{k})$.
For one-parametric families of maps $f$ the number
of zeroes is equal to $\ {\rm deg}_c[r_{n,k}]$.
Each zero describes
some merging of the elementary $\sigma$-domains.
The number of zeroes is
related to the number of $\alpha$-parameters.

\bigskip

All these numbers are related by sum rules:
\be
\sum_{p=0}^\infty {\cal N}_n^{(p)} = {\rm deg}_c [G_n(w_c,c)], \nn \\
\sum_{p=0}^\infty q(p){\cal N}_n^{(p)} = {\rm deg}_c [d_n], \nn \\
\sum_{p=1}^\infty {\cal N}_n^{(p)} =
\sum_{k|n}{\rm deg}_c[r_{n,k}].
\label{sumrule}
\ee

\bigskip

{\bf Example.}
For $f$ which is a polynomial of degree $d$, $f(x) = x^d + c$, we have:

\bigskip

$\bullet$ $n=1:$

\# of orbits: $d$,

\# of "potentially-stable" orbits: $S(1;x^d+c) = 1$.

Since $G_1(0;f) = f(0) = a_0 = c$, the number of solutions
to $G_1(0;c) = 0$ is $1$.

Discriminant $d(G_1) = D(G_1) = D(F_1)$ vanishes when simultaneously
$x^d + c = x$ and $dx^{d-1} = 1$: there are $d-1$ solutions, thus
discriminant has $d-1$ zeroes.

\bigskip

$\bullet$ $n=2:$

\# of orbits: $\frac{d^2-d}{2}$,

\# of "potentially-stable" orbits: $S(2;x^d+c) = d-1$.

Since $G_2(0;c) = c^{d-1}+1 = 0$, the number of solutions
to $G_2(0;c) = 0$ is $1$.

Discriminant $d(G_2) = \sqrt{\frac{D(G_2)}{R(G_1,G_2)}}$ has
$(d-1)(d-2)$ zeroes. Each of the $d-1$ elementary stability domains for
orbits of order two has $d-2$ cusps.

\bigskip

We list much more data of this kind for particular families of maps
in Tables in s.\ref{Exa2} below.
Similar analysis can be performed for bounded grand orbits.

\subsection{Shapes of Julia and Mandelbrot sets}

\subsubsection{Generalities}

The shape of Mandelbrot can be investigated with the
help of stability constraints (\ref{stabcon}):
\be
\left\{ \begin{array}{c}
|F_n'(x;f) + 1| = 1,  \\
G_n(x;f) = 0.
\end{array} \right.
\label{stabco}
\ee
Exclusion of $x$ from these equations provides a
real-\-codimension-\-one subspace in
$\mu\subset{\cal M}$, which defines the boundary $\partial S_n$ of
stability domain $S_n$. Its connected components
are boundaries of the elementary domains $\sigma$.
Enumerating all $n$ we can obtain in this way the entire
boundary of Mandelbrot space $M(\mu)$.

Zeroes of all discriminants $D(G_n)$ belong to this boundary.
Indeed, if $D(G_n) = 0$ there is a point $x_n \in \bb{X}$ which
is common root of $G_n$ and $G_n'$: $G_n(x_n) = G'_n(x_n) = 0$
(actually, there is entire set of such points, labeled by additional
$\alpha$-index).
Then, since $F_n = \prod_{k|n} G_k = G_n \tilde F_n$, we
have $F_n'(x_n) = G_n(x_n) \tilde F_n'(x_n) +
G_n'(x_n) \tilde F_n(x_n) = 0$ and thus both equations
(\ref{stabco}) are satisfied. Furthermore, since according to
(\ref{Dd}), a zero of every resultant $R(G_n,G_{n/m})$
is also a zero of $D(G_n)$, we conclude that
zeroes of all such resultants also belong to $\partial M$.
In the world of iterated maps only resultants
of this kind are non-trivial, $R(G_n,G_k) = const$ unless
$k|n$ or $n|k$.

One possibility to define the shape of $M(\mu)$ is just to
{\bf plot many enough zeroes of discriminants and resultants}
and -- since they are dense in $\partial M(\mu)$ -- this provides
approximation with any desired accuracy.
Advantage of this approach is that it is pure algebraic and --
once formulated -- does not contain any reference to stability
equations and to the notion of stability at all.
The disadvantage is that -- at least in presented form --
it does not separate points from different elementary domains:
zeroes of each particular resultant are distributed between
many elementary domains, which can even belong to different
disconnected components $M_{k\alpha}$ of $M$. Also, it does not
describe directly the boundaries of elementary
domains, which -- outside a few cusps, located at zeroes of
reduced discriminants $d(G_n)$,-- are smooth curves of peculiar
(multi-cusp) cardioid-like form. Finally, it does not explain the
similarity of various components $M_{k\alpha}$, which -- for
reasonably chosen families $\mu$ -- belong to a set of
universality classes labeled by $\bb{Z}_{d-1}$-symmetries.

Another option is to use equations (\ref{stabco}) more intensively
and transform them into less transcendental form, making
reasonable reparametrizations and approximations. The
choice of parameterization, however,
can impose restrictions on the families $\mu$, but instead one can
move much further in explicit description of constituents of
Mandelbrot and even Julia sets.

In the case of Julia sets the situation with non-algebraic
approaches is somewhat more difficult. These
sets do not have elementary smooth constituents, except for
exactly at the bifurcation points, i.e. at the boundary points of
Mandelbrot set, but even on $\partial M$ the decomposition
of $J(f)$ depends discontinuously on the
point at the boundary. As soon as one goes inside Mandelbrot set,
infinitely many different smooth structures interfere and only
some traces of them can be
seen approximately. The structure of Julia sets is pure
algebraic, not smooth (or, if one prefers, consistent with $d$-adic
rather than with $\bb{C}$-topology). The boundary of Julia set for
given $f$ is formed by solutions of all the equations $F_{n,s} = 0$
for all $n$ and $s$, with exclusion of a few
orbits, which are stable for this $f$, -- and
this remains the best existing {\it constructive} definition of
Julia set in general situation. Also, approximate methods
can be used to analyze some features of Julia sets.

\subsubsection{Exact statements about $1$-parametric families
of polynomials of power-$d$}

Assuming that $\mu \subset {\cal P}_d$ and ${\rm dim}_{\bb{C}}\mu = 1$,
we can label the maps $f$ in the family by a single parameter $c$:
$\mu = \{f_c\}$.
For given $c \in M_1(\mu)$ define:
\be
F'_n(x;c) + 1 = e^{i\varphi (d-1)}.
\label{varphipar1}
\ee
This relations maps the roots of $G_n(x;c)$ (the periodic orbits of
order $n$) into the unit circle, parameterized by the angle $\varphi$.
Considering all possible $n$ and taking the closure, one can
extend this map to entire boundary of Julia set.
For $c \in M_1$ this provides a one-to-one correspondence
between $\partial J_c$ and the unit circle.
If $c \in M_{k\alpha}$, analogous expression,
$$
F'_{nk}(x;c) + 1 = e^{i\varphi (d-1)},
$$
relates a {\it part} of $\partial J_c$ and unit circle.
Other parts are mapped onto additional circles.
Parameterization (\ref{varphipar1}) is adjusted to describe
the boundary of Mandelbrot set: it explicitly solves the first
equation in (\ref{stabco}) and the function $x(\varphi)$ can be
substituted into $G_n(x;c) = 0$ to obtain $c(\varphi)$. The
map $c(\varphi)$ has several branches, associated with different
elementary domains $\sigma$ from stability domain $S_n$.
This program can be realized in certain approximation.

Before going to approximations, let us give an example of exact
statement. The map $S^1 \rightarrow \partial \sigma$ is
singular (there are cusps on the boundary $\partial\sigma$)
whenever
\be
\left\{\begin{array}{c}
\left|\frac{\partial c}{\partial \varphi}\right| =
(d-1)\left|\frac{G_n'}{H_n}\right| = 0,  \\
G_n = 0.
\end{array}\right.
\label{cuspeq}
\ee
Here $\dot G_n = \partial G_n/\partial c$ and
$$
H_n = \{F_n', G_n\} =
\dot F_n'G_n' - F_n'' \dot G_n.
$$
Solutions to this system are all zeroes of the resultant
\be
R(G_n,G_n') \sim D(G_n) = d^n(G_n) \prod_{k|n} r^{n-k}(G_n,G_k)
\label{RGG'D}
\ee
which are {\it not} simultaneously zeroes of another resultant,
$R(G_n',H_n)$, or, what is much simpler to check, $R(G_n,H_n)$.
Actually excluded are roots of all reduced resultants $r(G_n,G_k)$,
and the cusps of $\partial\sigma$ are at zeroes
of reduced discriminants $d(G_n)$.
Indeed, whenever $r(G_n,G_k) = 0$ for $k|n$, there is a common zero
$\tilde x$ of the two functions $G_n(\tilde x) = G_k(\tilde x) = 0$.
Also, because of (\ref{RGG'D}), $G'_n(\tilde x) = 0$, therefore
$H_n(\tilde x) = -F_n''\dot G_n(\tilde x)$.
Both $G_n$ and $G_k$ enter the product (\ref{rnk}) for $F_n$,
$F_n \sim G_nG_k$ and two derivatives in $F_n''$ are not enough to
eliminate $G_n$, $G_k$ and $G_n'$ from the product, therefore also
$F_n''(\tilde x) = 0$ and $H_n(\tilde x) = 0$.

For polynomial maps elementary domains of  Mandelbrot
set $M(c)$ belong to universality classes, which
are represented by multi-cusp cardioids (cycloids),
described by the function $\varepsilon(\varphi)$
in complex $\varepsilon$-plain
\be
C_0:\ \ \ \varepsilon = e^{i\varphi}, \nn \\
C_{d-1}:\ \ \ \varepsilon = e^{i\varphi} - \frac{1}{d}e^{i\varphi d},
\ \ \ {\rm for}\ d>1. \nn
\ee
The real curve $C_{d-1}$ has discrete symmetry
$$
\bb{Z}_{d-1}:\ \ \varphi \rightarrow \varphi + \theta,\ \
\varepsilon \rightarrow e^{i\theta}\varepsilon,
$$
it is singular,
$\left|\partial\varepsilon/\partial\varphi\right| = 0$, provided
$\varphi = \frac{2\pi ik}{d-1}$, $k = 0,1,\ldots,d-2$, i.e. the
number of cusps is $d-1$.

Generically, the boundaries of  elementary domains belong to
the class $C_1$ for level $p=0$ and to $C_0$ for all other levels
$p\geq 1$. However, for special families, like
$f_c = x^d + c$, the situation will be different: at level $p=0$
the elementary domains $\sigma^{(0)}$ can belong to the class
$C_{d-1}$ (has $d-1$ cusps), while at higher levels it decreases to
$C_{d-2}$ (has $d-2$ cusps). If symmetry is broken by adding
lower powers of $x$ with small coefficients to $f_c$, the small-size
components $M_{k\alpha}$ fall into generic $C_0\oplus C_1$ class,
and the bigger the symmetry-breaking coefficients, the bigger
are components that switch from $C_{d-2}\oplus C_{d-1}$
to $C_0 \oplus C_1$ class.

\subsubsection{Small-size approximation
\label{ssa}}

In order to explain how cardioids arise in description of elementary
domains, we continue
with the example of one-parameter family of
polynomials of degree $d$ and
describe an {\it approximation}, which can be used to explain the
above mentioned results. It is based on expansion of equations
(\ref{stabco}) around the point $(x,c) = (w,c_n)$, where
$w_c$ is a critical point of $f_c(x)$ and $c_n$
is the "center" of an elementary domain from $S_n$, defined from
solution of the equations
\be
\left\{ \begin{array}{c}
f_c'(w_c)=0,  \\
G_n(w_c;c) = 0.
\end{array} \right.
\label{cripog}
\ee
We now substitute into (\ref{stabco}) $x = w + \chi$ and
$c = c_n + \varepsilon$, expand all the functions in
powers of $\varepsilon$ and $\chi$, and leave the first
relevant approximation. Typically $|\chi| < d^{-n}$ and
$|\varepsilon| < d^{-n}$, actually there is much stronger
damping for elementary domains belonging to "remote" $M_{k\alpha}$,
thus approximation can be numerically very good in most situations.

One can even promote this approach to
alternative description of Mandelbrot set as a blow up of
the system of points (\ref{cripog}): each point gets surrounded
by its own domain $\sigma$. Some of these domains get big enough
to touch each other -- and form the connected components
$M_{k\alpha}$,-- some remain disconnected topologically, but get
instead connected by {\it trails}. Consistency of such
approach and the one, based on resultant's zeroes, requires that
the number of solutions to (\ref{cripog}) with given $n$
coincides with the total number of zeroes of $r_{nm}(c)$ with
all $m < n$ plus the number of components $M_{n\alpha}$
growing from the elementary domain $\sigma^{(0)}[n\alpha]$,
\be
{\rm deg}_c G_n(w_c,c) = |\nu_n^{(0)}| +
\sum_{m|n} {\rm deg}_c r_{n,m}(c)
\ \ \ \forall {\rm \ 1-parametric\ families\ } \mu
\label{sumrule1}
\ee
where $\# F(c) = {\rm deg}_c F(c)$ denotes the number of roots of
$F(c)$, which for polynomial $F(c)$ is equal to its degree
(for multiparametric families this is the relation between the
degrees of algebraic varieties).
A similar sum rule related the number of roots of reduced
determinant $d_n(c)$ and the numbers ${\cal N}_n^{(p)}$ of solutions to
(\ref{cripog}) with $q(p)$ cusps:
\be
\sum_p q(p){\cal N}_n^{(p)} = {\rm deg}_c d_n(c).
\label{sumrule2}
\ee

\subsubsection{Comments on the case of $f_c(x) = x^d + c$
\label{xdExa}}

In this case there is a single critical point $w_f = 0$.

Stability domain $S_1$ for orbits of order one
consists of the single elementary domain,
$S_1 = \sigma[1]$ and is defined by a system
\be
d|x^{d-1}| = 1, \nn \\
x^d - x + c =0.
\label{boundd1}
\ee
Its boundary can be parameterized as follows:
$x = d^{-1/(d-1)}e^{i\phi}$, and
\be
\partial S_1:\ \
c = x - x^d = \frac{1}{d^{1/(d-1)}}
\left(e^{i\phi} - \frac{1}{d}e^{i\phi d}\right)
\label{boundd}
\ee
This is a curve with a cusp at $c = {d-1}/{d^{d/(d-1)}}$
and the symmetry group $\bb{Z}_{d-1}$, the shift $\phi \rightarrow
\phi + \frac{2\pi}{d-1}$ multiplies $c$ by $e^{\frac{2\pi i}{d-1}}$
(so that there are actually $d-1$ cusps).
Julia sets have another symmetry, $\bb{Z}_d$.

\bigskip

We use this example to show how approximate similarity
of different components $M_{k\alpha}$ of Mandelbrot set can be
explained by "perturbation theory" from the previous
subsection \ref{ssa}. We discuss just a very
particular part of the story: the central domains
$\sigma^{(0)}[k\alpha]$, obtained by
the procedure from s.\ref{summsim}.
The procedure implies that we solve the equations (\ref{stabcon})
\be
\left\{ \begin{array}{c}
F_n(x;c) = 0,  \\
|F'_n(x;c) + 1| = 1
\end{array} \right.
\label{boueq}
\ee
by expansion near the point $(x,c) = (0,c_0)$, where
$c_0$ is a root of the equation $F_k(0,c_0) = 0$. This means that
we now write $x = 0 + \chi$, $c = c_0 + \varepsilon$ and assume
that $\chi$ and $\varepsilon$ are small -- and this will be justified
{\it a posteriori}.
In this approximation
$$
F_n(\chi,c) = c_n(c) - \chi + b_n\chi^d + O(\chi^{2d}),
$$
and
\be
c_{n+1} = c_n^d + c, \nn \\
b_{n+1} = b_nc_n^{d-1} d. \nn
\ee
At $c=c_0$ the $c_k(c_0) = 0$ and $c_k(c_0 + \varepsilon) =
\dot c_k \varepsilon$ (dot denotes derivative with respect to $c$).
Substituting $F_k = \dot c_k(c_0) \varepsilon -\chi + b_k(c_0)\chi^d
+ O(\chi^{2d})$
into (\ref{boueq}), we get in this approximation:
\be
db_k|\chi^{d-1}| = 1, \nn \\
b_k\chi^d - \chi + \dot c_k\varepsilon = 0, \nn
\ee
i.e. $\chi = \rho_k e^{i\phi}$,
$\rho_k = (b_kd)^{-1/(d-1)}$ and
\be
\partial \sigma_k: \ \
\varepsilon = \frac{x - b_kx^d}{\dot c_k} =
\frac{\rho_k}{\dot c_k}
\left(e^{i\phi} - \frac{1}{d}e^{i\phi d}\right)
\label{bounddk}
\ee
Thus such
$\partial \sigma_k$ are approximately similar to
$\partial S_1 = \partial\sigma_1$, described by eq.(\ref{boundd}).

Approximation works because the radius $\rho_k$ is numerically
small. Presumably it can be used to find Feigenbaum indices and their
$d$-dependence.
The same approximation can be used in the study of Julia sets
for $c \in M_{k\alpha}$ with $k>1$.
From this calculation it is also clear that the shape of $M_{k\alpha}$
is dictated by the symmetry of the problem: for families $\mu$
of maps, which do not have any such symmetries one should expect all
the higher $M_{k\alpha}$ to be described by (\ref{bounddk}) with
$d=2$. Obviously, Feigenbaum parameters depend only on effective value of
$d$. Also, one can easily find families $\mu$, for which above
approximated scheme will not work: then one can expect the
breakdown of self-similarity of the Mandelbrot set.

However, if one tries to be more accurate, things get more sophisticated.
Precise system (\ref{stabco}) contains the condition $G_n(x;c) = 0$
rather than $F_n(x;c) = 0$ in (\ref{boueq}). In variance with $F_n(x)$,
which has a gap between $x$ and $x^d$, $G_n(x)$ contains all powers of $x$,
for example, for $n=2$
$$
G_2 = 1 + \sum_{k=0}^{d-1}c^{d-k-1}\chi^{k} +
\sum_{k=0}^{d-2}(d-k-1)c^{d-k-2}\chi^{d+k} + \ldots  +
\chi^{d^{k-1}}.
$$
This makes analysis more subtle. We leave this issue, together with
approximate description of Julia sets to the future work.

\bigskip

\subsection{Analytic case \label{analyt}\label{cont}}

Before ending discussion of approaches to
creation of theory of the universal Grand
Mandelbrot (discriminant)
variety and Julia sets and proceeding to examination
of various examples, in the remaining two subsections
we explain how the relevant
notions look in the case of generic analytic functions,
not obligatory polynomial. As already mentioned, the
theory of discriminant, resultant and
Mandelbrot varieties can be made pure topological
and free of any algebraic structure, however in this case
one can loose other interesting -- pure algebraic -- examples
(like $\bb{X} = \bb{F}_p$, to begin with) as well as the
skeleton machinery, based on the use of functions $F_n$
and $G_k$
(to define them one needs subtraction operation and entire
ring structure respectively) and ordinary
discriminant calculus (which is again algebraic).
Thus in the remaining subsections we accept that these structures
are important and explain instead, why we do {\it not} think that
restrictions to polynomials is needed anywhere.
Actually, polynomials play distinguished role in
particular examples: this is because the vocabulary for
iterations of other functions -- even trigonometric --
was never developed and we do not have any adequate language
to discuss, say
$$
G_2(x; \sin\omega x) =
\frac{\sin(\omega\sin \omega x) - x}{\sin \omega x - x}
$$
or
$$
G_{1,1}(x; \sin\omega x) =
\frac{\sin(\omega\sin \omega x) - \sin\omega x}{\sin \omega x - x}
= G_2(x; \sin\omega x) - 1
$$
or there no-\-less-\-interesting {\it trigonometric} counterparts
$$
\tilde G_2(x; \sin\omega x) =
\frac{\sin(\omega\sin \omega x) - \sin x}{\sin \omega x - \sin x},
$$ $$
\tilde G_{1,1}(x; \sin\omega x) =
\frac{\sin(\omega\sin \omega x) - \sin\omega x}
{\sin \omega x - \sin x} = \tilde G_2(x; \sin\omega x) - 1
$$
While iterated polynomial is always a polynomial, just
of another degree, even iterated exponential
or iterated sine do not have names.
This could be not a too
important restriction if one agrees to rely
more on computer experiments than on theoretical considerations,
but even here existing symbolic-calculus programs are
better adjusted to work with polynomials...
Still, non-polynomial singularities play increasingly important
role in modern theories, they appear already in the
simplest examples of $\tau$-functions of KdV and KP equations,
nothing to say about more general partition functions. Thus
we find it important to emphasize that discussion of
phase transitions and their hidden algebraic structure in the
present paper is by no means restricted to polynomials.

\bigskip

Take an arbitrary (locally) analytic function $f(x)$. By the same argument
as in the polynomial case $f$ maps roots of $F_n(x)$ into roots, giving us
the a representation of $\bb{Z}_n$ on the set of zeros of
$F_n(x)$,\footnote{
This is because if $x_0$ is a root of $F_n(x) = f^{\circ n}(x) - x$, then
$f(x_0)$, $f^{\circ 2}(x_0)$ etc are also roots, and $f^{\circ n}(x_0) =
x_0$.
}
which is the analogue of the Galois group action for polynomials. The orders
of orbits of this actions are also divisors of $n$, but the set of those
orbits now is infinite for each given $n$. Although for series there is no
notion of divisibility, still
if $n$ is divisible by $k$, then $F_k(x)=0$ implies $F_n(x)=0$.
Indeed, if $n=km$ and $f^{\circ k}(x_0) = x_0$, then
$f^{\circ n}(x_0) = f^{\circ(km)}(x_0) =
f^{\circ (k(m-1))}(f^{\circ k}(x_0)) = f^{\circ (k(m-1))}(x_0)
= \ldots = x_0$, i.e. every root of $F_k(x)$
is obligatory the root of $F_n(x)$.

\bigskip

{\bf Example:}
Take $f(x)=\sin(x)+cx+bx^4$. Fig.\ref{sin4} shows real roots of
$F_4(x)$ and $F_2(x)$  for $c=2,\ b=-0.01$.
The root $x\approx 2.39$ is the
root of $F_2(x)$ and its orbit has order $2$.
The root $x\approx 2.03$ is
the root of $G_4$ (independent of $F_2(x)$)
and its orbit has order $4$.

\FIGEPS{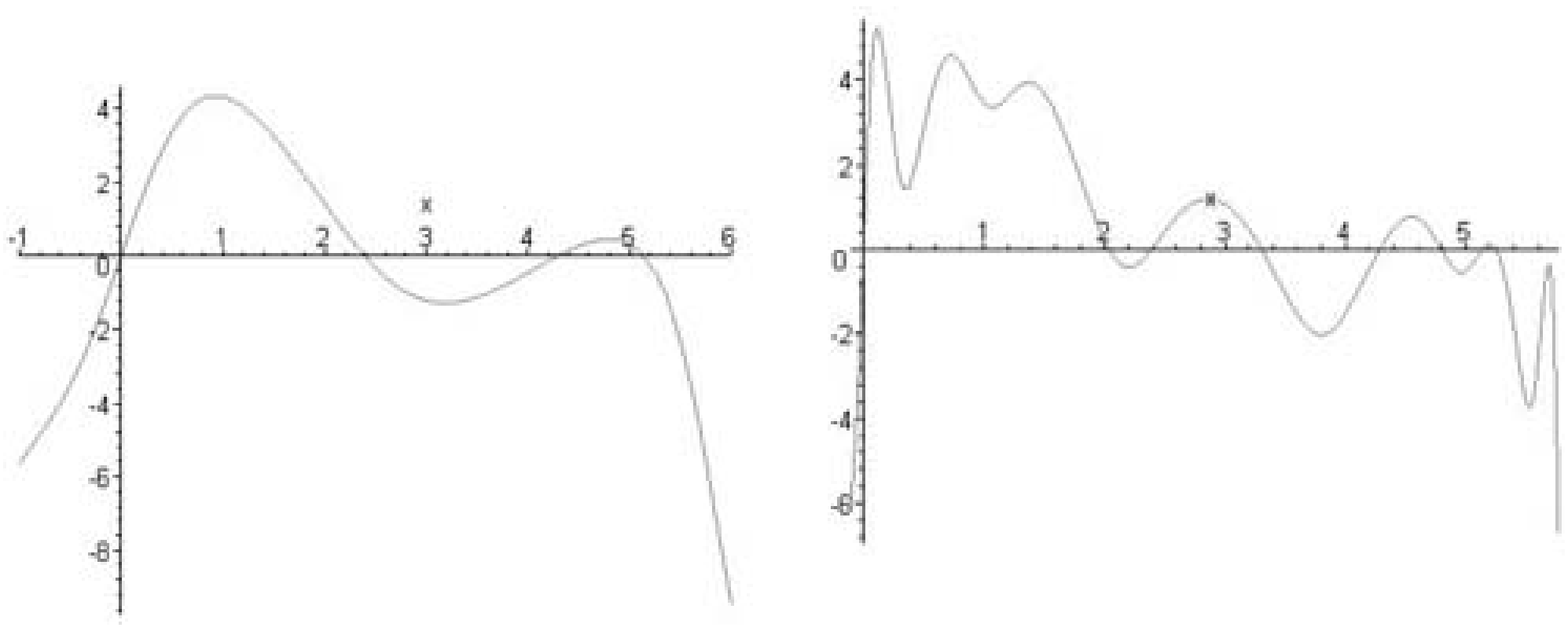}{}
{409,162}
{Real roots of $F_2(x;f)$ and $F_4(x;f)$ for the map
$f(x) = \sin(x) + 2x -0,01x^4$.}

\bigskip

If $U$ is the definition domain (the set of regular points) of $f$,
then $f\circ f$ and $F_2$ are defined in $U_{\circ 2} = U\cap f(U)$
and so on. Denote by
$$
U^\infty(f) = \cap_{n=0}^\infty f^{\circ n}(U)
$$
the inverse limit of the sequence
$$
U\hookleftarrow
U_{\circ 2}\hookleftarrow\dots\hookleftarrow
U_{\circn}\hookleftarrow\dots
$$
For polynomial $f(x)$ the set  $U^\infty(f)=\bb{C}$
is the whole complex
plane.

The $\bb{Z}_n$ representation on the roots of $F_n$ is also a
representation of $\bb{Z}$, which we denote by $V_n$.
In $n$ is divisible by $k$, then every root of $F_k$ is the root of $F_n$
and we have an embedding, which is actually a
morphism of $\bb{Z}$-representations $V_k\to V_n$.
This provides a directed set of
$\bb{Z}$-representations
\be
\begin{array}{ccccccc}
&&V_4&\to&\vdots&&\\
&\nearrow&&\searrow&&&\\
V_2&&&&V_{12}&\to&\vdots\\
&\searrow&&\nearrow&&&\\
&&V_6&\to&\vdots&&\\
&\nearrow&&&&&\\
V_3&\to&\vdots&&&&\\
\dots&&&&&&
\end{array}
\nn
\ee
Its direct limit $V_\infty(f)=\cup_n V_n$ is a representation of
$\bb{Z}$ which is a pertinent characteristic of the map $f$.
All roots of every $F_n$ satisfy $x=f^{\circ n}(x) = f^{\circ(nm)}(x)$
for any $m$, thus they belong to $U^\infty$, i.e.
each orbit of $V_\infty$ totally belongs to $U^\infty$.
Thus we have the
following hierarchy of subsets, characterizing a holomorphic function $f$:
$$
V_\infty(f) \subset U^\infty(f) \subset\ldots\subset U_{\circ 2}(f)
\subset U(f) \subset\bb{C}
$$
The closure of $V_\infty$-space is called
the {\bf algebraic Julia set} $J_A$ of $f$.

\bigskip

{\bf Hypothesis:} This definition coincides with the one
in sec.\ref{graob}.

\bigskip

{\bf Example:}
For $f(x)=x^d$ the Julia set $J(f)$ is the unit circle $S^1\subset
\bb{C}$ and the $V_\infty$ space is an everywhere dense subset in this
circle.

\bigskip

In the general case, when $f$ is a series with complex coefficients,
there is no distinguished $f$-invariant subsets analogous to
$\bb{R}\subset\bb{C}$. But the existence of $f$-orbits
implies, that
a deformation of coefficients of $f(x)$, which makes any two
roots of some
$F_n(x;f)$ lying on different orbits in
$V_\infty(f) \subset U^\infty(f)$ coincide,
results in simultaneous pairwise merging of {\it all}
the roots on those orbits,
so the resulting orbit acquires multiplicity.

For given $n$ take the subset $C_n$ in the space of coefficients $a$ of $f$
where some of the orbits in the set of zeros of $F_n$ merge,
i.e. $f$ belongs to the $I_n$-pullback of the discriminant component
${\cal D}_{\circn}$.
The closure of $C_\infty:=\cup_n C_n$ is the boundary $\partial
M$ of the {\bf Mandelbrot set} of $f$.

\bigskip

{\bf Hypothesis:}
For any finite-parameter family $\mu \subset {\cal M}$ of maps
$\overline{C_\infty}(\mu)$ is a subset of {\it real}
codimension $1$ in the family $\mu$).

\bigskip

Provided this is true,
$\partial M$ separates the space of complex coefficients
into disjoint components.
Note, that every particular $C_n$ has real codimension $2$
(since the two equations
$F_n(x)=0$ and $F_n'(x)=0$ can be used to define a complex
common root $x$ and
one complex relation between the coefficients of $F_n(x)$ --
and thus between the
coefficients of $f$).

\subsection{Discriminant variety ${\cal D}$ \label{disc}}

This section reminds the standard definitions of discriminants
and resultants for the case of polynomial functions and briefly
comments on the possible ways of generalization to the case of
arbitrary analytic functions.

\subsubsection{Discriminants of polynomials}

In the polynomial case
${\cal D}({\cal P}_n)$ is defined \cite{Lang} as an algebraic variety
in ${\cal P}_n$ by an equation
\be
D(F) = 0
\label{disceq}
\ee
in the space of coefficients.
Here $D(F)$ is the square of the Van-der-Monde product,
for $F(x) = v_n\prod_{k=1}^n (x-\alpha_k)$
\be
D(F) = v_n^{2n-2}\prod_{k<l}^n (\alpha_k - \alpha_l)^2
\label{discVdM}
\ee
Remarkably, it can be also expressed as a {\bf resultant} of
$F(x)$ and its derivative $F'(x)$
\be
D(F) = (-)^{n(n-1)/2}v_n^{-1} R(F(x),F'(x)),
\label{discres}
\ee
which is a polynomial of the coefficients of $F(x)$
(while the individual roots $\alpha_k$ can be sophisticated
functions of these coefficients, not even expressible in radicals),
so that the equation (\ref{disceq}) is indeed algebraic.

The resultant of two polynomials
$F(x) = \sum_{k=0}^n v_k x^{k} = v_n\prod_{k=1}^n (x-\alpha_k)$ and
$G(x) = \sum_{k=0}^m w_k x^{k} = w_m\prod_{l=1}^m (x-\beta_l)$ is
defined as a double product
\be
R(F,G) = v_n^m w_m^n \prod_{k,l} (\alpha_k - \beta_l) =
v_n^m \prod_{k=1}^n G(\alpha_k) =
(-)^{nm} w_m^n \prod_{l=1}^m F(\beta_l),
\label{resulroots}
\ee
depends on symmetric functions of the roots $\{\alpha\}$ and/or
$\{\beta\}$, thus -- according to Vieta formula -- depends only on
the coefficients of $F$ and/or $G$,
and is representable as determinant
\be
R(F,G) = \det_{(m+n)\times (m+n)}
\left(
\begin{array}{ccccccc}
v_0 & v_1 & \ldots & v_n & & & \\
    & v_0 & v_1 & \ldots & v_n & & \\
&&&\ldots&&& \\
&&& v_0 & v_1 & \ldots & v_n \\
w_0 & w_1 & \ldots & w_m & & & \\
    & w_0 & w_1 & \ldots & w_m & & \\
&&&\ldots&&& \\
&&& w_0 & w_1 & \ldots & w_m
\end{array}
\right)
\label{discf}
\ee
The resultant obviously vanishes when $F(x)$ and $G(x)$ have a common
root $x_0$, accordingly the matrix in (\ref{discf})
annihilates the column-vector $(1,x_0,x_0^2,\ldots,x_0^{n+m})$.

Directly from the definitions it follows that discriminant of a product
of two polynomials is decomposed in the following way
\be
D(FG) = D(F)D(G)R^2(F,G).
\label{resfact}
\ee

If $F(x) - H(x)$ is divisible by $G(x)$,
$F = H\ {\rm mod}\ G$ (for example, $H$ is the residual
from $F$ division by $G$), then (\ref{resulroots})
implies that
\be
R(F,G) \sim \prod_{\beta: \ G(\beta) = 0} F(\beta)
\sim \prod_{\beta: \ G(\beta) = 0} H(\beta) \sim\
R(H,G)\ \sim \prod_{\gamma: \ H(\gamma) = 0} G(\gamma)
\label{resresu}
\ee
where proportionality signs imply the neglect of factors
like $v_n$, $w_m$ and minus signs in (\ref{resulroots}),
in this form the relation is applicable to any analytic
functions, not obligatory polynomials.
(Note that the products at the r.h.s. of (\ref{resresu})
over the roots of $H$ and, say, of $H+\kappa G$ will be
the same -- up to above-mentioned rarely essential factors.)

\subsubsection{Discriminant variety in entire ${\cal M}$}

To define the discriminant in analytic (non-polynomial) case one can make
use of both definitions (\ref{discVdM}) and (\ref{discres})
and either take the limit of a
double-infinite product over the roots or handle determinant of an infinite
matrix. It is an open question, when exactly this can be done and when the
two limits can coincide.

A possible approach in the case of determinantal representation
 can make use of the
following recursive procedure.
Discriminant of a polynomial $a_0+a_1x+\ldots+a_{d+1}x^{d+1}$ may be
written as (pre-factor $a_{d+1}^{-1}$ in (\ref{discres}) is important here)
$$
D_{d+1}\{a_0,\ldots,a_{d+1}\}
=a_d^2\cdot
D_d\{a_0,\ldots,a_d\}+\sum_{k=1}^d b_{d+1,k}\{a_0,\ldots,a_d\}\cdot
a_{d+1}^k,
$$
where $D_d\{a_0,\ldots,a_d\}$ is discriminant of
$a_0+a_1x+\ldots+a_{d}x^{d}$ of the previous degree
and $b_{d+1,k}\{a_0,\ldots,a_d\}$ are
polynomials of coefficients $a_0, \ldots, a_d$.
Then for an analytic function locally represented by series $f(x)=\sum
a_ix^i$ with numerically given coefficients $\{a_i\}$ we have a sequence
\be
\check D_1=1,\ \ \check D_2=a_1^2-4a_0a_2,\nn \\
\check D_3 = a_2^2(a_1^2-4a_0a_2) +
(18a_0a_1a_2-4a_1^3)a_3 - 27a_0^2a_3^2, \nn \\
\dots,\ \check D_d,\dots \nn
\ee
of numbers, equal to
the values of the corresponding polynomials. The numeric limit of this
sequence may be regarded as the value of the infinite expression
$$\dots(a_d^2\cdot\dots(a_3^2\cdot(a_2^2\cdot(a_1^2\cdot
1+b_{21}\{a_0,a_1\}a_2)$$
$$+b_{31}\{a_0,a_1,a_2\}a_3+b_{32}\{a_0,a_1,a_2\}a_3^2)$$
$$+b_{41}\{a_0,a_1,a_2,a_3\}a_4+
b_{42}\{a_0,a_1,a_2,a_3\}a_4^2+b_{43}\{a_0,a_1,a_2,a_3\}a_4^3)$$
$$+\dots)+\sum_{i=1}^d b_{{d+1},i}\{a_0,\dots,a_d\}a_{d+1}^i )\dots$$
(this computation is an algebraic counterpart of the chain fractions)
which may be called the {\bf discriminant} $D(f)$
of the analytic function $f$.

Since in the case of polynomial $f$ the coefficients
of all functions $F_n$ are polynomials $\{a^{(n)}_k(a)\}$
of the coefficients $\{a_k\}$ of the series $f$, such procedure can
work simultaneously for $f$ and $F_n(f)$.
The possibility of generalizations from polynomial to analytic case can be
formulated in the form of the following

\bigskip

{\bf Hypothesis:}
$\{a_i\}\in C_n$ iff $D(F_n)=0$, i.e. some orbits within the radius of
convergence of series $f=\sum a_ix^i$ merge iff the above defined
discriminant $D(F_n)$ of series $F_n=\sum a^{(n)}_k(a)x^k$ is zero.

\bigskip

{\bf Example:} For $f_c(x) = e^{x^2} - c$ we have
$\check D_1 = 1,\ \check D_2 = 4(c-1),\ \check D_3 = 4(c-1),\
\check D_4 = 8(c-1)(2c-1)^2,\ \ldots$. It is easy to check that all higher
$\check D_k$ will be also proportional to $c-1$, so that discriminant
$D\left(e^{x^2} - c\right)$, defined as numerical limit of this sequence
vanishes at $c=1$, where the two real roots of $f_c(x)$ merge (and go to
complex domain).

\subsection{Discussion}

The above considerations suggest the following point of view.

Given a (complex) analytic function $f(x)$ we get in the domain
$U^\infty (f) \subset \bb{X}$
a set of discrete $f$-invariant indecomposable subsets. These subsets are
orbits of a $\bb{Z}$-action on $U^\infty(f)$, generated by $f$.
The union of periodic orbits gives a $\bb{Z}$-representation
$V_\infty(f)$ which may be regarded as an intrinsic characteristic of $f$.
Each periodic orbit belongs to
a set of roots of some $G_n(x;f)$ of minimal degree $n$, which can be
assigned to the orbit as its degree. As the shape of $f$ changes, i.e.
as $f$ moves in the moduli space ${\cal M}$, the
$f$-orbits move in $\bb{X}$. The dynamics of this motion
is worth studying. For $f$'s of certain shapes some orbits can merge.
The corresponding subset in ${\cal M}$ can be regarded as
inverse image (pullback) ${\cal D}^{*}$ of the discriminant
variety ${\cal D}\subset {\cal M}$, induced by the functions $G_n(x;f)$.
We believe that discriminants and resultants can
be well-defined not only for polynomials, but also for analytic functions,
while the variety ${\cal D}^{*}$ makes sense in even more general --
topological -- setting.
We identify the closure of the union
$\cup_{m,n=1}^\infty {\cal R}_{mn}^{*} =
\cup_{n=1}^\infty {\cal D}_n^*$
as the boundary of the Mandelbrot set, which separates the moduli
space ${\cal M}$
into disjoint components, what can by used to classify analytic functions.
Even more structures (Julia sets, secondary and Grand Mandelbrot sets)
arise in a similar way from consideration of pre-orbits and grand orbits of
$f$ and of their change with the variation of $f$.

Even in multi-dimensional case for every map
$\vec f(\vec x) \in {\cal M}$ one can take its iterations
$\vec F_n(\vec x;\vec f) = \vec{f^{\circ n}}(\vec x) - \vec x$
and construct the real-codimension-two subsets ${\cal D}^*_n$ and
${\cal R}^*_{mn}$ in ${\cal M}$, made out of zeroes of reduced discriminant
$d_n(\vec f)$ and resultant $r_{mn}(\vec f)$ functions.\footnote{
The function $d_n$ can be defined as irreducible constituent of
$$
D_n = \left\{\begin{array}{c} \vec F_n(\vec x;\vec f) = 0 \\
\det_{ij} \frac{\partial F_n^i}{\partial x_j}(\vec x;\vec f) = 0,
\end{array}\right.
$$
and $r_{mn}=0$ whenever $\vec F_m(\vec x;\vec f) =
\vec F_n(\vec x;\vec f) = 0$ (though seemingly
overdefined, this system defines a real-codimension-two subspace
in ${\cal M}$ if $m|n$ or $n|m$). Stability set $S_n$ can be defined as
$$
S_n = \left\{\begin{array}{c} \vec F_n(\vec x;\vec f) = 0 \\
\left|{\rm e.v.}\left(\delta^i_j +
\frac{\partial F_n^i}{\partial x_j}(\vec x;\vec f)\right)\right| < 1,
\end{array}\right.,
$$
where ${\rm e.v.}(A^i_j)$ denote eigenvalues of matrix $A$.
}
A union $\bigcup_{m=1}^\infty {\cal R}^*_{nm} = \partial S_n$ has
real codimension one in ${\cal M}$ and can be considered as a boundary of
a codimension-zero {\it stability set} $S_n \subset {\cal M}$, so that
$d_n \in \partial S_n$ and
$\partial S_n \cap \partial S_m = {\cal R}^*_{mn}$. What is non-trivial,
the intersection of stability domains, not only their boundaries, is just
the same (and has real codimension two!):
$\partial S_n\cup \partial S_m = {\cal R}^*_{mn}$.
This makes the structure of {\it universal Mandelbrot set} $M = \cup_n S_n$
non-trivial, and requires description in terms of projections to
{\it multipliers trees} and representation theory, as suggested in this paper.
Sections of the Mandelbrot set by real-dimension-two manifolds (by
one-complex-parametric map families $\mu_c \subset {\cal M}$) help to
visualize some of the structure, but special care should be taken to
separate intrinsic properties of $M$ from peculiarities
of particular section $M(\mu) = M\cup \mu$. For example, the tree structure
is universal, while the number of vertices and links is not.

\newpage

\section{Map $f(x) = x^2 + c$: from standard example to general
conclusions
\label{Exa}}

In this section we use the well publicized example to illustrate
our basic claims. Namely:

1) The pattern of periodic-orbit bifurcations in {\it real} case is well
described in the language of discriminants.

2) Discriminants carry more information in two aspects:
there are more bifurcations in real case than captured by the
period-doubling analysis and there are even more bifurcations
in complex domain which are not seen in real projection.

3) Fractal Mandelbrot set is in fact a union of well-defined
domains with boundaries, which (i) are described by algebraic
equations, (ii) are smooth almost everywhere (outside zeroes of
associated discriminants), (iii) are densely populated
by zeroes of appropriate resultants. When resultant vanishes,
a pair of smooth domains touch and the touching point
serves as a single "bridge", connecting the two domains.

4) Julia set is related to inverse limit of almost any grand orbit:
its infinitely many branches originate in the vicinities of
infinitely many unstable orbits, which constitute the
boundary of the algebraic Julia set.

5) Changes of stability (bifurcations of Julia set)
always occur when something happens to periodic orbits, but
these orbits that matter are not necessarily stable.
Unstable orbits can be best studied with the help of
grand orbits, because even pre-images of stable orbits carry
information about unstable ones: all grand orbits originate
in the vicinity of unstable orbits (grand orbit's periodic branches
are more-or-less in one-to-one correspondence with unstable
orbits). At the same time, everything what one can wish to know about
grand orbits is again describable in terms of discriminants
and resultants.

6) If $f\in M_1$, the fractal Julia set $J(f)$ is in fact a
continuous (but not necessarily smooth) deformation of a
unit disc.
The unit circle -- the boundary of the disc --
is densely covered by grand orbits, associated
with unstable periodic orbits, and one stable periodic orbit,
together with its grand orbit, lies inside
the disc. The vanishing of the relevant
resultant implies that the stable grand orbit from inside
comes to the boundary, intersects and exchange stability with a particular
unstable grand orbit, which becomes stable, quits the boundary and immerses
into the Julia set. This results in merging the corresponding points of
the boundary and changes the topology of Julia set -- a bifurcation occurs.
Topology of the Julia set $J(f)$ is defined by position of the
map $f$ in the Mandelbrot set and by the path through "bridges", which
connects $f$ with the central domain of the Mandelbrot set
(i.e. by the branch of the powerful tree $T_1$).

7) If $f \in M_{k\alpha}$ with $k>1$ the disc/ball is substituted
by $k$-dependent collection of disjoint discs/balls,  grand orbits are jumping
between the discs, but return to original one every $n$ times, if
the orbit is of order $n$. All bifurcations when $f$ moves between
different components inside the given $M_{k\alpha}$ are described in the
same way as in 6). If $f$ leaves Mandelbrot set, $f\notin M$,
there are no stable orbits left and Julia set looses its "body"
(interior of the disc), only the boundary survives, which contains all
the bounded grand orbits.

\subsection{Map $f(x) = x^2 + c$. Roots and orbits, real and complex}

$f_c(x) = x^2 + c$ is a well known example, examined in numerous papers
and textbooks. This makes it convenient for illustration of
general arguments.

\subsubsection{Orbits of order one (fixed points)}

$\bullet$ ${\cal S}_1(f_c)$  -- is the set of roots of
$$F_1(x;f_c) = x^2-x+c = G_1(x;f_c).$$
Since the order of polynomial $f_c$ is $d=2$, the number of roots
of $F_1=G_1$ is two, each of the two roots,
\be
{\cal S}_1(f_c) = \left\{ \frac{1}{2} \pm \frac{\sqrt{1-4c}}{2}\right\},
\label{S1}
\ee
is individual orbit of order one.

The orbits = roots are real, when $c \leq \frac{1}{4}$.

The action of $\hat f(1)$ leaves each of the orbits = roots intact.
The action of $\hat f(1)$ can be lifted to that of entire Abelian
group $\bb{Z}$, the same for all periodic orbits of arbitrary order
(see s.\ref{cont}). This action, however, does not distinguish
between conjugate and self-conjugate orbits,
to keep this information an extra $\bb{Z}_2$ group is needed.
If we denote by $v_q$ representation of $\bb{Z}$ when the group acts
as cyclic permutations on the sequence of $q$ elements,
and label the trivial and doublet representations of $\bb{Z}_2$
by superscripts $0$ and $\pm$, then
the  representation of $\bb{Z}\otimes \bb{Z}_2$
associated with $F_1(x)$ is $2v^0_1 = v_1^+ \oplus v_1^-$.

Discriminant $d(G_1) = D(G_1) = D(F_1(f_c)) = 1 - 4c$,
its only zero is at $c = \frac{1}{4}$,
where the two order-one orbits intersect. At intersection
point the orbits are real, $x = \frac{1}{2}$ and remain real for
real $c < \frac{1}{4}$. On the real line it looks like the two real
roots are "born from nothing" at $c=\frac{1}{4}$.

Since $f'(x) = 2x$ the plus-orbit is unstable for all real $c$,
the minus-orbit is stable on the segment $-3/4 < c < 1/4$.

In the domain of complex $c$ stability region $S_1$ is bounded by the curve
$|2x| = 1$ where $x$ is taken from (\ref{S1}), i.e.
\be
\partial S_1: \ \ \ \left.
\begin{array}{c}
| 1 \pm \sqrt{1-4c} | = 1,\ \ {\rm or} \\
c = {1}/{4} + e^{i\varphi}\sin^2{\varphi}/{2}.
\end{array}\right.
\label{cur1}
\ee
In polar coordinates $(r,\varphi)$
with the center at $\frac{1}{4}$ the curve is given by
$r = \sin^2\frac{\varphi}{2}$, see Fig.\ref{lem1}.
Our expectation is that the points on this curve (an everywhere dense
countable subset in it) will be zeroes of the resultants
$R(F_1,F_n)$, or $r(G_1,G_{n})$ to be precise,
with all possible $n$.

The critical point of the map $f_c = x^2+c$, i.e. solution to $f_c'(x)=0$
is $w_c=0$. Equation $F_1(w_c;f_c) = F_1(0;f_c) = f_c(0) = 0$ has a single
solution, $c=0$, which lies inside the single elementary component of
$S_1 = \sigma^{(0)}[1]$.

\subsubsection{Orbits of order two}

$\bullet$ ${\cal S}_2(f_c)$  --
the roots of
\be
F_2(x;f_c) = (x^2-x+c)(x^2+x+c+1) = G_1(x;f_c)G_2(x;f_c).
\label{S2}
\ee
$F_2$ is divisible by $F_1$ since $1$ is a divisor of $2$,
the ratio
$$G_{2}(x;f_c) = F_2/F_1 = x^2+x+c+1.$$

The $d^2=4$ roots of $F_2$ form two conjugate order-one orbits and one
($\frac{d^2-d}{2}=1$) self-conjugate order-two orbit.
The new orbit -- in addition to the two order-one
orbits inherited from ${\cal S}_1(f_c)$
-- is of order $2$ and consists of two points $-\frac{1}{2} \pm
\frac{\sqrt{-3-4c}}{2}$ -- the zeroes of $G_2$.

The action of $\hat f(2)$ interchanges these two points.
Representation of $\bb{Z}\otimes \bb{Z}_2$,
associated with $F_2$, is $2v_1\oplus v_2 = (v_1^+ \oplus v_1^-)
\oplus v_2^0$.

Discriminant
\be
D(F_2) = (4c-1)(4c+3)^3 = \nn \\ =
D(G_1)D(G_{2})R^2(G_{2},G_1) =
d(G_1)d^2(G_2)r^3(G_2,G_1),\nn
\ee
and
\be
D(G_1) = d(G_1) = 1-4c, \nn \\
D(G_{2}) = d^2(G_2)R(G_2,G_1) = -3-4c, \nn \\
d(G_2) = -1, \nn \\
R(G_{2},G_1) = r(G_2,G_1) = 3+4c.\nn
\ee
The only new zero
is at $c = -\frac{3}{4}$. Since it is a zero of $R(G_{2},G_1)$,
the new orbit, associated with the roots of $G_{2}(x;f_c)$,
intersects an old one, related to zeroes $G_1(x;f_c)$.
Indeed, at $c=-3/4$
the stable order-one orbit intersects the order-$2$ orbit
and loses stability, while the order-$2$ orbit becomes stable
and stays real for all real $c < -\frac{3}{4}$.

Since at intersection point $c = -\frac{3}{4}$
the two roots of $G_{2}$ merge,
$D(G_{2})$ has a simple zero. Similarly, below all zeroes of
$R(G,\tilde G)$ will be also zeroes of either $D(G)$ or $D(\tilde G)$
(in particular, $D(G)$ are usually reducible polynomials over any field
$\bb{X}$), moreover their multiplicities are dictated by the properties
of intersecting orbits.

Stability criterium for the order-$2$ orbit of $f_c(x)$ is
$$
\left\{ \begin{array}{c}
|f'(x)f'(f(x))| = |4xf_c(x)| = |4x(x^2+c)| < 1 \\
G_2(x;f_c) = x^2+x+c+1 = 0.
\end{array}\right.
$$
The boundary of stability region $S_2$ in complex plane $c$ is obtained
by changing inequality for equality.
Then $x(x^2+c) = -x(x+1) = c+1$, and the boundary is just a circle
\be
\partial S_2:\ \ \ |c+1| = \frac{1}{4}
\label{cur2}
\ee
with center at $c=-1$ and radius $\frac{1}{4}$, see Fig.\ref{lem1}.
It intersects with the curve (\ref{cur1}) at $c=-\frac{3}{4}$, which
is the zero of $R(G_{2},G_1) = 3 + 4c$.

Eq. $G_2(w_c;f_c) = G_2(0;c) = c+1 = 0$ now implies that $c=-1$, and
this point lies inside the single elementary domain of $S_2 =
\sigma^{(1)}[2|1]$.

\FIGEPS{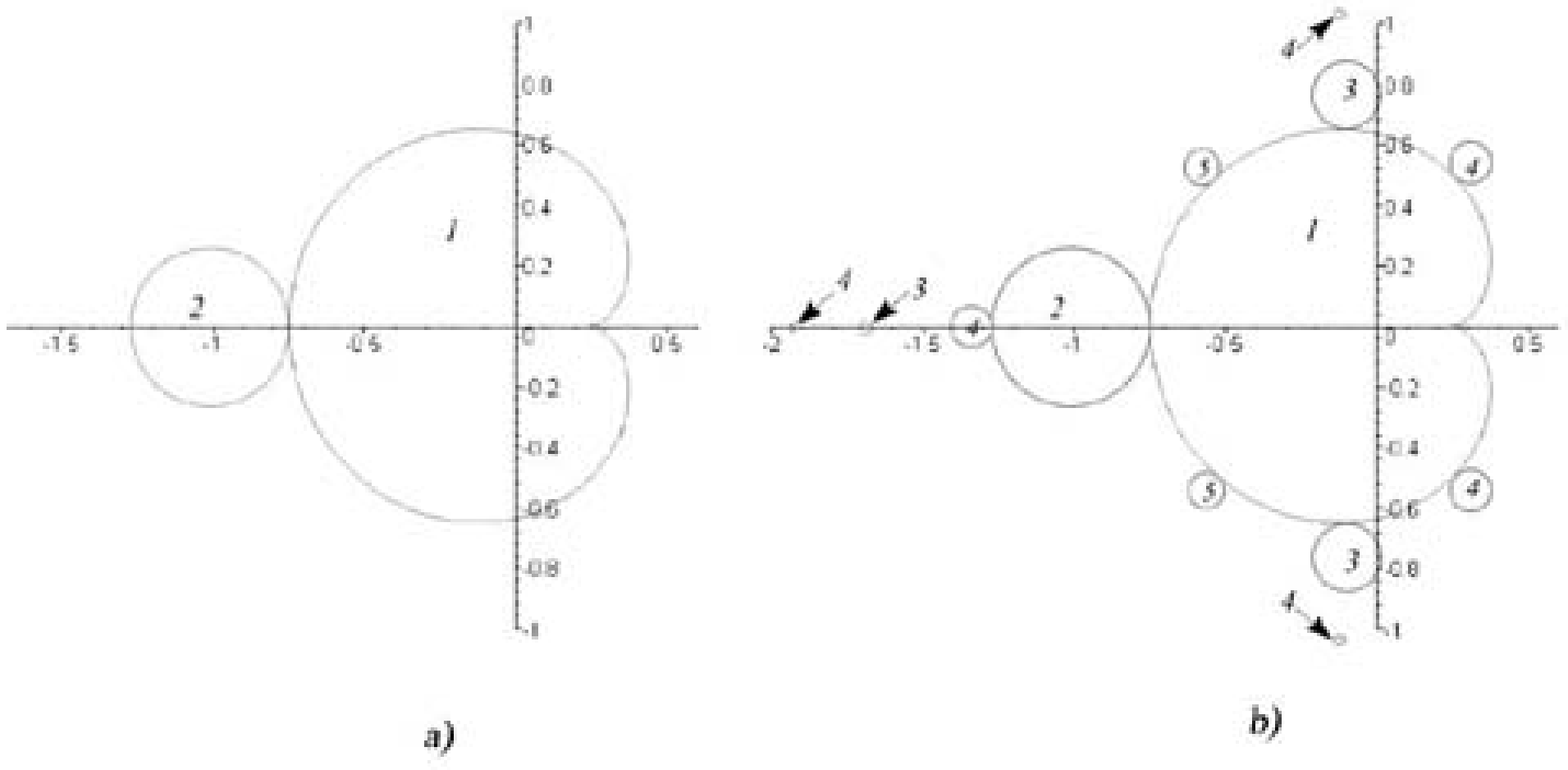}{}
{420,215}
{Stability domains in the plane of complex $c$:
a) for orbits of order one and two, b) for orbits of orders up to four.}

\subsubsection{Orbits of order three}

$\bullet$ ${\cal S}_3(f_c)$  --
the roots of
\be
F_3(x;f_c) = (x^2-x+c)\left(x^6
+ x^5
+ x^4( 3c + 1 )
+ x^3( 2c + 1 )
+ \phantom{5^{5^5}} \right. \nn \\ \left. \phantom{5^{5^5}}
+ x^2( 3c^2 + 3c + 1 )
+ x( c^2 + 2c+1)
+ ( c^3+2c^2+c+1 )\right)
= \nn \\
= G_1(x;f_c)G_{3}(x;f_c).\nn
\ee
$F_3(x)$ is divisible by $F_1(x) = G_1(x)$ since $1$ is a divisor of $3$,
the ratio $G_{3}(x) = F_3(x)/F_1(x)$.

The $d^3 = 8$ roots of $F_3$ form two conjugate order-one orbits
and two ($\frac{d^3-d}{3}=2$) conjugate order-$3$ orbits.

Representation of $\bb{Z}\otimes \bb{Z}_2$, associated with $F_3$ is
$2v_1 \oplus 2v_3 = (v_1^+\oplus v_1^-) \oplus (v_3^+ \oplus v_3^-)$.

Discriminant
\be
D(F_3) = (4c-1)(4c+7)^3(16c^2+4c+7)^4 = \nn \\ =
D(G_1)D(G_{3})R^2(G_{3},G_1) =
d(G_1)D^3(G_3)r^4(G_3,G_1) \nn
\ee
and
\be
D(G_1) = d(G_1) = 1-4c, \nn \\
D(G_{3}) = d^3(G_1)r^2(G_3,G_1) =  -(4c+7)^3(16c^2+4c+7)^2, \nn \\
d(G_3) = -(4c+7), \nn \\
R(G_{3},G_1) = r(G_3,G_1) = 16c^2+4c+7.\nn
\ee
The new -- as compared to $F_1$ (but not $F_2$, which is not a divisor
of $F_3$) -- zeroes of $D(F_3)$ are at $c = -\frac{7}{4}$ and at
$c = -\frac{1}{8} \pm \frac{3\sqrt{3}i}{8}$.
The latter two values are zeroes of $R(G_{3},G_1)$ and they describe
intersection of two order-$3$ orbits with the stable order-$1$
orbit (one of the two roots of $G_1$) at essentially complex $c$.
No trace of these intersections
is seen in the plane of real $x$ and $c$.
The two order-$3$ orbits intersect at the remaining zero of $D(F_3)$,
which is real, $c = -\frac{7}{4}$, and they (orbits) remain real
at all real $c < -\frac{7}{4}$. However, since $c = -\frac{7}{4}$
is actually a root of $d(G_{3})$, it has nothing to do with $G_1$, and the
new real order-$3$ orbits look -- from the {\it real} point of view --
"born from nothing", not at the position of any other
"previously existing" orbits.
In complex domain the orbits always exist, and just intersect
(coincide) get real at
$c=-\frac{7}{4}$.

Stability criterium for an order-$3$ orbit of $f_c(x)$ is
\be
\left\{ \begin{array}{cc}
|f'(x)f'(f(x))f'(f^{\circ 2}(x))| = |8xf_c(x)f_c^{\circ 2}(x)| =
\\ =
|8x(x^2+c)(x^4 + 2c x^2 + c^2+c)| < 1, \\
G_3(x;c) = 0.
\end{array}\right.
\nn
\ee
This system is hard to solve explicitly. Solving instead
$$ G_3(0,c) = c^3+2c^2+c+1 = 0 $$
we obtain three points, lying
inside the three elementary domains of
$$
S_3 = \sigma^{(0)}[3] \bigcup
\sigma^{(1)}\left[\left.\begin{array}{c} 3\\1 \end{array}\right| 1 \right]
  \bigcup \sigma^{(1)}
\left[\left.\begin{array}{c} 3\\2 \end{array}\right| 1 \right].
$$
This is the first example when we observe the appearance of
$\alpha$-parameters.

\subsubsection{Orbits of order four}

$\bullet$ ${\cal S}_4(f_c)$  --
the roots of
\be
F_4(x;f_c) = (x^2-x+c)(x^2+x+c+1)\left(x^{12} +
6cx^{10}
+ x^9
+ \phantom{5^{5^5}} \right. \nn \\ \left. \phantom{5^{5^5}}
+ x^8(15c^2 + 3c)
+ 4c x^7
+ x^6(20c^3 + 12c^2 + 1)
+ x^5(6c^2 + 2c )
+ \phantom{5^{5^5}} \right. \nn \\ \left. \phantom{5^{5^5}}
+ x^4(15c^4 + 18c^3 + 3c^2 + 4c )
+ x^3(4c^3 + 4c^2 + 1)
+ \phantom{5^{5^5}} \right. \nn \\ \left. \phantom{5^{5^5}}
+ x^2(6c^5 + 12 c^4 + 6c^3 + 5c^2 + c )
+ x(c^4+ 2c^3+c^2+2c)
+ \phantom{5^{5^5}} \right. \nn \\ \left. \phantom{5^{5^5}}
+ (c^6 + 3c^5 + 3c^4 + 3c^3 + 2c^2 + 1)\right)
= \nn \\
= F_2(x)G_{4}(x) = G_1(x)G_{2}(x)G_{4}(x).\ \ \
\nn
\ee
$F_4$ is divisible by $F_1$ and $F_2$ since $1$ and $2$ are divisors
of $4$, the ratio $G_{4} = F_4/F_2$ is a polynomial of order $12$.

The $d^4 = 16$ roots of $F_4$ form two conjugate order-one orbits
(inherited from $G_1$), one self-conjugate order-$2$ orbit (from
$G_2$) and three ($\frac{d^4-d^2}{4}=3$)
new order-$4$ orbits, two conjugate and one self-conjugate.

Representation of $\bb{Z}\otimes \bb{Z}_2$, associated with $F_4$, is
$$
2v_1 \oplus v_2 \oplus 3v_4 =
(v_1^+\oplus v_1^-) \oplus v_2^0
\oplus v_4^0 \oplus (v_4^+ \oplus v_4^-).
$$

Discriminant
\be
D(F_4) =
(4c-1)(4c+3)^3\cdot \nn \\ \cdot
(4c+5)^6 (16c^2-8c+5)^5 (64c^3+144c^2+108c+135)^4 = \nn \\ =
D(G_1)D(G_{2})D(G_{4})R^2(G_{2},G_1)R^2(G_{4},G_1)
R^2(G_{4},G_{2}) = \nn \\ =
d(G_1)d^2(G_2)d^4(G_4) r^3(G_2,G_1)r^5(G_4,G_1)r^6(G_4,G_2)
\nn
\ee
where
\be
D(G_1) = 1-4c, \nn \\
D(G_{2}) = -3-4c, \nn \\
D(G_{4}) = (4c+5)^2(16c^2-8c+5)^3
(64c^3+144c^2+108c+135)^4, \nn \\
R(G_{2},G_1) = 3+4c, \nn \\
R(G_{4},G_1) = (16c^2-8c+5), \nn \\
R(G_{4},G_{2}) = (4c+5)^2
\nn
\ee
and
\be
d(G_1) = 1-4c, \nn \\
d(G_2) = -1, \nn \\
d(G_4) = 64c^3+144c^2+108c+135, \nn \\
r(G_2,G_1) = 3+4c, \nn \\
r(G_4,G_1) = 16c^2-8c+5, \nn \\
r(G_4,G_2) = 4c+5.
\nn
\ee

The roots of $D(F_4)$ describe the following phenomena:

$c=\frac{1}{4}$ and $c=-\frac{3}{4}$ were already examined in
connection with $F_1$ and $F_2$.

$c=-\frac{5}{4}$ is a zero of $D(G_{4})$,
thus it could describe either intersection
of two order-$4$ orbits or merging of roots of a single
order-4 orbit. In the latter case that orbit would degenerate into
the one of order $2$ or $1$. Since this $c$ is simultaneously the
zero of $R(G_{4},G_{2})$, the right choice is
intersection of self-conjugate order-$4$ orbit with the previously
existing (inherited from $G_{2}$) self-conjugate order-$2$ orbit.
Since at intersection point two pairs of roots of $G_{4}$ merge,
the order of zero of $D(G_{4})$ is two.
For lower real $c<-\frac{5}{4}$ the self-conjugate order-$4$ orbit
remains real.

$c = \frac{1}{4} \pm \frac{i}{2}$ are complex zeroes of
$R(G_{4},G_1)$ and describe intersection of conjugate
order-$1$ orbits with the conjugate
order-$4$ orbits in complex domain. No traces of this intersection
are seen in the plane of real $x$ and $c$.
Since at intersection points all the four roots of $G_{4}$
merge together, $D(G_{4})$ has cubic zeroes.

The remaining three roots of $D(F_4)$ are roots of $d(G_{4})$
alone and describe entirely the world of order-$4$ orbits:
the $C^2_3 = 3$ intersections of $3$ orbits. Namely,

$c = -\frac{3}{4}(1 + 2^{2/3})$ -- two conjugate order-$4$ orbits
intersect, get real and stay real for lower real values of $c$.
Look as "born from nothing" from the real perspective. Since
$4$ pairs of roots need to merge simultaneously, the order of
zero of $D(G_{4})$ is four.

$c = -\frac{3}{4} + \frac{3}{8} 2^{2/3}(1 \pm i\sqrt{3})$ --
self-conjugate order-$4$ orbit intersects with one or another
of the two conjugate order-$4$ orbits in complex domain. No traces
at real plane. $D(G_{4})$ has quadruple zero.

Stability criterium for an order-$4$ orbit of $f_c(x)$ is
$$
\left\{ \begin{array}{cc}
|f'(x)f'(f(x))f'(f^{\circ 2}(x))f'(f^{\circ 3}(x))| =
|2^4xf_c(x)f_c^{\circ 2}(x)f_c^{\circ 3}(x)| = \nn \\ =
|8x(x^2+c)(x^4 + 2c x^2 + c^2+c)(x^8 + \ldots )| < 1, \\
G_4(x;c) = 0.
\end{array}\right.
$$
Solving instead of this system
$$
G_4(0,c) = c^6 + 3c^5 + 3c^4 + 3c^3 + 2c^2 + 1 = 0
$$
we obtain six points, lying
inside the six elementary domains of
$$
S_4 = \left(\bigcup_{\alpha=1}^3
\sigma^{(0)}[4, \alpha]\right)
\bigcup
\sigma^{(1)}\left[\left.\begin{array}{c} 4\\1 \end{array}\right| 1\right]
  \bigcup \sigma^{(1)}\left[\left.\begin{array}{c} 4\\2 \end{array}\right| 1\right]
\bigcup \sigma^{(2)}[2\ 2 | 1].
$$

\subsubsection{Orbits of order five}

$\bullet$ ${\cal S}_5(f_c)$  --
the roots of
$$
F_5(x;f_c) = (x^2-x+c)(x^{30} + \ldots ) = G_1(x)G_{5}(x)
$$

There are $\frac{d^5-d}{5}=6$ order-$5$ orbits.

\be
D(G_{5}) = d^5(G_5)r^4(G_1,G_{5}), \nn \\
R(G_1,G_{5}) = r(G_1,G_5) = 256c^4 + 64c^3 + 16c^2-36c + 31, \nn \\
d(G_5) = -(4194304c^{11} + 32505856c^{10} + \nn \\
+ 109051904c^9 + 223084544c^8 + 336658432c^7 + \nn \\
+ 492464768c^6 + 379029504c^5 + 299949056c^4 + 211327744c^3 + \nn \\
+ 120117312c^2 + 62799428c + 28629151) = \nn \\
= -\left((4c)^{11} + 31(4c)^{10} + 416(4c)^9 + \ldots\right)
\label{DG5.2}
\ee
The powers $4$ and $5$ appear here because when an order-$5$ orbit
intersects with the order-$1$ orbit, the $5$ different
roots of $F_5$ should
merge simultaneously, thus discriminant has a zero of order $4=5-1$,
while at intersection of any two order-$5$ orbits, the $5$ pairs of roots
should coincide pairwise, so that discriminant has a zero of order $5$.

$G_5(0;c)$ is polynomial of order $2\cdot 8 -1 = 15$ in $c$, thus equation
$G_5(0;c) = 0$ has $15$ solutions. These should describe the would be
$C^2_6 = 15$ pairwise intersections of $6$ order-$5$ orbits.
However, since the power of $c$ in $d(G_5)$ is only $11$, we conclude that
only $11$ of such intersections really occur.
Instead there are
$4$ intersections between order-$1$ and order-$5$ orbits (since
$r(G_1,G_5)$ is of order $4$ in $c$).
$$
15 = 11 + 4,
$$
and stability domain $S_5$ consists of $15$ elementary domains.

\subsubsection{Orbits of order six}

$\bullet$ ${\cal S}_6(f_c)$  --
the roots of
\be
F_6(x;f_c) = G_1(x)G_{2}(x)G_{3}(x)G_{6}(x) = \nn \\
(x^2-x+c)(x^2 + x + c + 1)(x^6 + \ldots)(x^{54} + \ldots)
\nn
\ee
Here the degree-$54$ polynomial $G_{6} = F_6F_1/F_2F_3 =
F_6/F_1(F_2/F_1)(F_3/F_1)$
since $6$ has three divisors $1$,$2$ and $3$.
There are $\frac{d^6-d^2-d^3+d}{6}=9$ order-$6$ orbits.

\be
R(G_1,G_{6}) = r(G_1,G_6) = 16c^2 - 12c +3, \nn \\
R(G_{2},G_{6}) = r^2(G_2,G_6) = (16c^2 + 36c + 21)^2, \nn \\
R(G_{3},G_{6}) = r^3(G_3,G_6) = (64c^3+ 128c^2 + 72c +81)^3, \nn \\
D(G_{6}) = -d^6(G_6) R^5(G_1,G_{6})
R^2(G_{2},G_{6})R(G_{3},G_{6}) = \nn \\ =
d^6(G_6)r^5(G_6,G_1)r^4(G_6,G_2)r^3(G_6,G_3), \nn \\
d(G_6) =
1099511627776c^20+10445360463872c^19+44873818308608c^18+\nn \\
+121736553037824c^17+245929827368960c^16+399107688497152c^15+\nn\\
+535883874828288c^14+617743938224128c^13+631168647036928c^12+\nn\\
+576952972869632c^11+484537901514752c^10+376633058918400c^9+\nn\\
+263974525796352c^8+173544017002496c^7+104985522188288c^6+\nn\\
+58905085704192c^5+33837528259584c^4+15555915962496c^3+\nn\\
+8558772746832c^2+1167105374568c+3063651608241 = \nn \\
=(4c)^{20} + \ldots
\label{dg6}
\ee

$G_6(0,c)$ is polynomial of degree $32-1-1-3 = 27$ in $c$: there are
$27$ solutions to $G_6(0,c) = 0$.
On the other hand there could be up to $C^2_9 = 36$
pairwise intersections between
$9$ order-$6$ orbits, however there are only $20$, instead there are
$2$ intersections between order-$6$ and order-$1$,
$2$ -- between order-$6$ and order-$2$, and $3$ - between order-$6$
and order-$3$ orbits (all these numbers are read from the powers of
$c$ in the reduced discriminant $d(G_5)$
and reduced resultants $r(G_6,G_i)$).
We have
$$
27 = 20 + 2 + 2 + 3,
$$
stability domain $S_6$ has $27$ elementary components.

\bigskip

In this subsection we used the example of $f_c(x) = x^2 + c$
to demonstrate the application of discriminant analysis.
We explained in what sense the standard
period-doubling bifurcation tree in Fig.\ref{lemMa} is a part of a more
general pattern. Even on the plane of real $x$ and $c$ much more
is happening. Period doublings do not exhaust all possible
bifurcations, other orbits of various orders are "born from nothing"
-- not at the positions of previously existing orbits --
if we move from higher to lower $c$ or "merge and disappear"
if we move in the opposite direction.
If we go away from real $c$ and $x$, period doubling gets
supplemented by tripling, quadrupling and so on.
In complex domain no orbits
can be born, merge or disappear -- they can only intersect in
different ways.
Moreover, such intersections
do not need to leave any traces on the real $c-x$ plane.
Specifics of the iterated maps $F_n$ is that whenever they hit the
discriminant variety the whole orbits intersect: many different
roots coincide simultaneously. As $n$ increases, the pattern
becomes more and more sophisticated and begins looking chaotic as
$n \rightarrow \infty$. Still the whole story is no more no less
than that of the intersecting algebraic varieties (infinite-dimensional,
if we speak about infinitely-large $n$).
We used the standard example in order to formulate
the appropriate language
for discussion of these -- in fact, pure algebraic -- phenomena.

\subsection{Mandelbrot set for the  family $f_c(x) = x^2 + c$
\label{man2}}

Fig.\ref{lem1} shows  stability domains
for different orbits (roots of the polynomials $F_n$ and $G_{n}$)
in the complex $c$ plane. Fig.\ref{lemMa} shows the
full picture, where one can easily recognize the standard Mandelbrot
set \cite{defs}. We see that it is nothing but the union of
stability domains for orbits of different orders, which touch
at single points:
at zeroes of associated resultants. The picture also shows in what
sense the boundary of every particular stability domain is formed
by the zeroes of resultants (the zeroes constitute  countable
dense subsets in these boundaries). Note also that the two domains
touch only if the order of one orbit divides another,
otherwise the resultants do not have zeroes.
For example, $R(G_{2},G_{3}) \equiv 1$, and
it is easy to see that  indeed these two $r$-polynomials can not have
a common zero, since $G_{3}(x;f_c) - (x^4+2cx^2+x+c^2+c)G_{2}(x;f_c)
= 1$. Similarly, $R(G_{4},G_{6}) \equiv 1$ etc.

\FIGEPS{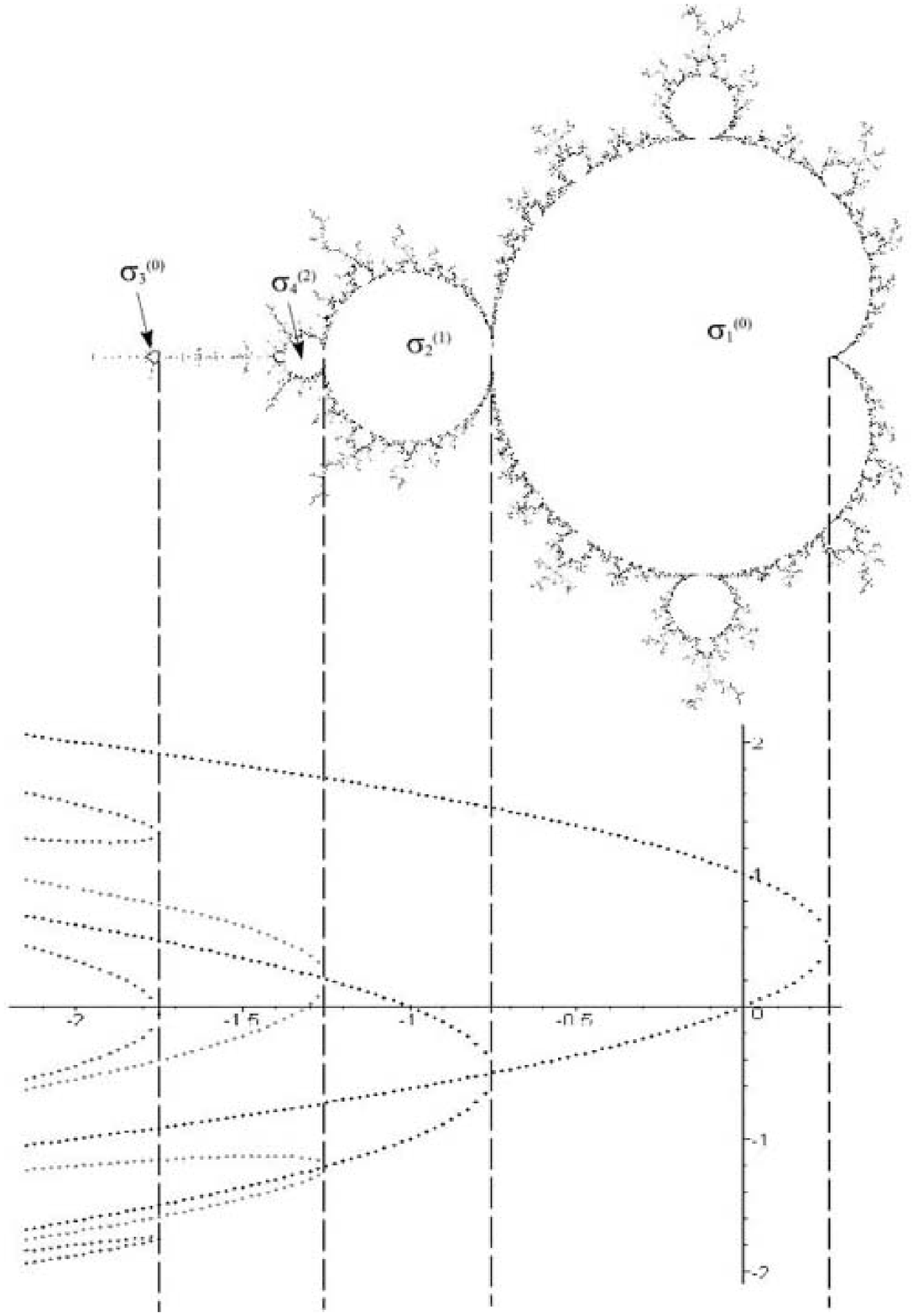}{}
{311,445}
{Mandelbrot set = union of all stability domains in the
plane of complex $c$ for the family of maps $f_c(x) = x^2 + c$.
Boundary of the white region is densely filled by collection of all
zeroes of all resultants = discriminant variety.
The picture below shows the real section of the "Julia sheaf",
by the plane of real $x$ and real $c$: for each $c$ on the real line
shown are the real orbits of different orders, stable and unstable.}

According to this picture every point in the Mandelbrot set
(i.e. every map $f$ from a given family -- not obligatory one-parametric)
belongs to one particular stability domain, and there is a uniquely
defined sequence of "bridges" (tree structure), which should
be passed to reach the "central" stability domain -- the one with
a stable fixed point. As we shall see below, this sequence (path)
is the data, which defines the structure of the Julia set $J(f)$.
As will be demonstrated in sec.\ref{Exa2}, from the point of view
of this picture there is nothing special in the maps $f_c(x) = x^2+c$,
neither in quadraticity of the map, nor in unit dimension of the family:
the {\it tree-forest-trail} structure of Mandelbrot set is always the same.

However, particular numbers of vertices (elementary domains $\sigma$),
links (zeroes of $r_{mn}$'s), cusps (zeroes of $d_n$'s) and trails
are {\it not} universal -- depend on the choice of the family $\mu$ of maps
(on the section of the universal Mandelbrot set). Moreover, cyclic
ordering of links in a vertex of a tree can also change, if $\mu$ crosses
a singularity of the resultant variety (i.e. if $R(r_{kl},r_{mn}) = 0$):
the fat-graph structure of the forest is preserved only locally:
under {\it small} variations of the 1-parametric families $\mu$.

\subsection{Map $f(x) = x^2 + c$. Julia sets, stability and preorbits
\label{Juse}}

In this subsection we describe a few examples of orbits and pre-orbits
to illustrate that normally the orbits tend
to particular stable periodic orbits
"in the future" and originate from entire set of unstable periodic orbits
"in the past". We suggest to call the closure of this latter set
$\partial J_A(f) = \overline{{\cal O}_-(f)}$
the boundary of algebraic Julia set of $f$.
We demonstrate that there is no difference between the behaviour of
{\it preorbits} of different points $x \in \bb{X}$:
for $x$ one can take invariant point (i.e. orbit of order one) of $f$,
stable or unstable, a point of any periodic orbit or a point with
unbounded grand orbit,-- despite behaviour of the {\it orbits} is absolutely
different in these cases. Often the boundary of
Julia set is defined as pre-image of
unstable invariant point, but our examples seem to demonstrate that this
is unnecessary restriction.

\bigskip

Now, for given $c \in M(\mu)$ we can switch to what happens in the
complex $x$ plane (provided $\bb{X} = \bb{C}$).
In this plane we have infinitely many periodic orbits. Their
points form the set ${\cal O}(f_c) = {\cal O}_+(f_c)\cup {\cal O}_-(f_c)$,
the union of all zeroes of all $F_n(x;f_c)$.
There are  a few stable periodic orbits and all the rest are unstable.
For polynomial $f_c(x)$,
all other stable orbits are actually located at $x=\infty$,
unstable orbits separate them from the finite stable orbits.
Generic grand orbit approaches either one of the stable orbits
from ${\cal O}_+(f_c)$ or goes to infinity.
The orbits which do {\it not} tend to infinity form {\bf Julia set}
$J(f_c) \subset \bb{X}$. The boundary of Julia set $\partial J(f_c)$
has at least three possible interpretations:

(i) it is a closure of the set of all unstable orbits, Fig.\ref{J1-1},
$$\partial J(f_c) = \overline{{\cal O}_-(f_c)};$$

(ii) it is a closure of a bounded grand orbit, associated with {\it any}
single unstable periodic orbit (for example, with an orbit of order
one, Fig.\ref{J1-4} -- an unstable fixed point of $f_c$, -- or any other,
Fig.\ref{J1-5} and Fig.\ref{iii2}),
$$\partial J(f_c) = \overline{GO_1^-};$$

(iii) it is the set of end-points of
all branches (not obligatory unstable, Fig.\ref{J1-2}, or
periodic, Fig.\ref{J1-3})
of any generic grand orbit (this set is already full,
no closure is needed).

Hypothetically all the three definitions describe the same object,
moreover, for $f \in M$, i.e. when the map belongs to the Mandelbrot
set, this object is indeed a boundary of some domain -- the
Julia set. If $f$ quits the Mandelbrot set, the object is still
well defined, but is not a boundary of anything: the "body" of Julia set
disappears, only the "boundary" survives.

\bigskip

For the family $f_c = x^2+c$ we know that there is exactly one
stable orbit for all
$c \in M(x^2+c)$, see Fig.\ref{lemMa}; for $c \notin M(x^2+c)$
{\it all} orbits (not just some of them) go to $x = \infty$, so that
$J(x^2+c) = \emptyset$.

$\bullet$ $c\in M_1$. The set ${\cal O}_+$ consists of a
single stable invariant (fixed) point (orbit of order one)
$x_+^{(1)} = \frac{1}{2} - \frac{\sqrt{1-4c}}{2}$.

\FIGEPS{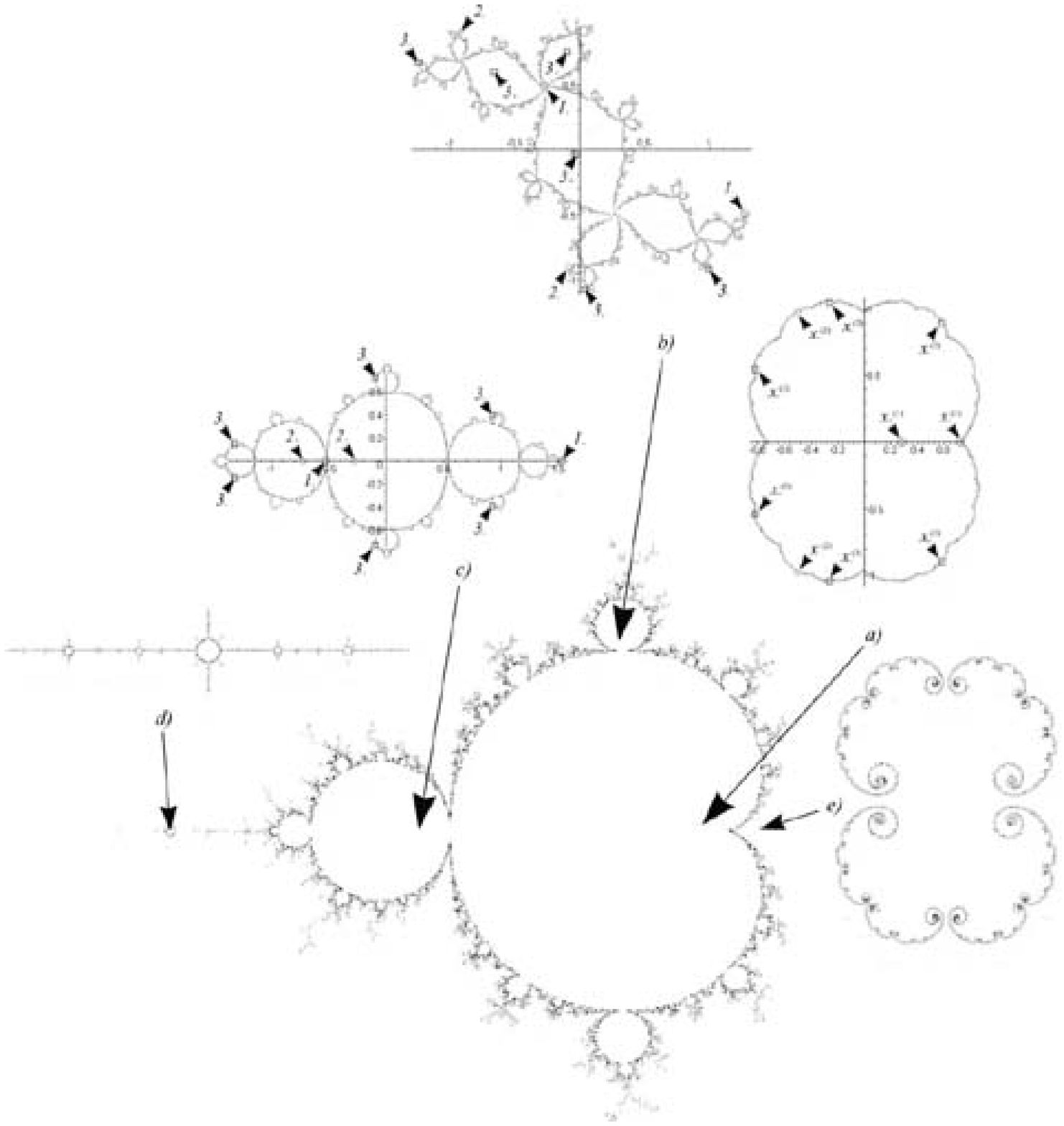}{}
{500,507}
{The few lowest (of orders 1,2 and 3) periodic orbits.
If all higher-order orbits are included, the closure
provides the boundary $\partial J$ of Julia set
for the family $f_c(x) = x^2+c$ by definition (i).
The same orbits are shown on different fibers of Julia sheaf:
a) $c=0.2$,  b) $c =-0.1+0.75i$,  c) $c =-0.8$, d) $c=-1.76$, e) $c=0.3$.
For a)-c) values of $c\in M$ there is exactly one
stable orbit {\it inside} the Julia set and all others are
at the boundary.}

\FIGEPS{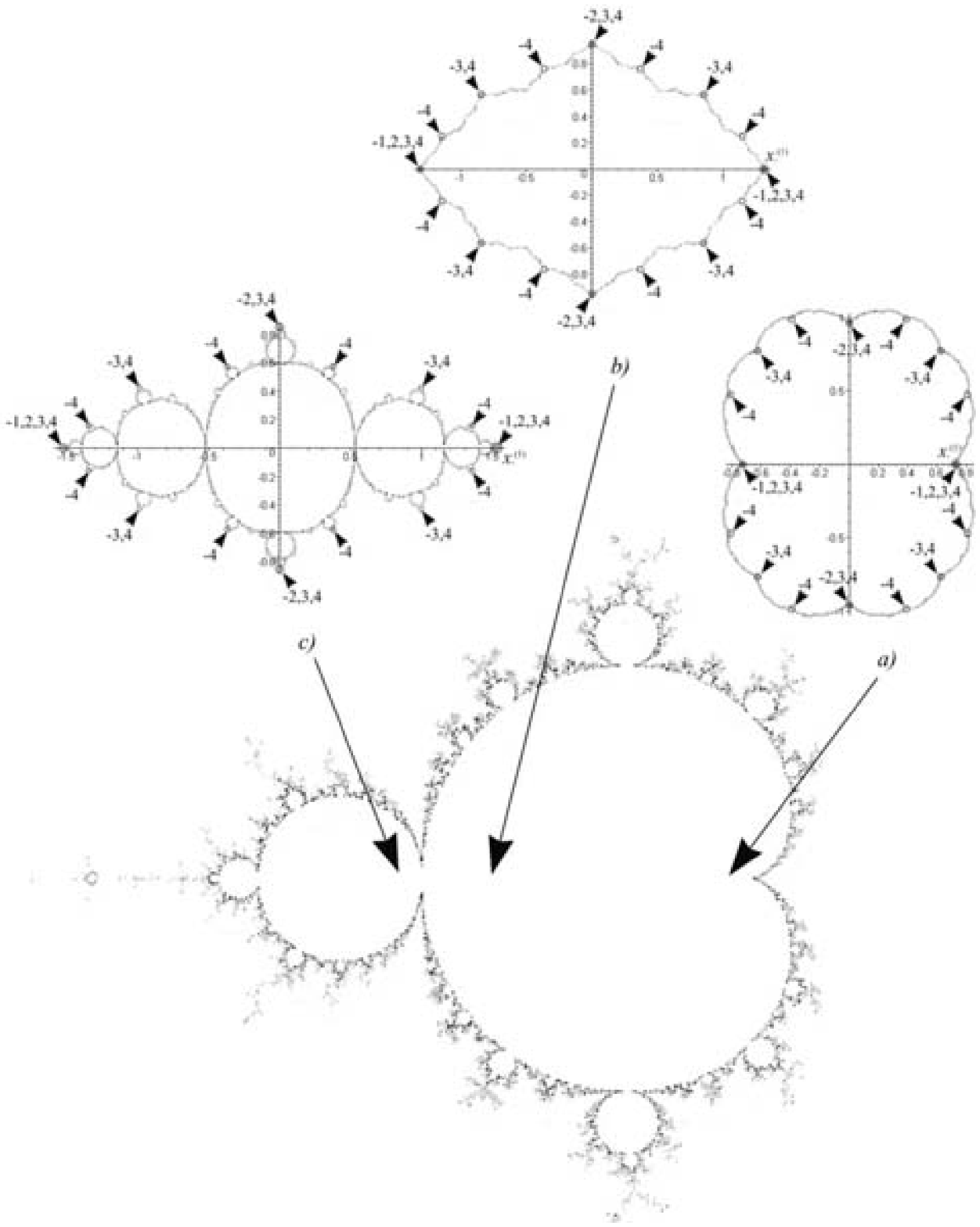}{}
{400,478}
{Grand orbit (circles) of unstable fixed point
$x_-^{(1)} = \frac{1}{2} + \frac{\sqrt{1-4c}}{2}$.
Preimages of orders from 1 to 4 are shown.
Its closure provides the boundary $\partial J$ of Julia set
for the family
$f_c(x) = x^2 + c$ by definition (ii). The same grand orbit is shown
for different fibers of Julia sheaf:
a) $c=0.2$,  b) $c =-0.4$,  c) $c =-0.8$.
The grand orbit of unstable fixed point consists of
all the "extreme" points ("spikes") of Julia set for
${\rm Re}\ c < 0$, while these are the "pits"
for $0<{\rm Re}\ c < 1/4$. The pits
are approaching the points of grand orbit of a stable fixed point,
which lie inside the Julia set, and merge with them at $c=1/4$,
see Fig.\ref{Jdeco}.
Both "spikes" and "pits" are dense everywhere in the boundary
$\partial J$, if pre-images of all levels are included.}

\FIGEPS{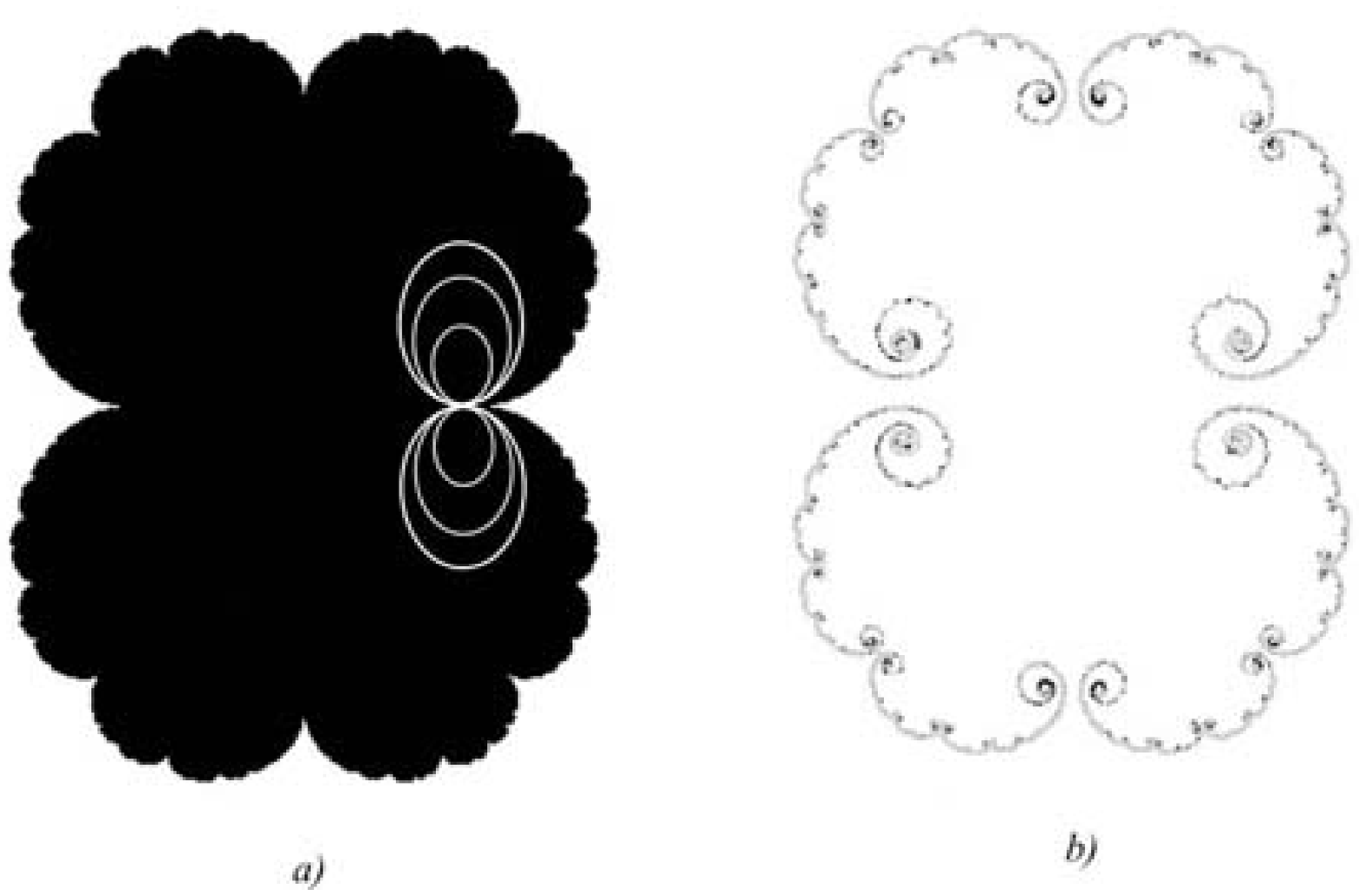}{}
{410,251}
{Decomposition of Julia set at $c=1/4$, where
the two fixed points, stable and unstable,-- and thus
their entire grand orbits -- coincide.
\p
a) The pattern is shown as obtained by moving from {\it inside} the
Mandelbrot set. White lines show peculiar circular $f$-flows in the
vicinity of the stable fixed point (similar flow patterns occur in
the vicinities of all other cusps).
When the boundary of Mandelbrot set is crossed, all the
cusps (located at all the points of the
coincident grand orbits of two merged fixed points,
which densely populate $\partial J$) dissociate into
pairs of points, thus $\partial J$ acquires holes (almost everywhere)
and fails to be a boundary of anything: the "body" of Julia
set "leaks away" through these holes and disappears.
\p
b) The resulting pattern as obtained by moving from {\it outside}
the Mandelbrot set. Shown is unlarged piece of the Julia set,
which now consists ony of the everywhere discontinuous boundary.
The holes grow with increase of ${\rm Re}\ c - \frac{1}{4} > 0$.
The spiral structure inside the former Julia set is formed by
separatrices between the flows, leaking through the holes and
remaining "inside". Infinitely many separatrices (at infinitely
many cusps) are formed by (infinitely many) branches of the
grand orbit of the fixed point. At the same time each point
on the grand orbit is a center of its own spiral.
Spiral shape is related to the nearly circular
shape of the $f$ flows in the vicinity of the stable fixed point
and its pre-images just before the formation of holes
(for ${\rm Re}\ c$ slightly smaller than $\frac{1}{4}$).
If the boundary of Mandelbrot set is crossed
at another point, which is a zero of some other $d_n$
(and belongs to some $\sigma_n^{(p)}$), the same role
is played by grand orbits of the intersected orbits of order $n$:
holes are formed at all points of these intersecting grand orbits.}

\FIGEPS{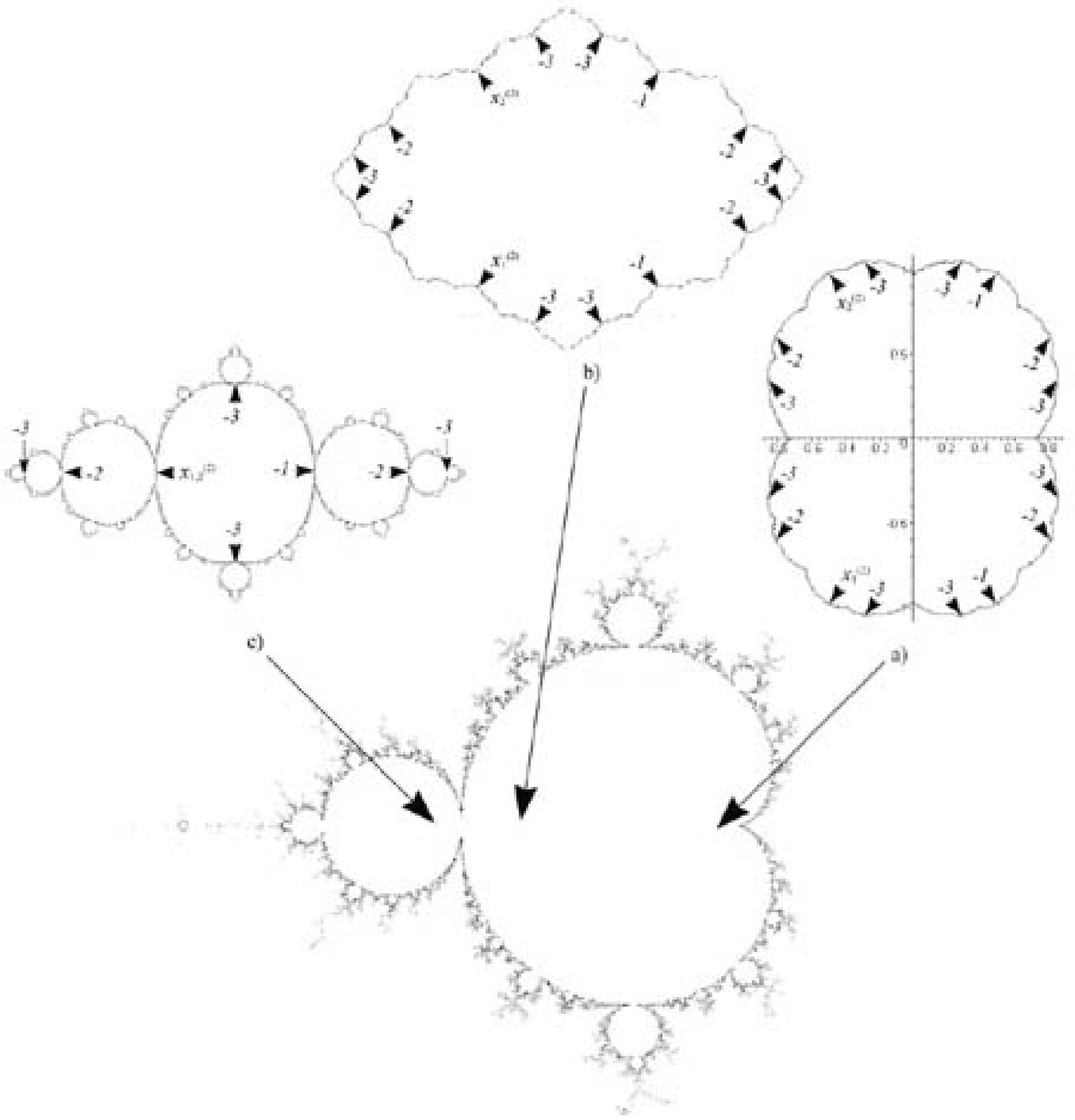}{}
{400,394}
{The analogue of Fig.\ref{J1-4} for another grand orbit:
the one of unstable orbit of order $2$, formed by
the points $x^{(2)}_\pm = -\frac{1}{2} \pm \frac{\sqrt{-3-4c}}{2}$.
Again, the same grand orbit is shown for three different fibers of Julia sheaf:
a) $c=0.2$,  b) $c =-0.4$,  c) $c =-0.8$. This grand orbit
also fills the boundary $\partial J$ densely, in accordance with
the definition (ii).}

\FIGEPS{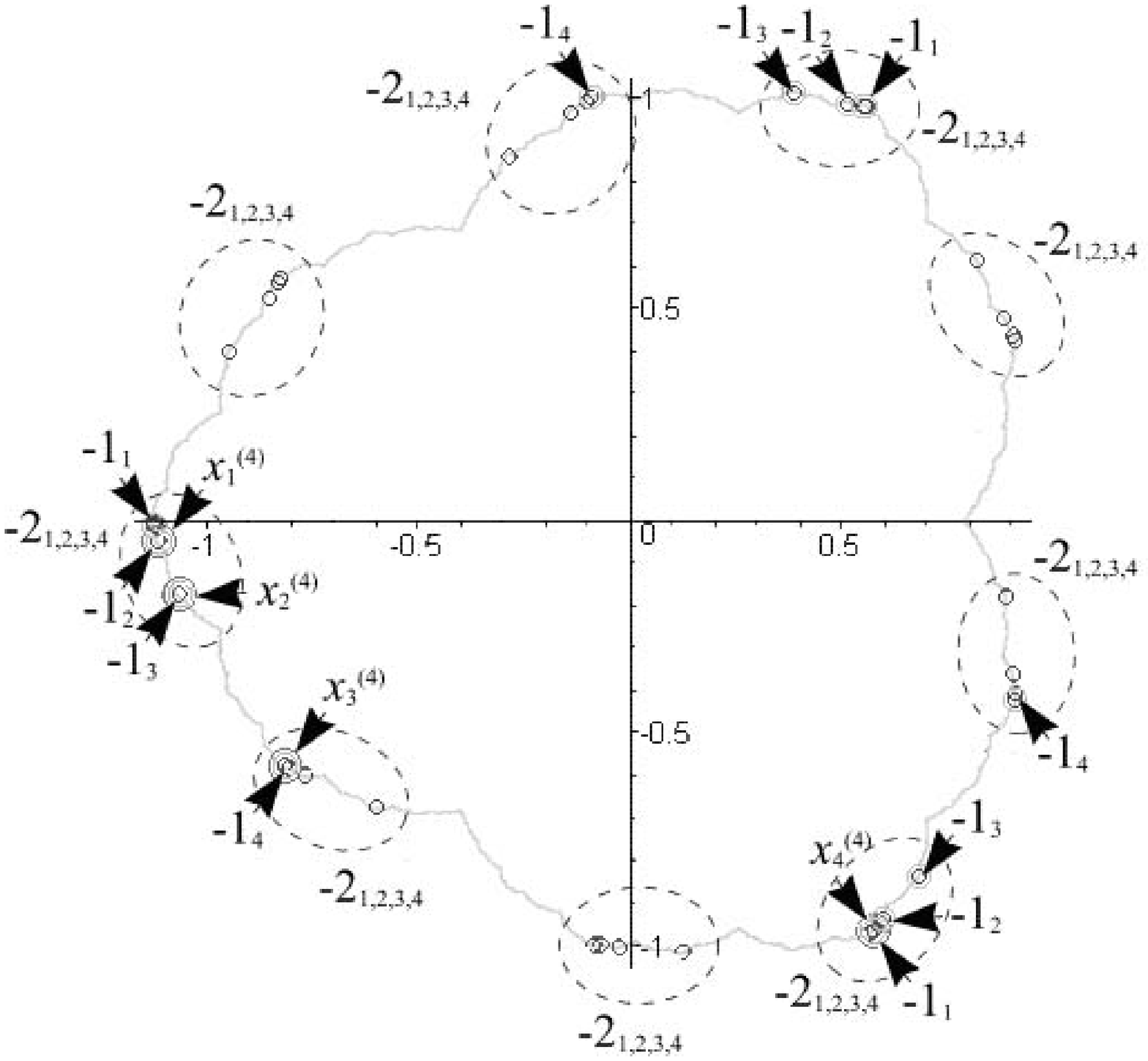}{}
{259,235}
{Non-degenerate bounded grand orbit of order $4$
for cubic $f(x)$, $d=3$, $c=0.3$ -- one more illustration
of the definition (ii). The first two preimages are shown.
The dashed lines enclose each of the 9 second preimages of the whole orbit.}

\FIGEPS{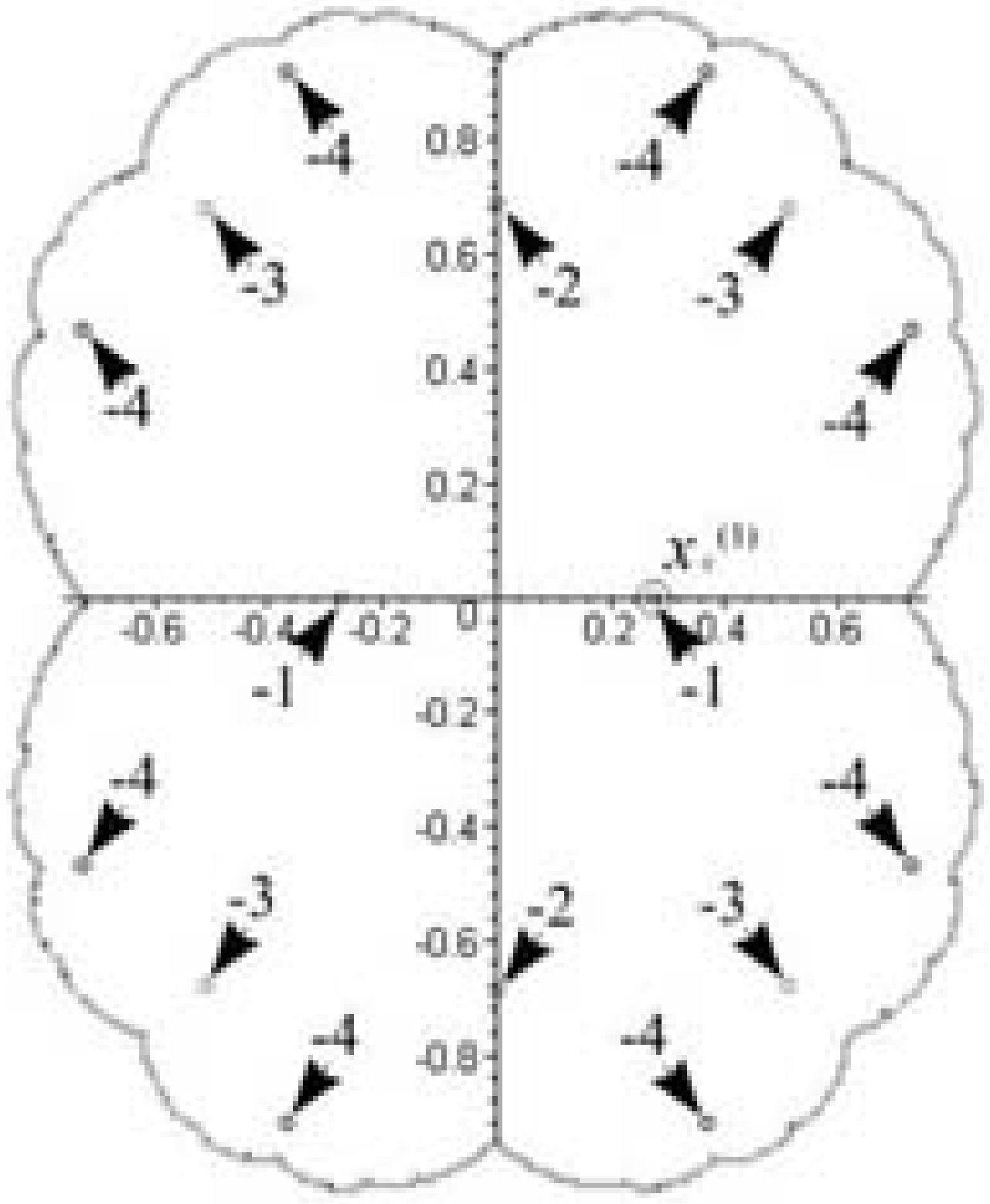}{}
{274,302}
{Grand orbit of a stable fixed point $x_+^{(1)}$, $c=0.2$.
Its end- (or, better, starting-) points approach the boundary $\partial J$ of
Julia set, if one goes against the $f$ flow
and fill it densely, according to the definition (iii).
The first four preimages are shown.}

\FIGEPS{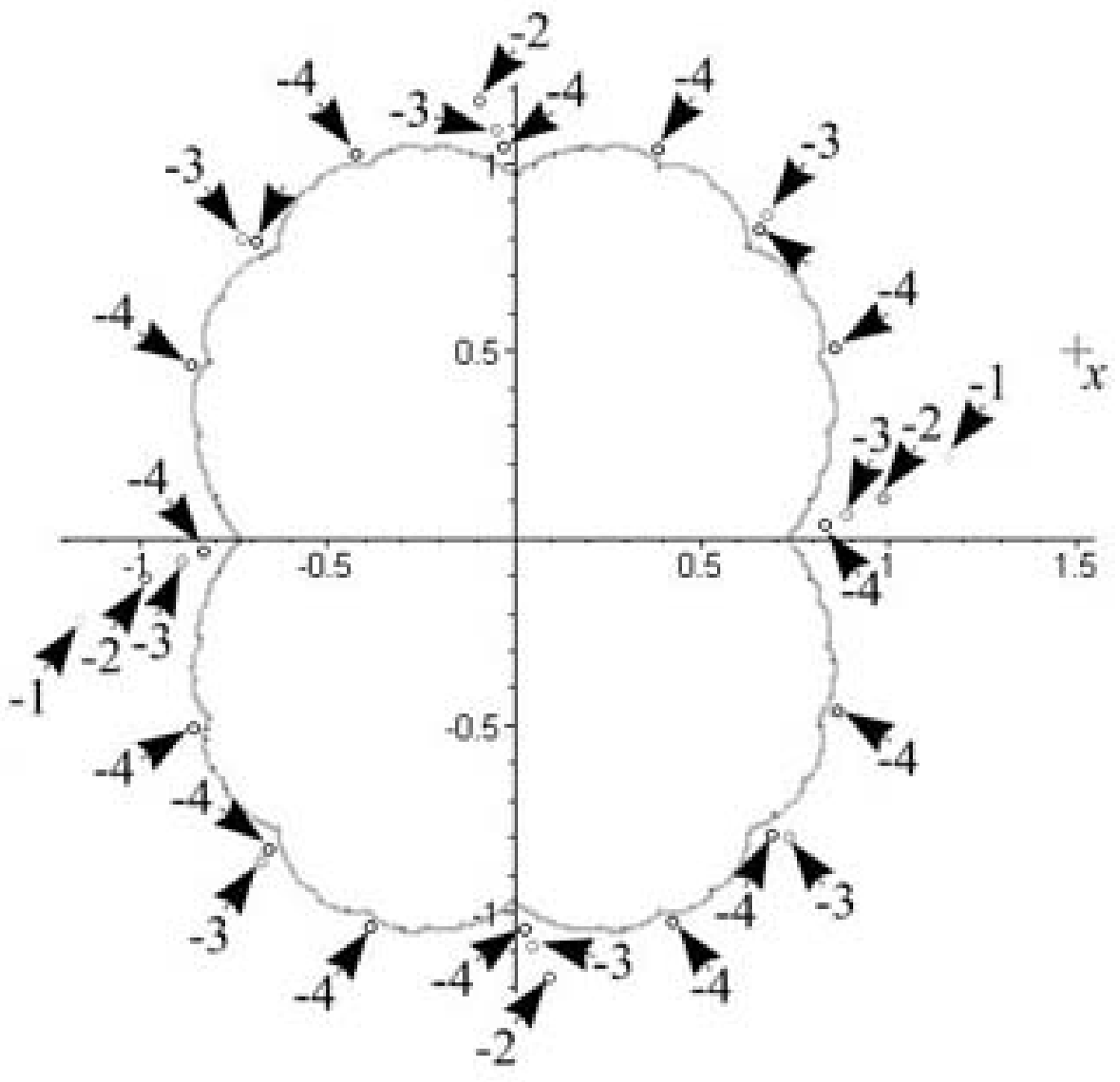}{}
{253,233}
{Typical unbounded grand orbit at the same value of $c=0.2$.
Again, its origin fill densely the boundary of the same
Julia set, $\partial J$, according to the definition (iii).}

$\bullet$ $c\in M_2$. The set ${\cal O}_+$ consists of a
single stable orbit of order two formed by the points
$x^{(2)}_\pm = -\frac{1}{2} \pm \frac{\sqrt{-3-4c}}{2}$.

\bigskip

We now elaborate a little more on the definition (iii) of the Julia set
boundary. Generic grand orbit is a tree with $d$ links entering and
one link exiting every vertex. Bounded (non-generic) grand orbit
of order $n$ ends in a loop, with $d-1$ trees
attached to every of the $n$ vertices in the loop.
Degenerate grand orbits (bounded or unbounded) have some branches
eliminated, see the next subsection \ref{secMa}. In the rest of
this subsection we consider non-degenerate grand orbits.

Going {\it backwards} along the tree we need to chose between $d$
possible ways at every step. Particular path (branch) is therefore
labeled by an infinite sequence of integers modulo $d$, i.e.
by the sequence of elements of the ring $\bb{F}_d$ (if $d=p$ the ring
is actually a field and
the sequences are $p$-adic numbers). According to definition (iii)
these sequences form the boundary of Julia set -- this can be
considered as a map between real line $\bb{R}$ (parameterized by
the sequences) and $\partial J_A$, provided by a given grand orbit.
Different grand orbits provide different maps. These maps deserve
investigation. Here we just mention that among all the sequences
the countable set of periodic ones is well distinguished, and they
{\it presumably} are associated with periodic unstable orbits,
which are dense in $\partial J_A$ according to the definition (i).

\bigskip

{\bf Important example:} The map $f(x) = x^2$
(the $c=0$ case of $x^2+c$)
has a stable fixed point $x=0$ and unstable orbits of order $n$
consisting of points on unit circle,
which solve equation $x^{2^n} = x$ and do not
lie on lower-order orbits:
\be
\exp\left( 2\pi i\frac{2^k+j}{2^n-1}\right),\ \ \
k=0,\ldots,n-1.
\ee
The Julia set -- the attraction domain of $x=0$ --
is the interior of the unit circle (the unit disc)
and all periodic unstable
orbits fill densely its boundary, as requested by (i).
Unstable fixed point is $x=1$ and its grand orbit consists of
all the points $\exp \left(\frac{2\pi i l}{2^m}\right)$,
$l = 0,1,\ldots,2^m-1$ of its
$m$-th pre-images, which fill densely the unit circle, as requested
by (ii).
The grand orbit of any point $x = r e^{i\varphi}$,
which does no lie on a unit
circle, $|x|=r \neq 1$, consists of the points
$x^{2^k}$, with all integer $k$. For negative $k = -m$ the choice of one
among $2^m$ phase factors $\exp \left(\frac{2\pi i l}{2^m}\right)$
should be made in order to specify one particular branch of the
grand orbit.
The branch can be also labeled by the sequence of
angles $\{\varphi_m\}$, such that
$\varphi_{m+1} = \frac{1}{2}(\varphi_{m} + 2\pi s_m)$,
where $\{s_m\}$ is a sequence with entries $0$ or $1$. If the sequence is
periodic with the period $n$, i.e. $s_{m+n} = s_m$, then
$\varphi_{m+n} = \frac{1}{2^n}(\varphi_m + 2\pi p_m^{(n)},$
where $p_m^{(n)} = \sum_{i=0}^{n} 2^{i}s_{m+i}$.
Then the origin of the corresponding branch is a periodic orbit
of the order $n$, containing the point
on the unit circle with the phase
$\varphi_m^{(n)} = \frac{2\pi p_n}{2^n-1}$,
which solves the equation
$\varphi_m^{(n)} = \frac{1}{2^n}(\phi^{(n)} + 2\pi p_m^{(n)})$.
Different $m \in (m_0+1,\ldots, m_0+n)$ describe $n$ different elements
of the periodic orbit, and $m_0$ labels
the place where the sequence becomes periodic.
For pure periodic sequences one can put $m_0=0$.

For $n=1$ we have two options:
\be
s_1=0,\  \ p_1=0, \ \  \varphi^{(1)} = 0\ \ \ {\rm and} \nn \\
s_1=1,\  \ p_1=1, \ \  \varphi^{(1)} = 2\pi
\ee
which have equivalent limits: the two branches originate at the
single unstable orbit of order one, i.e. at the point $x=1$.

For $n=2$ there are two options for the $s$-sequences:
\be
s = \{0,1\},\ \ p_2 = 2,\ \ \varphi^{(2)} = \frac{4\pi}{3}, \nn \\
s = \{1,0\},\ \ p_2 = 1,\ \ \varphi^{(2)} = \frac{2\pi}{3},
\ee
which describe the single existing orbit of order $2$ as the origin for
two different branches. The sequences $s = \{0,0\}$ and $s = \{1,1\}$
are actually of period one, not two

For $n=3$ there are six options for the $s$-sequences:
\be
s = \{0,0,1\},\ \ p_3 = 4,\ \
\varphi^{(3)} = \frac{8\pi}{7}, \nn \\
s = \{0,1,0\},\ \ p_3 = 2,\ \
\varphi^{(3)} = \frac{4\pi}{7}, \nn \\
s = \{0,1,1\},\ \ p_3 = 6,\ \
\varphi^{(3)} = \frac{12\pi}{7}, \nn \\
s = \{1,0,0\},\ \ p_3 = 1,\ \ \varphi^{(3)} = \frac{2\pi}{7}, \nn \\
s = \{1,0,1\},\ \ p_3 = 5,\ \
\varphi^{(3)} = \frac{10\pi}{7}, \nn \\
s = \{1,1,0\},\ \ p_3 = 3,\ \
\varphi^{(3)} = \frac{6\pi}{7},
\ee
which describe the two orbits,
$(2,4,8)\times\frac{\pi}{7}$ and $(6,12,10)\times\frac{\pi}{7}$,
of order $3$ as the origins for
six different branches.

Continuation to higher $n$ is straightforward.

For more general map $f(x) = x^d$ we have
\be
\varphi_{m+1} = \frac{1}{d}(\varphi_{m} + 2\pi s_m),
\ \ \ s_m \in (0,1,\ldots,d-1) = G_d; \nn \\
\varphi_m^{(n)} = \frac{2\pi p_n}{d^n-1}, \ \ \
p_n = \sum_{i=0}^{n-1} {s_i}{d^{i}}.
\ee

\bigskip

In order to understand what happens when $c\neq 0$ we need to study
the deformation of grand orbits with the variation of $c$.
The deformation deforms the disc, but preserves the mutual
positions of grand orbits on its boundary, i.e.
the structure, partly shown in Fig.\ref{disc1}. As soon as we reach
the boundary of the domain $M_1(\mu)$ in Mandelbrot set, the
attractive fixed point loses stability and this means that its
grand orbit approaches the boundary of the disc -- and intersects there
with an orbit of order $n$ (if $f$ approaches the point
$M_1\cap M_n \in \partial M_1$). Every point of $GO_1$
merges with $n$ points of $GO_n$, which are located at different
points of the disc boundary, so that the disc gets separated into
sectors -- the Julia set decomposes into a fractal-like structure,
see Figs.\ref{J12} and \ref{J13}.
When $f$ enters the domain $M_n$, the orbit $O_n$ gets stable and
its grand orbit fills densely the interior of the disc, while $O_1$
remains unstable and lies on the boundary, keeping it contracted (glued)
at infinitely many points.
Moreover, the angles $\alpha$ between the $n$ components merging at
contraction points increase smoothly from $\alpha = 0$ at $M_1\cap M_n$
to $\alpha = \frac{2\pi}{n}$ when the next bifurcation point
$M_n\cap M_l$ is reached. At $M_n\cap M_l$ new contractions are added
-- at the points, belonging to grand orbit $GO_l$,
which become merging points of $l/n$ sectors of our disc. And so on.

\FIGEPS{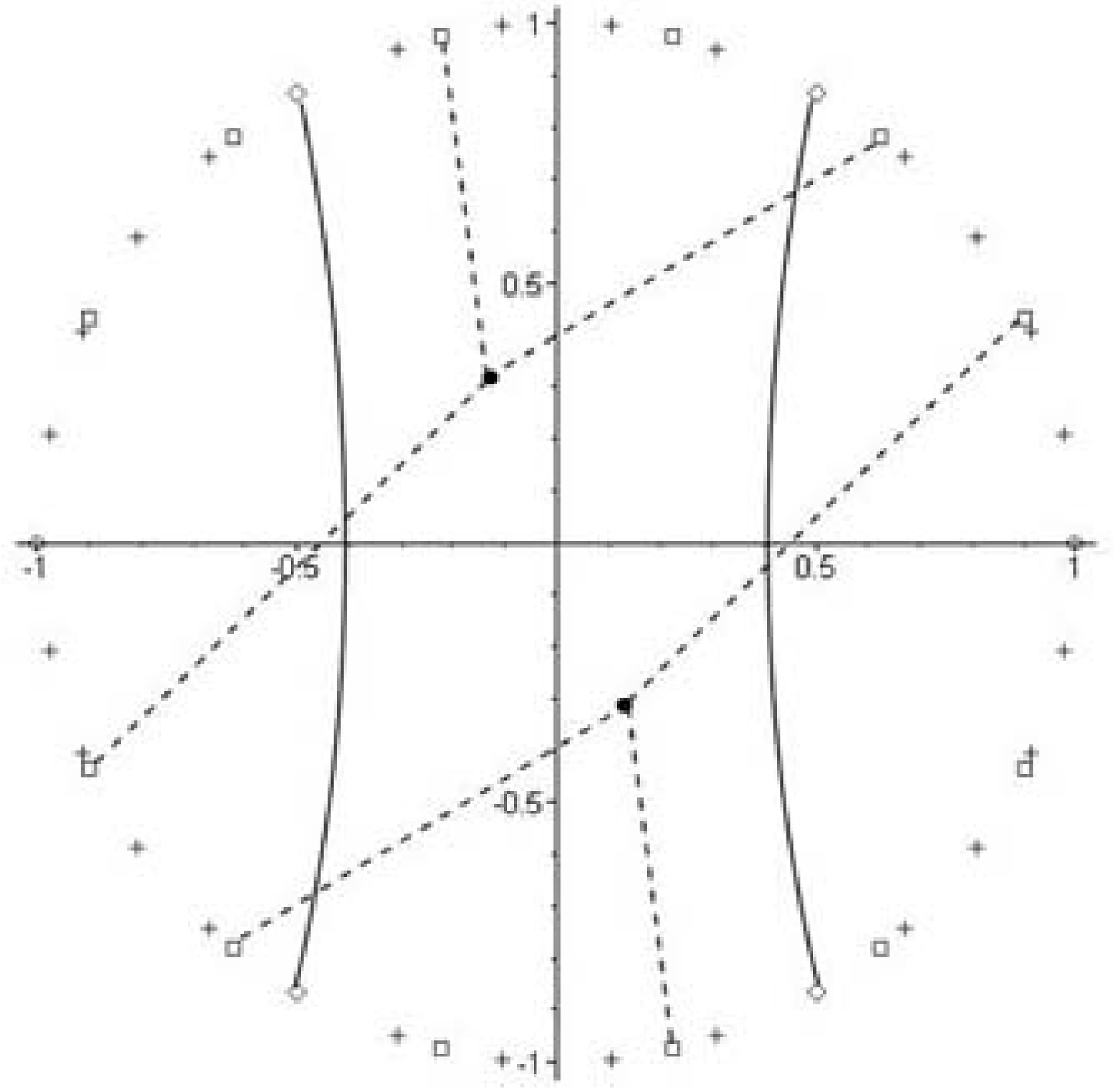}{}
{266,249}
{Unit disc (Julia set at $c=0$) with location of the first
preimages (orbit and the first preorbit) of unstable orbits
of orders one (circles), two
(diamonds), three (boxes) and four (crosses). Locations of
the points are shown for $c=0$, but mutual positions of grand
orbits remain the same for all complex $c$ (though particular
points can change order: a dot can pass through a cross -- when
$c = M_1\cap M_2$, -- but another dot will pass through the
same cross in the opposite direction. Shown are also the lines,
connecting the two points of the orbit
of order two and the two points of its first pre-image -- they
will be among the contracted separatrices in Fig.\ref{J12},--
and analogous lines for an order-three orbit -- to be contracted
in Fig.\ref{J13}.}

\FIGEPS{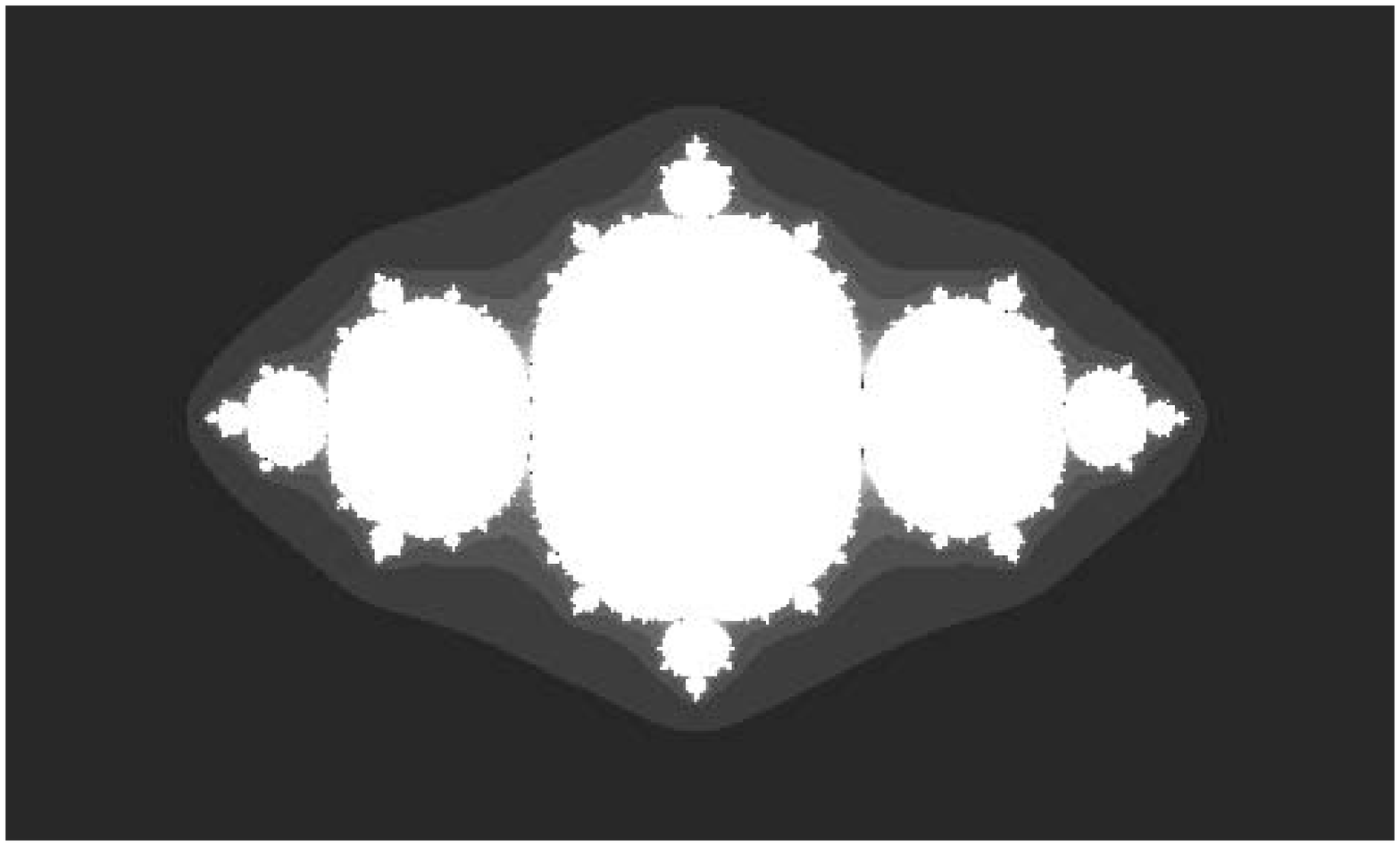}{}
{241,145}
{Julia set at $c=-\frac{3}{4}$, i.e.
when $c = \sigma_1^{(1)}\cap \sigma_2^{(2)}$. See also Fig.\ref{Julia12to1}.
}

\FIGEPS{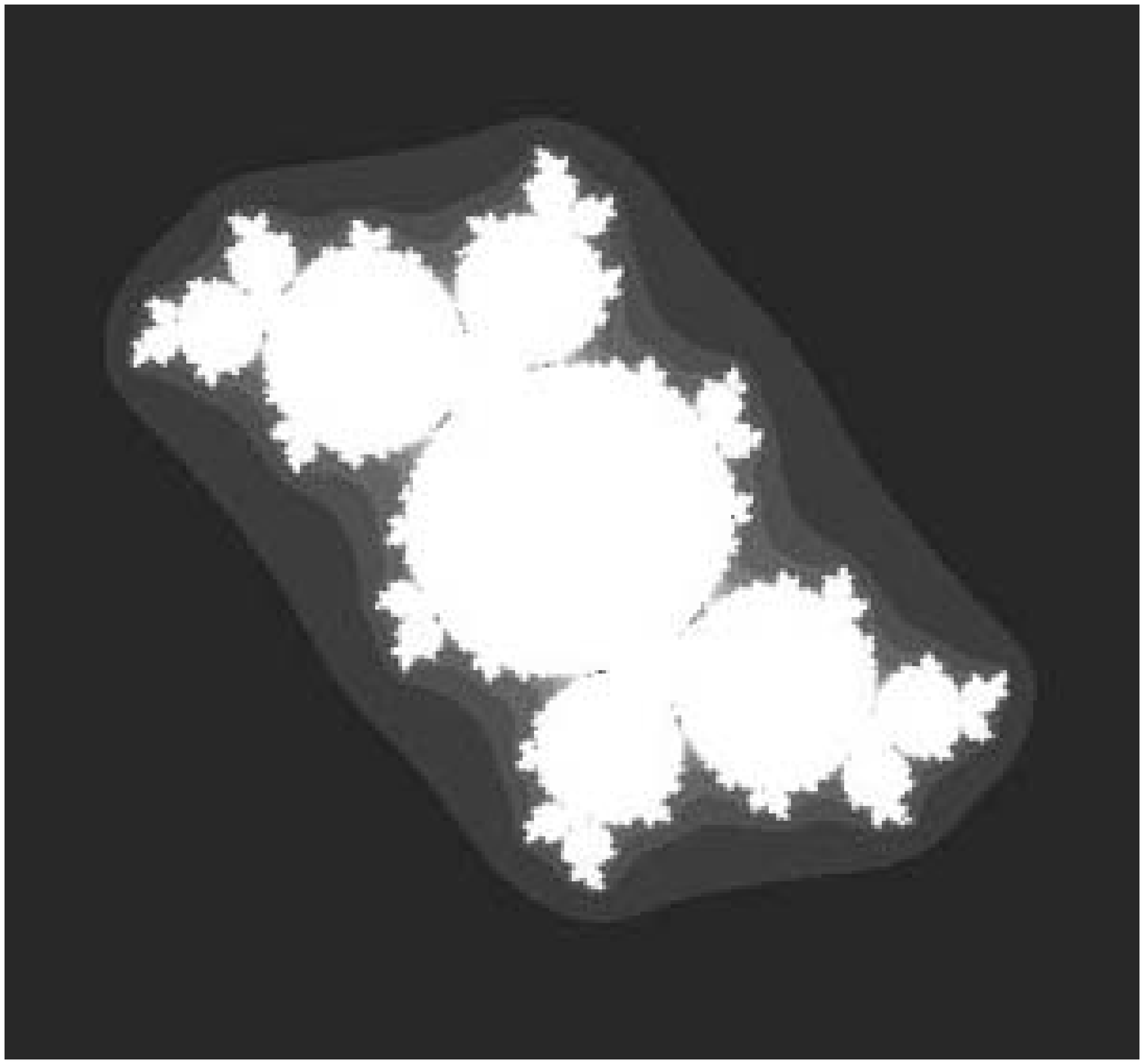}{}
{186,173}
{Julia set at $c=-\frac{1}{8}+i\frac{3}{8}\sqrt{3}$, i.e.
when $c = \sigma_1^{(1)}\cap \sigma_3^{(2)}$. See also Fig.\ref{Julia13to1}.
}

\FIGEPS{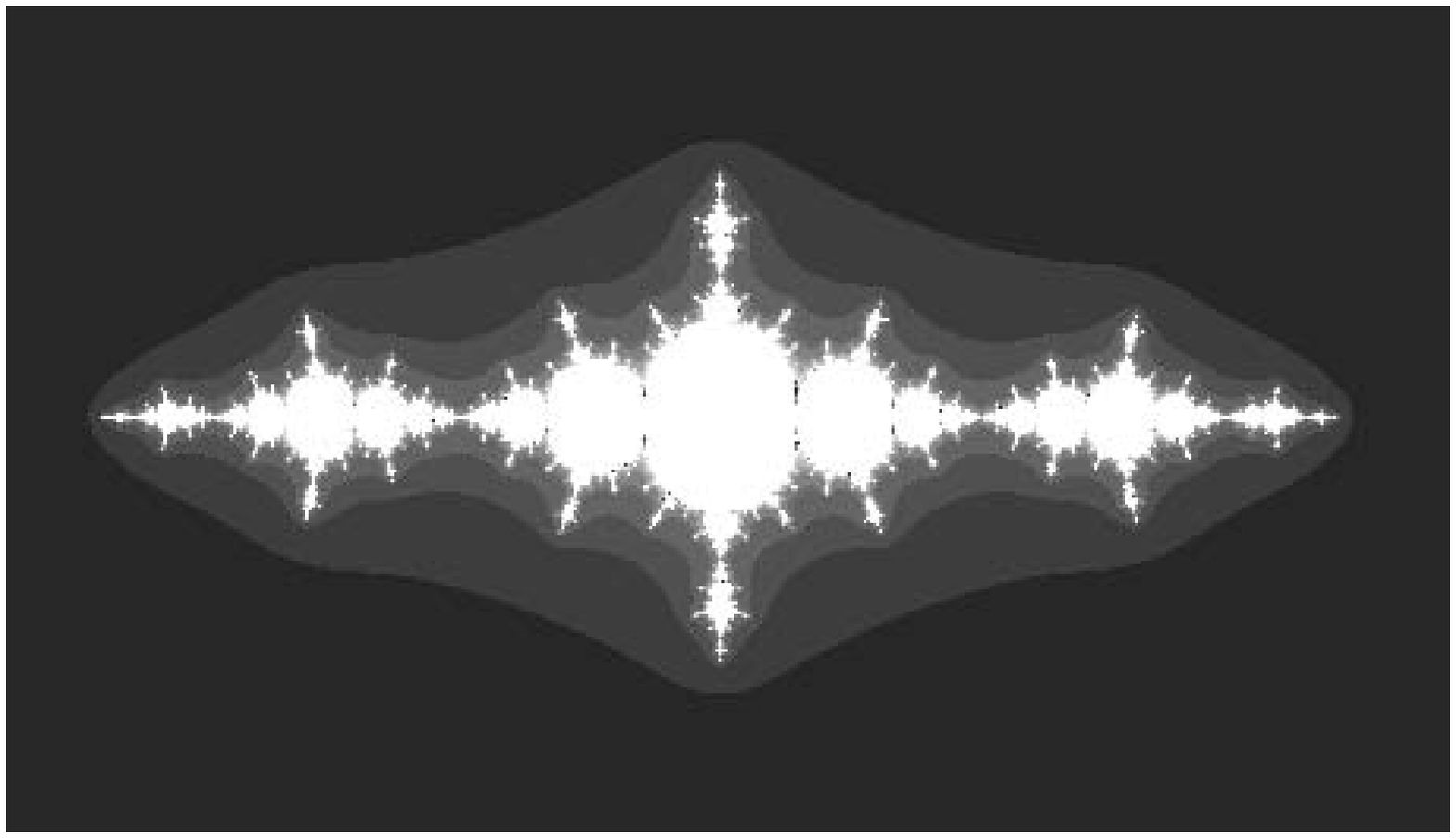}{}
{229,131}
{Julia set at $c=-5/4$, i.e.
when $c = \sigma_2^{(2)}\cap \sigma_4^{(3)}$.
This time the pairs of points from the grand orbit of order two
should be strapped, as well as the quadruples
of points from a grand orbit of order four.}

\FIGEPS{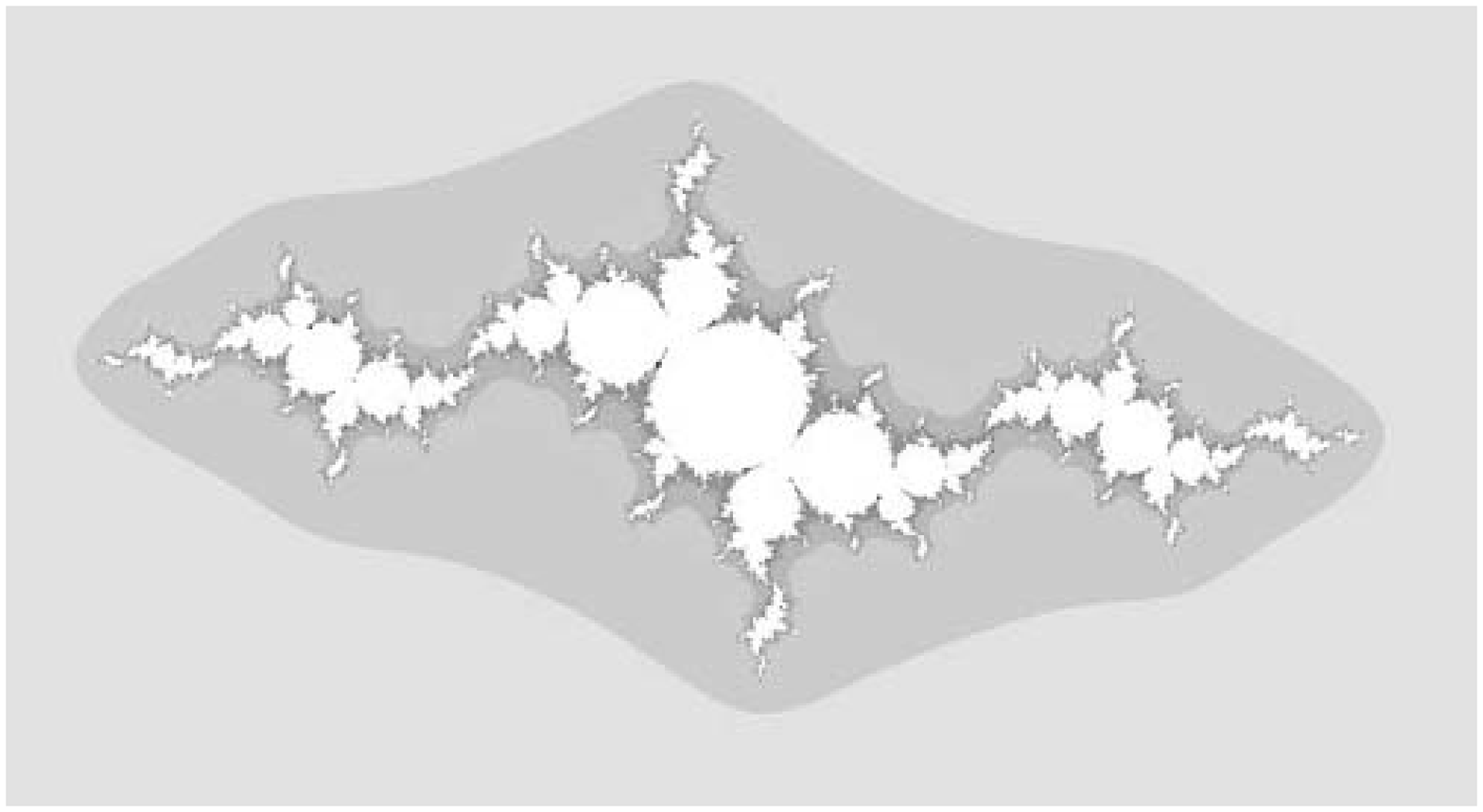}{}
{351,191}
{Julia set at $c=-\frac{9}{8}+i\frac{\sqrt{3}}{8}$, i.e.
when $c = \sigma_2^{(2)}\cap \sigma_6^{(3)}$.
This time the pairs of points from the grand orbit of order two
should be strapped, as well as the sextets
of points from a grand orbit of order six.}

\subsection{Map $f(x) = x^2 + c$. Bifurcations of Julia set and Mandelbrot
sets, primary and secondary
\label{secMa}}

The real slices of the lowest orbits and pre-orbits for
the family $f_c(x) = x^2 + c$ are shown in Fig.\ref{bif3}.
In order to understand what happens beyond real section,
one can study Fig.\ref{bif1s}.

In this case $z(c) = c$ -- this is the point with degenerate pre-image,
which consists of one rather than two points; this is ramification
point of the Riemann surface of functional inverse of the
analytic function $x^2 + c$. Actually, $z(c) = c$ for all the families
$f_c(x) = x^d + c$, but then it has degeneracy $d-1$. This $z(c)$
is $f$-image of the critical point $w(c) = 0$ of multiplicity $d-1$.

\FIGEPS{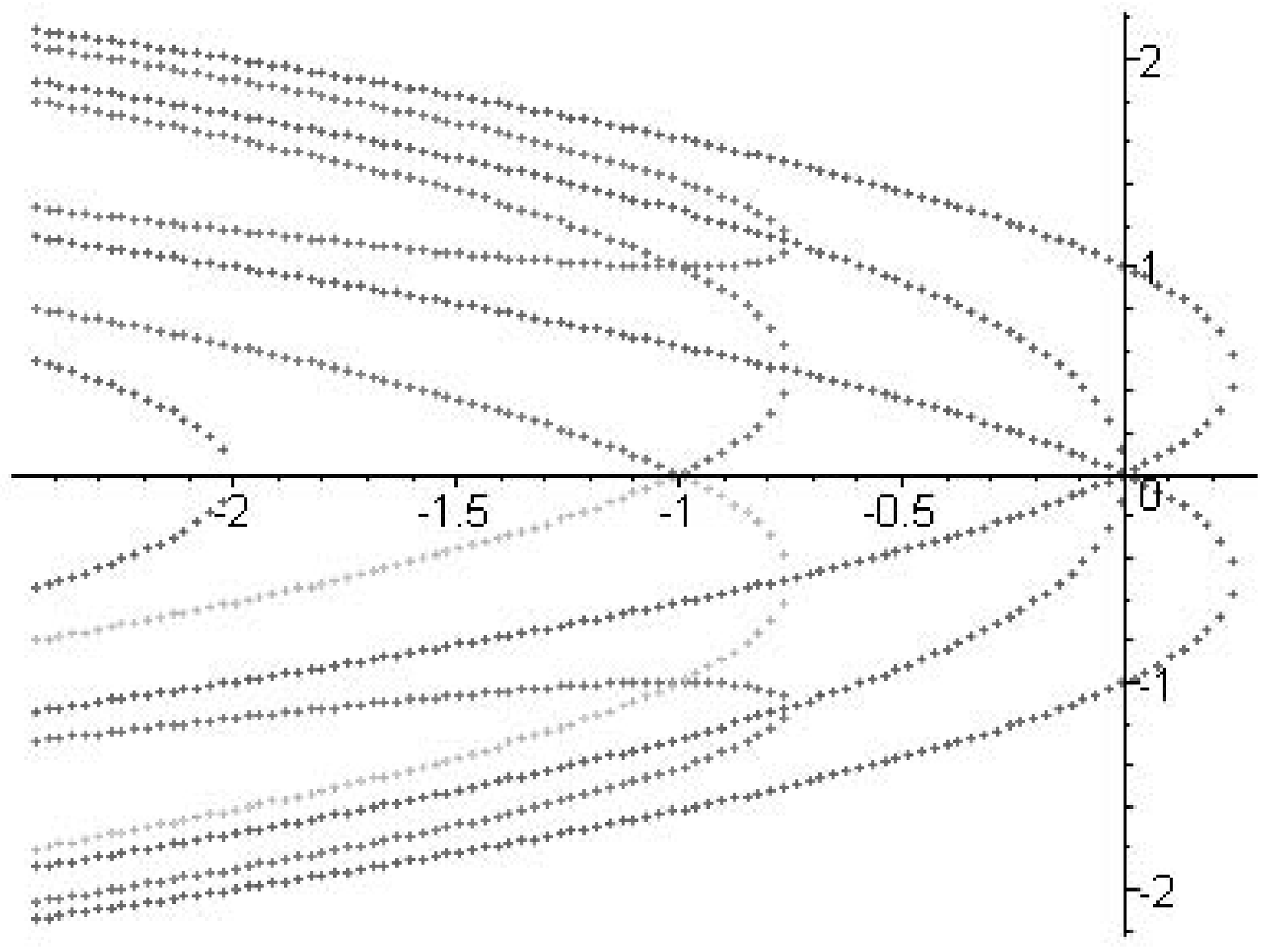}{}
{199,144}
{One more kind of section of Julia sheaf.
The lowest two orbits of $f_c(x) = x^2 + c$:
of orders one ($G_1(x;c)=0$) and two  ($G_2(x;c)=0$)
together with their first two pre-orbits ($G_{1,1}(x;c)=0$,
$G_{1,2}(x;c)=0$ and $G_{2,1}(x;c)=0$,
$G_{2,2}(x;c)=0$). The section ${\rm Im}\ c =0$, ${\rm Im}\ x=0$
of entire 2-complex-dimensional pattern is shown, real $c$ is
on horizontal line, real $x$ -- on the vertical one.
Distinguished points on the horizontal line, where
the roots of various polynomials $G_{n,s}$ coincide, are: $c=1/4$ --
the zero of $d_1(c) = 1-4c$ and thus also of $D(G_{1,1})\sim d_1$ ;
$c=0$ -- the zero of $w_1(c) = c$ and thus also
a zero of $W_{1,2}(c)=w_1(c)w_{1,2}(c)$ and of $D(G_{1,2})
\sim W_{1,2}$;
$c=-3/4$ -- the zero of $r_{12}(c) = 3+4c$, thus vanishing
at this point are also discriminants
$D(G_2)$, $D(G_{2,1})$, $D(G_{2,2})$ and resultants
$r_{1,1|1,2}$, $r_{1,1|2,1}$, $r_{1,2|2,2}$, all proportional to
$r_{12}$; $c=-1$ -- the zero of
$w_2(c) = c+1$ and of the resultants $r_{2|2,1}$, $r_{2|2,2}$,
$r_{2,1|2,2}$, all proportional to $w_2$;
$c=-2$ -- the second zero of $w_{1,2}(c)=c+2$ and thus of
$D(G_{1,2}) \sim W_{1,2}$.
The zeroes of $w_{2,2} = c^2+1$, where the two branches
of $G_{2,2}=0$ intersect, are pure imaginary and not seen
in this real section. $w_{1,1}$ and $w_{2,1}$ are identically unit
for $d=2$.}

\FIGEPS{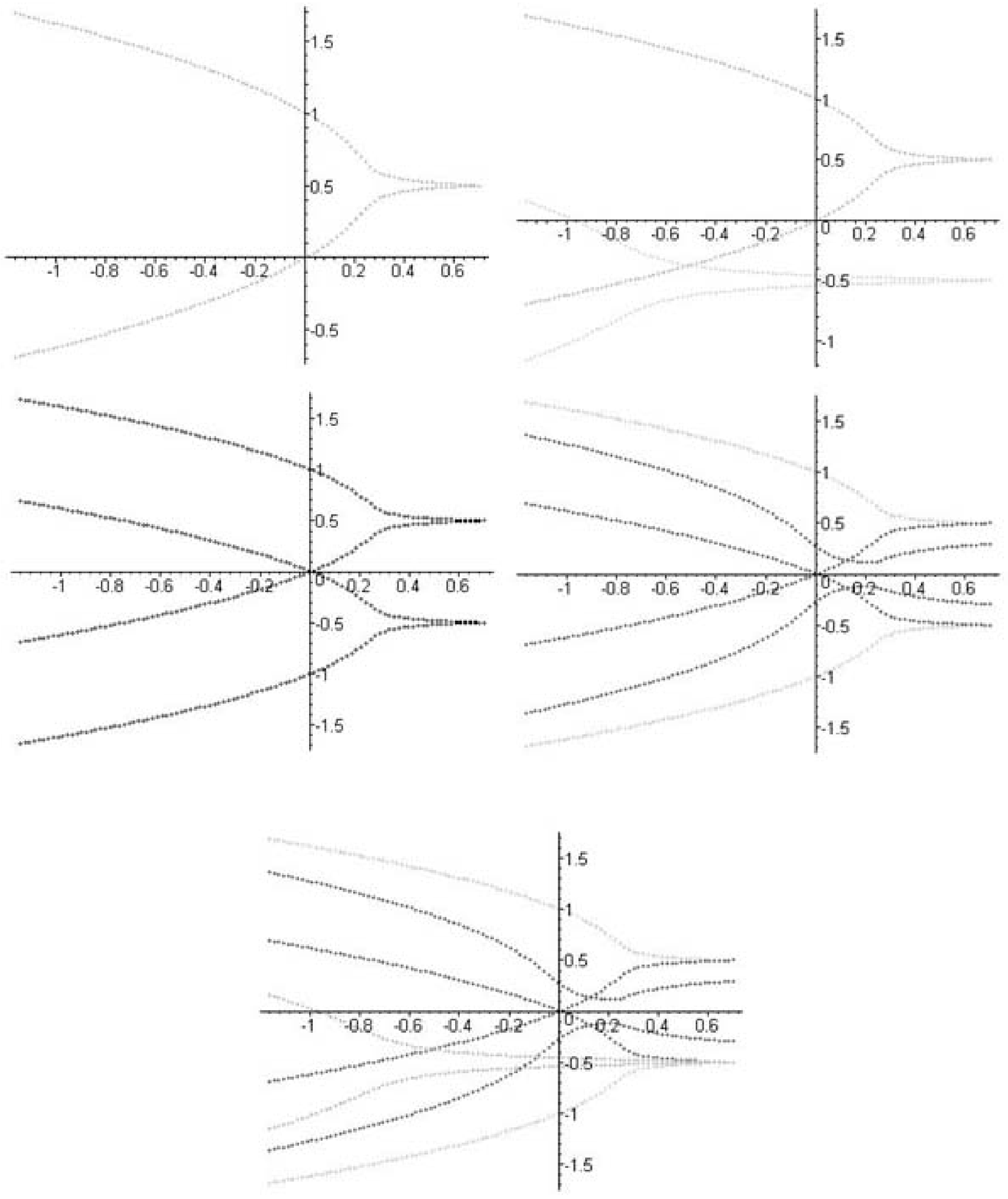}{}
{400,500}
{Instead of the {\it section} of the two-complex-dimensional
pattern by the plane ${\rm Im}\ c =0$, ${\rm Im}\ x=0$, shown in
Fig.\ref{bif3},
this picture represents {\it projection} onto this plane,
with ${\rm Re}\ c$ and ${\rm Re}\ x$ on the horizontal and
vertical axes respectively. If it was
projection of the same section, the only difference
from Fig.\ref{bif3} would be attaching
of horizontal lines to the right of the bifurcation points,
corresponding, say, to
${\rm Re}\left(\frac{1}{2} \pm \frac{\sqrt{1-4c}}{2}\right) = \frac{1}{2}$
for $c>1/4$ (while there are no traces of these complex roots
in the {\it section}, they are well seen in {\it projection}).
If we now slightly change the section, from
${\rm arg}(c) = 0\ {\rm mod}\pi$ to a small, but non-vanishing value
(i.e. consider a section of Mandelbrot set not by a real axis, but
by a ray, going under a small angle), then in projection we get
the curves, shown in these pictures. a) The fixed points
$x = \frac{1}{2} \pm \frac{\sqrt{1-4c}}{2}$. b) The same fixed points
together with the orbit of order two,
$x = -\frac{1}{2} \pm \frac{\sqrt{-3-4c}}{2}$.
c) Fixed points together with their first pre-images,
$x = -\frac{1}{2} \mp \frac{\sqrt{1-4c}}{2}$.
d) The same with the second pre-orbit (zeroes of $G_{1,2}$) added.
e) The same with the orbit of order two added.}

For this family of maps one can straightforwardly study the pre-orbits
and compare with the results of the resultant analysis -- and this
is actually done in Figs.\ref{bif3},\ref{bif1s}.
Shown in these pictures are:

\bigskip

$\bullet$ the two fixed points (orbits of order one), i.e. zeroes of
$G_1(x) = x^2 - x + c$:
$$x = \frac{1}{2} \pm \frac{\sqrt{1-4c}}{2};$$

\bigskip

$\bullet$ their first pre-images, i.e. zeroes of
$G_{1,1}(x) = x^2 + x + c$:
$$x = -\frac{1}{2} \mp \frac{\sqrt{1-4c}}{2}$$
(the lower branch of $G_{1,1}=0$ is mapped by $f$ onto the upper branch
of $G_{1}=0$, $(c,x)=(0,0)$ is a fixed point of a family $f_c(x)$);

\bigskip

$\bullet$ second pre-images, i.e. zeroes of
$G_{1,2}(x) = x^4 + (2c+1) x^2 + c(c+2)$:
$$x = \pm\sqrt{-\frac{1+2c}{2} \mp \frac{\sqrt{1-4c}}{2}};$$

\bigskip

$\bullet$ the orbit of order two, i.e. zeroes of
$G_2(x) = x^2 + x + c + 1$:
$$x = -\frac{1}{2} \pm \frac{\sqrt{-3-4c}}{2};$$

\bigskip

$\bullet$ its first pre-image, i.e. zeroes of
$G_{2,1}(x) = x^2 - x + c + 1$:
$$\frac{1}{2} \pm \frac{\sqrt{-3-4c}}{2}$$.
(this time the lower branch of $G_{2,1}=0$ is mapped by $f$
onto the lower branch of $G_{1}=0$, $(c,x)=(-1,0)$ is {\it not}
a fixed point of a family $f_c(x)$);

\bigskip

$\bullet$ its second pre-images, i.e. zeroes of
$G_{2,2}(x) = x^4 + (2c-1) x^2 + c^2+1$.
$$x = \pm\sqrt{\frac{1-2c}{2} \pm \frac{\sqrt{-3-4c}}{2}}.$$

\bigskip

From these pictures it is clear that along with the zeroes of
ordinary discriminants and resultants (points $c=1/4$ and $c=-3/4$),
well reflected in the structure of universal Mandelbrot set,
new characteristic points emerge: the zeroes of $w_n$ and $w_{n,s}$
(like points $c=0$, $c=-1$ and $c=-2$). Some of these points are
shown in Fig.\ref{Mandwithw} together with the Mandelbrot set.
By definition these additional points are associated with the
boundary of Grand Mandelbrot set. It seems clear from Fig.\ref{Mandwithw}
that there is also an intimate relation between some of these
points with the {\it trail structure} of the Mandelbrot set
itself. In Fig.\ref{Juliaw} we show what happens to Julia set at
zeroes of $w_n$ and $w_{n,s}$. Clearly, zeroes of $w_{n,s}$
mark the places where Julia sets change the number of connected
components.

\FIGEPS{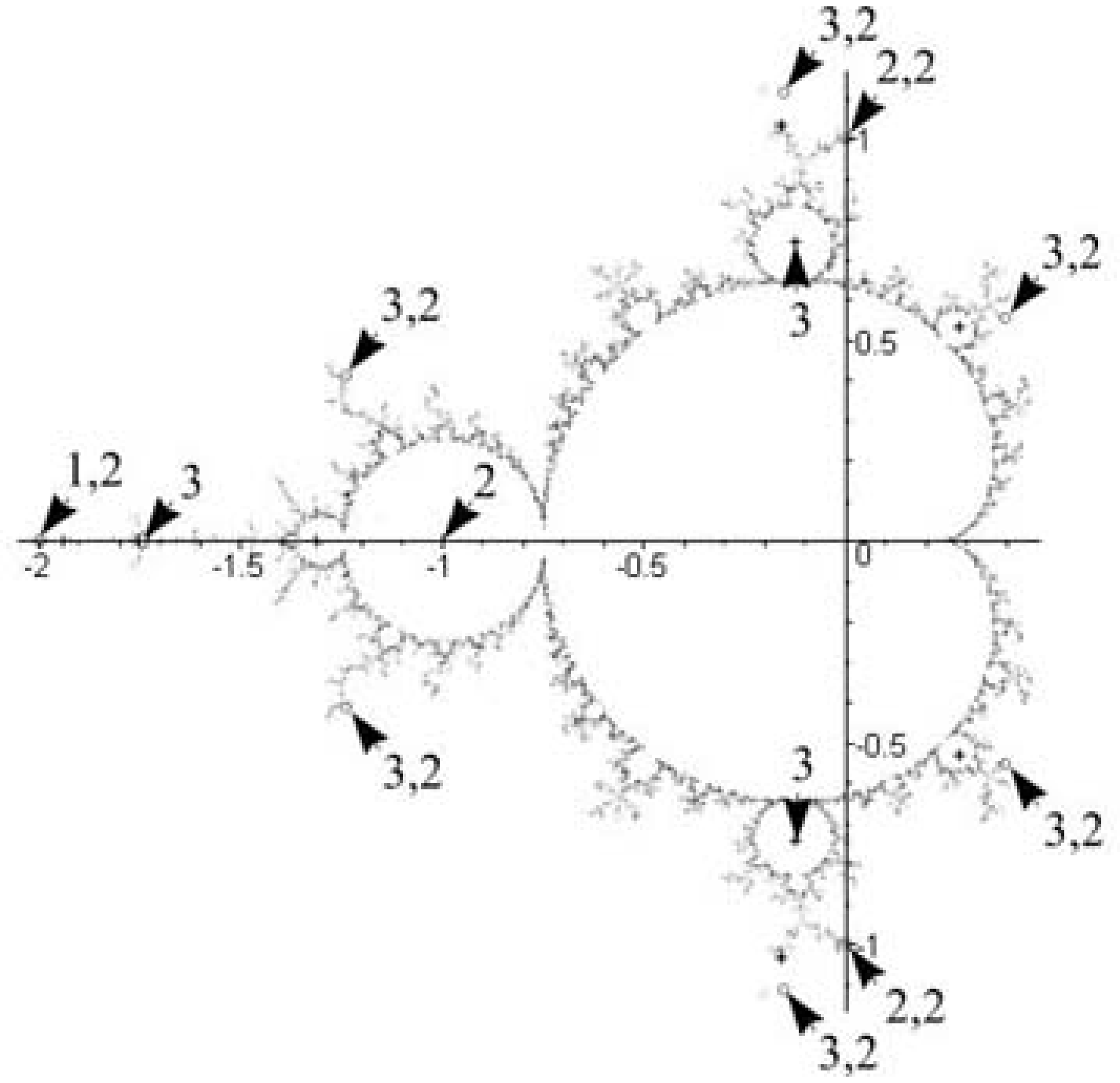}{}
{400,275}
{Mandelbrot set, shown together with the zeroes
of some $w_n$ (crosses) and $w_{n,1}$ (circles), associated with the more
general structure
of Grand Mandelbrot set.}

\FIGEPS{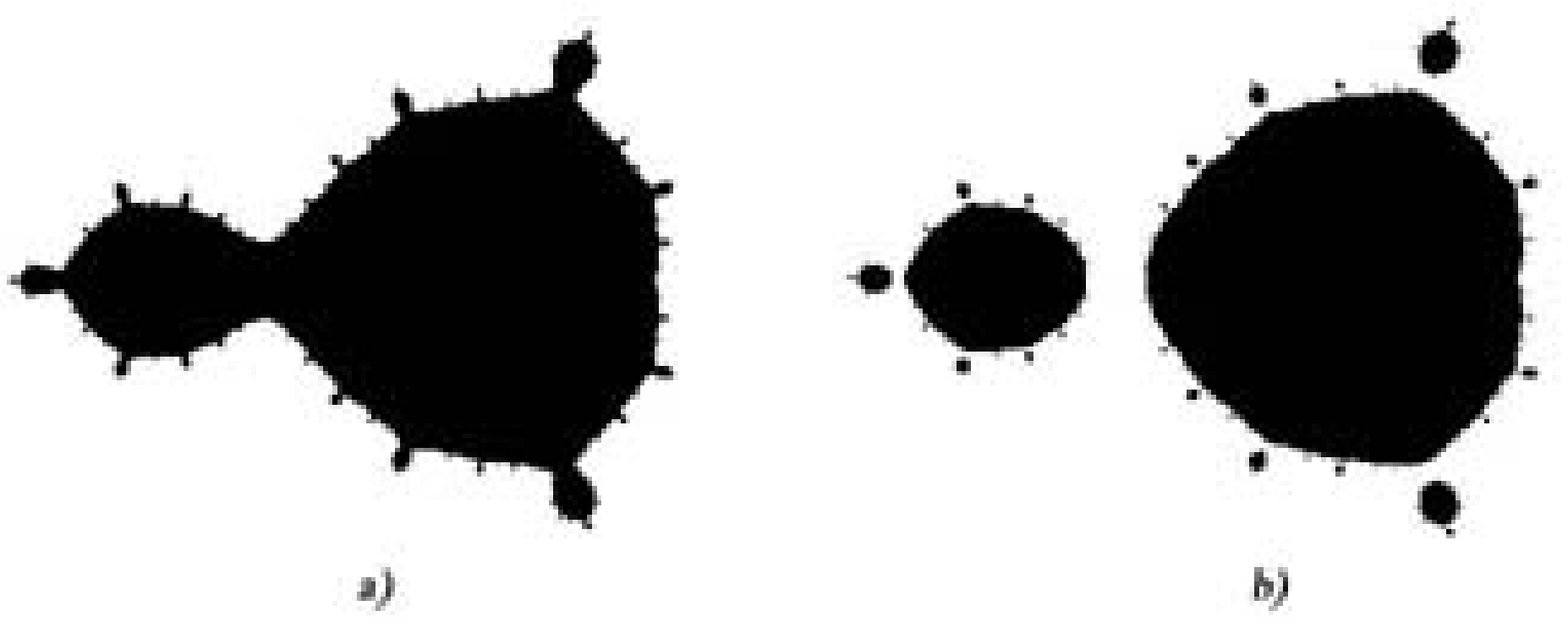}{}
{380,150}
{This picture illustrates how the Julia set gets disconnected
in the vicinity of zero $c=4/9$ of $w_{1,1}$ for the (one-parametric)
family $cx^3+x^2$. Shown are two views: before and after the
decay phase transition.}

Discriminants
for lowest pre-orbits are given by the following formulas
(note that$d_2=1$ and all $w_{n,1}=1$ for $d=2$):

{\footnotesize
$$
\begin{array}{ccccc}
ns&&D(F_{ns}) &&D(G_{ns}) \\
&&&& \\
11&& 2^4d_1^2 w_1^2 &
& d_1 \\
12&& 2^{16}d_1^4w_1^7w_{12} &
& 2^4d_1^2w_1w_{12} \\
13&& 2^{48}d_1^8w_1^{18}w_{12}^3w_{13} &
& 2^{16}d_1^4w_1^3w_{12}^2w_{13}\\
14&& 2^{128}d_1^{16}w_1^{41}w_{12}^7w_{13}^3w_{14} &&
     2^{48}d_1^8w_1^7w_{12}^4w_{13}^2w_{14} \\
15&& 2^{320}d_1^{32}w_1^{88}w_{12}^{15}w_{13}^7w_{14}^3w_{15} &&
     2^{128}d_1^{16}w_1^{15}w_{12}^8w_{13}^4w_{14}^2w_{15}\\
&&&& \\
21&& 2^8d_1^2r_{12}^6w_1^2w_2^2 &
& -r_{12} \\
22&& 2^{32}d_1^4 r_{12}^{12}w_1^7w_{12}w_2^6w_{22} &
& 2^4r_{12}^2w_{22} \\
23&& 2^{96}d_1^8r_{12}^{24}w_1^{18}w_{12}^3w_{13}
     w_2^{15}w_{22}^3w_{23} &&
     2^{16}r_{12}^4w_2w_{22}^2w_{23} \\
24&& 2^{256}d_1^{16}r_{12}^{48}w_1^{41}w_{12}^7w_{13}^3w_{14}
     w_2^{33}w_{22}^7w_{23}^3w_{24} &&
     2^{48}r_{12}^8w_2^2w_{22}^4w_{23}^2w_{24} \\
25&& 2^{640} d_1^{32}r_{12}^{96}w_1^{88}w_{12}^{15}w_{13}^7
     w_{14}^3w_{15}w_2^{70}w_{22}^{15}w_{23}^7w_{24}^3w_{25} &&
     2^{128}r_{12}^{16}w_2^5w_{22}^8w_{23}^4w_{24}^2w_{25}\\
&&&& \\
31&& 2^{16}d_1^2d_3^6r_{13}^8w_1^2w_3^2 &
& d_3^3r_{13}^2 \\
32&&&& 2^{12}d_3^6r_{13}^4w_{32} \\
33&&&& 2^{48}d_3^{12}r_{13}^8w_{32}^2w_{33} \\
34&&&& 2^{144}d_3^{24}r_{13}^{16}w_3w_{32}^4w_{33}^2w_{34} \\
35&&&& 2^{384}d_3^{48}r_{13}^{32}w_3^2w_{32}^8w_{33}^4w_{34}^2w_{35} \\
&&&&\\
41&&&& d_4^4r_{14}^3r_{24}^2 \\
42&&&& 2^{24}d_4^8r_{14}^6r_{24}^4w_{42} \\
43&&&& 2^{96}d_4^{16}r_{14}^{12}r_{24}^8w_{42}^2w_{43} \\
&&&&\\
51&&&& d_5^5r_{15}^4 \\
52&&&& 2^{60} d_5^{10}r_{15}^8w_{52} \\
&&&&\\
61&&&& d_6^6r_{16}^5r_{26}^4r_{36}^3
\end{array}
$$
}

\subsection{Conclusions about the structure of the "sheaf" of
Julia sets over moduli space (of Julia sets and their dependence on
the map $f$)
\label{Juexa}}

\noindent

{\bf 1.} {\bf Julia set} $J(f)$ is defined for a map $f_{\vec c}
\in \mu \subset {\cal M}$. If $f \in \sigma_n(\mu)$, $J(f)$ is an
attraction domain of the stable $f$-orbit of order $n$.
Unstable periodic orbits and all bounded grand orbits (including
pre-orbit of the stable orbit)
lie at the boundary $\partial J(f)$. Every bounded
grand orbit is dense in $\partial J(f)$. This property
allows to define $\partial J(f)$ even for $f \notin M$, i.e.
outside the Mandelbrot set: as a closure of any bounded grand orbit
(or, alternatively, as that of a union of all periodic orbits).
Just outside Mandelbrot set this closure fails to be a boundary of
anything -- there are no stable periodic orbits and
Julia set $J(f)$ itself does not exist.

\bigskip

{\bf 2.}
If $f \in \sigma_n(\mu) \subset M_1(\mu)$, the Julia set
$J(f)$ is actually a
deformation of unit disc. Its exact shape is defined by
grand orbits of unstable periodic orbits, in particular all its
"exterior points"/spikes are preimages of unstable fixed
point(s). Its topology is even more transparent: dictated by
position of $f$ on the tree-powerful of the Mandelbrot set --
as explained at the end of sec.\ref{Juse}.

\bigskip

In more detail:
$J(f)$ is a union of real domains in $\bb{X}$ (sectors of the
unit disc):
\be
J(f) = \cup_{\nu}^\infty J_\nu(f).
\ee
In variance with the Mandelbrot set decomposition into elementary domains
$\sigma_n$'s, the constituents $J_\nu$ (decomposition of the unit disc)
and even their labelings $\nu$ are not uniquely defined.

{\bf 2.1.}
At particular point (map) $f \in \sigma_k^{(p)}\cap \sigma_n^{(p+1)} =
\partial \sigma_k^{(p)}\cap \partial \sigma_n^{(p+1)}$, where
$R(G_k,G_n) = 0$ and two orbits $O_k$ and $O_n$ of orders $n = mk$
intersect, every point of $O_k$ coincides with exactly
$m = n/k$ merged points of $O_n$. Then every point of grand orbit
$GO_k$ is a singular point of $\partial J(f)$, where exactly $m$
components from the set $\{J_\nu[O_n,O_k]\}$ "touch" together at
zero angles, $\alpha = 0$,
see examples of $(k,n)=(1,2)$ in Fig.\ref{J12},
$(k,n)=(1,3)$ in Fig.\ref{J13},
$(k,n)=(2,4)$ in Fig.\ref{J24} and $(k,n)=(2,6)$ in Fig.\ref{J26}.
There are no other singularities of $\partial J(f)$ at
$f \in \sigma_k^{(p)}\cap \sigma_n^{(p+1)}$.

\FIGEPS{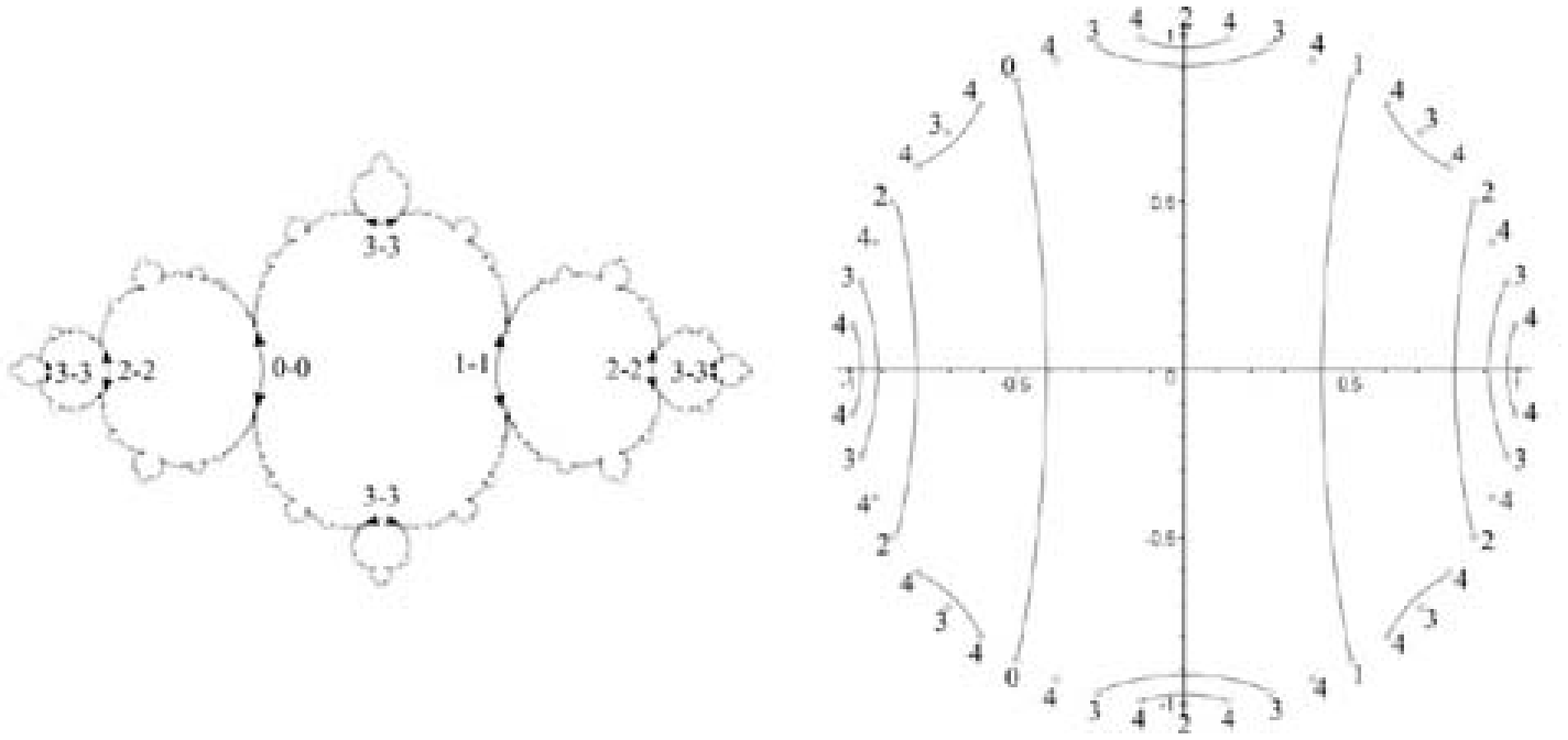}{}
{400,178}
{The deformation of Fig.\ref{J12}:
Julia set in the vicinity of intersection
point $c=\sigma_1^{(1)}\cap \sigma_2^{(2)}$ with
$c \in \sigma_1^{(1)}$. It is obtained from the unit
disk by strapping the {\it pairs} of points on the
boundary, which belong to grand orbit of the orbit
of order $2$. Domains $J_\nu$ overlap and separatrices
connect the points of grand orbit of order $2$, which are
strapped at the intersection (phase transition) point. The length
of separatrices can serve as an order parameter in this phase
($\sigma_1^{(1)}$).
Actually, there are infinitely many different order parameters
inside $\sigma_1^{(1)}$, associated with transitions to different
$\sigma^{(2)}_n$. Separatrices are closures of peculiar
(not-bounded) $f$-orbit and of its pre-images.}

\FIGEPS{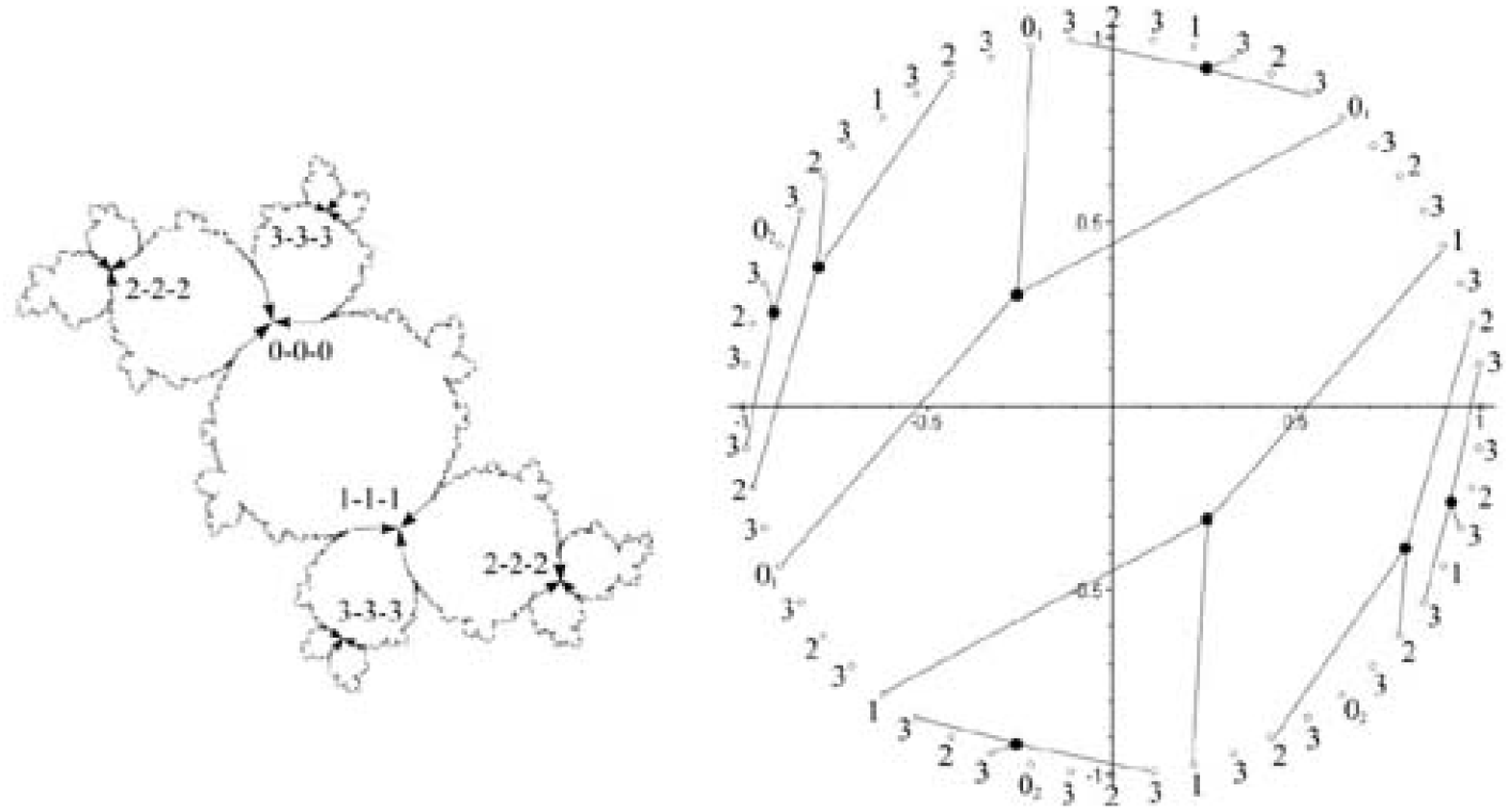}{}
{400,209}
{Analogous deformation of Fig.\ref{J13}:
Julia set in the vicinity of intersection
point $c=\sigma_1^{(1)}\cap \sigma_3^{(2)}$ with
$c \in \sigma_1^{(1)}$. It is obtained by strapping the
{\it triples} of points on the boundary of the disk,
belonging to a grand orbit of the order $3$. }

\FIGEPS{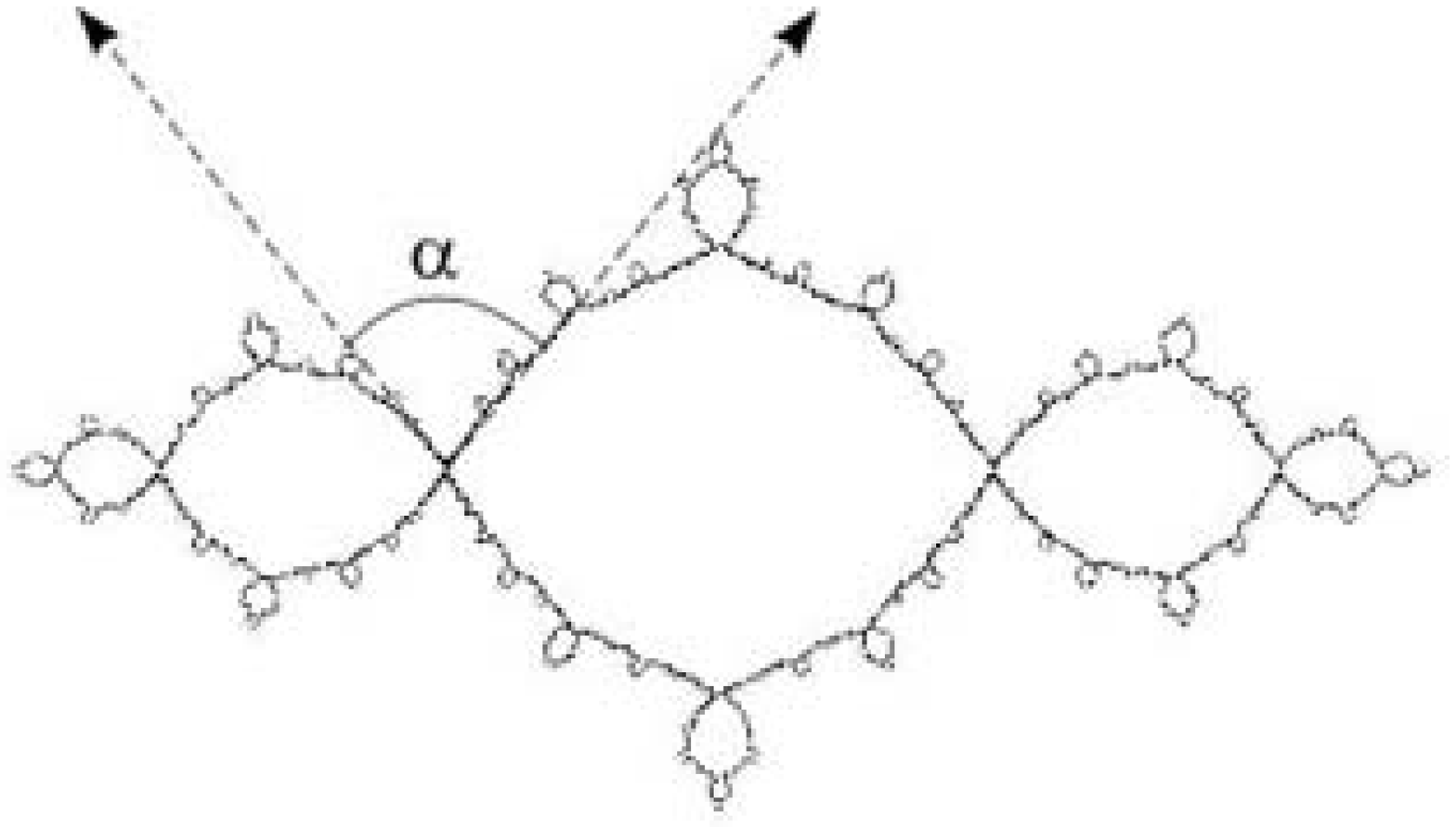}{}
{300,168}
{Deformation of Fig.\ref{J12}
in another direction: Julia set in the vicinity of intersection
point $c=\sigma_1^{(1)}\cap \sigma_2^{(2)}$ with
$c \in \sigma_2^{(2)}$. In this phase the order parameter is
the angle $\alpha > 0$ between the merging domains. It
increases from $\alpha = 0$ at transition point
$c=\sigma_1^{(1)}\cap \sigma_2^{(2)}$ to the maximal value
$\alpha = 2\pi/m = \pi$ when $c$ reaches other points of
the boundary $\partial \sigma^{(2)}_2$, i.e. when the next
phase transition occurs, e.g. when
$c=\sigma_2^{(2)}\bigcap \sigma_4^{(3)}$ (Fig.\ref{J24})
or $c=\sigma_2^{(2)}\bigcap \sigma_6^{(3)}$ (Fig.\ref{J26}).}

\FIGEPS{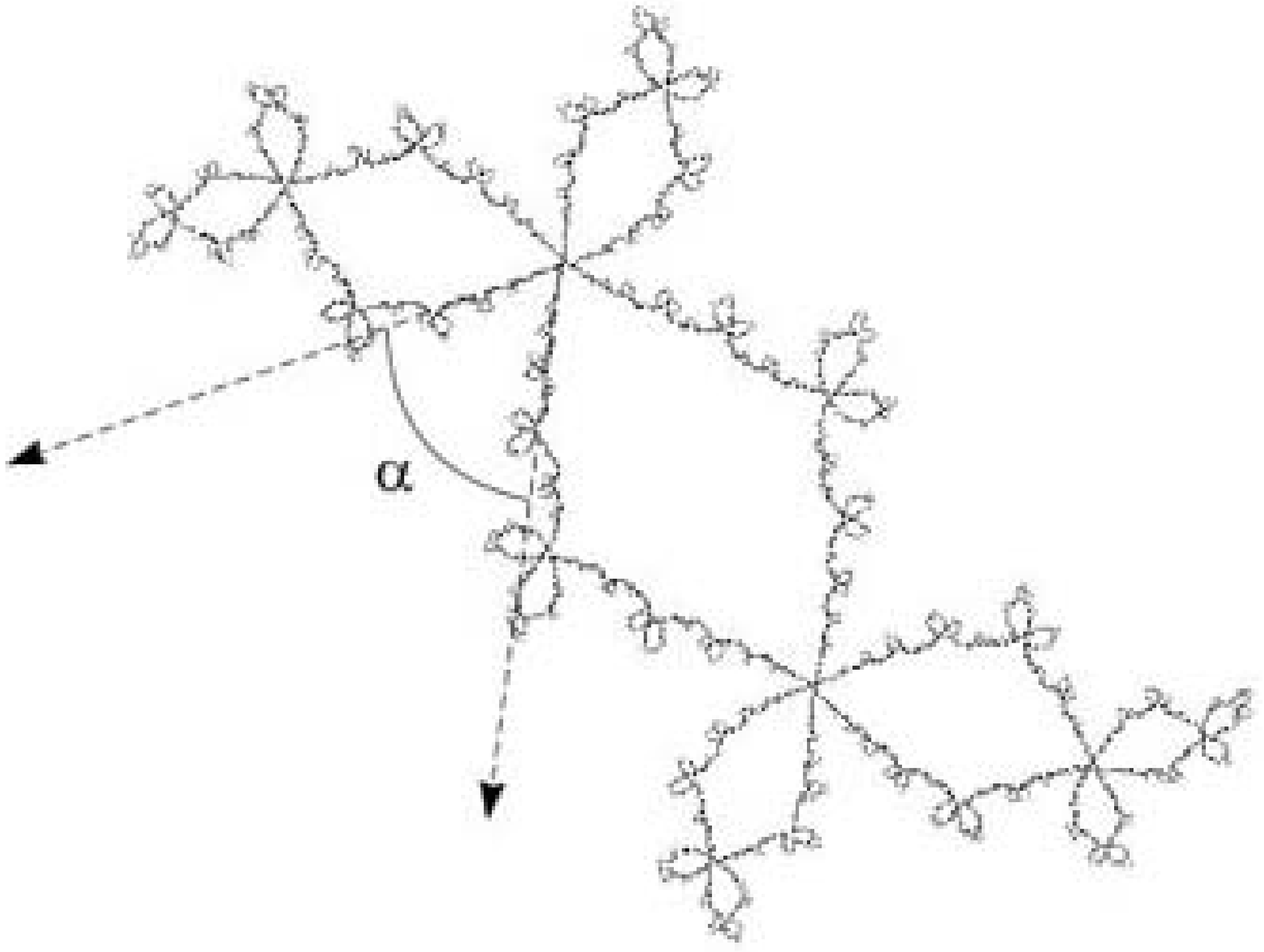}{}
{338,230}
{Analogous deformation of Fig.\ref{J13}:
Julia set in the vicinity of intersection
point $c=\sigma_1^{(1)}\cap \sigma_3^{(2)}$ with
$c \in \sigma_3^{(2)}$. In this phase the angles $\alpha$
change from $\alpha = 0$ at $c=\sigma_1^{(1)}\cap \sigma_3^{(2)}$ to
$\alpha = 2\pi/m = 2\pi/3$ at any other point of the boundary
$\partial \sigma^{(2)}_3$.}

{\bf 2.2.} If we start moving $f$ inside $\sigma_k^{(p)}$
($k < n$), every
$m$-ple from $\{J_\nu[O_n,O_k]\}$, which was {\it touching}
at a single point for $f\in \sigma_k\cap \sigma_n$, acquires common
boundary segments and
the singularities get partly resolved: a point turns into $m$
points and the full non-degenerate orbit $O_n$ is restored, see
Figs.\ref{Julia12to1} and \ref{Julia13to1}.
On this side of the $(k,n)$-"phase transition" the "order parameter"
is the length of the common segments (separatrices):
the deeper in $\sigma_k$ the longer
the segments. These segments can be also considered as
peculiar unbounded grand orbit which densely fills "separatrices"
between the merging orbits $O_n$ (on $\partial J(f)$) and $O_k$
(inside $J(f)$) when we approach the transition point from below --
from the side of $\sigma_k$ ($k < n$).

{\bf 2.3.} If we start moving $f$ inside $\sigma_n^{(p+1)}$
($n > k$), every
$m$-ple from $\{J_\nu[O_n,O_k]\}$, which was {\it touching}
at a single point for $f\in \sigma_k\cap \sigma_n$, continues to intersect
at a single point -- an orbit $O_k$ is now unstable and should
stay at the boundary $\partial J(f)$,-- but angle $\alpha$
(the "order parameter" above the $(k,n)$-"phase transition")
is no longer vanishing, but increases as we go deeper into $\sigma_n$,
see Figs.\ref{Julia12to2} and \ref{Julia13to3}.
This is how the orbits $O_k$ and $O_n$
exchange stability: $O_n$ leaves $\partial J(f)$ while $O_k$
emerges at $\partial J(f)$. Since $n>k$ this is not so trivial to
achieve and the problem is resolved by creating highly singular
merging-points of $m$ varieties at all the points of
the grand orbit $GO_k$, when it emerges at $\partial J(f)$.

{\bf 2.4.} If, while moving inside $\sigma_n^{(p+1)}$ ($n>k$),
the map $f$ reaches a
new intersection point $\sigma_n^{(p+1)}\cap \sigma_N^{(p+2)}$,
$N>n$, our $(k,n)$-structure
($n/k$-merged points at $GO_k$) feels this!
Namely, when it happens, the angles
$\alpha_{k,n}$ reach their maximal values of
$\frac{2\pi}{m}$, our $m=n/k$
merging varieties turn into needles in the vicinity of
the merging points (i.e.
at points of $GO_k$), see Figs.\ref{J24} and \ref{J26}.

{\bf 2.5.} All this describes the behaviour of $J_\nu$ sets,
associated with the subvariety
$\sigma_k^{(p)}\cap \sigma_n^{(p+1)} \subset \partial M(\mu)$
at the boundary between smooth constituents of
Mandelbrot set. However, if we are at some
point {\it inside} $\sigma_k$, there are many different structures of this
type, coming from different boundary points (intersections with
different $\sigma_{km}$ with all integer $m$),
and no one is distinguished. Thus the
decomposition $J(f) = \cup_\nu J_\nu$ is ambiguous for $f$ inside
$\sigma_k$ and gets really well defined only
exactly at $\partial \sigma_k$ -- though is
an everywhere discontinuous function on this boundary, changes
abruptly as $f$ moves along $\partial \sigma_k$.

{\bf 2.6.} If $f$ crosses the boundary $\partial M(\mu)$   and goes
outside the Mandelbrot set then there are no stable orbits left
and nothing can fill the disc: only the boundary remains,
the Julia set has no "body" -- it becomes a collection of
disconnected curves, which are no longer boundaries of anything
(but are still densely filled by bounded grand orbits and/or unstable
periodic orbits).
This dissociation of the boundary occurs at discriminants points,
where the two orbits of the same order, one stable and another
unstable on the Mandelbrot-set side cross and become unstable
beyond the Mandelbrot set. When they diverge after crossing,
the orbits (and their entire grand orbits)
pull away their pieces of the boundary of Julia set, and
the "body" of Julia set "leaks away" through emerging holes,
see Fig.\ref{Jdeco}.

\bigskip

Despite it is not exhaustive, this decomposition structure  carries
significant information, in  fact it is information
about {\it topology} of Julia sets and singularity structure
of their boundaries. A remote region $\sigma_n^{(p)}$ in
the main component of Mandelbrot set $M_1(\mu)$
can be reached from $\sigma_1^{(1)}$ by a path which lies entirely
inside $M(\mu)$ and passes through a sequence of
points $\sigma_k\cap \sigma_l$
connecting different components. Passage through every such
separation point changes Julia set in above-described way,
drastically changes its topology and angles $\alpha$.
However, the change of the {\it shape} (i.e. geometry
rather than just topology) of $J(f)$
when $f$ is varied {\it inside} every particular component $\sigma_n$
is not described in equally exhaustive manner.

In other words, the suggested theory gives a
complete description of the following characteristics of Julia set:

-- "Extreme" points (like "spikes" and "pits" in Fig.\ref{J1-4})
-- they belong to grand orbits
of unstable fixed points (orbits of order one);

-- Merging points of sectors and needles -- they belong
to grand orbits, which became unstable on the path of $f$ from inside
the region $\sigma_1^{(1)}(\mu)$ in Mandelbrot set $M_1(\mu)$,
where the order-one orbit
was stable. Every merging point is associated to particular "bridge
crossing" at the point
$\sigma_n\cap \sigma_k$ in $M_1(\mu)$, $k|n$, where the stable
orbit $O_k$ "exchanges stability" with $O_n$.
The number of merging components is equal to $m=n/k$.
The angles between merging components of Julia set depend on how far
we are from the corresponding "bridge crossing", and change from
$\alpha = 0$ (touching) at the bridge  to $\alpha = \frac{2\pi}{m}$
(needle) when the next bridge is reached.

{\bf 3.} The shape of $\partial J(f)$ is defined pure algebraically,
because all the bounded grand orbits lie densely inside $\partial J(f)$.
However, in variance with the boundary $\partial M$ of Mandelbrot
sets, $\partial J(f)$ can not be decomposed into smooth constituents
(like $\partial \sigma_n$ which are smooth almost everywhere):
no countable set of points from bounded grand orbits fill densely
any smooth variety (like the zeroes of $R(G_{k},G_{km})$ did for given
$k$ and arbitrary $m$ in the case of $\partial \sigma_k$).

{\bf 4.} Unbounded (generic) grand orbits lie entirely either inside $J(f)$
or beyond it. Each unbounded grand orbit originates in the vicinity of
entire $\partial J(f)$: for every point of $\partial J(f)$
and for every grand orbit there is a
branch which comes (with period $d$) close enough to this point.
If grand orbit lies in $J(f)$, it tends to a stable orbit --
of order $n$ if $f \in \sigma_n$. If grand orbit lies outside of $J(f)$,
it tends to infinity.
One can say that infinity is a "reservoir" of stable orbits: if the
degree $d$ of polynomial map $f$ is increased, the new stable orbits
come from infinity.

{\bf 5.} For $f \in M_{n\alpha}$ lying in other
components of Mandelbrot set, disconnected from $M_1$
(but linked to it by densely populated
{\it trails}), the structure of Julia set is similar, with the only
exception: the starting point is not a single disc, but a collection
of discs, formed when the map $f$ leaves $M_1$ and steps onto one
or another trail leading to the other components $M_{n\alpha}$.
Both periodic orbits and pre-orbits jump between the boundaries
(if unstable) or interiors (if stable) of these discs.
With this modification, the same description of Julia sheaf is valid,
with "monodromies" described by
the same procedure of interchanging stable grand orbits inside with
unstable grand orbits on the boundary, which causes fractalization
of the boundary.

{\bf 6.} If $f \notin M(\mu)$ gets outside the Mandelbrot set, the
Julia set is no longer a disc or collection of discs: it looses its
"body" (which would be occupied by the stable bounded grand orbit)
and consists only of the boundary (where unstable periodic orbits and
their grand orbits live). Its bifurcations can still be described
algebraically, but now the intersections of unstable orbits will be
the only ones to play the role. These bifurcations are controlled by the
Grand Mandelbrot set.

\newpage

\section{Other examples
\label{Exa2}\label{moreexa}}

In this section we present some more examples of the few-parametric
families of maps $f_c(x)$ in order to show that our considerations
and conclusions are in no way restricted to the peculiar case of
$f_c(x) = x^2 + c$. For every family $\mu\subset {\cal M}$
we describe the Mandelbrot set $M(\mu)$ as the union of stability
domains $M_n(\mu)$, its boundary $\partial M(\mu)$ as the closure
of appropriate discriminant variety, the structure of grand orbits and
of algebraic Julia sets.

We are going to illustrate and support the following claims:

$\bullet$ Consideration of various $1$-parametric families shows
universality of the tree structure of Mandelbrot sets and
non-universality of particular numbers: shows the difference between
$n,k,m$- and $\alpha$-parameters.

$\bullet$ Particular $1$-parametric families
(like $\{f_c(x) = x^d + c\}$ or $\{f(x) = cx^d+x^2\}$)
are not in generic position -- even among the maps of given degree --
and the corresponding sections of Mandelbrot/discriminant variety
are not fully representative: have additional special features,
symmetry properties and accidental degeneracies.

$\bullet$ $2$-parametric families can be considered as families
of $1$-parametric families. Their analysis helps to understand
how the non-universal components of the description change and
helps to understand their nature and relation to next-level algebraic
structures like singularities of resultant varieties, controlled
by the resultants of the higher order. Clearly, non-universality
is no more but the corollary of our restriction to ordinary resultants
and further work will provide unique interpretation to all
properties of particular sections/families $\mu$.

In variance with the previous section \ref{Exa} we do not give
exhaustive treatment of every particular example below.
Each example serves to illustrate one or another particular aspect
of the problem and a particular line of further development.

\subsection{Equivalent maps}

Different maps $f$ and $\tilde f$
can be equivalent from the point of view
of our considerations if they are related by a diffeomorphism
$\phi$ of $\bb{X}$, $f\circ \phi = \phi \circ \tilde f$, or
\be
f(\phi(x)) = \phi(\tilde f(x)).
\label{diffeo}
\ee
Then, obviously, $f^{\circ n} \circ \phi = \phi \circ
\tilde f^{\circ n}$,
$F_n\circ \phi = \phi\circ \tilde F_n$, $G_n \circ \phi =
\phi \circ \tilde G_n$ and so on.
If $f$ and $\tilde f$ are polynomials of the same degree -- what
will be the most important application of equivalence below --
than $\phi$ should be a linear transformation, $\phi(x) = ax+b$.

\subsection{Linear maps
\label{linear}}

\subsubsection{The family of maps $f_{\alpha\beta}
= \alpha + \beta x$}

In this case
$$
F_n(x;f_{\alpha\beta}) =
(\beta^n - 1)\left(x + \frac{\alpha}{\beta - 1}\right)
$$
and decomposition formula (\ref{rnk}) reads:
\be
F_n = \prod_{k|n}^{\tau(n)} G_k, \nn \\
G_1(x;f_{\alpha\beta}) = (\beta - 1)x + \alpha, \nn \\
G_k(x;f_{\alpha\beta}) = g_k(\beta),\ \ k>1.
\label{lindeco}
\ee
The functions $g_k(\beta)$ are circular polynomials:
$$
\beta^n - 1 = \prod_{k|n}^{\tau(n)} g_k(\beta)
$$
with
\be
g_1(\beta) = \beta - 1,\nn \\
g_2(\beta) = \beta + 1 = -g_1(-\beta), \nn \\
g_3(\beta) = \beta^2 + \beta + 1,\nn \\
g_4(\beta) = \beta^2 + 1,\nn \\
g_5(\beta) = \beta^4 + \beta^3 + \beta^2 + \beta + 1,\nn \\
g_6(\beta) = \beta^2 - \beta + 1 = g_3(-\beta),\nn \\
\ldots
\label{gpol}
\ee
For simple $p>2$
\be
g_p(\beta) = \frac{\beta^p -1}{\beta - 1} =
\beta^{p-1} + \ldots + \beta + 1, \nn \\
g_{2p}(\beta) = \frac{\beta^p +1}{\beta + 1} =
\beta^{p-1} - \beta^{p-2} + \ldots - \beta + 1,\nn \\
g_{p^k}(\beta) = \frac{\beta^{p^k} -1}{\beta^{p^{k-1}} - 1} =
\beta^{p(p-1)} + \ldots + \beta^p + 1 \nn
\ee
and so on.

All $G_k(x;f_{\alpha\beta})$ in (\ref{lindeco})
except for $G_1$ do not
depend on $x$. This means that there are no orbits of order
$k>1$ unless $\beta$ is appropriate root of unity --
when there are infinitely many such orbits.
For example, when $\beta = -1$ any pair $(x,\alpha - x)$
form a second-order orbit of the map $x \rightarrow \alpha - x$;
and when $\beta = \pm i$ any quadruple $(x,\pm ix+\alpha,
-x + (1\pm i)\alpha, -ix \pm i\alpha)$ is an orbit of order
four etc.

Of course, the family of linear maps is
a highly degenerate example, still this
peculiar section of Mandelbrot/discriminant variety should be
reproduced in particular limits of more general families of
maps below.

\subsubsection{Multidimensional case}

We use the chance to illustrate the new ingredients of the
theory which arise in multidimensional situation.

Take $\bb{X} = \bb{C}^r$ and consider the linear map
$f:\ x \rightarrow Bx$ where $B$ is an $r\times r$ matrix.
The iterated map remains linear, $f^{\circ n}(x) = B^nx$
and $F_n(x) = (B^n-I)x$. Irreducible components $G_k = g_k(B)$
are independent of $x$ for $k>1$, just as in the one-dimensional
situation. However, for $r>1$ degeneration pattern is richer:
of interest are all situations when a pair of eigenvalues
coincide, matrices acquire Jordan form and some two eigenvectors
become collinear.

{\bf The case of $r=2$.} The eigenvector $x = (x_1,x_2)$ of
$B = \left(\begin{array}{cc} b_{11}&b_{12}\\b_{21}&b_{22}
\end{array}\right)$
is a quadric in $\bb{CP}^1$ in homogeneous coordinates, i.e. satisfies
\be
\frac{1}{2} x_iQ_{ij}(B)x_j =
b_{12}x_2^2 + (b_{11}-b_{22}) x_1x_2 - b_{21}x_1^2 = 0.
\label{quadri}
\ee
The linear map $B$ is degenerate when determinant of the
matrix $Q(B)$ (or, what is the same, discriminant of the polynomial
$x_iQ_{ij}x_j/x_2^2$ in $x_1/x_2$ vanishes),
$$
\det_{2\times 2} Q(B) = (b_{11}-b_{22})^2 + 4b_{12}b_{21} = 0.
$$
We call this expression discriminant of the linear map $B$.
For $r>2$ consideration is a little more sophisticated and we
do not discuss it here, see \cite{Doldis}.
The resultant of two maps $A$ and $B$ vanishes when some
eigenvector of $A$ gets collinear to some eigenvector of $B$.
Of course, different iterations of the same linear map have
the same eigenvectors, and resultant analysis
for linear maps remain trivial.
Still, it is important to remember about these structures in
analysis of non-linear multidimensional maps.

\subsection{Quadratic maps}

This class of examples helps to demonstrate, that even
equivalent maps (related by diffeomorphism in ${\cal M}$)
can have differently {\it looking} Mandelbrot sets.
Of course, this is not a big surprise: non-linear change of
variables (say, $c \rightarrow c^2$) in ${\cal M}\cap \mu$
changes the shape and even the number of domains in Fig.\ref{lemMa}.

\subsubsection{Diffeomorphic maps}

Consideration of generic 3-parametric family of quadratic maps
$f_{\alpha\beta\gamma}(x) = \alpha + \beta x+\gamma x^2$
can actually be reduced to that of $f_c(x) = x^2 + c$ by the
rule (\ref{diffeo}), because
$$
\gamma f_{\alpha\beta\gamma}(x) + \frac{\beta}{2} =
f_c(\gamma x + \frac{\beta}{2})
$$
provided
\be
c = \alpha\gamma + \frac{\beta}{2} - \frac{\beta^2}{4}
\label{cabg}
\ee
Then $\gamma f_{\alpha\beta\gamma}^{\circ 2}(x) + \frac{\beta}{2} =
f_c(\gamma f_{\alpha\beta\gamma}(x) + \frac{\beta}{2}) =
f_c(f_c(\gamma x + \frac{\beta}{2})) =
f_c^{\circ 2}(\gamma x + \frac{\beta}{2})$
and, in general,
$$
\gamma f_{\alpha\beta\gamma}^{\circ n}(x) + \frac{\beta}{2} =
f_c^{\circ n}(\gamma x + \frac{\beta}{2}).
$$
In other words, $f_{\alpha\beta\gamma}:\ {\cal M} \rightarrow
{\cal M}$ is diffeomorphic to $f_c:\ {\cal M} \rightarrow {\cal M}$.
Discriminants and resultants for $f_{\alpha\beta\gamma}$
can be obtained from those for $f_c$ by a substitution
(\ref{cabg}). However, since (\ref{cabg}) is non-linear
(quadratic) transformation, it can change the numbers of zeroes,
and the numbers of elementary domains in sections of
${\cal D}^*$ can be different for different
$1$-parametric sub-families of $\{f_{\alpha\beta\gamma}\}$.

For two 2-parametric families to be analyzed below we have
equivalences:
$$
x^2+px+q: \ \ \ p=\beta,\ q= \alpha\gamma,\ \ c=\frac{4q+2p-p^2}{4}
$$
and
$$ \gamma x^2 + (b+1)x:\ \ \ p = b+1, q = 0,\ \ c = \frac{1-b^2}{4}.
$$

\subsubsection{Map $f = x^2 + c$}

First of all, we summarize the discussion of Mandelbrot set in the
previous section \ref{Exa} in the form of a table:
{\footnotesize
\be
\thispagestyle{empty}
\begin{tabular}{|c|ccccccccc|}
\multicolumn{10}{c}{}\\
\multicolumn{10}{c}{}\\
\hline
\multicolumn{10}{|l|}{}\\
\multicolumn{10}{|l|}{\ \ \ $f = x^2 + c$}\\
\multicolumn{10}{|l|}{}\\
\hline
&\ &\ \ &&&&&&& \\
n                            &&    1&2&3&4&5&6&& \ldots \\
&&&&&&&&& \\
\hline
&&&&&&&&& \\
\#\ {\rm of\ orbits}       &&    2&1&2&3&6&9&& \\
= ${\rm deg}_x [G_n]/n$ &&&&&&&&& \\
&&&&&&&&& \\
\#\ {\rm of\ el.\ domains} &&    1&1&3&6&15&27&& \\
= ${\rm deg}_c [G_n(w_c,c)]$   &&    &&&&&&& \\
&&&&&&&&& \\
$d_n$ && 1-4c &$i$&-7-4c&$64c^3+144c^2+$&
    {\rm see\ eq.}(\ref{DG5.2})&
    {\rm see\ eq.}(\ref{dg6})&&  \\
{\rm total}\ \#\ {\rm of\ cusps}&&  &  &     &   +108c+135 &&&& \\
= ${\rm deg}_c [d_n]$             && 1&0&1&3&11&20&& \\
&&&&&&&&& \\
\hline
&&&&&&&&& \\
$r_{n,n/m}$ {\rm \ and}&&&&&&&&& \\
{\rm total}\ \#\ {\rm of}\ (n,n/m)&&&&&&&&& \\
{\rm touching\ points}&&&&&&&&& \\
= ${\rm deg}_c[r_{n,n/m}]$ \ {\rm for}&&&&&&&&& \\
&&&$r_{12}$&&$r_{24}$&&$r_{36}$&& \\
m=2&&-&3+4c&-&4c+5&-&$64c^3+128c^2+$&& \\
&&&&&&&+72c+81&& \\
&&-&1&-&1&-&3&& \\
&&&&&&&&& \\
&&&&$r_{13}$&&&$r_{26}$&& \\
m=3&&-&-&$16c^2+4c+7$&-&-&$16c^2+36c+21$&& \\
&&-&-&2&-&-&2&& \\
&&&&&$r_{14}$&&&& \\
m=4&&-&-&-&$16c^2-8c+5$&-&-&& \\
&&-&-&-&2&-&-&& \\
&&&&&&$r_{15}$&&& \\
m=5&&-&-&-&-&$256c^4+64c^3+$&-&& \\
&&&&&&$+16c^2-36c+31$&&& \\
&&-&-&-&-&4&-&& \\
&&&&&&&$r_{16}$&& \\
m=6&&-&-&-&-&-&$16c^2-12c+3$&& \\
&&-&-&-&-&-&2&& \\
&&&&&&&&& \\
\hline
&&&&&&&&& \\
${\cal N}_n^{(p)}$: \
\#\ {\rm of}\ {\rm el.\ domains}&&&&&&&&& \\
$\sigma^{(p)}_n\subset S_n$&&&&&&&&& \\
{\rm with}\ $q(p)$\ {\rm cusps}&&&&&&&&& \\
&&&&&&&&& \\
$p=0,\ q=1$&& 1 & 0 &1 &3&11&20&& \\
&&&&&&&&& \\
$p=1,\ q=0$&& 0&1&2&2&4&3&& \\
&&&&&&&&& \\
$p=2,\ q=0$&& 0 &0&0&1&0&4&& \\
&&&&&&&&& \\
$p=3,\ q=0$&&0&0&0&0&0&0&& \\
&&&&&&&&& \\
\ldots&&&&&&&&& \\
&&&&&&&&& \\
\hline
\end{tabular} \nn
\ee
}
$$
r_{17} = (4c)^6 + (4c)^5 + (4c)^4 + (4c)^3 +
15\cdot(4c)^2 - 17\cdot(4c) + 127,
$$
$$r_{18} = 256c^4 + 32c^2 - 64c + 17,$$
and so on. Stars in the table stand for too long expressions.
${\cal N}^{(p)}_n$ is the number of domains $\sigma^{(p)}_n$
of given level $p$ and order $n$, which still differ by
the values of parameters $m_i$ and $\alpha_i$ that are
ignored in this table. For one-parametric families the number $q$
of cusps often depends only on the level $p$. Taking this into
account, we can write down -- and check in this and other
examples -- the set of {\it sum rules} for the numbers
${\cal N}^{(p)}_n$:
\be
\sum_{p=0}^\infty {\cal N}_n^{(p)} = {\rm deg}_c [G_n(w_c,c)], \nn \\
\sum_{p=0}^\infty q(p){\cal N}_n^{(p)} = {\rm deg}_c [d_n], \nn \\
\sum_{p=1}^\infty {\cal N}_n^{(p)} =
\sum_{m}{\rm deg}_c[r_{n,n/m}].
\label{sumrules}
\ee
Actually, for every given $n$ only finitely many values of $p$
contribute: $p$ is the number of non-unit links in the branch of the
multipliers tree, thus $p\leq \log_2 n$.
From (\ref{sumrules}) one can deduce an identity
which does not include sums over $p$:
\be
{\cal N}_n^{(0)} + \sum_{m}{\rm deg}_c[r_{n,n/m}] =
{\rm deg}_c [G_n(w_c,c)],
\label{sumrules1}
\ee
and in many cases (including the families $x^d +c$) the r.h.s. is just
${\cal N}_n(d)$.
Also in many cases (again including $x^d+c$) $q(0) = d-1$ and $q(p)
= d-2$ for $p>0$. In such situations
\be
(d-1){\cal N}_n^{(0)} + (d-2)
\sum_{m}{\rm deg}_c[r_{n,n/m}] =  {\rm deg}_c [d_n],
\label{sumrules2}
\ee
and we also have a consistency condition between (\ref{sumrules1})
and (\ref{sumrules2}):
\be
{\rm deg}_c [d_n] + \sum_{m}{\rm deg}_c[r_{n,n/m}] =
(d-1){\rm deg}_c [G_n(w_c,c)].
\label{sumrules12}
\ee
Coming back to our particular family $x^2+c$, one can easily check
that consistency condition is satisfied by the data in the table,
and also ${\cal N}_n^{(0)} =  {\rm deg}_c [d_n]$, as requested by
(\ref{sumrules2}).

For analysis of Julia sheaf above table should be supplemented by the
data concerning pre-orbits and associated resultants, but this remains
beyond the scope of the present paper.

\subsubsection{Map $f_{\gamma\beta 0} =
\gamma x^2 + \beta x = \gamma x^2 + (b+1)x$
\label{adresqualin}}

According to (\ref{cabg}) we have
\be
c = \frac{\beta}{2} - \frac{\beta^2}{4} = \frac{1}{4} -
\frac{(\beta -1)^2}{4} = \frac{1}{4}(1-b^2),
\label{cb}
\ee
and one can expect that quadratic cusp for domain
$\sigma_1$ at $c=\frac{1}{4}$ disappears after such
transformation. In fact, transformation (\ref{cb})
makes from cardioid $\sigma_1$ in $c$-plane a pair
of unit discs in $b$-plane with centers at $b=0$
and $b=2$, which touch each other at the point $b=1$,
$S_1 = \sigma^{(0)}[1+]\cup \sigma^{(0)}[1-]$:
this is a good example, demonstrating that
links with $m=1$ in multiples tree sometime contribute
to the structure of Mandelbrot set.

The boundary $\partial S_1$ is described by a system
of equations (\ref{stabco}):
$$
\partial S_1:\ \ \left\{ \begin{array}{c}
|2\gamma x+\beta| = 1 \\
\gamma x^2+\beta x = x
\end{array}\right.
$$
i.e. $x=0$ or $x = \frac{1-\beta}{\gamma}$,
$|\beta|=1$ or $|\beta-2|=1$
and $|1\pm b| = 1$.
Accordingly the resultants $r_{1n}(b)$ should have roots
lying on these two unit circles. Actually, they are all made
from products of circular polynomials (\ref{gpol}):
\be
r_{1n}(b) = \gamma^{N_n} g_n(\beta)g_n(2-\beta) =
\gamma^{N_n} g_n(1+b)g_n(1-b),\nn \\
r_{2,2n} = \gamma^{N_{2n}} g_{n}(5-b^2),\nn \\
\ldots \nn
\ee
in particular (for $\gamma = 1$)
\be
\gamma^{-2}r_{12}(b) = 3+4c =
(\beta+1)(3-\beta) = (2+b)(2-b), \nn \\
\gamma^{-6}r_{13}(b) = 16c^2 + 4c + 7 =
(\beta^2+\beta+1)(\beta^2-5\beta +7) =
(3+3b+b^2)(3-3b+b^2),\nn \\
\gamma^{-12}r_{14}(b) = 16c^2-8c +5 =
(\beta^2+1) (\beta^2-4\beta+5)=
(2+2b+b^2)(2-2b+b^2),\nn \\
\gamma^{-30}r_{15}(b) = 256c^4 + 64c^3 + 16c^2-36c + 31 = \nn \\ =
(\beta^4+\beta^3+\beta^2+\beta + 1)
(\beta^4-9\beta^3+31\beta^2-49\beta + 31) = \nn \\ =
(5+10b+10b^2+5b^3+b^4)(5-10b+10b^2-5b^3+b^4) ,\nn \\
\gamma^{-54}r_{16}(b) = 16c^2-12c+3 =
(\beta^2-\beta+1)(\beta^2-3\beta +3)=
(1+b+b^2)(1-b+b^2), \nn \\
\ldots \nn \\
\gamma^{-12}r_{24}(b) =  4c+5 = 6-b^2 = 1+(5-b^2), \nn \\
\gamma^{-54}r_{26}(b) = 16c^2+36c+21 = b^4 - 11b^2 + 21 =
(5-b^2)^2 + (5-b^2) + 1, \nn \\
\ldots \nn \\
\gamma^{-108}r_{36}(b) =
64c^3 + 128c^2 + 72c + 81 = -b^6 + 11b^4 - 37b^2 + 108,\nn \\
\ldots \nn
\ee
The last expression, for $r_{36}$, has nothing to do with circular
polynomials: the map (\ref{cb}) transforms $\sigma^{(0)}_1$,
which is quadratic cardioid in $c$-plane, into a bouquet of two
unit discs in $b$-plane; it converts $\sigma^{(1)}_2$,
which is a disc in $c$-plane,
into domain $|b^2-5|<1$ in $b$-plane; but it does not simplify
higher $\sigma_n$, which are complicated in $c$-plane and remain
complicated in $b$-plane.

What is interesting in this example,
dependence on $\gamma$ is trivial in
$b$-plane, and the pattern of Mandelbrot set (its section by quadratic
maps) remain the same in the limit of $\gamma \rightarrow 0$.
In this limit the biggest part of Mandelbrot set is associated with
orbits, lying at very big $x \sim \gamma^{-1}$. Naive limit $\gamma = 0$,
considered in above s.\ref{linear}, ignores such orbits and only
the unit disc $|\beta|<1$ (domain $\sigma^{(0)}_1$) is seen by
examination of linear maps. However, the other parts are actually present
(nothing happens to them in the limit of $\gamma = 0$), just not
revealed by consideration of strictly linear maps. Similarly,
consideration of quadratic maps alone can ignore (overlook) other
pieces of Mandelbrot set, and so does every restriction to maps of a
given degree.

Powers of $\gamma$ in above formulas
for resultants characterize intersection of orbits
at infinity. Since in this particular case all periodic orbits
except for a single fixed point tend to $x=\infty$ as $\gamma
\rightarrow 0$, the asymptotic of resultants is defined by
the following rule (see eq.(\ref{resa2}) below):
$$
R(G_k,G_l) = r_{kl}^l \sim \gamma^{N_k(N_l-\delta_{l,1})},
\ \ k>l,
$$
where $N_l = N_l(d=2)$ is the degree of the polynomial $G_l(x;f)$
for quadratic map $f$, and $N_l/l$ is the number of periodic orbits
of order $l$.

We omit the table for this family, because all the numbers
in it are obtained by doubling the corresponding numbers in
the table for $x^2+c$ (provided degrees of all polynomials are
counted in terms of $b$ rather than $c$).

\subsubsection{Generic quadratic map and $f = x^2+px+q$}

Expressions for generic quadratic map $\gamma x^2 + \beta x + \alpha$
can be obtained from formulas below by a trivial change of variables:
$x \rightarrow \gamma x$, $p = \beta$, $q=\alpha\gamma$.
$$
G_1 = F_1 = x^2 + (p-1)x + q,
$$ $$
G_2 = \frac{F_2}{G_1} = x^2 + (p+1)x + (p+q+1),
$$ $$\ldots  $$ $$
G_{1,1} = \frac{F_{1,1}}{G_1} = G_2 - 1 =
x^2 + (p+1)x + (p+q+1),
$$ $$
G_{1,2} = \frac{F_{1,2}}{G_1G_{1,1}} =
x^4 + 2p x^3 + (2q + p^2+ p +1)x^2 +
$$ $$
\ \ \ + (2pq+p^2+p)x +
(q^2+pq+2q+p),
$$ $$
G_{2,1} = \frac{F_{2,1}}{G_1G_2G_{1,1}} =
x^2+(p -1)x + (q+1),
$$ $$
G_{2,2} = \frac{F_{2,2}}{G_1G_2G_{1,1}G_{1,2}G_{2,1}} =
$$ $$
= x^4 + 2p x^3 + (p^2 + 2q + p -1)x^2 +
(p^2 + 2pq -p)x + (q^2+pq+1)
$$ $$\ldots $$

The critical point is $w = -p/2$,
the "irreversible point" $z = f(w) = \frac{4q-p^2}{4q}$.
Accordingly, for generic map $w = -\beta/2\gamma$  and
$z = \frac{4\alpha\gamma - \beta^2}{4\gamma}$.

$$w_1 = q + \frac{p}{2}-\frac{p^2}{4},$$
$$w_2 = 1 + w_1,$$
$$w_3 = 1 + 2pq + q + q^3 + \frac{p}{2} + \frac{p^2}{4}  -
\frac{3p^2q^2}{4} + \frac{3pq^2}{2} - $$ $$
- \frac{p^2q}{2}  -\frac{3qp^3}{4} - \frac{p^4}{16}
-\frac{3p^3}{8} - \frac{p^6}{64} +
\frac{3p^4q}{16} + \frac{3p^5}{32},$$
$$\ldots$$
$$w_{11} = 1,$$
$$w_{12} = 2 + w_1,$$
$$\ldots$$

\subsubsection{Families as sections
\label{faasse}}

We now use these examples to briefly discuss the geometric
interpretation of the theory. The Mandelbrot set in Fig.\ref{lemMa}
is obtained as a union of infinitely many elementary domains
$\sigma$ in the plane of complex $c$, glued together in a very
special way along the discriminant variety (which has {\it real}
codimension one and is a closure of a union of complex varieties
${\cal R}^*_{k,n}$ of {\it complex} codimension one).
Each domain $\sigma$ can be described as a real domain,
defined by some equation of the form $\Sigma(c) < 1$
with a real function $\Sigma(c)$ of a complex variable $c$.

Switching to the two-parametric family $f = x^2+px+q$, we
get the Mandelbrot set, which is a similar union of domains
$\tilde\sigma$ in the space $\bb{C}^2$ of two
complex variables $p$ and $q$, with each $\tilde\sigma$
defined by the equation $\tilde\Sigma(p,q) < 1$, where
essentially $\tilde\Sigma(p,q) = \Sigma\left(c = q+\frac{1}{4}
-\frac{1}{4}(p-1)^2\right)$.

Since we can not draw pictures in
$\bb{C}^2$, we restrict the whole pattern to real $p$ and $q$:
this badly spoils the nice picture, but preserves the main
property which we are going to discuss.
After restriction to real $c$, Fig.\ref{lemMa} turns into
collection of segments (implicit in the lower part of that
figure), see Fig.\ref{reMa1}. Instead we can now extend the
picture in another direction: to two-dimensional plane ($p,q$).
Segments turn into parabolic strips, see Fig.\ref{reMa2},
where different families, like $x^2+c$ and $x^2 + (b+1)x$,
are represented as different sections. It is clear from the
picture why one and the same Mandelbrot set looks very different
if restricted to different families. Another version of the
same pattern is presented in Fig.\ref{reMa3}, where Mandelbrot
consists of straight rather than parabolic strips (after
complexification strips become "cylindrical" or, better,
toric domains). However, beyond quadratic family such ideal
"torization" is not possible, different resultant
varieties can not be made exactly "parallel", because sometime
they intersect, see eqs.(\ref{highdege}) for the simplest
example -- at level of cubic families. The best one can try
to achieve is the representation in terms of somewhere-strapped
tori, represented by chains of sausages, see Fig.\ref{sausages}
at the end of the paper. It is an interesting question, whether
Mandelbrot set can be represented as a toric variety.

\FIGEPS{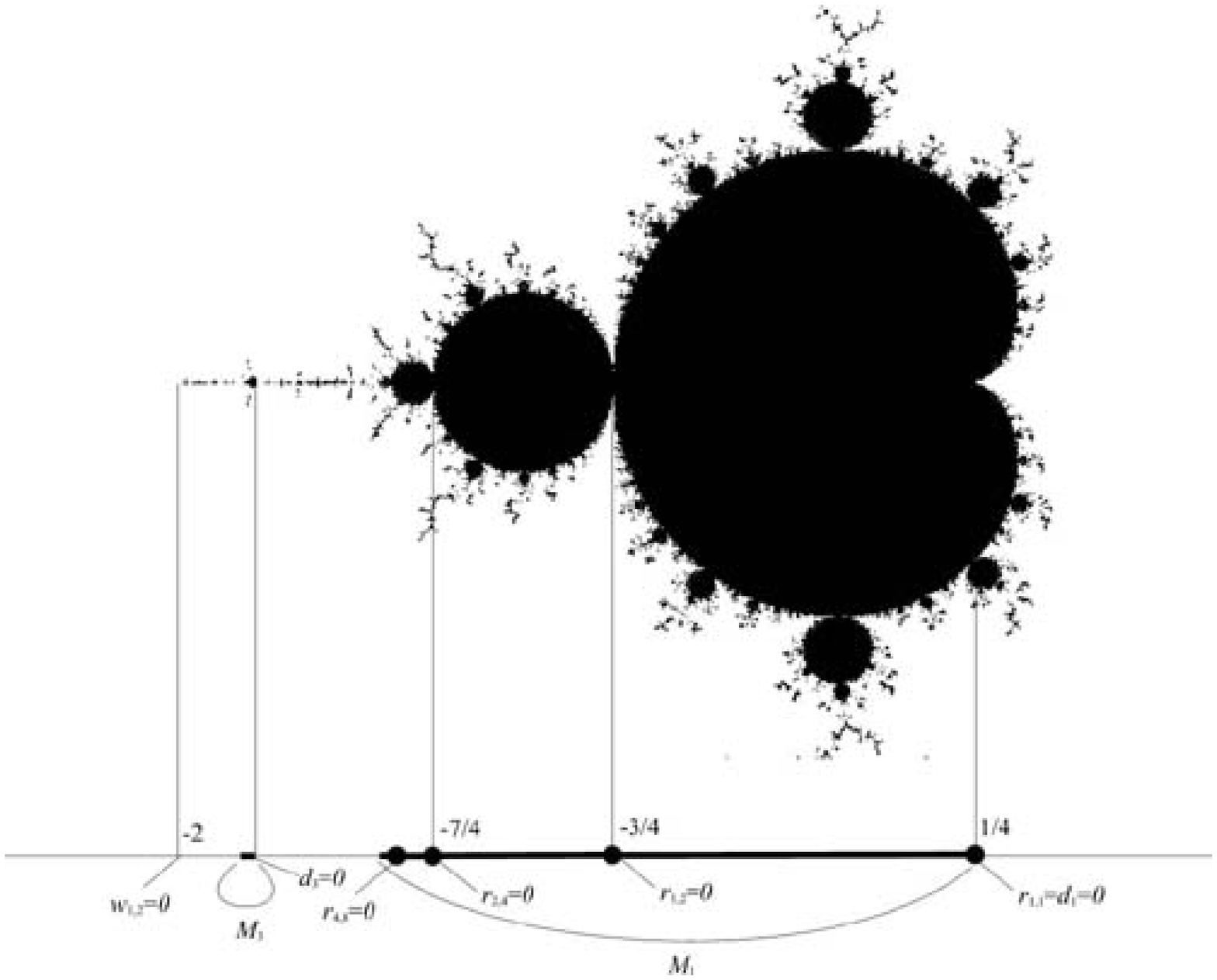}{}
{400,321}
{The Mandelbrot set for the family $x^2+c$ shown
together with its section by a real line ${\rm Im}\ c = 0$,
where it turns into collection of segments. These segments form
disconnected groups -- sections of different components
$M_n$ (actually, only the big group, associated with $M_1$,
and a tiny one, associated with $M_3$, are seen in the picture;
if the section was not by a real line, a single $M_1$ would
provide many disconnected groups of segments).
Within each group the segments, corresponding to different
$\sigma^{(p)}_k$ touch by their ends, real zeroes of resultants,
and free ends are real zeroes of discriminants (perhaps,
those of pre-orbit polynomials, i.e. zeroes of $d_n$ or $w_{n,s}$).}

\FIGEPS{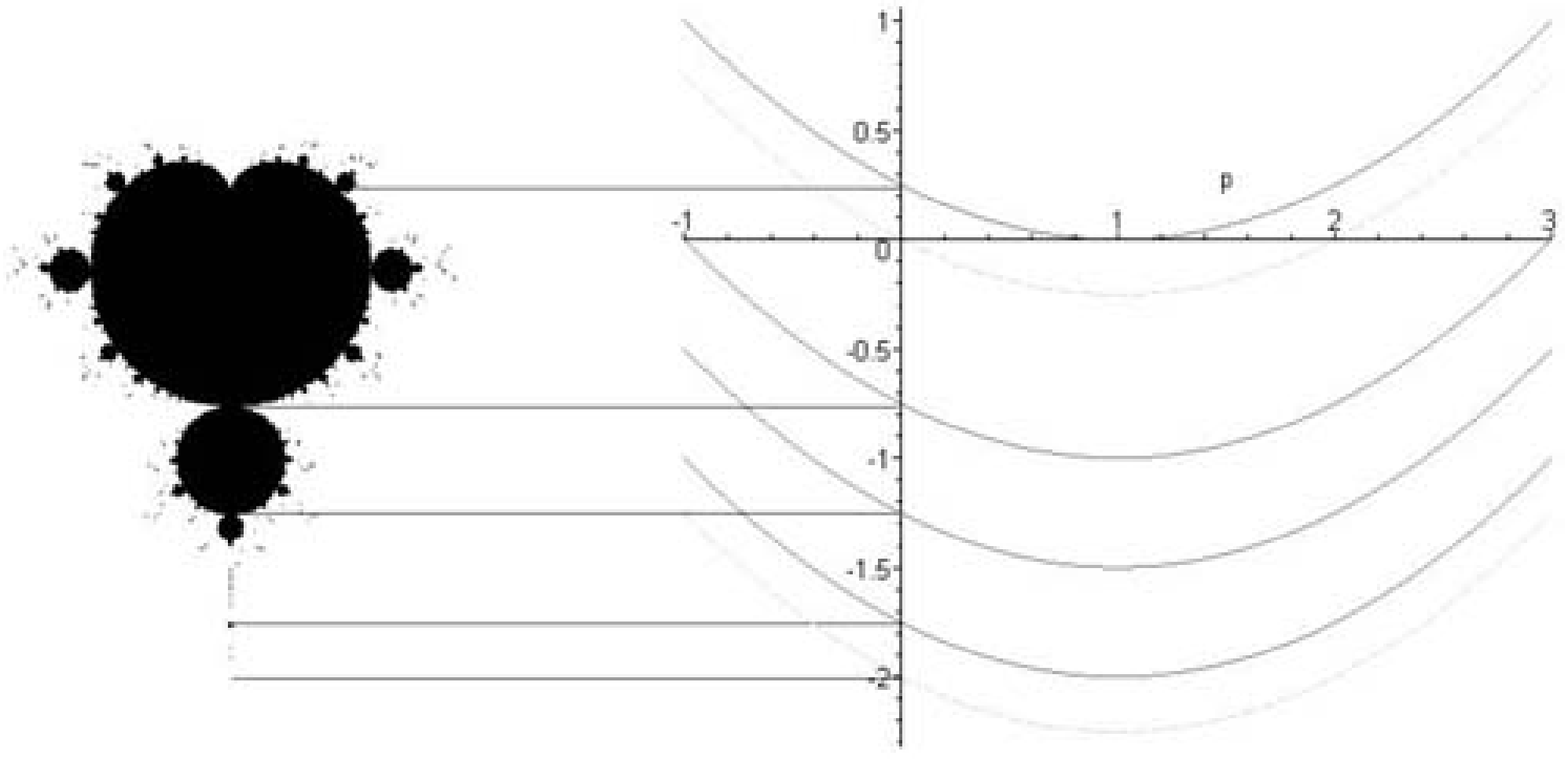}{}
{450,220}
{The same picture in the {\it real} $(p,q)$ plane, $p$ is plotted along
the absciss axis, $q$ -- along ordinate.
This section of Mandelbrot set represents it as a collection of
domains bounded by parallel parabolas, some domain touch (have
common boundaries, which consist of zeroes of the resultants --
now they are not points, but codimension-one parabolas).
}

\FIGEPS{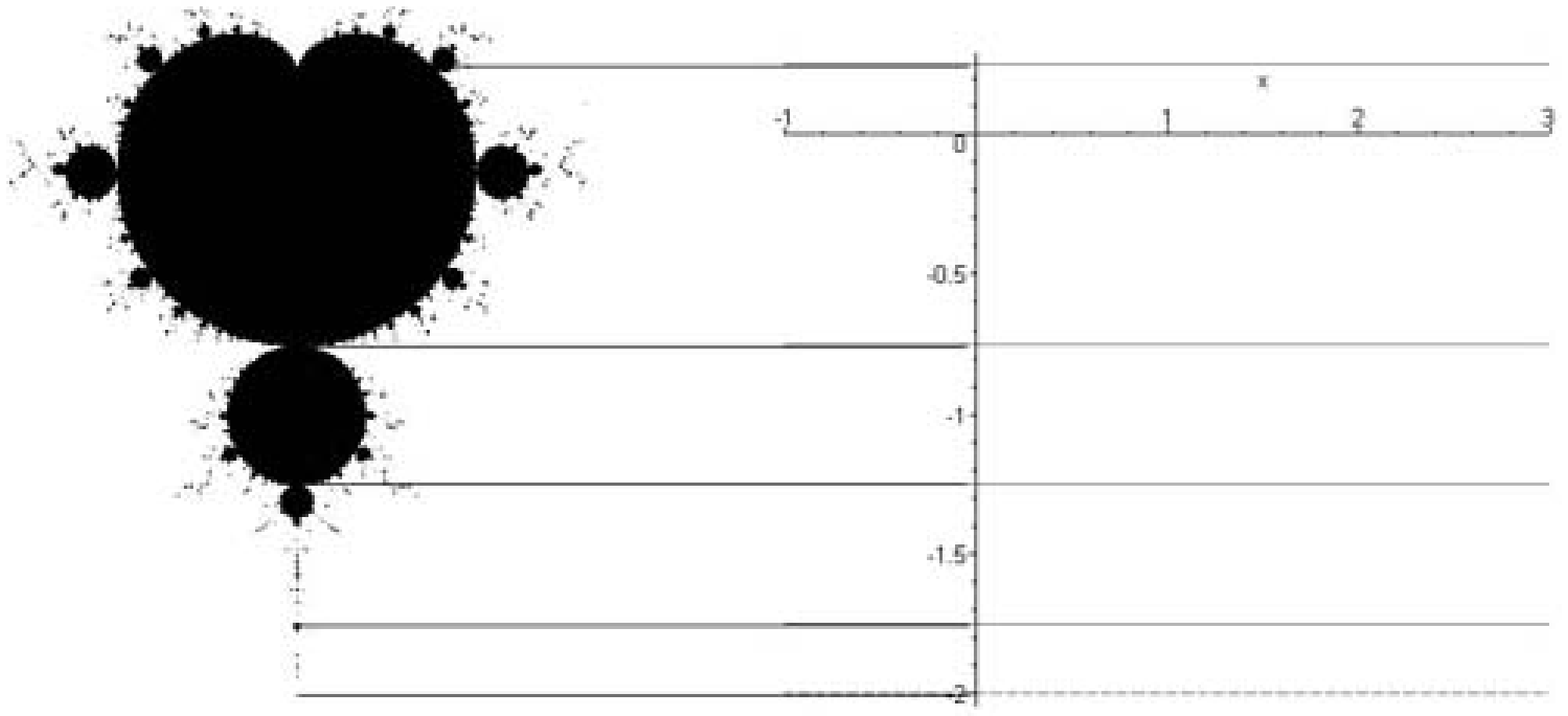}{}
{447,220}
{The same picture in the real ($c,p$)-plane: $p$ plotted on
absciss axis, $c$ -- on ordinate. The Mandelbrot
set is represented by collection of straight horizontal strips.
Resultants and discriminants depend only on $c$ and not on $p$
thus their zeroes form vertical lines, separating the strips.
In general situation
(beyond quadratic families) the Mandelbrot set can not be made
"cylindrical": the obstacle is provided by intersections of
different resultant and discriminant varieties, controlled by
higher resultants and discriminants, see (\ref{highdege})
for examples. The best one can hope to achieve is a "sausage-chain"
representation, see Fig.\ref{sausages} at the end of this paper.}

\subsection{Cubic maps}

Generic cubic map -- the 4-parametric family
$f_{\alpha\beta\gamma\delta}(x) =
\alpha + \beta x + \gamma x^2 + \delta x^3$
-- is diffeomorphic to
a 2-parametric family $f_{p,q}(x) = x^3 + px + q$,
\be
p = \beta - \frac{\gamma^2}{3\delta}, \nn \\
q = \frac{2\gamma^3 + 9(1-\beta)\gamma\delta + 27\alpha\delta^2}
{27\delta\sqrt{\delta}} \nn
\ee

In addition to some initial information
about this 2-parametric family
we provide a little more details about several
1-parametric sub-families of cubic maps,
$$f_c(x) = x^3+c\ \ {\rm with}\ \ (p,q)=(0,c),$$
$$f_c(x) = cx^3+x^2\ \ {\rm with}\ \
(p,q) = \left(-\frac{1}{3c},\frac{9c+2}{27c\sqrt{c}}\right),$$
$$f_c(x) = ax^3 + (1-a)x^2 + c\ \ {\rm with} \ \
(p,q) = \left(-\frac{(1-a)^2}{3a},
\frac{2(1-a)^3 + 27a^2c}{27a\sqrt{a}}\right)$$
respectively (in the last case $a$ is considered as an additional
parameter).

Our main purpose will be to demonstrate four (inter-related)
new phenomena, not seen at the level of quadratic maps:

-- Keeping degree of the maps fixed, we ignore (hide) the biggest part
of Julia set, which is detached from the "visible" part and is located
at $x = \infty$. Julia set decays at peculiar
"decay" bifurcation points, which are presumably are zeroes
of $w_{n,s}(f)$. See Fig.\ref{Juliaw}.

-- For essentially multiparametric families $\mu$ of maps (like the
2-parametric $f_{p,q}(x) = x^3 + px + q$), different components
$r_{kn}$ and $d_n$ of the resultant/discriminant variety intersect and
higher order singularities of this variety can be revealed in this way.
At these singularities particular
sections of universal Mandelbrot set, associated with 1-parametric
families, reshuffle, and new components $M_{n\alpha}$ can split
from previously existing. Thus these reshufflings are responsible for
formation of the trail structure of particular Mandelbrot sets.

-- If the $\bb{Z}_d$ symmetry of the map $f(x) = x^d + c$ is slightly,
but fully broken (e.g. by addition of a term $\beta x$
or $\gamma x^2$ with small non-vanishing $beta$ or $\gamma$),
then the $\bb{Z}_{d-1}$ symmetry of the section of Mandelbrot set gets
broken in peculiar way:
the central domain $M_1$ and the nearest (in a relevant topology,
actually, those with largest sizes $\rho_n$) components $M_{n\alpha}$
continue to possess the $\bb{Z}_{d-1}$ symmetric shape, while remote
components $M_{n\alpha}$ acquire the non-symmetric form (characteristic
for the Mandelbrot set for the family $x^2+c$). See Fig.\ref{symmebre}.
Detailed analysis of reshuffling process, shown in Figs.\ref{resh}
and \ref{zeropos},
reveals how the non-central components and trail structure of Mandelbrot
set is formed.

\FIGEPS{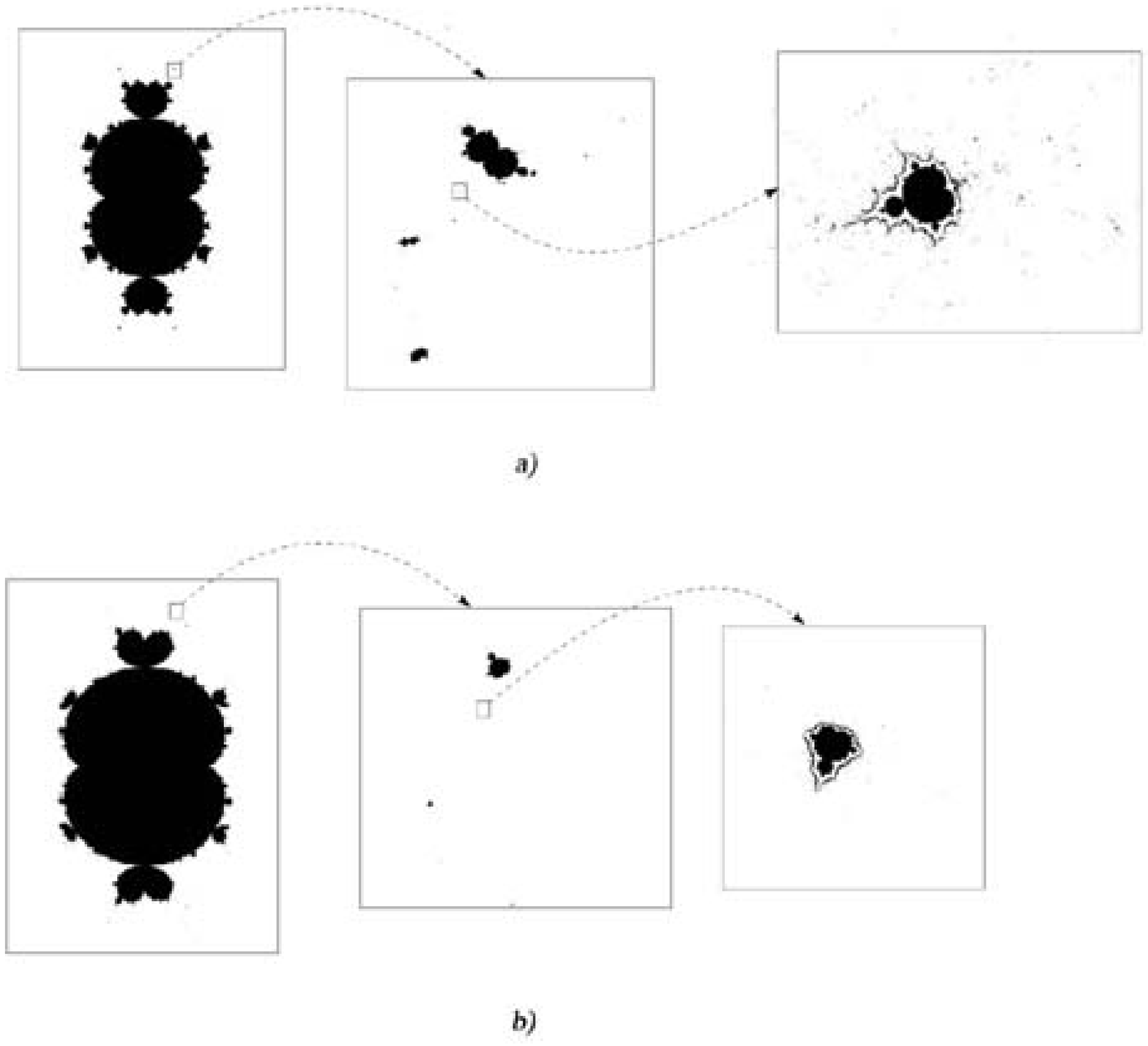}{}
{500,428}
{Different, close and remote, components
of the Mandelbrot set in the complex $c$ plane for the family
$ax^3+(1-a)x+c$ for different values of $a$
(considered as an additional parameter): a) $a=4/5$, b) $a=2/3$. One can see how
the symmetry of different domains is changing with the variation of $a$:
the smaller $a$ the larger is the low-symmetry domain.
See Fig.\ref{resh} for description of transition (reshuffling) processs
and s.\ref{acfa} for comments. Halo around the domain in the right
pictuires (at high resolution) is an artefact and should be ignored.}

\FIGEPS{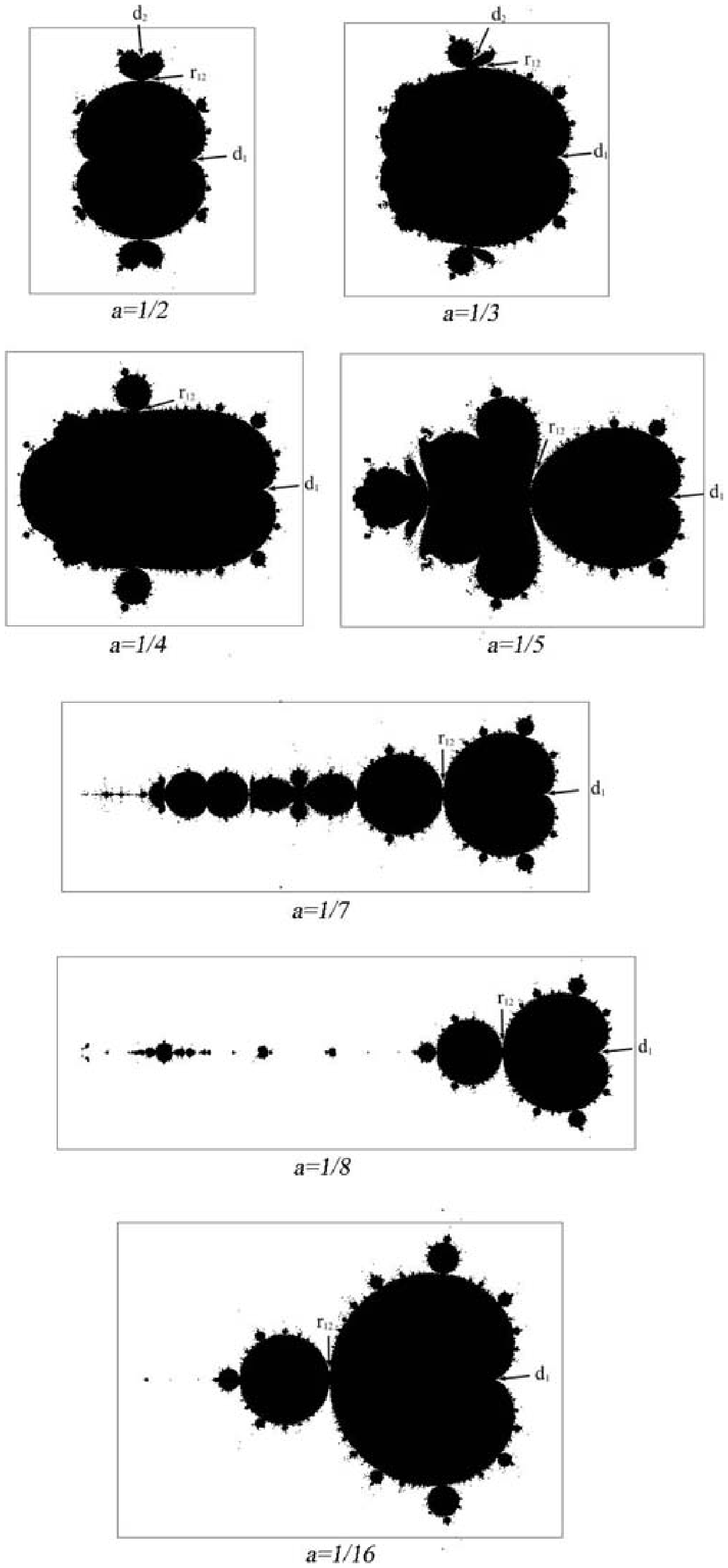}{}
{282,630}
{Shown is a sequence of views of Mandelbrot sets
for different values of $a$ in the 2-parametric family
$\{ax^3 + (1-a)x^2 + c\}$, interpolating between the 1-parametric
families $\{x^3+c\}$ at $a=1$ and $\{x^2+c\}$ at $a=0$.
See s.\ref{acfa} for additional comments and footnote \ref{FEcr}
in that section for {\bf precaution} concerning this particular Figure:
only the right parts of the pictures can be trusted, the left parts
should be reflections of the right. Arrows show positions of some
zeroes of $d_1$, $d_2$ and $r_{12}$. The full sets of these zeroes
are shown in fully reliable (though less picturesque) Fig.\ref{zeropos}.}

\FIGEPS{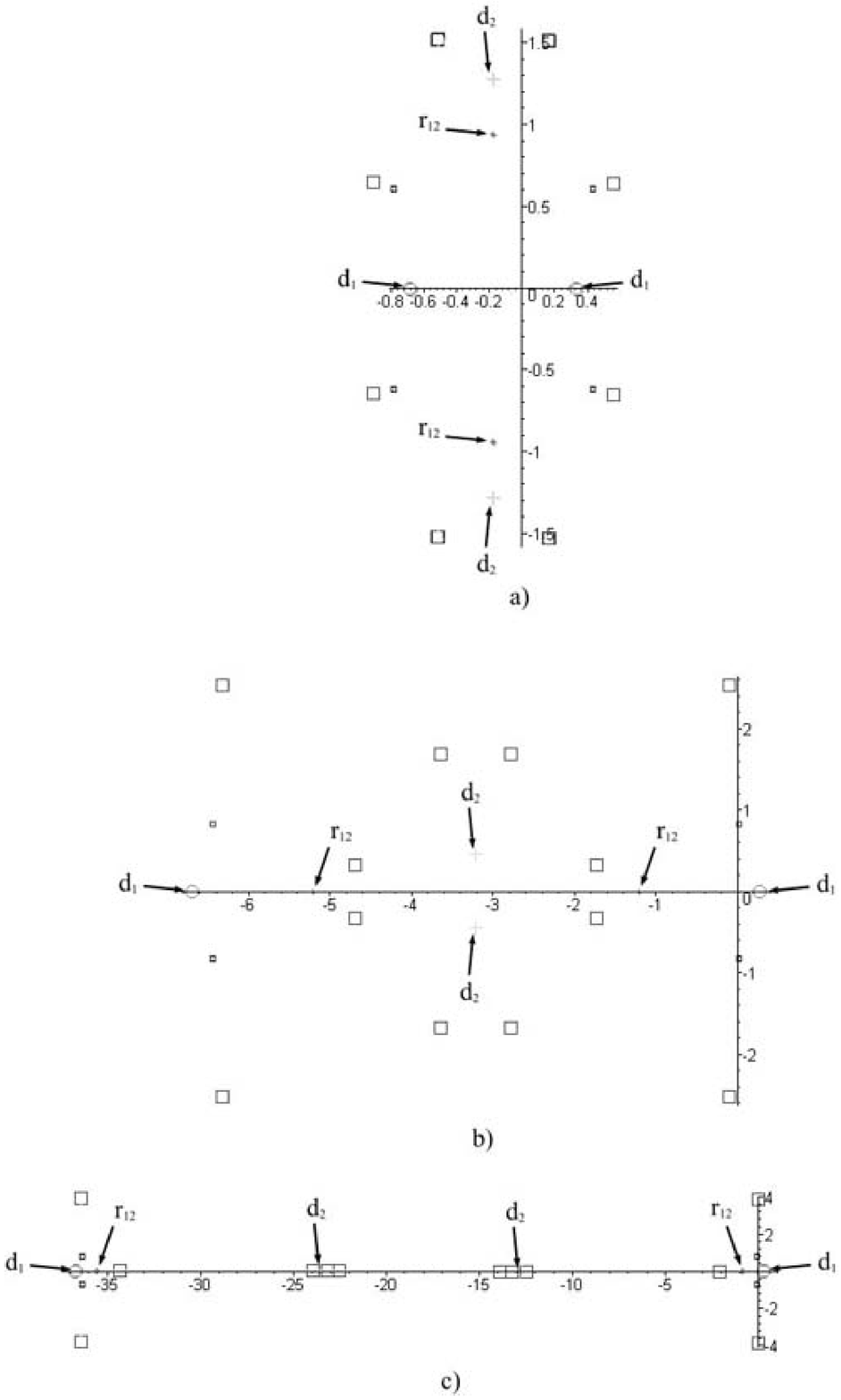}{}
{345,580}
{A sequence of pictures, showing positions in the
complex $c$-plane of zeroes
of $d_1(c)$ (big cirlces), $d_2(c)$ (big crosses), $d_3(c)$
(big boxes), $r_{12}(c)$ (small crosses), $r_{13}(c)$ (small boxes)
for the 2-parametric family
$\{ax^3 + (1-a)x^2 + c\}$, interpolating between the 1-parametric
families $\{x^3+c\}$ at $a=1$ and $\{x^2+c\}$ at $a=0$: a) $a=2/3$, b) $a=1/6$,
c) $a=1/15$. These are
particular special points in the patterns, presented in Fig.\ref{resh}.
Three arrows point to a zero of $d_1$, a zero of $r_{12}$ and a zero of $d_2$.
After the two roots of $r_{12}$ merge at the real-$c$ line,
this happens at $a=\frac{5-\sqrt{21}}{2} \approx 1/5$,
they start moving in opposite directions along this line. We arbitrarily
choose to point at the right of the two zeroes (because the left one is
not seen in the partly erroneous Fig.\ref{resh}).}

-- In generic situation a given map (at a given point in the Mandelbrot
space) can possess {\it several} stable periodic orbits
(not just one as it happens in the case of $x^2+c$ -- and this brings us
closer to continuous situation, where several stable points,
limiting cycles or other attractors can co-exist).
Bifurcations of Julia set occur whenever any of these
orbits exchange stability with some unstable one.
See Fig.\ref{multiorb}.

\FIGEPS{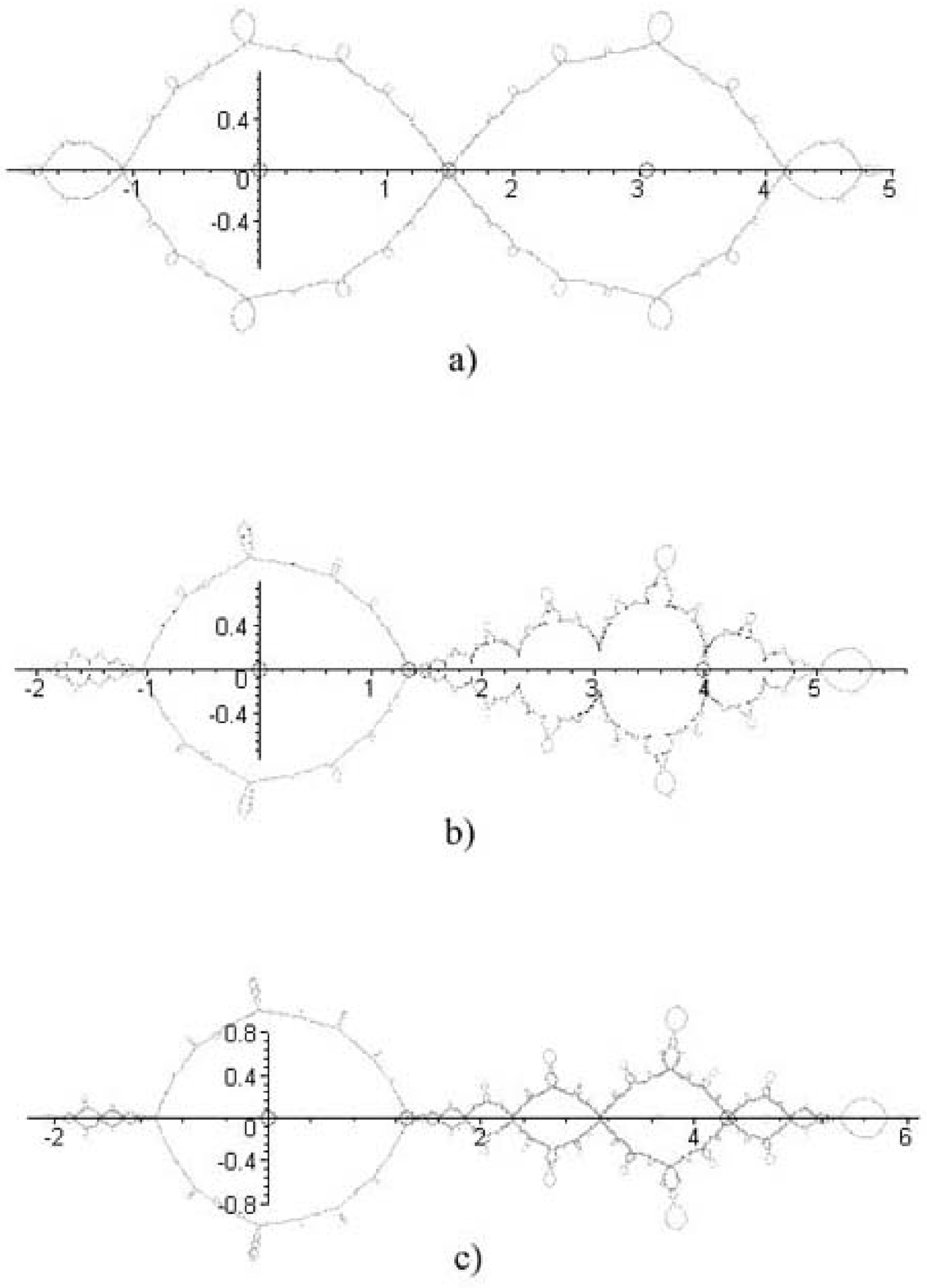}{}
{400,549}
{Coexisting stable orbits for the family $cx^3+x^2$: a) $c=-0.22$, b) $c=-188$, c) $c=-0.178$. Shown
are Julia sets, and stable orbits lie inside $J(f)$. In the left
hand part of the Julia set there is a stable fixed point. The stable orbit
inside the right hand part is different:
shown is the exchange of stability between {\it another} fixed point
(stable in the case (a)) and an orbit of order two (stable in the case (c)).
In (b) the transition point is shown, where the fixed point and the
orbit of order two intersect and Julia set changes topology.}

\subsubsection{Map $f_{p,q}(x) = x^3 + px + q$
\label{x3pxq}}

$$
G_1(x) =   x^3  + (p-1)x + q,
$$ $$
G_2(x) = x^6  + (2p+1) x^4  + 2qx^3 +
(p^2+p+1)x^2  + ((2p+1)q-1)x + p + q^2 + 1,
$$ $$
G_{1,1}(x) = G_2-1,
$$
The two critical points are $w = \pm \sqrt{-p/3}$, or
$p = -3w^2$.
Here are some irreducible components of pre-orbit
resultants and discriminants (they are all products
of two factors, associated with $+w$, and $-w$,
such products are polynomials in integer powers of $p$;
we list expressions which are not too long):
{\footnotesize
\be
w_1 = (q-2w^3-w)(q+2w^3+w) = \frac{1}{27}
(27q^2+4p^3-12p^2+9p),
\nn \\
w_2 = (4 w^6  - 2 w^4  - 4 w^3 q - 2 w^2  + w q + q^2 + 1)
(4 w^6  - 2 w^4  + 4 w^3 q - 2 w^2  - w q + q^2 + 1),
\nn \\
w_{1,1} = (q -2w^3+2w) (q +2w^3-2w) = \frac{1}{27}
(27q^2+4p^3+24p^2+36p),
\nn \\
w_{2,1} = (q^2+1 - 2wq + w^2 - 4w^3q + 4w^4 + 4w^6)
(q^2+1 + 2wq + w^2 + 4w^3q + 4w^4 + 4w^6),
\nn \\
\ldots\nn
\ee
}
Some discriminants and resultants:
$$
d_1 = -(27q^2+ 4p^3 - 12p^2 +12p - 4),
$$ $$
r_{12} = 27q^2+4p^3-12p^2 + 16,
$$ $$
d_2 = 27q^2+4p^3 + 24p^2 + 48p + 32,
$$ $$
r_{1|1,1} = 27(q + w + 2 w^3 ) (q - w-2w^3) = 27w_1,
$$ $$
d_{1,1} = \frac{D(F_{1,1})}{d_1^3 r_{1|1,1}^2} =
\frac{D(G_{1,1})}{d_1^2}= $$ $$\ \ \ =
-27 (q - 2 w + 2 w^3 ) (q + 2 w  - 2 w^3) =
-27w_{1,1},
$$ $$
r_{1|1,2} = 27(q + w + 2 w^3 ) (q - w-2w^3) = 27w_1,
$$ $$
r_{1,1|1,2} = R(G_{1,1},G_{1,2})^{1/2} =
27(q + w + 2 w^3 ) (q - w-2w^3) = 27w_1,
$$ $$ \ldots $$

\bigskip

Zeroes of $w_{1,1}(f)$ form a submanifold, where the Julia set
can decay into two disconnected components ($(p,q)=(-3/4,3/4)$ is a
particular point on this submanifold, another point is
$(p,q) = (0,0)$, where Julia set turns into an ideal circle.)
As cubic $f$ degenerates into quadratic map, namely,
when $q\rightarrow 0$
with $p$ fixed, one of these two components travels to $x = \infty$.
See Fig.\ref{Juliaw}.

\bigskip

Since in this subsection
we are dealing with a two-parametric family, we can observe
intersections of various components $r_{k,n}$ and $d_n$ of the universal
discriminant variety (these intersections lie in complex codimension two).
They are seen in the section by complex-dimension-two family
$f = x^3+px+q$ as particular points: zeroes of the resultants of
$d(p,q)$ and $r(p,q)$, considered as polynomials of $q$ for a fixed $p$
or vice versa. For example:
\be
R(r_{12},d_1|q) = 16\cdot 3^6 (3p-5)^2,
\nn \\
R(r_{12},d_2|q) = 16\cdot 3^6 (3p+2)^2,
\nn \\
R(d_1,d_2|q) = 16\cdot 3^{10} (p^2+p+1)^2;
\nn \\  \\
R(r_{12},d_1|p) = -2^{10}(729q^2+32),
\nn \\
R(r_{12},d_2|p) = 2^8(729q^2 + 256),
\nn \\
R(d_1,d_2|p) = 2^83^9(27q^4+ 16).
\nn
\label{highdege}
\ee

\subsubsection{Map $f_c = x^3 + c$
\label{x3c}}

$$G_1 = x^3-x+c,$$
$$G_2 = x^6 + x^4 + 2cx^3 + x^2 + cx + c^2 + 1,$$
$$G_{1,1} = G_2 - 1,$$
$$\ldots$$

The double stable point is $w=0$, and
$$w_1 = c^2,$$
$$w_2 = (1+c^2)^2,$$
$$w_3 = (1+c^2+3c^4+3c^6+c^8)^2,$$
$$w_4 = c^8,$$
$$\ldots$$
$$w_{11} = c^2,$$
$$w_{12} = c^2(c^4+3c^2+3)^2,$$
$$\ldots$$
$$w_{21} = (1+c^2)^2,$$
$$w_{22} = 1+3c^4+2c^6+3c^8+3c^{10}+c^{12},$$
$$\ldots$$
$$w_{31} = (1+c^2+3c^4+3c^6+c^8)^2,$$
$$\ldots$$
These quantities are obviously factorizable -- in seeming contradiction
with our claim that they are all irreducible. In fact, factorization
is accidental and is lifted by infinitesimal deformation, for example,
by taking $p\neq 0$, see the previous subsection \ref{x3pxq}.
Such deformation also breaks the accidental coincidences,
like that of $w_2$ and $w_{21}$.

{\footnotesize
\be
\begin{tabular}{|c|ccccccc|}
\hline
\multicolumn{8}{|l|}{}\\
\multicolumn{8}{|l|}{\ \ \ $f = x^3 + c$}\\
\multicolumn{8}{|l|}{}\\
\hline
&\ &\ \ &&&&& \\
n                            &&    1&2&3&4&5& \ldots \\
&&&&&&& \\
\hline
&&&&&&& \\
\#\ {\rm of\ orbits}       &&  3 & 3 & 8 & 18 & 48 &\\
= ${\rm deg}_x [G_n]/n$ &&&&&&& \\
&&&&&&& \\
\#\ {\rm of\ el.\ domains} &&    &&&&& \\
= ${\rm deg}_c [G_n(w_c,c)]$   &&   1&2&8&24&80& \\
&&&&&&& \\
$d_n$ &&$4-27c^2$&$(27c^2+32)i$&*&*&?& \\
{\rm total}\ \#\ {\rm of\ cusps}&& &&&&& \\
= ${\rm deg}_c [d_n]$             && 2&2&12&40&152& \\
&&&&&&& \\
\hline
&&&&&&& \\
$r_{n,n/m}$ {\rm \ and}&&&&&&& \\
{\rm total}\ \#\ {\rm of}\ (n,n/m)&&&&&&& \\
{\rm touching\ points}&&&&&&& \\
= ${\rm deg}_c[r_{n,n/m}]$ \ {\rm for}&&&&&&& \\
&&&&&&& \\
&&&$r_{12}$&&$r_{24}$&& \\
m=2&&-&$27c^2+16$&-&$729c^4+1620c^2+1000$&-& \\
&&-&2&-&4&-& \\
&&&&$r_{13}$&&& \\
m=3&&-&-&$729c^4+27c^2+169$&-&-& \\
&&-&-&4&-&-& \\
&&&&&$r_{14}$&& \\
m=4&&-&-&-&$729c^4-324c^2+100$&-& \\
&&-&-&-&4&-& \\
&&&&&&$r_{15}$& \\
m=5&&-&-&-&-&$3^{12}c^8+3^9 4c^6+$& \\
&&&&&&$+3^6 46c^4 -7\cdot 27\cdot 263 c^2 + 11^4$& \\
&&-&-&-&-&8& \\
&&&&&&& \\
\ldots &&&&&&& \\
&&&&&&& \\
\hline
&&&&&&& \\
${\cal N}_n^{(p)}$: \
\#\ {\rm of}\ {\rm el.\ domains}&&&&&&& \\
$\sigma^{(p)}_n\subset S_n$&&&&&&& \\
{\rm with}\ $q(p)$\ {\rm cusps}&&&&&&& \\
&&&&&&& \\
p=0,\ q=2&&1&0&4&16&72& \\
&&&&&&& \\
p=1,\ q=1&&0&2&4&4&8& \\
&&&&&&& \\
p=2,\ q=1&&0&0&0&4&0& \\
&&&&&&& \\
p=3,\ q=1&&0&0&0&0&0& \\
&&&&&&& \\
\ldots&&&&&&& \\
&&&&&&& \\
\hline
\end{tabular} \nn
\ee
}
{\footnotesize
$$
d_3 =
3^{18}c^{12} + 2\cdot 3^{19}c^{10} + 3^{13}3571 c^8 +
2^4 3^9 11\cdot 19\cdot 107 c^6+ $$ $$ + 2^5 3^8 13^2 137c^4 +
2^63^413^417c^2 + 2^8 13^{16}.
$$
\be
d_4=(150094635296999121c^{24}+1200757082375992968c^{22}+
\nn \\ +
4203267461712259335c^{20}+8399740516065395253c^{18}+
\nn \\ +
10909964351274746418c^{16}+10526401881511556976c^{14}+
\nn \\ +
8522156414444085612c^{12}+5544611719418268000c^{10}+
\nn \\ +
2750472027922567500c^8+1314354779366400000c^6+
\nn \\ +
459901255680000000c^4+167772160000000000)\cdot \nn \\
\cdot(282429536481c^{16}+1757339338104c^{14}+4642459719687c^{12}+
\nn \\ +
6806074010589c^{10}+6891783220746c^8+5994132959232c^6+
\nn \\ +
4118269132800c^4+1739461754880c^2+1073741824000) = \nn \\
= 3^{36}c^{24} + \ldots = (27c^2)^{12} + \ldots
\nn
\ee
}
$d_4$ appears factorizable, but this is an accidental factorization,
lifted by any deformation of the symmetric family $x^3+c$.

The sum rules (\ref{sumrules}) now imply:
$$
{\cal N}_n^{(0)} + \sum_{m}{\rm deg}_c[r_{n,n/m}] =
{\rm deg}_c [G_n(w_c,c)],
$$
$$
2{\cal N}_n^{(0)} +
\sum_{m}{\rm deg}_c[r_{n,n/m}] =  {\rm deg}_c [d_n],
$$
$$
\sum_{p\geq 1}{\cal N}_n^{(p)} = \sum_{m}{\rm deg}_c[r_{n,n/m}]
$$
and the consistency condition
$$
{\rm deg}_c [d_n] + \sum_{m}{\rm deg}_c[r_{n,n/m}] =
2{\rm deg}_c [G_n(w_c,c)],
$$
all obviously satisfied by the data in the table
(question marks near some numbers in the table indicate, that
they were found not independently, but with the help
of the sum rules).

\bigskip

With these explicit expressions we can illustrate the
small general theorem,
formulated and proved after eq.(\ref{RGG'D}).
Namely, solutions to the system
$$
\left\{\begin{array}{c} G'_n/H_n = 0 \\ G_n = 0 \end{array}\right.
$$
where $H_n = \{F'_n,G_n\}$,
are given by zeroes of {\it irreducible} discriminant $d_n$.

For $n=1$ we have: $G'_1 = 3x^2-1$, the system has solutions, i.e.
$G'_1$ and $G_1$ possess a common zero, when their resultant $D(G_1) =
4-27c^2$ vanishes, and $H_1 = -6x$ never has common zeroes with $G'_1$.
Thus solutions of the system are zeroes of the irreducible discriminant
$d_1$, which in this case coincides with $D(G_1)$.

For $n=2$ $G'_2 = 6x^5 + 4x^3 + 6cx^2 + 2x + c$ and
$H_2 = -36x(x^3+x+c)(x^3+c)(x^3-x+c)$ both depend on $c$ and have
a common root whenever their resultant $R(H_2,G'_2) =
-2^{10}3^{13}c^4(27c^2-16)(27c^2+16)(27c^2+8)$ vanishes.
At the same time, the system can have solution only when $G'_2$ and
$G_2$ possess common zeroes, i.e. when $D(G_2) = (27c^2+16)(27c^2+32)^2$
vanishes. We see that a pair of the roots, $27c^2+16$ is actually
eliminated, because they are also zeroes of $R(H_2,G'_2)$ and
what remains are the roots of the irreducible discriminant $d_2 =
27c^2+32$.

\bigskip

The first few pre-orbit discriminants for the family $x^3+c$ are:
$$
\begin{array}{ccc}
ns&&D(G_{ns}) \\
&& \\
11&& -3^3d_1^2w_{11}  \\
12&& -3^{27}d_1^6w_1w_{11}^3w_{12} \\
13&& -3^{135}d_1^{18}w_1^4w_{11}^{9}w_{12}^3w_{13}\\
14&& -3^{567}d_1^{54}w_1^{13}w_{11}^{27}w_{12}^9w_{13}^3w_{14}\\
&& \\
21&& 3^6d_2^4r_{12}^2w_{21} \\
22&& 3^{54}d_2^{12}r_{12}^6w_{21}^3w_{22} \\
&& \\
31&& 3^{24}d_3^6r_{13}^4w_{31}
\end{array}
$$

The Julia sheaf for this family is shown in Fig.\ref{Jush3}.

\FIGEPS{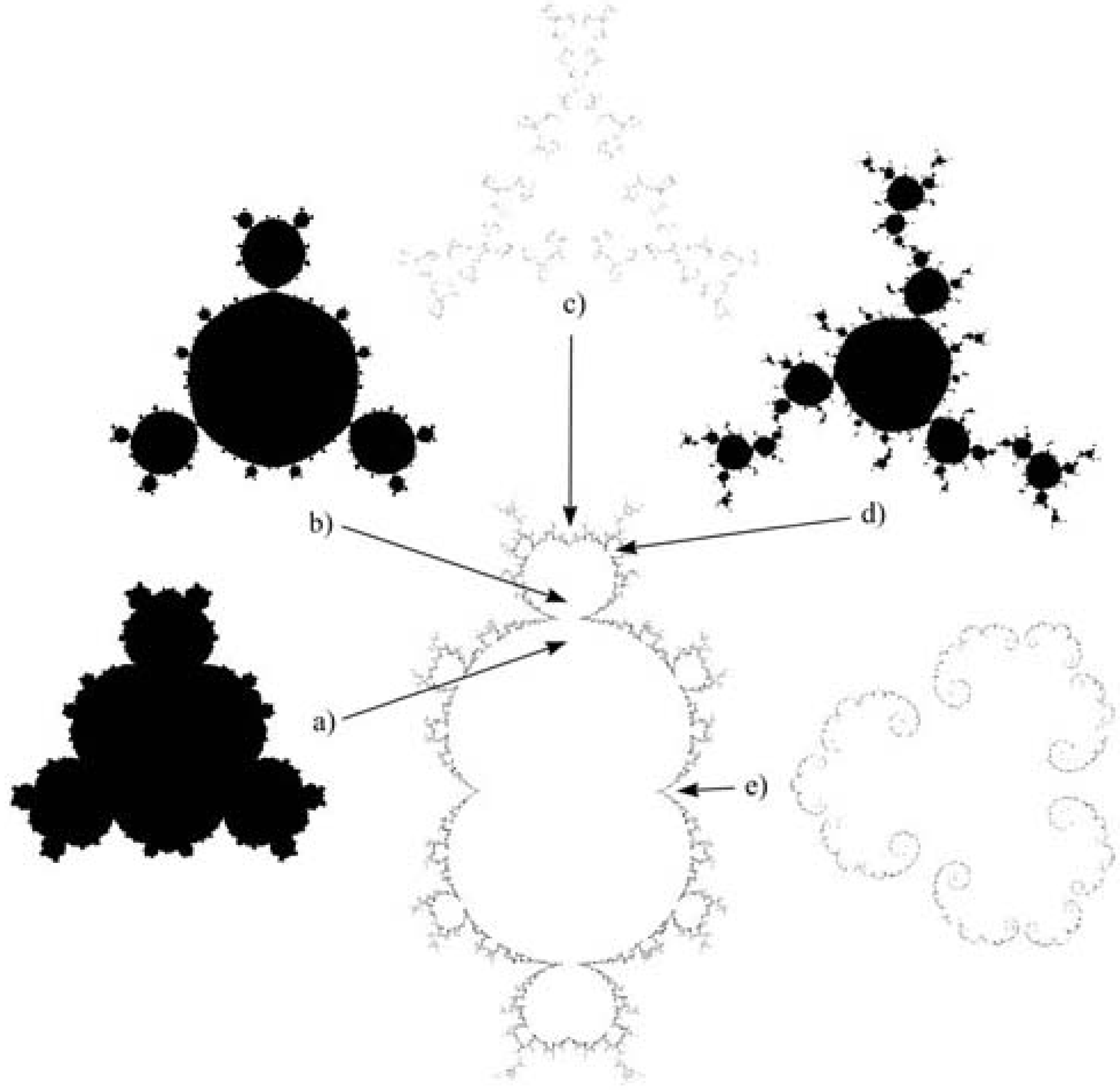}{}
{410,398}
{A fragment (fibers shown at several points only)
of the Julia sheaf for the family $x^3+c$ -- the analogue of
Figs.\ref{J1-1}, \ref{J1-4} for $x^2+c$ and \ref{Juexa} for $x^4+c$,
a) $c=0+0.7i$, b) $c=0+0.9i$, c) $c=0.2+1.08i$, d) $c=1.14i$, e) $c=0.45$.}

\subsubsection{Map $f_c(x) = cx^3+x^2$
\label{cx3x2}}

$$G_1= cx^3+x^2-x,$$
$$G_2 = c^3x^6+2c^2x^5+c(c+1)x^4+2cx^3+(c+1)x^2+x+1,$$
$$G_{1,1} = G_2 - 1,$$
$$\ldots$$

Critical points $w=\{0, -\frac{2}{3c}\}$.
Contribution of critical point $w=0$ can be ignored in most applications,
and we define in this case:
\be
w_1=G_1(w)=F_1(w) = \frac{2}{3c^2}\left(c+\frac{2}{9}\right), \nn \\
F_2(w) = \frac{2}{3c^5}\left(c+\frac{2}{9}\right)
\left(c^3 - \frac{2}{9}c^2 + \left(\frac{2}{9}\right)^2c +
2\left(\frac{2}{9}\right)^3\right),\nn \\
F_{1,1}(w) = F_2(w)-F_1(w) =
-\frac{4}{27c^5}\left(c-\frac{4}{9}\right)
\left(c+\frac{2}{9}\right)^2, \nn \\
G_{1,1}(w) = \frac{F_{1,1}(w)}{F_1(w)}
= -\frac{2}{9c^3}\left(c-\frac{4}{9}\right)
\left(c+\frac{2}{9}\right), \nn \\
w_{1,1}=\frac{G_{1,1}(w)}{G_1(w)} =
-\frac{1}{3c}\left(c-\frac{4}{9}\right).
\nn
\ee
$c=4/9$ means that $(p,q)=(-3/4,3/4)$.

{\footnotesize
\be
\begin{tabular}{|c|cccccc|}
\hline
\multicolumn{7}{|l|}{}\\
\multicolumn{7}{|l|}{\ \ \ $f = cx^3+x^2$}\\
\multicolumn{7}{|l|}{}\\
\hline
&\ &\ \ &&&& \\
n                            &&    1&2&3&4& \ldots \\
&&&&&& \\
\hline
&&&&&& \\
\#\ {\rm of\ orbits}       &&    &&&& \\
= ${\rm deg}_x [G_n]/n$ &&3&3&8&18& \\
&&&&&& \\
\#\ {\rm of\ el.\ domains} &&    &&&& \\
= ${\rm deg}_c [G_n(\varepsilon,c)]$   &&    1&3&12&36& \\
&&&&&& \\
$d_n$ &&$4c+1$&$ic^{5/2}(4-13c+32c^2)$&*&*& \\
{\rm total}\ \#\ {\rm of\ cusps}&& &&&& \\
= ${\rm deg}_c [d_n]$             && 1&2&10&31& \\
&&&&&& \\
\hline
&&&&&& \\
$r_{n,n/m}$ {\rm \ and}&&&&&& \\
{\rm total}\ \#\ {\rm of}\ (n,n/m)&&&&&& \\
{\rm touching\ points}&&&&&& \\
= ${\rm deg}_c[r_{n,n/m}]$ \ {\rm for}&&&&&& \\
&&&&&& \\
&&&$r_{12}$&&$r_{24}$& \\
m=2&&-&$c^5(6c+3)$&-&$c^{105}(100c^3-220c^2+47c+20)$& \\
&&-&1&-&3& \\
&&&&$r_{13}$&& \\
m=3&&-&-&$c^{22}(169c^2+68c+7)$&-& \\
&&-&-&2&-& \\
&&&&&$r_{14}$& \\
m=4&&-&-&-&$c^{70}(100c^2+44c+5)$& \\
&&-&-&-&2& \\
&&&&&& \\
\hline
&&&&&& \\
${\cal N}_n^{(p)}$: \
\#\ {\rm of}\ {\rm el.\ domains}&&&&&& \\
$\sigma^{(p)}_n\subset S_n$&&&&&& \\
{\rm with}\ $q(p)$\ {\rm cusps}&&&&&& \\
&&&&&& \\
p=0,\ q=1&&1&2?&10?&31?& \\
&&&&&& \\
p=1,\ q=0&&0&?&?&?& \\
&&&&&& \\
p=2,\ q=0&&0&?&?&?& \\
&&&&&& \\
p=3,\ q=0&&0&?&?&?& \\
&&&&&& \\
\ldots&&&&&& \\
&&&&&& \\
\hline
\end{tabular} \nn
\ee
}

{\footnotesize
\be
d_3 =c^{66}(1235663104c^10-765891776c^9+315356704c^8-\nn \\
-107832976c^7+33146817c^6-9493768c^5+2585040c^4+\nn\\
+233040c^3-123072c^2+42752c+7168) = \nn \\
= c^{66}(2^8 13^6 c^{10} + \ldots)
\nn
\ee
\be
d_4 = c^{500}(+180143985094819840000000000000c^{31}
-39631676720860364800000000000c^{30}
+ \nn \\
+8693002811987722240000000000c^{29}
-13044663126539108352000000000c^{28}
+ \nn \\
+1173757802691870851072000000c^{27}-
251521821595369943859200000c^{26}
+ \nn \\
+571799816500523292950528000c^{25}
+31829281120908656668016640c^{24}
+ \nn \\
+4498587759541291509802744c^{23}
-22393560229455792230880276c^{22}
- \nn \\
-4844558197092071004917119c^{21}-391359614494908199135640c^{20}
+ \nn\\
+847227219438292501416048c^{19}
+319824545156577533637200c^{18}
+ \nn \\
+43548434837473605814592c^{17}
+33835232982782581502464c^{16}
+ \nn\\
+6794824688429177434112c^{15}
-295875264228762906624c^{14}
-183114267457753161728c^{13}
+ \nn \\
+165660797805589659648c^{12}
+93048634733415038976c^{11}
+19011408301729447936c^{10}
+ \nn \\
+25910347327455363072c^9+16153160956010037248c^8+4359984836267474944c^7
+ \nn\\
+933600163325280256c^6
+314124845396787200c^5+94391968968212480c^4
+ \nn \\
+17132514753642496c^3
+1782088231550976c^2
+99381248262144c
+2319282339840) =
\nn \\
= c^{500}(2^{67}5^{13}c^{31}
+\ldots),
\nn
\ee
$$
r_{15} = c^{236}(14641c^4+12506c^3+4021c^2+576c+31) =
c^{236}((11c)^4 + \ldots)
$$
$$ r_{23} = c^{72},$$
$$r_{25} = c^{720},$$
$$r_{34} = c^{864},$$
$$\ldots$$
}

It is instructive to reproduce the powers of $c$ in these expressions
by direct analysis of roots of the polynomials $G_n(x)$
to demonstrate the strong correlation between different roots of iterated
polynomials. We consider here only the simplest case:
the periodic orbits of orders one and two in the limit $c\rightarrow 0$.
In this limit all polynomials $G_n(x;f)$ for cubic map $f$ should turn
into the same polynomials, but for quadratic map $f$, thus the powers
of the polynomials should decrease appropriately: from
$N_n(d=3)$ to $N_n(d=2)$. In particular, the degrees of $G_1(x)$
and $G_2(x)$ should change from $3$ to $2$ and from $6$ to $2$
respectively. For cubic map $G_1(x;f)$ has three roots, one of them
grows as $\frac{1}{c}$, another two remain finite, so that
$G_1(x) \sim c\left(x+\frac{1}{c}\right)x(x-1) + O(c)$.
(In this case the roots can be found exactly, they are: $x = 0$
and $x = \frac{1}{2c}(-1 \pm \sqrt{1+4c})$.)
$G_2(x)$ has six roots, four grow as $c \rightarrow 0$, and two
remain finite. However, the four roots can not all grow as $\frac{1}{c}$,
because then $G_2(x) = c^3 \prod_{i=1}^6(x - \rho_i^{(2)})$
would grow as $c^{3-4} = c^{-1}$ instead of turning into finite quadratic
polynomial $x^2+x+1 +O(c)$. Actually, only two of the four routs
grow as $\frac{1}{c}$, while the other two -- only as $\frac{1}{\sqrt{c}}$.

Still, this is only the beginning. Unless the coefficients in front
of the singular terms are carefully adjusted, we will not reproduce
correct $c$-asymptotics of $d_1$, $d_2$ and $r_{12}$. Actually,
the first discriminant is simple:
$$d_1 = D(G_1) \sim c^{2\cdot 3 - 2}
\left(\frac{1}{c}\right)^{2\cdot 2} \sim c^0 + O(c),$$
where the first factor is the standard $\alpha_d^{2d-2}$
from discriminant definition, and the second comes from squared
differences between the singular root of $G_1(x)$ and two finite ones.

Similarly we can naively estimate:
$$r_{12} \sim c^{3\cdot 3 + 1\cdot 6}
 \left(\frac{\ldots}{c} - \frac{\ldots}{c}\right)^2
 \left(\frac{\ldots}{c} - \frac{\ldots}{\sqrt{c}}\right)^2
 \left(\frac{\ldots}{c} - \rho^{(2)}\right)^2
 \left(\rho^{(1)} - \frac{\ldots}{c}\right)^4
\cdot $$ $$ \cdot
 \left(\rho^{(1)} - \frac{\ldots}{\sqrt{c}}\right)^4
\left(\rho^{(1)} - \rho^{(2)}\right)^4
\stackrel{?}{\sim}
c^{15-2-2-2-4-2}(1+O(c)) \sim c^3(1+O(c)),$$
where the first factor is $\alpha_d^{\frac{1}{2}((N_1-1)N_2 + N_1N_2)}$
(see s.\ref{gemaded}), the three next come from differences between the
singular root of $G_1$ and the roots of $G_2$, while the remaining
three terms come from the differences of two finite roots of $G_1$
with the roots of $G_2$ (here $\rho^{(1)}$ and $\rho^{(2)}$ denote
{\it finite} roots of $G_1$ and $G_2$, there are two and two).
However, here we run into a problem: the asymptotics $c^3$ is wrong,
actually $r_{12} \sim c^5(1+O(c))$. The reason for this is that the
factor $ \left(\frac{\ldots}{c} - \frac{\ldots}{c}\right)^2$
is actually finite, $\sim c^0$, rather than $\sim \frac{1}{c^2}$
as we naively assumed, because the singular
roots of $G_1$ and $G_2$ are strongly correlated:
the root of $G_1$ is $-\frac{1}{c} + O(1)$, while the two most singular
roots of $G_2$ have exactly the same asymptotics:
$-\frac{1}{c} + O(\sqrt{c})$: coefficients in front of $\frac{1}{c}$
are the same for all these three roots!

For $D(G_2)$ we have:
{\footnotesize
$$
D(G_2) = d_2^2 r_{12} \sim $$ $$
\sim (c^3)^{2\cdot 6 -2}
\left[\left(\frac{\ldots}{c} - \frac{\ldots}{c}\right)
 \left(\frac{\ldots}{c} - \frac{\ldots}{\sqrt{c}}\right)^4
\left(\frac{\ldots}{\sqrt{c}} - \frac{\ldots}{\sqrt{c}}\right)
 \left(\frac{\ldots}{c} - \rho^{(2)}\right)^4
 \left(\frac{\ldots}{\sqrt{c}} - \rho^{(2)}\right)^4
\left(\rho^{(2)}_1 - \rho^{(2)}_2\right)\right]^2 \sim $$ $$
\sim c^{30}\left(\frac{\ldots}{c} - \frac{\ldots}{c}\right)^2
\left(\frac{1}{c^{4+1/2+4+2}}\right)^2(1+O(c))
\sim c^9\left(\frac{\ldots}{c} - \frac{\ldots}{c}\right)^2(1+O(c))
$$ }
We know already, that the remaining difference is not $\frac{1}{c}$,
since both most singular roots of $G_2$ are  $\sim -\frac{1}{c}$
with identical coefficients $(-1)$. However, there is more that:
they are actually $-\frac{1}{c} \pm i\sqrt{c} + \ldots$, so that
the difference is actually $\sim \sqrt{c}$, and $D(G_2)\sim c^{10}$.
There is no similar cancellation between the two less singular roots,
because they are actually $\sim \pm\frac{i}{\sqrt{c}}$ with
{\it opposite} rather than equal coefficients. Thus, since we know
asymptotics of $D(G_2)$ and $r_{12}$, $d_2^2 \sim c^5(1+O(c))$.

\bigskip

In this case the Mandelbrot set has no $\bb{Z}_2$ symmetry
(like for the family $x^3+c$), therefore we expect that
all the cusps belong only to the zero-level components
$\sigma_n^{(p)}$, i.e.
$q(p)=1$ for $p=0$ and $q(p)=0$ for $p>0$ (like for
the family $x^2+c$).
The sum rules (\ref{sumrules})  imply in this case:
$$
{\cal N}_n^{(0)} + \sum_{m}{\rm deg}_c[r_{n,n/m}] =
{\rm deg}_c [G_n(\varepsilon,c)],
$$
$$
{\cal N}_n^{(0)} =  {\rm deg}_c [d_n],
$$
$$
\sum_{p\geq 1}{\cal N}_n^{(p)} = \sum_{m}{\rm deg}_c[r_{n,n/m}]
$$
and the consistency condition
$$
{\rm deg}_c [d_n] + \sum_{m}{\rm deg}_c[r_{n,n/m}] =
{\rm deg}_c [G_n(\varepsilon,c)],
$$
all obviously satisfied by the data in the table, provided
the number of elementary domains is counted as the $c$-degree
of $G_n(x,c)$ with small $c$-independent $x=\varepsilon$
(exactly at critical point $x=0$ all $G_n(x,c)$ vanish, and at
another critical point $x=-2/3c$ they have negative powers in
$c$).

\subsubsection{$f_\gamma = x^3 + \gamma x^2$}

The family $cx^3 + x^2$ is diffeomorphic (i.e. equivalent) to
$x^3 + \gamma x^2$ with $\gamma = 1/\sqrt{c}$, which is much simpler
from the point of view of resultant analysis.

This time the map has two different critical points,
$\{w_f\} = \{0,-2\gamma/3\}$. Therefore one could expect that
the number of elementary domains is counted by a sum of two
terms:
$$ \# \ {\rm of\ elementary\ domains}\  \stackrel{?}{=}
{\rm deg}_\gamma [G_n(0,\gamma)] +
{\rm deg}_\gamma \left[G_n\left(-\frac{2\gamma}{3},\gamma\right)\right]
$$
However, $G_n(x=0,\gamma)$ vanishes identically because of the
high degeneracy of the map $f = x^3+\gamma x^2$ at $x=0$ and does
not contribute. Also two out of three roots of
$G_1\left(-\frac{2\gamma}{3},\gamma\right)$ for $n=1$
coincide (and equal zero),
and label one and the same elementary domain. Thus, actually,
$$ \# \ {\rm of\ elementary\ domains}\ =
{\rm deg}_\gamma \left[G_n\left(-\frac{2\gamma}{3},\gamma\right)\right]
-\delta_{n,1}.
$$

{\footnotesize
\be
\begin{tabular}{|c|cccccc|}
\hline
\multicolumn{7}{|l|}{}\\
\multicolumn{7}{|l|}{\ \ \ $f = x^3+\gamma x^2$}\\
\multicolumn{7}{|l|}{}\\
\hline
&\ &\ \ &&&& \\
n                            &&    1&2&3&4& \ldots \\
&&&&&& \\
\hline
&&&&&& \\
\#\ {\rm of\ orbits}       &&    &&&& \\
= ${\rm deg}_x [G_n]/n$ &&3&3&8&18& \\
&&&&&& \\
\#\ {\rm of\ el.\ domains} &&    &&&& \\
= ${\rm deg}_\gamma [G_n\left(-\frac{2\gamma}{3},\gamma\right)]
-\delta_{n,1}$
&&    2&6&24&72& \\
&&&&&& \\
$d_n$ &&$\gamma^2+4$&$i(4\gamma^4-13\gamma^2+32)$&*&*& \\
{\rm total}\ \#\ {\rm of\ cusps}&& &&&& \\
= ${\rm deg}_\gamma [d_n]$             && 2&4&20&62 & \\
&&&&&& \\
\hline
&&&&&& \\
$r_{n,n/m}$ {\rm \ and}&&&&&& \\
{\rm total}\ \#\ {\rm of}\ (n,n/m)&&&&&& \\
{\rm touching\ points}&&&&&& \\
= ${\rm deg}_\gamma[r_{n,n/m}]$ \ {\rm for}&&&&&& \\
&&&&&& \\
&&&$r_{12}$&&$r_{24}$& \\
m=2&&-&$3\gamma^2+16$&-&$20\gamma^6+47\gamma^4-220\gamma^2+1000$& \\
&&-&2&-&6& \\
&&&&$r_{13}$&& \\
m=3&&-&-&$7\gamma^4 + 68\gamma^2 + 169$&-& \\
&&-&-&4&-& \\
&&&&&$r_{14}$& \\
m=4&&-&-&-&$5\gamma^4+44\gamma^2+100$& \\
&&-&-&-&4& \\
&&&&&& \\
\hline
&&&&&& \\
${\cal N}_n^{(p)}$: \
\#\ {\rm of}\ {\rm el.\ domains}&&&&&& \\
$\sigma^{(p)}_n\subset S_n$&&&&&& \\
{\rm with}\ $q(p)$\ {\rm cusps}&&&&&& \\
&&&&&& \\
p=0,\ q=1&&2&4 ?&20 ?&62 ?& \\
&&&&&& \\
p=1,\ q=0&&?&?&?&?& \\
&&&&&& \\
p=2,\ q=0&&?&?&?&?& \\
&&&&&& \\
p=3,\ q=0&&?&?&?&?& \\
&&&&&& \\
\ldots&&&&&& \\
&&&&&& \\
\hline
\end{tabular} \nn
\ee
}
{\footnotesize
\be
d_3 = 7168a^{20}+42752a^{18}-123072a^{16}+233040a^{14}+ \nn \\
+2585040a^{12}-9493768a^{10}+33146817a^8-107832976a^6+ \nn \\
+315356704a^4-765891776a^2+1235663104
\nn
\ee
\be
d_4 =
2319282339840\gamma^{62} +99381248262144\gamma^{60}
+1782088231550976\gamma^{58}+ \nn \\
+17132514753642496\gamma^{56}+94391968968212480\gamma^{54}
+314124845396787200\gamma^{52} + \nn \\
+933600163325280256\gamma^{50}+4359984836267474944\gamma^{48}
+16153160956010037248\gamma^{46} + \nn \\
+25910347327455363072\gamma^{44}+19011408301729447936\gamma^{42}
+93048634733415038976\gamma^{40} + \nn \\
+165660797805589659648\gamma^{38}-183114267457753161728\gamma^{36}
-295875264228762906624\gamma^{34} + \nn \\
+6794824688429177434112\gamma^{32}
+33835232982782581502464\gamma^{30}
+43548434837473605814592\gamma^{28} + \nn \\
+319824545156577533637200\gamma^{26}
+847227219438292501416048\gamma^{24}
-391359614494908199135640\gamma^{22} - \nn \\
-4844558197092071004917119\gamma^{20}
-22393560229455792230880276\gamma^{18}
+4498587759541291509802744\gamma^{16}+ \nn \\
+31829281120908656668016640\gamma^{14}
+571799816500523292950528000\gamma^{12}
-251521821595369943859200000\gamma^{10}+ \nn \\
+1173757802691870851072000000\gamma^8
-13044663126539108352000000000\gamma^6
+8693002811987722240000000000\gamma^4- \nn \\
-39631676720860364800000000000\gamma^2
+180143985094819840000000000000
\ \ \ \ \ \ \ \ \ \ \ \ \ \ \ \ \ \
\nn
\ee
}

As already mentioned in the previous subsection \ref{cx3x2},
in this case the Mandelbrot set has no $\bb{Z}_2$ symmetry
(like for the family $x^3+c$), therefore we expect that
all the cusps belong only to the zero-level components
$\sigma_n^{(p)}$, i.e.
$q(p)=1$ for $p=0$ and $q(p)=0$ for $p>0$ (like for
the family $x^2+c$).
The sum rules (\ref{sumrules})  imply in this case:
$$
{\cal N}_n^{(0)} + \sum_{m}{\rm deg}_\gamma[r_{n,n/m}] =
{\rm deg}_\gamma
\left[G_n\left(-\frac{2\gamma}{3},\gamma\right)\right] - \delta_{n,1},
$$
$$
{\cal N}_n^{(0)} =  {\rm deg}_\gamma [d_n],
$$
$$
\sum_{p\geq 1}{\cal N}_n^{(p)} = \sum_{m}{\rm deg}_\gamma[r_{n,n/m}]
$$
and the consistency condition
$$
{\rm deg}_\gamma [d_n] + \sum_{m}{\rm deg}_\gamma[r_{n,n/m}] =
{\rm deg}_\gamma
\left[G_n\left(-\frac{2\gamma}{3},\gamma\right)\right]-
\delta_{n,1},
$$
all obviously satisfied by the data in the table.

\bigskip

Comparing the data in tables in this section and in the previous
s.\ref{cx3x2} we see, that the numbers of elementary domains
differ by two -- despite the two maps are equivalent (diffeomorphic)
in the sense of eq.(\ref{diffeo}). This is the phenomenon which we
already studied at the level of quadratic maps (which are all
diffeomorphic) in s.\ref{faasse}: different families of maps
(even diffeomorphic) are different sections of the same entity --
universal discriminant variety, but what is seen in the section
depends on its (section's) particular shape.
Relation between $cx^3 + x^2$ and $x^3+\gamma x^2$ is such that
$c = 1/\gamma^2$, so it is not a big surprise that in the
$\gamma$ plane we see each domain from the $c$ plane twice.

It is instructive to see explicitly how it works, at least
for the simplest case of $\sigma_1^{(0)}$. This (these)
domain(s) is (are) the (parts of) stability domain $S_1$,
defined by relations (\ref{stabco}),
$$\left\{
\begin{array}{c}
cx^3 + x^2 - x = 0 \\
|3cx^2 + 2x| < 1
\end{array}\right.
$$
for the family $cx^3+x^2$ and
$$\left\{
\begin{array}{c}
x^3 + \gamma x^2 - x = 0 \\
|3x^2 + 2\gamma x| < 1
\end{array} \right.
$$
for the family $x^3+\gamma x^2$.
The first of these systems is easily transformed to
$$\left\{
\begin{array}{c}
|3-x| < 1, \ {\rm or}\ x=3 - e^{i\phi} \\
c = \frac{1}{x^2} - \frac{1}{x}
\end{array}\right.
$$
while the second one -- to
$$ \left\{
\begin{array}{c}
|3-\gamma x| < 1,  \ {\rm or}\ x = \frac{3 - e^{i\phi}}{\gamma} \\
\gamma = \frac{1}{x} - x
\end{array}\right.
$$
The critical point (single) of the first system,
where ${\partial c}/{\partial\phi} = 0$,
is at $x_{cr} = 2$, thus $c_{cr} = -\frac{1}{4}$ -- the zero
of discriminant $d_1(c) = 4c+1$.
The critical points of the second system,
where ${\partial \gamma}/{\partial\phi} = 0$,
are at $x_{cr} = \pm i$, thus $c_{cr} = -2\mp i$ -- the zeroes
of discriminant $d_1(\gamma) = \gamma^2 + 4$.
These points are the positions of the single cusp of a single elementary
domain $\sigma_1^{(0)}$ in the first case and the two cusps of
two elementary domains $\sigma_{1,\pm}^{(0)}$ (one cusp per domain)
in the second case.

\subsubsection{Map $f_{a;c} = ax^3 + (1-a)x^2 + c$
\label{acfa}}

We now use another possibility, provided by the study of 2-parametric
families: we look at the situation with broken $\bb{Z}_d$ symmetries
and at the interpolation between $\bb{Z}_{d}$ and $\bb{Z}_{d-1}$.
In the family $\{ax^3 + (1-a)x^2 + c\}$ the $\bb{Z}_3$ symmetry of Julia
set and $\bb{Z}_2$ symmetry of Mandelbrot set occurs at $a=1$, while
at $a=0$ it is reduced to $\bb{Z}_2$ for Julia sets and to nothing for
Mandelbrot set. Accordingly, the number of cusps of elementary components
of Mandelbrot set should decrease as $a$ changes from $1$ to $0$.
It is instructive to see, how continuous change of parameter $a$
causes change of discrete characteristic, like the number of cusps
(i.e. that of zeroes of irreducible discriminants $d_n(c)$).
Part of the answer is given by Fig.\ref{symmebre}: at $a=1$ all
the zero-level components $\sigma_n^{(0)}$ have the $2$-cusp shape;
when $a$ is slightly smaller
than $1$, only remote components $\sigma_n^{(0)}$ change their shape
from $2$-cusp to $1$-cusp type, while central domains stay in the $2$-cusp
shape;
and the smaller $a$ the closer comes the boundary between $2$-cusp
and $1$-cusp shapes. Analytically, this means that positions of zeroes
of $d_n(c)$ depend on $a$ in a special way.

Transition reshuffling process ("perestroika") is a separate story,
it is shown in Fig.\ref{resh} and
it repeats itself with all the components on Mandelbrot set, though
the moment of transition is different for different components, as
clear from Fig.\ref{symmebre}. Fig.\ref{resh} demonstrates that
what was a single central component $M_1$ with the multipliers-tree
internal structure for the family $\{x^3+c\}$ at $a=1$,
deforms and splits into many such components as $a$ decreases from
$1$ to $0$. Most of these detached components disappear at $c=\infty$
as $a\rightarrow 0$, but some remain, including the central domain
$M_1$ and, say, the $M_3$-domain for the family $\{x^2+c\}$.
This picture clearly shows how the trail structure between $M_1$ and $M_3$
(and its continuation further, towards the end-point $c=-2$) is
formed for $\{x^2+c\}$ from a single domain $M_1$ for $\{x^3+c\}$.
Our analytic method allows to describe and analyze this process by tracking
the motion of zeroes of resultants and discriminants with changing $a$.
For our 2-parametric family\footnote{Alternative technical approach
(especially effective for $a$ close to $1$), can substitute our family by
a diffeomorphic one, $\{x^3 + \gamma x^2 + \alpha\}$
with $\gamma = \frac{1-a}{\sqrt{a}}$, $\alpha = c\sqrt{a}$.
For this family expressions for discriminants and resultants are
somewhat simpler, for example
$$d_1 = -27\alpha^2-(4\gamma^3+18\gamma )\alpha+(\gamma^2+4),$$
$$r_{12} = 27\alpha^2+(4\gamma^3+18\gamma )\alpha+(3\gamma^2+16),$$
$$d_2 = i(27\alpha^2+(4\gamma^3+18\gamma )\alpha+
(4\gamma^4-13\gamma^2+32)),$$
$$r_{13} = 729\alpha^4 + (216\gamma^3+972\gamma )\alpha^3 +
(16\gamma^6 + 144\gamma^4+351\gamma^2 + 27)\alpha^2 +
$$ $$+(4\gamma^5 + 22\gamma^3 18\gamma )\alpha +
(7\gamma^4+68\gamma^2+169),$$
$$\ldots$$
Instead interpretation in the region of $a$ near $0$
is less transparent.
}
$$d_1 = (1+a)^2 +2(a-1)(a+2)(2a+1)c-27a^2c^2,$$
$$r_{12} = a^5\left[(a+3)(3a+1) -2(a-1)(a+2)(2a+1)c + 27a^2c^2\right],$$
$$d_2 = a^{5/2}i\left[(4a^4-29a^3+82a^2-29a+4) -2a(a-1)(a+2)(2a+1)c
+27a^3c^2\right],$$
$$r_{13} = a^{22}(7+40a+4c-4a^5c+48a^5c^2+15a^4c^2+216a^2c^3+16c^2+
$$ $$+40a^3+48ac^2+16a^6c^2-324a^4c^3+2ca-216a^5c^3+75a^2+7a^4+729a^4c^4+
-131a^3c^2-2a^4c-8a^2c+8a^3c+15a^2c^2+324a^3c^3),$$
$$\ldots$$
$$w_{1,1} = \frac{27q^2+4p^3+24p^2+36p}{27} =
\frac{4(2a^4-8a^3+3a^2+10a-7)-12a^2(a-1)^3c+81a^4c^2}{81a^3}$$
$$\ldots$$
and positions of zeroes are shown in Fig.\ref{zeropos}.
As $a\rightarrow 0$, of two zeroes of $d_1(c)$ one tends to
$\frac{1}{4}+O(a)$
(position of the cusp for the family $\{x^2+c\}$), another
grows as $-\frac{4}{27a^2}-\frac{2}{9a} - \frac{1}{36}+O(a)$;
of two zeroes of $r_{12}(c)$
one tends to $-\frac{3}{4}+O(a)$
(position of the intersection $\sigma_1^{(0)}\cap \sigma_2^{(1)}$ for
the family $\{x^2+c\}$), another grows as
$-\frac{4}{27a^2}-\frac{2}{9a} - \frac{1}{36}+1+O(a)$;
both zeroes of $d_2(c)$ grow -- as $-\frac{1}{a}+2+O(a)$ and
$-\frac{4}{27a^2}+\frac{7}{9a}-\frac{16}{9}+O(a)$.
On their way, the pairs of complex (at $a=1$) roots of $r_{12}$ and $d_2$
reach the real line ${\rm Im}\ c = 0$
and merge on it when the higher discriminants
$$D(r_{12}|c) \sim (a-1)^2(a+2)^2(2a+1)^2-27a^2(a+3)(3a+1) =
$$ $$=4a^6+12a^5-84a^4-296a^3-84a^2+12a+4 =
4(a^2-5a+1)(a^2+4a+1)^2 = ((2a-5)^2-21)((a+2)^2-3)^2$$ and
$$D(d_2|c) \sim a^2(a-1)^2(a+2)^2(2a+1)^2-27a^3(4a^4-29a^3+82a^2-29a+4)=
=4a^2(a^2-8a+1)^3$$
vanish. Actually, this happens at $a = \frac{5-\sqrt{21}}{2} \approx
0.208712152$ (i.e. $\gamma = \sqrt{3}$)
and $a= 4-\sqrt{15} \approx 0,127016654$ respectively.
After that the roots diverge again and
move separately along the real line towards
their final values at $c=-\frac{3}{4}$ and $c=\infty$. Similar is
the behaviour of zeroes of $r_{13}$, $d_3$ and of higher resultants.
The "main" splitting, shown in the central picture
(with $a=1/7$) of Fig.\ref{resh}, takes place at
$a \approx 0.141$ -- it corresponds to
the splitting of "end-point" of the component
$M_1$ of Mandelbrot set (which is a limiting point at
$n\rightarrow \infty$ of the sequence of zeroes of resultants $r_{n,2n}$).
The "final splitting", shown in the next picture (with $a=1/8$) of
Fig.\ref{resh}, takes place at
$a \approx 0.121$, which is the zero of the discriminant $D(w_{1,2}|c)$ --
this is obvious from the fact that the "end-point" $c=-2$ of Mandelbrot set
for $\{x^2+c\}$ is a zero of $w_{1,2}(c)$, and two such end-points
can be seen to touch at this cadre of Fig.\ref{resh}. Of course, there is
nothing special in $w_{1,2}$ -- except for that $1$ and $2$ are small
numbers and that $w_{1,2}(c)$ has real zeroes, -- other stages of
reshuffling (similar catastrophes, involving smaller
components of Mandelbrot set) take place when other $D(w_{n,s}|c)=0$.

From Fig.\ref{zeropos} it is clear, that for small $a$ an exact
-- only mirror-reflected -- copy
of the Mandelbrot set for the family $\{x^2+c\}$
(which is located near $c=0$) is formed in the vicinity of
$c=\infty$ (far to the left of Fig.\ref{zeropos}). More evidence
to this statement is provided by analysis of relation (\ref{stabco}).
For example, for stability domain $S_1$:
$$
\left\{\begin{array}{c}
|3ax^2 + 2(1-a)x| < 1, \\
c = x - (1-a)x^2 - ax^3.
\end{array} \right.
$$
At small $a$ this domain decays into two identical parts:
$$
\left\{\begin{array}{c}
2|x| < 1 + O(a), \\
c = x - x^2 + O(a),
\end{array} \right.
$$
and, for $x = -\frac{2(1-a)}{3a} - \tilde x$,
$$
\left\{\begin{array}{c}
\left|2(1-a)\tilde x
\left(1+\frac{3a}{2(1-a)}\tilde x\right)\right| < 1, \\
 c = x - (1-a)x^2 - ax^3 = \left(-\frac{4}{27a^2} - \frac{2}{9a} +
\frac{2}{9}\right) - (\tilde x -\tilde x^2) + O(a).
\end{array} \right.
$$
In general, the map $x \rightarrow ax^3 + (1-a)x^2 + c$
is equivalent to $\tilde x \rightarrow a\tilde x^3 + (1-a)\tilde x^2
+ \tilde c$ with
$$\tilde c = -c - \frac{4(1-a)^3}{27a^2} - \frac{2(1-a)}{3a} =
\frac{2(a-1)(a+2)(2a+1)}{27a^2}-c.$$
This is the peculiar $\bb{Z}_2$
symmetry of the Mandelbrot set for our family.\footnote{\label{FEcr}
To avoid possible confusion, note that $\bb{Z}_2$
symmetry, discussed in this paragraph, is not
respected in Fig.\ref{resh}, obtained with the help of
{\it Fractal Explorer} program \cite{FE}.
The program obviously misinterprets some properties of Mandelbrot sets in
the case of non-trivial families, like $\{ax^3 + (1-a)x^2 + c\}$.
Still it seems to adequately describe some qualitative
features even for such families and we include Fig.\ref{resh}
for illustrative purposes, despite it is not fully correct.
Actually, the right parts of the pictures seem rather reliable,
while the left parts should be obtained by reflection.
Fully reliable is Fig.\ref{zeropos}, but it is less detailed:
it contains information only about several points of the Mandelbrot
sets. Some of these points (located in the right parts of the pictures)
are shown by arrows in both Figs.\ref{resh} and \ref{zeropos} to help
comparing these pictures.
}

{\footnotesize
\be
\begin{tabular}{|c|cccccc|}
\hline
\multicolumn{7}{|l|}{}\\
\multicolumn{7}{|l|}{\ \ \ $f = ax^3+(1-a)x^2+c$,\ \ \
$\begin{array}{c}{\rm small}\ a\\
\left[a\ {\rm near}\ 1\right]\end{array}$ }\\
\multicolumn{7}{|l|}{}\\
\hline
&\ &\ \ &&&& \\
n                            &&    1&2&3&4& \ldots \\
&&&&&& \\
\hline
&&&&&& \\
\#\ {\rm of\ orbits}       &&    &&&& \\
= ${\rm deg}_x [G_n]/n$ &&3&3&8&18& \\
&&&&&& \\
\#\ {\rm of\ el.\ domains} &&    2&4&16&48& \\
= $2{\rm deg}_c [G_n(0;c)]$&&    [1]&[2]&[8]&[24]& \\
&&&&&& \\
{\rm total}\ \#\ {\rm of\ cusps}&& &&&& \\
= ${\rm deg}_c [d_n]$             && 2&2&12&40& \\
&&&&&& \\
\hline
&&&&&& \\
{\rm total}\ \#\ {\rm of}\ (n,n/m)&&&&&& \\
{\rm touching\ points}&&&&&& \\
= ${\rm deg}_c[r_{n,n/m}]$ \ {\rm for}&&&&&& \\
&&&&&& \\
m=2&&-&2&-&4& \\
&&&&&& \\
m=3&&-&-&4&-& \\
&&&&&& \\
m=4&&-&-&-&4& \\
&&&&&& \\
\hline
&&&&&& \\
${\cal N}_n^{(p)}$: \
\#\ {\rm of}\ {\rm el.\ domains}&&&&&& \\
$\sigma^{(p)}_n\subset S_n$&&&&&& \\
{\rm with}\ $q(p)$\ {\rm cusps}&&&&&& \\
&&&&&& \\
$p=0,\ q=1$&&2&2&12&40& \\
\ \ \ \ \ [q=2]&&[1]&[0]&[4]&[16]& \\
&&&&&& \\
$p=1,\ q=0$&&0&2&4&?& \\
\ \ \ \ \ [q=1]&&[0]&[2]&[4]&[4]& \\
&&&&&& \\
$p=2,\ q=0$&&0&0&0&?& \\
\ \ \ \ \ [q=1]&&[0]&[0]&[0]&[4]& \\
&&&&&& \\
$p=3,\ q=0$&&0&0&0&0& \\
\ \ \ \ \ [q=1]&&[0]&[0]&[0]&[0]& \\
&&&&&& \\
\ldots&&&&&& \\
&&&&&& \\
\hline
\end{tabular} \nn
\ee }
Every map of our family has two critical points, $x=0$ and $\tilde x = 0$,
i.e. $x = \frac{2(1-a)}{3a}$, which are both independent of $c$.
Therefore the number of elementary domains,
$\sum_{w_c} {\rm deg}_c [G_n(w_c;c)] = 2{\rm deg}_c [G_n(0;c)]$.
In this table most data without square brackets corresponds to the case
of small $a$, i.e. to the small deformation of the family $\{x^2+c\}$.
However, even a minor deformation causes immediate switch of
degrees of $d_n(c)$ and $r_{n,k}(c)$ to their values,
characteristic for the cubic map family $\{x^3 + c\}$.
At the same time, the numbers $q(p)$ of cusps remain the same
as they were for the quadratic map family $\{x^2+c\}$.
The table shows what this means for the numbers of elementary
domains. Numbers in square brackets count elementary domains
in the vicinity of the $\{x^3+c\}$
family, when some of components got merged (see Fig.\ref{resh})
and the numbers of cusps increased. Thus the table
shows transition between cubic and quadratic families in
terms of discrete numbers. Of course, at least one of the sum rules
and consistency condition also change under the transition,
since so do the numbers $q(p)$:
$$
{\rm deg}_c [d_n]
= q(1){\rm deg}_c [G_n(w_c,c)] + {\cal N}_n^{(0)},
$$
and
$$
{\rm deg}_c [d_n] + \sum_{m}{\rm deg}_c[r_{n,n/m}] =
q(0){\rm deg}_c [G_n(w_c,c)],
$$
while two other relations remain the same for all values of $a$:
$$
{\cal N}_n^{(0)} + \sum_{m}{\rm deg}_c[r_{n,n/m}] =
{\rm deg}_c [G_n(w_c,c)],
$$
$$
\sum_{p\geq 1}{\cal N}_n^{(p)} = \sum_{m}{\rm deg}_c[r_{n,n/m}].
$$

One more warning should be made: since transition
occurs at different values of $a$ for different components of
Mandelbrot set (see Fig.\ref{symmebre}), the meaning of "close"
(when we mention $a$ close to $1$ or to $0$) depends on considered
component (actually, on the value of $n$ in the table).

\subsection{Quartic maps}

\subsubsection{Map $f_c = x^4 + c$}

This section of Julia sheaf was used as an illustration in
Figs.\ref{0Man4} and \ref{JuliaMix4}.

In this case
$$G_1 = x^4-x+c,$$
$$G_2 = x^{12} + x^9 + 3cx^8 + x^6 + 2cx^5 + 3c^2x^4 + x^3 + cx^2 + c^2x +
c^3 + 1,$$
$$G_{1,1} = G_2-1 = (x^4+x+c)(x^8 + 2cx^4 + x^2 + c^2).$$
Factorization of the last polynomial is accidental and is lifted by
infinitesimal variation of the family $x^4+c$, e.g. provided by additional
term $px$. The same happens with accidental factorizations in the
following formulas.

The triple critical point is $w=0$.

$$w_1 = c^3,$$
$$w_{11} = c^6,$$
$$w_{12} = c^6(c^3+2)^3(c^6+2c^3+2)^3,$$
$$\ldots$$

{\footnotesize
\be
\begin{tabular}{|c|cccccc|}
\hline
\multicolumn{7}{|l|}{}\\
\multicolumn{7}{|l|}{\ \ \ $f = x^4 + c$}\\
\multicolumn{7}{|l|}{}\\
\hline
&\ &\ \ &&&& \\
n                            &&    1&2&3&4& \ldots \\
&&&&&& \\
\hline
&&&&&& \\
\#\ {\rm of\ orbits}       &&    &&&& \\
= ${\rm deg}_x [G_n]/n$ &&4&6&20&60& \\
&&&&&& \\
\#\ {\rm of\ el.\ domains} &&    &&&& \\
= ${\rm deg}_c [G_n(w_c,c)]$   &&    1&3&15&60& \\
&&&&&& \\
$d_n$ &&$-27+256c^3$&$65536c^6+$&*&?& \\
   &&&$+152064c^3+91125$&&& \\
{\rm total}\ \#\ {\rm of\ cusps}&& &&&& \\
= ${\rm deg}_c [d_n]$    && 3&6&39&165& \\ \\
&&&&&& \\
\hline
&&&&&& \\
$r_{n,n/m}$ {\rm \ and}&&&&&& \\
{\rm total}\ \#\ {\rm of}\ (n,n/m)&&&&&& \\
{\rm touching\ points}&&&&&& \\
= ${\rm deg}_c[r_{n,n/m}]$ \ {\rm for}&&&&&& \\
&&&&&& \\
&&&$r_{12}$&&$r_{24}$& \\
m=2&&-&$256c^3+125$&-&$16777216c^9+53411840c^6+$& \\
&&&&&$59113216c^3+24137569$& \\
&&-&3&-&9& \\
&&&&$r_{13}$&& \\
m=3&&-&-&$65536c^6-2304c^3+9261$&-& \\
&&-&-&6&-& \\
&&&&&$r_{14}$& \\
m=4&&-&-&-&$65536c^6-24064c^3+4913$& \\
&&-&-&-&6& \\
&&&&&& \\
\ldots &&&&&& \\
&&&&&& \\
\hline
&&&&&& \\
${\cal N}_n^{(p)}$: \
\#\ {\rm of}\ {\rm el.\ domains}&&&&&& \\
$\sigma^{(p)}_n\subset S_n$&&&&&& \\
{\rm with}\ $q(p)$\ {\rm cusps}&&&&&& \\
&&&&&& \\
p=0,\ q=3&&1&0&9&45& \\
&&&&&& \\
p=1,\ q=2&&0&3&6&6& \\
&&&&&& \\
p=2,\ q=2&&0&0&0&9& \\
&&&&&& \\
p=3,\ q=2&&0&0&0&0& \\
&&&&&& \\
\ldots&&&&&& \\
&&&&&& \\
\hline
\end{tabular} \nn
\ee
}

{\footnotesize
\be
d_3=(1099511627776c^{15}+4367981740032c^{12}
+6678573481984c^9+\nn\\
+4811867160576c^6+1590250910976c^3+1312993546389)\cdot\nn\\ \cdot
(18446744073709551616c^{24}+149591565222738395136c^{21}+\nn\\ +
525955697232230481920c^{18}+1038531305524659486720c^{15}+\nn\\ +
1235398275557786386432c^{12}+894978163534410547200c^9+\nn\\ +
419572901219568058368c^6+181980149245232679936c^3
+79779353642425058769)
\nn
\ee
}

The sum rules (\ref{sumrules}) now imply:
$$
{\cal N}_n^{(0)} + \sum_{m}{\rm deg}_c[r_{n,n/m}] =
{\rm deg}_c [G_n(w_c,c)],
$$
$$
3{\cal N}_n^{(0)} +
2\sum_{m}{\rm deg}_c[r_{n,n/m}] =  {\rm deg}_c [d_n],
$$
$$
\sum_{p\geq 1}{\cal N}_n^{(p)} = \sum_{m}{\rm deg}_c[r_{n,n/m}]
$$
and the consistency condition
$$
{\rm deg}_c [d_n] + \sum_{m}{\rm deg}_c[r_{n,n/m}] =
3{\rm deg}_c [G_n(w_c,c)],
$$
all obviously satisfied by the data in above table.

\subsection{Maps $f_{d;c}(x) = x^d + c$}

Julia sets for these maps have symmetry $\bb{Z}_d$, because pre-image
$f^{\circ(-1)}(x)$ of any point $x$ is $\bb{Z}_d$-invariant set
$\left\{ e^{\frac{2\pi ik}{d}}x^{(-1)},\ k=0,\ldots,d-1\right\}$,
and thus entire grand orbits, their limits and closures are $\bb{Z}_d$
invariant. Sections of Mandelbrot set, associated with these families,
posses $\bb{Z}_{d-1}$ symmetry, because of invariance
of the equations $G_n(x;c) = 0$ under the transformations
$x \rightarrow e^{\frac{2\pi i}{d-1}}x,\
c \rightarrow e^{\frac{2\pi i}{d-1}}c$.

The critical point for all these maps is (multiple) $w_f=0$.
The values of polynomials $G_n$ with $n>1$ at this critical point,
$G_n(0;c)$, depend on $c^{d-1}$ only, therefore the systems of roots
$G_n(0;c) = 0$ possess $\bb{Z}_{d-1}$ symmetry (and the root
$c=0$ of $G_1(0,c) = c$ is a stable point of this symmetry).
This $\bb{Z}_{d-1}$ is the symmetry of entire Mandelbrot set.
\be
G_1(0,c) = c, \nn \\
G_2(0,c) = 1+c^{d-1}, \nn \\
G_3(0,c) = 1+c^{d-1}\left(1 + c^{d-1}\right)^d , \nn \\
G_4(0,c) = 1 + c^{d-1}
\frac{\left(1+c^{d-1}\left(1+c^{d-1}\right)^d\right)^d - 1}
{1 + c^{d-1}} = \nn \\ =
1 + \sum_{s=1}^d \frac{d!}{s!(d-s)!}c^{(d-1)(s+1)}
\left(1 + c^{d-1}\right)^{sd-1}, \nn \\
\ldots \nn
\ee
Degrees of these polynomials satisfy recurrent relation
$$
{\rm deg}_c [G_n(0,c)] =
d^{n-1} - \sum_{\stackrel{k|n}{k<n}}^{\tau(n)-1}
{\rm deg}_c [G_k(0,c)],
$$
which is similar to relation for powers of $G_n(x)$,
$$
N_n(d) = d^n - \sum_{\stackrel{k|n}{k<n}}^{\tau(n)-1} N_k(d).
$$
Therefore ${\rm deg}_c [G_n(0,c)] = N_n(d)/d$.

Note, that $F_n(f(0);c) = f(F_n(0;c)) - f(0)$,
and if $F_n(0;c)=0$, then $F_n(f(0);c)=0$.
For our maps $f(x) = x^d+c$, $f(0)=c$, and
$F_n(0;c)=0$ implies $F_n(c;c) = 0$,
actually $F_n(c;c) = F_n^d(0;c)$.

The sum rules (\ref{sumrules}) imply:
$$
{\cal N}_n^{(0)} + \sum_{m}{\rm deg}_c[r_{n,n/m}] =
{\rm deg}_c [G_n(w_c,c)],
$$
$$
{\rm deg}_c [d_n]
= (d-2){\rm deg}_c [G_n(w_c,c)] + {\cal N}_n^{(0)},
$$
$$
\sum_{p\geq 1}{\cal N}_n^{(p)} = \sum_{m}{\rm deg}_c[r_{n,n/m}].
$$
Consistency condition
$$
{\rm deg}_c [d_n] + \sum_{m}{\rm deg}_c[r_{n,n/m}] =
(d-1){\rm deg}_c [G_n(w_c,c)].
$$
All these relations are satisfied by the data in the
following table:

{\footnotesize
\be
\hspace{-1cm}
\begin{tabular}{|c|ccccccccc|}
\hline
\multicolumn{10}{|l|}{}\\
\multicolumn{10}{|l|}{\ \ \ $f = x^d + c$}\\
\multicolumn{10}{|l|}{}\\
\hline
&\ &\ \ &&&&&&& \\
n                            &&    1&2&3&4&5&6&& \ldots \\
&&&&&&&&& \\
\hline
&&&&&&&&& \\
$N_n(d)$  &&$d$&$d(d-1)$&$d(d^2-1)$&$d^2(d^2-1)$&$d(d^4-1)$&
$d(d^2-1)(d^3+d-1)$&& \\
&&&&&&&&& \\
\#\ {\rm of\ orbits}       &&    &&&&&&& \\
= ${\rm deg}_x [G_n]/n$ &&$d$&$\frac{d(d-1)}{2}$&
$\frac{d(d^2-1)}{3}$&$\frac{d^2(d^2-1)}{4}$&$\frac{d(d^4-1)}{5}$&
$\frac{d(d^2-1)(d^3+d-1)}{6}$&& \\
= $N_n(d)/n$ &&&&&&&&& \\
&&&&&&&&& \\
\#\ {\rm of\ el.\ domains} &&    &&&&&&& \\
= ${\rm deg}_c [G_n(w_c,c)]$
    &&1&$d-1$&$d^2-1$&$d(d^2-1)$&$d^4-1$&$(d^2-1)(d^3+d-1)$&& \\
= $N_n(d)/d$ &&&&&&&&& \\
&&&&&&&&& \\
{\rm total}\ \#\ {\rm of\ cusps}&&$d-1$&$(d-1)(d-2)$&
$(d-1)(d^2-3)$&$(d^2-1)(d^2-d-1)$&$(d-1)(d^4-5)$&
$(d-1)(d^5-2d^2-3d+2)$&& \\
= ${\rm deg}_c [d_n]$             && &&&&&&& \\
&&&&&&&&& \\
\hline
&&&&&&&&& \\
{\rm total}\ \#\ {\rm of}\ (n,n/m)&&&&&&&&& \\
{\rm touching\ points}&&&&&&&&& \\
= ${\rm deg}_c[r_{n,n/m}]$ \ {\rm for}&&&&&&&&& \\
&&&&&&&&& \\
&&&&&&&$(d-1)^3 + 2(d-1)^2$&& \\
m=2&&-&$d-1$&-&$(d-1)^2$&-&$=(d-1)^2(d+1)$&& \\
&&&&&&&&& \\
m=3&&-&-&$2(d-1)$&-&-&$2(d-1)^2$&& \\
&&&&&&&&& \\
m=4&&-&-&-&$2(d-1)$&-&-&& \\
&&&&&&&&& \\
m=5&&-&-&-&-&$4(d-1)$&-&& \\
&&&&&&&&& \\
m=6&&-&-&-&-&-&$2(d-1)$&& \\
&&&&&&&&& \\
\ldots &&&&&&&&& \\
&&&&&&&&& \\
\hline
&&&&&&&&& \\
${\cal N}_n^{(p)}$: \
\#\ {\rm of}\ {\rm el.\ domains}&&&&&&&&& \\
$\sigma^{(p)}_n\subset S_n$&&&&&&&&& \\
{\rm with}\ $q(p)$\ {\rm cusps}&&&&&&&&& \\
&&&&&&&&& \\
$p=0,\ q=d-1$&&1&0&$(d-1)^2$&$(d-1)^2(d+1)$&$(d-1)^2(d^2+2d+3)$&
$(d-1)d(d^3+d^2-2)$&& \\
&&&&&&&&& \\
&&&&&&&$2(d-1)+(d-1)^3$&& \\
$p=1,\ q=d-2$&&0&$d-1$&$2(d-1)$&$2(d-1)$&$4(d-1)$&$=(d-1)(d^2-2d+3)$&& \\
&&&&&&&&& \\
&&&&&&&$2(d-1)^2+2(d-1)^2$&& \\
$p=2,\ q=d-2$&&0&0&0&$(d-1)^2$&0&$=4(d-1)^2$&& \\
&&&&&&&&& \\
$p=3,\ q=d-2$&&0&0&0&0&0&0&& \\
&&&&&&&&& \\
\ldots&&&&&&&&& \\
&&&&&&&&& \\
\hline
\end{tabular} \nn
\ee
}
\noindent
Of the total of $(d-1)(d+1)(d^3+d-1)$ domains
$\sigma_6$ there are:

$\bullet$
$(d-1)d(d^3+d^2-2)$ zero-level components
$\sigma_6^{(0)}$ -- the centers of new isolated domains
$M_{6\alpha}$;

$\bullet$
$2(d-1)$ first-level components $\sigma_6^{(1)}$,
attached to the single $\sigma_1^{(0)}$ at $2(d-1)$ zeroes of
the resultant $r_{16}$, located on the boundary of $\sigma_1^{(0)}$,
one between every of $d-1$ cusps and $d-1$ zeroes of $r_{12}$ (where
the $\sigma_2^{(1)}$ components are attached to $\sigma_1^{(0)}$;

$\bullet$
$(d-1)^3$ first-level components $\sigma_6^{(1)}$, with
$d-1$ attached to each of the $(d-1)^2$ zero-level components
$\sigma_3^{(0)}$ at $(d-1)^3$ of $(d-1)^2(d+1)$ zeroes of $r_{36}$
in exactly the same way as $d-1$ components
$\sigma_2^{(1)}$ are attached to the central $\sigma_1^{(0)}$
at zeroes of $r_{12}$;

$\bullet$
$2(d-1)^2$ second-level components $\sigma_6^{(2)}$, with
$2(d-1)$ attached to each of the $d-1$ first-level components
$\sigma_2^{(1)}$ at zeroes of $r_{26}$
in exactly the same way as $2(d-1)$ components
$\sigma_3^{(1)}$ are attached to the central $\sigma_1^{(0)}$
at zeroes of $r_{13}$;

$\bullet$
$2(d-1)^2$ second-level components $\sigma_6^{(2)}$, with
$d-1$ attached to each of the $2(d-1)$ first-level components
$\sigma_3^{(1)}$ at remaining $2(d-1)^2$ zeroes of $r_{36}$
in exactly the same way as $2(d-1)$ components
$\sigma_2^{(1)}$ are attached to the central $\sigma_1^{(0)}$
at zeroes of $r_{12}$.

\bigskip

For fragments of Julia sheaf for various families
see Figs.\ref{J1-1} and \ref{J1-4} ($d=2$), \ref{Jush3} ($d=3$),
\ref{Juexa} ($d=4$), \ref{Jush5} ($d=5$) and \ref{Jush8} ($d=8$).

\FIGEPS{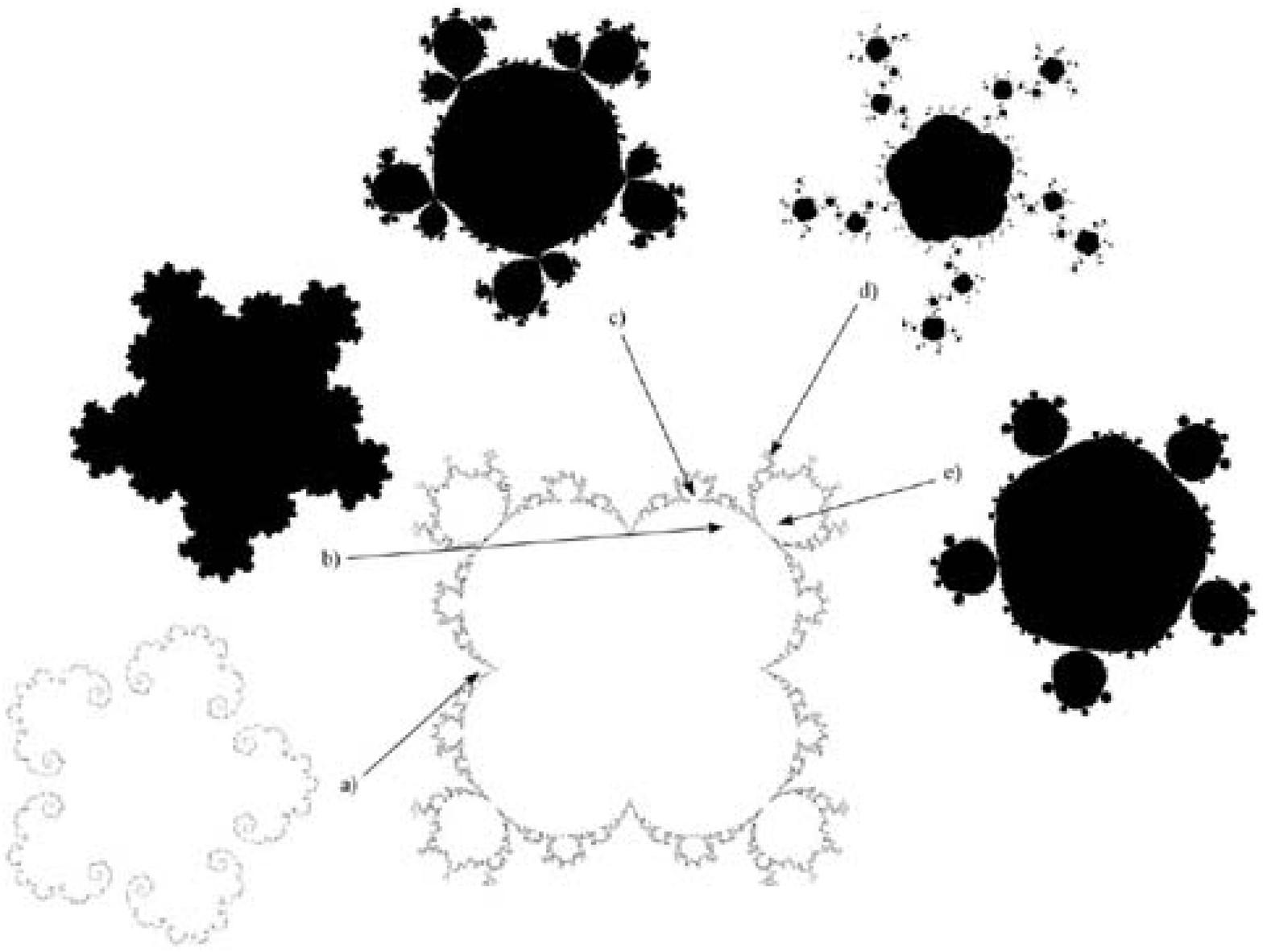}{A fragment (fibers shown at several points only)
of the Julia sheaf for the family $x^5+c$.
Julia sets have symmetry $\bb{Z}_5$, Mandelbrot set has
$\bb{Z}_4$ symmetry.}
{500,372}
{A fragment (fibers shown at several points only)
of the Julia sheaf for the family $x^5+c$.
Julia sets have symmetry $\bb{Z}_5$, Mandelbrot set has
$\bb{Z}_4$ symmetry: a) $c=-0.6$, b) $c=0.4+0.62i$,
c) $c=0.269+0.72i$, d) $c=5827+0.8834i$, e) $c=0.6+0.6i$.}

\FIGEPS{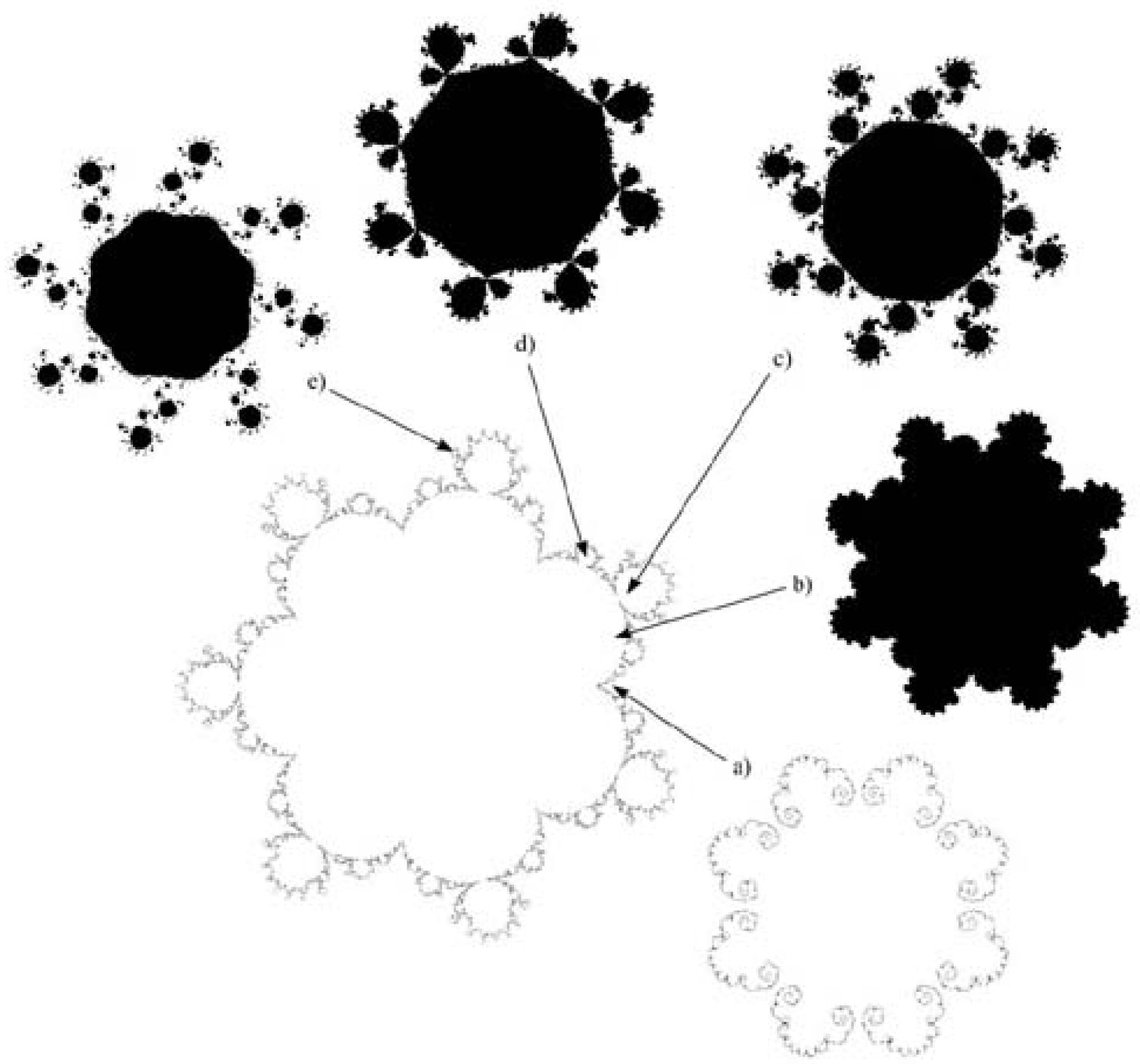}{A fragment (fibers shown at several points only)
of the Julia sheaf for the family $x^8+c$.
Julia sets have symmetry $\bb{Z}_8$, Mandelbrot set has
$\bb{Z}_7$ symmetry.}
{500,462}
{A fragment (fibers shown at several points only)
of the Julia sheaf for the family $x^8+c$.
Julia sets have symmetry $\bb{Z}_8$, Mandelbrot set has
$\bb{Z}_7$ symmetry: a) $c=0.7$, b) $c=0.73+0.25i$,
c) $c=0.068+0.9775i$, d) $c=0622+0.543i$, e) $c=0.786+0.504i$}

\subsection{Generic maps of degree $d$,
$f(x) = \sum_{i=0}^d \alpha_i x^i$
\label{gemaded}}

It is useful to introduce the following gradation:
\be
x \rightarrow \lambda^{-1} x, \nn \\
f(x) \rightarrow \lambda^{-1} f(x), \nn \\
\alpha_i \rightarrow \lambda^{i-1} \alpha_i, \nn \\
F_n(x) \rightarrow \lambda^{-1} F_n(x), \nn \\
G_1(x) \rightarrow \lambda^{-1} G_1(x), \nn \\
G_k(x) \rightarrow G_k(x)\ \ {\rm for}\ k>1 \nn
\ee
Then, since $G_k(x)$ is a polynomial of degree
$N_k(d)$ in $x$, it is clear that
$$
G_k(x) = \alpha_d^{\frac{N_k(d)-\delta_{k,1}}{d-1}}
\prod_{l=1}^{N_k(d)} (x-\rho^{(k)}_l)
$$
where the roots $\rho_l \rightarrow \lambda^{-1}\rho_l$
are sophisticated algebraic functions of the coefficients
$\alpha_i$.
Consequently, the resultant
\be
R(G_n,G_k) =
\alpha_d^{\frac{(N_n(d)-\delta_{n,1})N_k(d) +
N_n(d)(N_k(d)-\delta_{k,1})}{d-1}}
\prod_{l=1}^{N_n(d)}\prod_{l' = 1}^{N_k(d)}
(\rho^{(n)}_l-\rho^{(k)}_{l'})
\label{adres}
\ee
Assuming that $n>k$ we can simplify the expression for degree of
$\alpha_d$ to $\frac{2N_n(d)N_k(d) - \delta_{k,1}N_n(d)}{d-1}$.

{\bf Example.} As a simple application of these formulas,
take $d=2$. Then, as $a_2 \rightarrow 0$, all the roots grow
as $\alpha_2^{-1}$, $\rho^{(k)}_l \sim \alpha_2^{-1}$.
The only exception is {\it one}
of the two fixed points $\rho^{(1)}_+$ which remains finite,
but its differences with other roots are still growing as $\alpha_2^{-1}$.
Then from (\ref{adres}) we conclude that in this case
\be
R(G_n,G_k) \sim \alpha_2^{(2N_n(2)N_k(2) - \delta_{k,1}N_n(2))}
\alpha_2^{-N_n(2)N_k(2)} =
\alpha_2^{N_n(2)(N_k(2)-\delta_{k,1})}
\label{resa2}
\ee
We used this result in s.\ref{adresqualin} above.

When $\alpha_d \rightarrow 0$ for $d>2$, then $N_n(d-1)$ out of
the $N_n(d)$ roots of $G_n(x)$ (points on the orbits of order $n$)
remain finite, while the remaining $N_n(d)-N_n(d-1)$ grow as
negative powers of $\alpha_d$. However, different roots grow
differently, moreover, some asymptotics can coincide and
then differences of roots can grow slower than the roots themselves:
all this makes analysis rather sophisticated (see subsection
\ref{cx3x2} for a simplest example).

\newpage

\section{Conclusion}

In this paper we suggest to begin investigation of
Julia and Mandelbrot sets as algebraic varieties in
the phase space $\bb{X}$ and the moduli space of maps
${\cal M}$ respectively. We suggest to identify
the pure topological and/or pure algebraic entity --
the universal discriminant variety\footnote{
Since all resultants of iterated maps appear in
decompositions of their discriminants, see eq.(\ref{Dd}),
it is a matter of taste, which name, resultant
or discriminant variety to use.
However, to avoid confusion one should
remember that elementary irreducible
components are resultants ${\cal R}^*_{n,k}$, and
irreducible discriminants are rare (they can be identified
as regularized diagonal ${\cal R}^*_{nn}$.
}
${\cal D}^* = {\cal R}^*$
of {\it real} codimension one in ${\cal M}$ --
as the boundary $\partial M$ of Universal
Mandelbrot set $M$. The discriminant variety ${\cal R}^*$
naturally decomposes into strata of real codimension one,
$$
{\cal R}^* = \bigcup_{k=1}^\infty {\cal R}^*_k
$$
which, in turn, are made out of
{\it complex}-codimension-one algebraic varieties
${\cal R}^*_{k,mk}$,
$$
{\cal R}^*_k =
\overline{\bigcup_{m=1}^\infty {\cal R}^*_{k,mk}}.
$$
The intersection
$$
{\cal R}^*_n\bigcap {\cal R}^*_k = {\cal R}^*_{n,k}
$$
is non-empty only when $k$ is divisor of $n$ or vice versa.
This internal structure of the universal discriminant
variety is exhaustively represented by the {\it basic graph}:
a directed graph, obtained by identification of all
vertices of the multiples tree (see s.\ref{coo}) with
identical numbers at the vertices (this is nothing but
the graph, discussed in s.\ref{analyt} in relation to
representation theory of $\bb{Z}$-action on periodic orbits).
Irreducible resultants ${\cal R}^*_{k,mk}$ are
associated with the links of this graph and ${\cal R}^*_k$'s
-- with its vertices. Each ${\cal R}^*_n$ separates
the entire moduli space ${\cal M}$ into disjoint strata,
with one or another set of periodic orbits of order $n$ being
stable within each stratum. An intuitive image can be
provided by a chain of sausages: every sausage is stability
domain for particular set of orbits of given order,
every chain is bounded by particular ${\cal R}^*_k$,
two chains, labeled by $n$ and $k$ touch along a spiralling
line, represented by ${\cal R}^*_{n,k}$,
see Fig.\ref{sausages}.

\FIGEPS{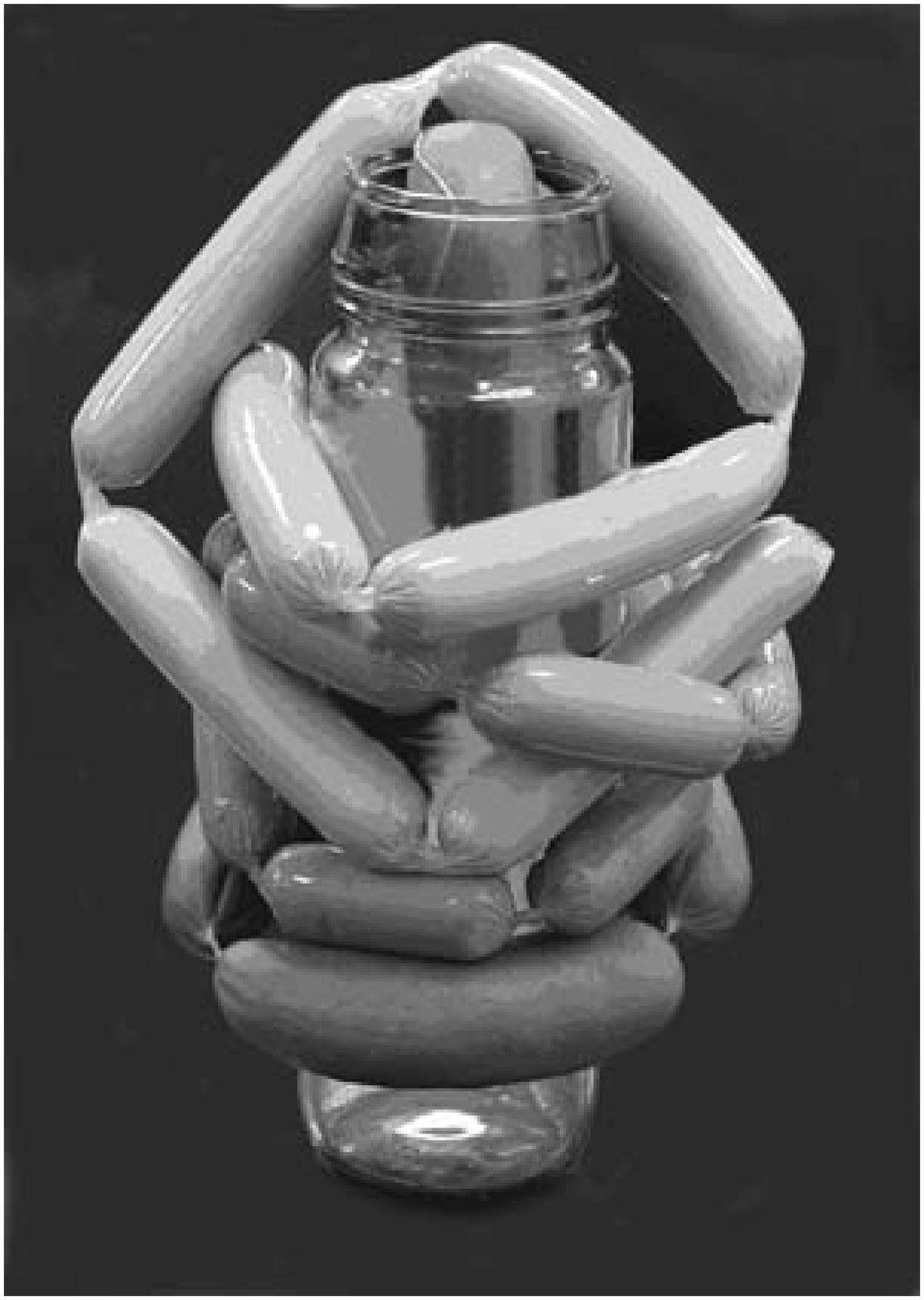}
{A symbolic picture of the universal Mandelbrot set.
A section can be obtained with the help of a knife.}
{1,2}
{A symbolic picture of the universal Mandelbrot set.
A section can be obtained with the help of a knife.}

Ordinary Mandelbrot sets, familiar from the literature,
are obtained as sections of this universal discriminant
variety by surfaces of low dimension, usually of complex
dimension one. In such section a single stratum and a single
irreducible resultant ${\cal R}^*_{n,mn}$ can look as
a set of circles and points respectively
(labeled by additional non-universal parameters $\alpha$
in s.3). Irreducibility of resultants guarantee that all
these points with given $n$ and $m$ belong to one and the
same entity.

While such consideration seems to be potentially exhaustive
for Mandelbrot sets
(though their trail structure, shapes, fractal dimensions and
Feigenbaum parameters still need to be better described
in algebraic terms),
it is not so clear about Julia sets: we did not identify them
as (infinite) unions of well defined domains, and we did not
establish full control over their geometry.
Instead, see s.\ref{Juexa}, we described Julia sets as
strapped discs in $\bb{X}$, with identified points at the
boundary. Identification of every group of points separates some
sectors from the disc, and since at every bifurcation infinitely
many groups of points are identified, the emerging structure
of the Julia set looks fractal, see Fig.\ref{strapping}.
Identified are points of various grand orbits,
and orbits involved in this procedure depend on the place
of the {\it map} in discriminant (Mandelbrot) variety: Julia sets
form a kind of a sheaf over universal discriminant variety.
We made just a few steps towards description of the structure
of this sheaf, its monodromy and singularity properties.
We came close to description of the underlying combinatorial
structure, which is a sheaf of rooted trees
(skeletons of Julia sets) over the basic graph (the skeleton
of the discriminant/Mandelbrot variety).

\FIGEPS{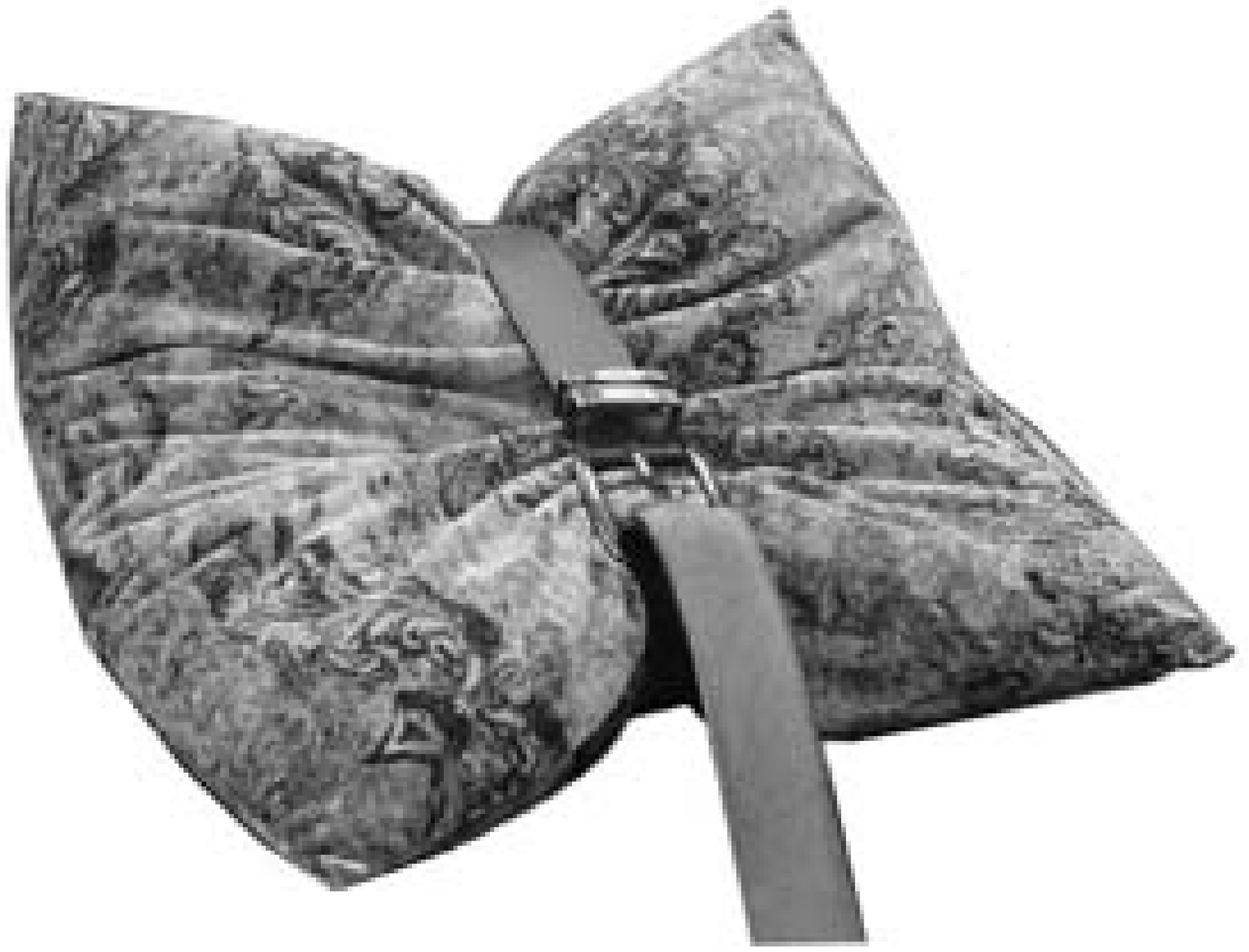}
{Getting Julia set by strapping a ball.
Actually there are double-infinity of strappings: going
in the direction of particular phase transition we strap
the points at the boundary, associated with different orbits;
and even if a direction is fixed, infinitely many points,
belonging to entire grand orbit, should be strapped.}
{1,2}
{Getting Julia set by strapping a ball.
Actually there are double-infinity of strappings: going
in the direction of particular phase transition we strap
the points at the boundary, associated with different orbits;
and even if a direction is fixed, infinitely many points,
belonging to entire grand orbit, should be strapped.}

Even more subtle is the issue of effective action and adequate
description of cyclic and chaotic RG flows.
Exact relation between $\tau$-functions, phase transitions
and Mandelbrot sets remains to be found. Generalizations to
arbitrary fields $\bb{X}$ and possible relations to multi-dimensional
flows are not explicitly worked out.
Last, but not the least, the obvious relations to {\bf landscape theory},
which studies the distributions of algebro-geometric quantities on
moduli spaces and their interplay with renormalization-group flows,
are left beyond the scope of the present paper.
We are just at the beginning of an interesting story.

\section{Acknowledgements}

Our work is supported by Federal Program of the Russian Ministry
for Industry, Science and Technology No 40.052.1.1.1112 and by
RFBR grants 04-02-17227 (V.D.) and 04-02-16880 (A.M.).

\end{document}